\documentclass{aa}  
\usepackage{graphicx}
\usepackage{txfonts}
\usepackage[colorlinks=true,citecolor=blue]{hyperref}
\usepackage{newtxtext,newtxmath}
\usepackage[T1]{fontenc}
\usepackage{amsmath}
\usepackage{amssymb}
\usepackage{color}
\usepackage{hyperref}
\usepackage{pdflscape}
\newcommand{\orcid}[1]{\href{https://orcid.org/#1}{\includegraphics[width=10pt]{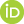}}}

\begin{document} 
\title{An updated measurement of the Hubble constant from near-infrared observations of Type Ia supernovae}
\titlerunning{An updated measurement of the $H_0$ from SNe Ia in the NIR}
\authorrunning{Galbany, de Jaeger, Riess, et al.}
\author{
Llu\'is Galbany\inst{1,2}\orcid{0000-0002-1296-6887}\thanks{E-mail: lgalbany@ice.csic.es (LG)},
Thomas de Jaeger,$^{3}$,
Adam G. Riess$^{4,5}$,
Tom\'as E. M\"uller-Bravo\inst{1,2}\orcid{0000-0003-3939-7167},
Suhail Dhawan$^{6}$,\\
Kim Phan$^{2,1}$,
M. D. Stritzinger\inst{7}\orcid{0000-0002-5571-1833},
E. Karamehmetoglu\inst{7}\orcid{0000-0001-6209-838X}
Bruno Leibundgut$^{8}$,
Chris Burns\inst{9}\orcid{0000-0003-4625-6629},
Erik Peterson$^{10}$,\\
W. D'Arcy Kenworthy$^{5}$,
Joel Johansson$^{11}$,
Kate Maguire$^{12}$,
Saurabh W. Jha$^{13}$.
}
\institute{
Institute of Space Sciences (ICE, CSIC), Campus UAB, Carrer de Can Magrans, s/n, E-08193 Barcelona, Spain. \and
Institut d’Estudis Espacials de Catalunya (IEEC), E-08034 Barcelona, Spain. 
\\ (\email{lgalbany@ice.csic.es})
\and
Institute for Astronomy, University of Hawaii, 2680 Woodlawn Drive, Honolulu, HI 96822, USA. \and
Space Telescope Science Institute, 3700 San Martin Drive, Baltimore, MD 21218, USA. \and
Department of Physics and Astronomy, Johns Hopkins University, Baltimore, MD 21218, USA. \and
Institute of Astronomy and Kavli Institute for Cosmology, University of Cambridge, Madingley Road, Cambridge CB3 0HA, UK. \and
Department of Physics and Astronomy, Aarhus University, Ny Munkegade 120, DK-8000 Aarhus C, Denmark. \and
European Southern Observatory, Karl-Schwarzschild-Strasse 2, D-85748 Garching, Germany.\and
Observatories of the Carnegie Institution for Science, 813 Santa Barbara St., Pasadena, CA 91101, USA. \and
Department of Physics, Duke University, Durham, NC 27708, USA. \and
The Oskar Klein Centre for Cosmoparticle Physics, Department of Physics, Stockholm University, SE-10691 Stockholm, Sweden. \and
School of Physics, Trinity College Dublin, College Green, Dublin 2, Ireland. \and
Department of Physics and Astronomy, Rutgers the State University of New Jersey, 136 Frelinghuysen Road, Piscataway, NJ 08854.
}
\date{Received September 7, 2022; accepted September 13, 2023}

\abstract{We present a measurement of the Hubble constant ($H_0$) using type Ia supernova (SNe Ia) in the near-infrared (NIR) from the recently updated sample of SNe Ia in nearby galaxies with distances measured via Cepheid period-luminosity relations by the SHOES project. We collect public near-infrared photometry of up to 19 calibrator SNe Ia and further 57 SNe Ia in the Hubble flow ($z>0.01$), and directly measure their peak magnitudes in the $J$ and $H$ band by Gaussian processes and spline interpolation. Calibrator peak magnitudes together with Cepheid-based distances are used to estimate the average absolute magnitude in each band, while Hubble-flow SNe are used to constrain the zero-point intercept of the magnitude-redshift relation. Our baseline result of $H_0$ is $72.3\pm1.4$ (stat) $\pm1.4$ (syst) km s$^{-1}$ Mpc$^{-1}$ in the $J$ band and $72.3\pm1.3$ (stat) $\pm1.4$ (syst) km s$^{-1}$ Mpc$^{-1}$ in the $H$ band, where the systematic uncertainties include  the standard deviation of up to 21 variations of the analysis, the 0.7\% distance scale systematic from SHOES Cepheid anchors, a photometric zeropoint systematic, and a cosmic variance systematic. Our final measurement represents a measurement with a precision of 2.8\% in both bands. Among all analysis variants the largest change in $H_0$ comes from limiting the sample to those SNe from the CSP and CfA programmes, noteworthy because these are the best calibrated, yielding $H_0\sim75$ km s$^{-1}$ Mpc$^{-1}$ in both bands. We explore applying stretch and reddening corrections to standardize SN Ia NIR peak magnitudes, and we demonstrate they are still useful to reduce the absolute magnitude scatter improving its standardization, at least up to the $H$ band. Based on our results, in order to improve the precision of the $H_0$ measurement with SNe Ia in the NIR in the future, we would need to increase the number of calibrator SNe Ia, be able to extend the Hubble-Lema\^itre diagram to higher redshift, and include standardization procedures to help reducing the NIR intrinsic scatter.}
\maketitle


\section{Introduction}

The expansion rate of the Universe parameterized by the Hubble-Lema\^itre parameter $H(z)$, has been a major endeavor in cosmology since the discovery of the expanding Universe \citep{1931MNRAS..91..490L,1929PNAS...15..168H}. $H(z)$ is not constant, but rather varies over cosmic time following the deceleration and acceleration of the Universe. In the last years, significant effort has been put forth to measure with high precision the local value of the Hubble-Lema\^itre parameter known as the Hubble constant ($H_0$, \citealt{2007LRR....10....4J,2010ARA&A..48..673F}), and today $H_0$ is estimated in the local Universe through the distance ladder technique with an uncertainty of $\sim$1 km s$^{-1}$ Mpc$^{-1}$ ($\lesssim$1.5\%, \citealt{2022ApJ...934L...7R}, hereafter R22). Perplexingly, these findings have revealed a dramatic discrepancy dubbed {\it the Hubble tension}: the estimation of $H_0$ from the local distance ladder is in strong disagreement (at 5$\sigma$ or 1 chance over $\sim$3.5 million) with the value inferred at high-redshift from the angular scale of fluctuations in the cosmic microwave background (CMB; \citealt{2020A&A...641A...6P}), possibly hinting towards new physics beyond the standard cosmological model. This discrepancy represents the most urgent puzzle of modern cosmology, and it is nowadays one of its hottest topics.

\begin{table*}\scriptsize
\centering
\caption{Properties of the 19 SNe Ia in galaxies with distances calibrated using Cepheids in \protect\cite{2022ApJ...934L...7R}.}
\label{tab:calibrators}
\begin{tabular}{llcccccccccc}
\hline
SN name             & Galaxy  & $\mu_{\rm Ceph}$& $m_J$        & $m_H$        & $M_J$           & $M_H$         & $A_J^{\rm MW}$&$A_H^{\rm MW}$& $K_J$    & $K_H$    & Ref.\\
\hline
2001el$^\dagger$ & NGC1448 & 31.287 (037)    & 12.821 (015) & 13.079 (045) & $-$18.466 (072) & $-$18.208 (077) & 0.010       & 0.006        & $-$0.010 & $-$0.007 & [1]     \\
2002fk$^\dagger$ & NGC1309 & 32.541 (059)    & 13.747 (009) & 13.957 (010) & $-$18.794 (085) & $-$18.584 (085) & 0.030       & 0.018        & $-$0.017 & $-$0.012 & [2]     \\
2003du$^\dagger$ & UGC9391 & 32.848 (067)    & 14.301 (117) & 14.628 (060) & $-$18.547 (148) & $-$18.220 (105) & 0.007       & 0.004        & $-$0.017 &    0.002 & [3]     \\
2005cf$^\dagger$ & NGC5917 & 32.363 (120)    & 13.770 (030) & 13.871 (058) & $-$18.593 (138) & $-$18.492 (141) & 0.072       & 0.044        & $-$0.018 & $-$0.006 & [4]     \\
2005df           & NGC1559 & 31.491 (061)    & 12.875 (013) & 13.120 (023) & $-$18.617 (087) & $-$18.371 (089) & 0.022       & 0.014        & $-$0.010 & $-$0.006 & [5]     \\
2006D            & Mrk1337 & 32.920 (123)    & 14.326 (010) & 14.610 (059) & $-$18.593 (137) & $-$18.310 (159) & 0.034       & 0.021        & $-$0.020 &    0.013 & [6,4]   \\
2006bh           & NGC7329 & 33.246 (117)    & 14.812 (007) & 15.014 (010) & $-$18.434 (132) & $-$18.232 (132) & 0.020       & 0.012        & $-$0.026 & $-$0.018 & [6]     \\
2007A            & NGC0105 & 34.527 (250)    & 15.685 (003) &   $\cdots$   & $-$18.842 (257) &    $\cdots$     & 0.054       &  $\cdots$    & $-$0.044 & $\cdots$ & [6]     \\
2007af$^\dagger$ & NGC5584 & 31.772 (052)    & 13.486 (006) & 13.634 (009) & $-$18.286 (080) & $-$18.138 (080) & 0.029       & 0.018        & $-$0.013 & $-$0.010 & [6]     \\
2008fv           & NGC3147 & 33.014 (165)    & 14.519 (251) &   $\cdots$   & $-$18.495 (306) &    $\cdots$     & 0.018       &  $\cdots$    & $-$0.024 & $\cdots$ & [7,4]   \\
2009Y            & NGC5728 & 33.094 (205)    & 14.473 (008) &   $\cdots$   & $-$18.621 (214) &    $\cdots$     & 0.075       &  $\cdots$    & $-$0.024 & $\cdots$ & [6]     \\
2011by$^\dagger$ & NGC3972 & 31.635 (089)    & 13.182 (044) & 13.408 (016) & $-$18.452 (116) & $-$18.227 (110) & 0.010       & 0.006        & $-$0.008 & $-$0.004 & [4]     \\
2011fe$^\dagger$ & M101    & 29.178 (041)    & 10.441 (023) & 10.723 (013) & $-$18.737 (076) & $-$18.456 (074) & 0.007       & 0.004        & $-$0.002 & $-$0.001 & [8]     \\
2012cg$^\dagger$ & NGC4424 & 30.844 (128)    & 12.280 (024) & 12.488 (066) & $-$18.564 (144) & $-$18.357 (156) & 0.015       & 0.009        & $-$0.003 & $-$0.002 & [9]     \\
2012fr           & NGC1365 & 31.378 (056)    & 12.731 (020) & 12.954 (018) & $-$18.647 (084) & $-$18.424 (084) & 0.015       & 0.009        & $-$0.012 & $-$0.009 & [10]    \\
2012ht           & NGC3447 & 31.936 (034)    & 13.437 (012) & 13.593 (016) & $-$18.499 (070) & $-$18.342 (071) & 0.022       & 0.013        & $-$0.008 & $-$0.007 & [11]    \\
2013dy           & NGC7250 & 31.628 (125)    & 13.676 (044) & 14.351 (097) & $-$18.846 (145) & $-$18.644 (168) & 0.114       & 0.070        & $-$0.009 & $-$0.007 & [12]    \\
2015F$^\dagger$  & NGC2442 & 31.450 (064)    & 13.157 (063) & 13.359 (038) & $-$18.293 (108) & $-$18.091 (096) & 0.150       & 0.093        & $-$0.011 & $-$0.009 & [11,13] \\
2017cbv          & NGC5643 & 30.547 (052)    & 11.786 (014) & 11.968 (001) & $-$18.761 (081) & $-$18.579 (080) & 0.125       & 0.077        & $-$0.010 & $-$0.006 & [14]    \\
\hline
\end{tabular}\\
\tablefoot{$^\dagger$Included in \protect\cite{2018AA...609A..72D}. 
Near-infrared photometry references: [1]~\cite{2003AJ....125..166K}; [2]~\cite{2014ApJ...789...89C}; [3]~\cite{2007AA...469..645S}; [4]~\cite{2015ApJS..220....9F}; [5]~\cite{2017RNAAS...1...36K}; [6]~\cite{2017AJ....154..211K}; [7]~\cite{2012AA...537A..57B}; [8]~\cite{2012ApJ...754...19M}; [9]~\cite{2016ApJ...820...92M}; [10]~\cite{2018ApJ...859...24C}; [11]~\cite{2018ApJ...869...56B}; [12]~\cite{2015MNRAS.452.4307P}; [13]~\cite{2017MNRAS.464.4476C}; [14]~\cite{2020ApJ...904...14W}.}
\end{table*}

The {\it Supernovae, H$_0$, for the Equation of State of Dark energy} (SHOES; R22) team has been leading the effort in the last two decades, building on the initial attempts to measure $H_0$ using the Hubble Space Telescope (HST) by the Type Ia Supernova HST Calibration Program \citep{2001ApJ...562..314S} and the HST Key Project \citep{2001ApJ...553...47F}. For that, they construct a {\it distance ladder}, which consists of three rungs. In the first and most nearby, the Cepheid period-luminosity relation \citep{1912HarCi.173....1L} is calibrated using galactic geometric distance anchors, such as parallaxes to those same Cepheids \citep{2021A&A...649A...4L}, detached eclipsing binaries (DEBs; \citealt{2019Natur.567..200P}), or water masers \citep{2019ApJ...885..131R}. This Cepheid calibration is used in turn in the second rung of the distance ladder to obtain distances to nearby galaxies hosting both Cepheids and type Ia supernovae (SNe Ia). The absolute magnitude of these SNe Ia in the second rung is calibrated using the distance obtained from this independent Cepheid method, and it is finally used in the third rung of the scale to calibrate the absolute magnitude of SN Ia host galaxies at larger distances. SHOES has recently provided the most precise direct measurement of $H_0$ in the late Universe (R22) by calibrating galactic Cepheids from Gaia EDR3 parallaxes, masers in NGC 4258, and DEBs in the Large Magellanic Cloud, and using the HST to measure distances to 38 galaxies hosting Cepheids and 42 SNe Ia. In their analysis, optical light-curves of 42 and 277 SNe Ia in the second and third rungs, respectively, were used.

Optical observations of SNe Ia have been widely used in the past decades to measure cosmological distances. SNe Ia are the most mature and well-exploited probe of the accelerating universe, and their use as standardisable candles provides an immediate route to measure dark energy \citep{1998AJ....116.1009R,1999ApJ...517..565P,2001ARA&A..39...67L,2011ARNPS..61..251G}. This ability rests on empirical relationships between SN Ia peak brightness and light-curve (LC) width \citep{1974PhDT.........7R,1977SvA....21..675P,1993ApJ...413L.105P}, and SN color \citep{1996ApJ...473...88R,1998A&A...331..815T}, which standardize the optical absolute peak magnitude of SNe Ia down to a dispersion of $\sim$0.12 mag ($\sim$6\% in distance; \citealt{2014A&A...568A..22B}). However, environmental dependences, such as the {\it mass step} (\citealt{2010MNRAS.406..782S,2010ApJ...715..743K,2010ApJ...722..566L}, but also with other global and local parameters such as the star-formation rate, metallicity, or age; see e.g. \citealt{2020A&A...644A.176R,2018MNRAS.476..307M,2011ApJ...740...92G}), have been found to contribute to the systematic uncertainty budget.

Increasing evidence suggests that SNe Ia are very nearly natural standard candles at maximum light at near-infrared (NIR) wavelengths, even without corrections for light-curve shape and/or reddening, yielding more precise distance estimates to their host galaxies than optical data alone \citep{1981ApJ...251L..13E,1985ApJ...296..379E,2000MNRAS.314..782M,2004ApJ...602L..81K,2008ApJ...689..377W,2014ApJ...784..105W,2015ApJS..220....9F,2019ApJ...887..106A}. Compared to the optical, SNe Ia in the NIR are relatively immune to the effects of extinction and reddening by dust (extinction corrections are a factor of $4-6$ smaller than in the optical $B$-band; \citealt{2018A&A...615A..45S}), and the correlation between peak luminosity and decline rate is much smaller (e.g., \citealt{2004ApJ...602L..81K}). For instance, in a sample of 15 SNe Ia located at $0.025 < z < 0.09$, \cite{2012MNRAS.425.1007B} found a scatter of 0.09 mag (4\% in distance) in the $H$ band without applying any corrections for host-galaxy dust extinction or K-corrections \citep{1968ApJ...154...21O}. However, SN Ia cosmology in the NIR is still less developed compared to the optical, for various reasons. Optical detectors were technologically simpler and put available before the more expensive NIR detectors. Moreover, SNe Ia are intrinsically fainter in the NIR, requiring bigger telescopes and longer integration times. 
Current efforts are focused on increasing the number of objects with NIR observations. 

\begin{table*}\footnotesize
\centering
\caption{Properties of the 57 SNe Ia in Hubble flow galaxies.}
\label{tab:hubbleflow}
\begin{tabular}{llcccccccccc}
\hline
SN name        &$z_{\rm helio}$&$z_{\rm cmb}$&$z_{\rm corr}$&$m_J$         &$m_H$     &$A_J^{\rm MW}$&$K_J$&$A_H^{\rm MW}$&$K_H$& Ref.\\
\hline
1999ee                     & 0.01115 & 0.01028 & 0.01074 & 14.723 (012) & 14.993 (012) & 0.015  & $-$0.027 & 0.009  & $-$0.020 & [15]\\ 
1999ek                     & 0.01752 & 0.01761 & 0.01758 & 15.752 (011) & 15.991 (032) & 0.412  & $-$0.042 & 0.254  & $-$0.030 & [16]\\ 
2004eo$^\dagger$           & 0.01517 & 0.01449 & 0.01491 & 15.476 (000) & 15.653 (029) & 0.080  & $-$0.040 & 0.049  & $-$0.020 & [6]\\
2004ey                     & 0.01583 & 0.01467 & 0.01526 & 15.449 (004) &   $\cdots$   & 0.103  & $-$0.040 &$\cdots$& $\cdots$ & [6]\\
2005M$^\dagger$            & 0.02484 & 0.02584 & 0.02609 & 16.494 (033) & 16.771 (029) & 0.023  & $-$0.057 & 0.014  & $-$0.026 & [6]\\
2005el$^\dagger$           & 0.01484 & 0.01482 & 0.01501 & 15.448 (009) & 15.672 (012) & 0.084  & $-$0.035 & 0.052  & $-$0.027 & [4,6]\\
2005eq$^\dagger$           & 0.02892 & 0.02833 & 0.02850 & 16.780 (023) & 17.155 (006) & 0.054  & $-$0.070 & 0.033  & $-$0.051 & [4,6]\\
2005hc                     & 0.04592 & 0.04496 & 0.04576 & 17.940 (001) &   $\cdots$   & 0.024  & $-$0.107 &$\cdots$& $\cdots$ & [6]\\
2005iq                     & 0.03404 & 0.03292 & 0.03361 & 17.241 (061) &   $\cdots$   & 0.016  & $-$0.083 &$\cdots$& $\cdots$ & [4,6]\\
2005kc$^\dagger$           & 0.01509 & 0.01386 & 0.01467 & 15.418 (012) & 15.600 (019) & 0.098  & $-$0.038 & 0.060  & $-$0.025 & [6]\\
2005ki$^\dagger$           & 0.01946 & 0.02066 & 0.02055 & 16.098 (018) & 16.239 (025) & 0.023  & $-$0.047 & 0.014  & $-$0.030 & [6]\\
2006ax$^\dagger$           & 0.01650 & 0.01774 & 0.01771 & 15.703 (013) & 15.931 (010) & 0.035  & $-$0.040 & 0.022  & $-$0.025 & [4,6]\\
2006et$^\dagger$           & 0.02240 & 0.02164 & 0.02204 & 16.153 (029) & 16.345 (027) & 0.014  & $-$0.057 & 0.009  & $-$0.032 & [6]\\
2006hx$^\dagger$           & 0.04539 & 0.04429 & 0.04466 & 17.761 (037) &   $\cdots$   & 0.022  & $-$0.104 &$\cdots$& $\cdots$ & [6]\\
2006kf                     & 0.02004 & 0.01955 & 0.01950 & 16.269 (028) &   $\cdots$   & 0.180  & $-$0.057 &$\cdots$& $\cdots$ & [6]\\  
2006le$^\dagger$           & 0.01744 & 0.01728 & 0.01854 & 15.888 (011) & 16.299 (039) & 0.302  & $-$0.042 & 0.186  & $-$0.023 & [4]\\ 
2006lf$^\dagger$           & 0.01319 & 0.01296 & 0.01250 & 14.878 (020) &   $\cdots$   & 0.702  & $-$0.032 &$\cdots$& $\cdots$ & [4]\\ 
2007S$^\dagger$            & 0.01386 & 0.01502 & 0.01509 & 15.353 (007) & 15.530 (008) & 0.019  & $-$0.034 & 0.012  & $-$0.024 & [4,6]\\
2007ai                     & 0.03176 & 0.03212 & 0.03297 & 17.001 (043) &   $\cdots$   & 0.246  & $-$0.085 &$\cdots$& $\cdots$ & [6]\\
2007as$^\dagger$           & 0.01757 & 0.01791 & 0.01817 & 15.881 (026) & 16.104 (037) & 0.106  & $-$0.046 & 0.065  & $-$0.022 & [6]\\
2007ba$^{\dagger\ddagger}$ & 0.03475 & 0.03863 & 0.03593 & 17.673 (019) & 17.516 (135) & 0.027  & $-$0.097 & 0.017  &    0.001 & [6]\\ 
2007bc                     & 0.02075 & 0.02187 & 0.02152 & 16.354 (018) &   $\cdots$   & 0.016  & $-$0.053 &$\cdots$& $\cdots$ & [6]\\
2007bd$^\dagger$           & 0.03044 & 0.03141 & 0.03144 & 17.066 (041) & 17.234 (101) & 0.025  & $-$0.075 & 0.015  & $-$0.037 & [6]\\
2007ca$^\dagger$           & 0.01406 & 0.01509 & 0.01457 & 15.689 (080) & 15.717 (011) & 0.049  & $-$0.036 & 0.030  & $-$0.024 & [4,6]\\
2008bc$^\dagger$           & 0.01509 & 0.01573 & 0.01564 & 15.519 (015) & 15.704 (024) & 0.194  & $-$0.038 & 0.119  & $-$0.019 & [6]\\
2008bf                     & 0.02342 & 0.02371 & 0.02473 & 16.381 (011) &   $\cdots$   & 0.025  & $-$0.058 &$\cdots$& $\cdots$ & [6]\\
2008gl                     & 0.03354 & 0.03312 & 0.03251 & 17.226 (022) & 17.422 (039) & 0.021  & $-$0.081 & 0.013  & $-$0.040 & [4,6]\\
2008gp                     & 0.03302 & 0.03292 & 0.03335 & 17.055 (038) & 17.459 (111) & 0.090  & $-$0.085 & 0.055  & $-$0.034 & [6]\\
2008hj                     & 0.03761 & 0.03643 & 0.03688 & 17.591 (037) &   $\cdots$   & 0.026  & $-$0.090 &$\cdots$& $\cdots$ & [6]\\
2008hs$^{\dagger\ddagger}$ & 0.01906 & 0.01665 & 0.01865 & 16.329 (047) & 16.619 (059) & 0.043  & $-$0.040 & 0.026  & $-$0.029 & [4]\\
2008hv$^\dagger$           & 0.01255 & 0.01360 & 0.01392 & 15.244 (145) & 15.491 (006) & 0.024  & $-$0.030 & 0.015  & $-$0.024 & [4,6]\\
2009aa                     & 0.02705 & 0.02827 & 0.02854 & 16.769 (029) & 16.987 (016) & 0.025  & $-$0.069 & 0.016  & $-$0.040 & [6]\\
2009ad$^\dagger$           & 0.02840 & 0.02834 & 0.02884 & 16.754 (018) & 16.992 (060) & 0.082  & $-$0.069 & 0.051  & $-$0.034 & [4,6]\\
2009al                     & 0.02207 & 0.02328 & 0.02327 & 16.437 (014) & 16.643 (026) & 0.018  & $-$0.056 & 0.011  & $-$0.039 & [4,6]\\
2009bv$^\dagger$           & 0.03667 & 0.03748 & 0.03882 & 17.512 (055) & 17.786 (025) & 0.007  & $-$0.094 & 0.004  & $-$0.065 & [4]\\ 
2010Y$^{\dagger\ddagger}$  & 0.01105 & 0.01115 & 0.01240 & 15.236 (023) & 15.110 (125) & 0.010  & $-$0.028 & 0.006  &    0.006 & [4]\\
2010ag$^\dagger$           & 0.03341 & 0.03367 & 0.03352 & 17.137 (050) &   $\cdots$   & 0.023  & $-$0.089 &$\cdots$& $\cdots$ & [4]\\ 
2010ai$^\dagger$           & 0.01827 & 0.01918 & 0.01800 & 16.459 (060) & 16.635 (100) & 0.007  & $-$0.047 & 0.004  & $-$0.023 & [4]\\ 
2010ju$^\dagger$           & 0.01524 & 0.01534 & 0.01523 & 15.544 (002) & 15.675 (041) & 0.310  & $-$0.038 & 0.191  & $-$0.017 & [4]\\ 
2010kg$^\dagger$           & 0.01613 & 0.01645 & 0.01602 & 15.698 (028) & 15.899 (022) & 0.112  & $-$0.039 & 0.069  & $-$0.030 & [4]\\ 
2011ao$^\dagger$           & 0.01074 & 0.01167 & 0.01210 & 14.829 (004) & 14.911 (026) & 0.015  & $-$0.027 & 0.009  & $-$0.016 & [4]\\ 
PTF10bjs$^\dagger$         & 0.03001 & 0.03055 & 0.03122 & 17.013 (027) & 17.159 (072) & 0.013  & $-$0.074 & 0.008  & $-$0.056 & [4]\\ 
PTF10hmv                   & 0.03200 & 0.03272 & 0.03465 &   $\cdots$   & 17.525 (033) &$\cdots$& $\cdots$ & 0.010  & $-$0.045 & [17,18]\\
PTF10mwb$^\dagger$         & 0.03136 & 0.03124 & 0.03138 & 16.951 (019) & 17.375 (276) & 0.022  & $-$0.075 & 0.014  & $-$0.059 & [4,17,18]\\ 
PTF10nlg                   & 0.05620 & 0.05612 & 0.05629 &   $\cdots$   & 18.575 (038) &$\cdots$& $\cdots$ & 0.008  & $-$0.100 & [17,18]\\
PTF10tce                   & 0.04095 & 0.03967 & 0.03971 & 18.035 (021) & 18.005 (010) & 0.033  & $-$0.097 & 0.021  & $-$0.072 & [17,18]\\
PTF10ufj$^\dagger$         & 0.07700 & 0.07620 & 0.07668 & 19.224 (084) & 19.242 (058) & 0.085  & $-$0.156 & 0.052  & $-$0.125 & [4,17,18]\\
PTF10wnm                   & 0.06560 & 0.06442 & 0.06449 & 18.867 (085) & 18.910 (040) & 0.025  & $-$0.141 & 0.015  & $-$0.116 & [17,18]\\
iPTF13asv                  & 0.03620 & 0.03640 & 0.03652 & 17.090 (027) & 17.249 (079) & 0.038  & $-$0.088 & 0.023  & $-$0.051 & [19]\\
iPTF13azs                  & 0.03383 & 0.03376 & 0.03564 & 17.278 (035) &   $\cdots$   & 0.016  & $-$0.078 &$\cdots$& $\cdots$ & [19]\\
iPTF13dge                  & 0.01586 & 0.01581 & 0.01586 & 15.632 (076) & 15.829 (281) & 0.067  & $-$0.037 & 0.041  & $-$0.028 & [19]\\
iPTF13duj                  & 0.01695 & 0.01586 & 0.01727 & 15.413 (057) & 15.927 (037) & 0.057  & $-$0.039 & 0.035  & $-$0.013 & [19]\\
iPTF13ebh                  & 0.01316 & 0.01238 & 0.01107 & 15.045 (006) & 15.217 (005) & 0.057  & $-$0.036 & 0.035  & $-$0.019 & [19]\\
iPTF14atg                  & 0.02129 & 0.02223 & 0.02513 & 16.721 (032) & 16.632 (097) & 0.008  & $-$0.059 & 0.005  &    0.017 & [19]\\
iPTF14bdn                  & 0.01558 & 0.01635 & 0.01597 & 15.213 (013) & 15.412 (001) & 0.008  & $-$0.038 & 0.005  & $-$0.026 & [19]\\
iPTF16abc                  & 0.02320 & 0.02415 & 0.02461 & 16.345 (030) & 16.613 (031) & 0.021  & $-$0.059 & 0.013  & $-$0.025 & [19]\\
iPTF16auf                  & 0.01502 & 0.01566 & 0.01538 & 15.515 (020) &   $\cdots$   & 0.011  & $-$0.038 &$\cdots$& $\cdots$ & [19]\\
\hline
\end{tabular}\\
\tablefoot{$^\dagger$Included in \protect\cite{2018AA...609A..72D}. $^\ddagger$Fast-decliner, removed from the baseline analysis.
Near-infrared photometry references: [4]~\cite{2015ApJS..220....9F}; [6]~\cite{2017AJ....154..211K}; [15]~\cite{2004AJ....127.1664K}; [16]~\cite{2004AJ....128.3034K}; [17]~\cite{2012MNRAS.425.1007B}; [18]~\cite{2013MNRAS.432L..90B}; [19]~\cite{2021ApJ...923..237J}.}
\end{table*}

Most of the NIR SN Ia data currently available at low redshift ($z<0.1$) comes from the Carnegie Supernova Project (CSP, \citealt{2017RNAAS...1...36K}) and the Center for Astrophysics (CfA,  \citealt{2015ApJS..220....9F}) follow-up programmes, with also significant contributions from smaller programmes (e.g. \citealt{2012MNRAS.425.1007B,2018A&A...615A..45S,2021ApJ...923..237J}). At intermediate redshift ($0.2<z<0.6$) the high-redshift subprogramme of the CSP \citep{2009ApJ...704.1036F} obtained $YJ$ imaging of 35 SNe Ia up to redshift 0.7 to build a $i$-band rest-frame Hubble diagram. More recently, The {\it Supernovae IA in the Near-InfraRed} (RAISIN, \citealt{2022ApJ...933..172J}) project has collected HST NIR observations of 45 SNe which, complemented with Pan-STARRS Medium Deep Survey (MDS;  \citealt{2016arXiv161205560C}) and Dark Energy Survey (DES, \citealt{2019ApJ...874..106B}) optical light-curves, are used to extend the Hubble-Lema\^itre diagram up to redshift 0.6 in rest-frame $Y$-band, the furthest rest-frame NIR Hubble-Lema\^itre diagram ever constructed. In the near future, the Nancy Roman Space Telescope supernova programme \citep{2021arXiv211103081R} with its F213 filter is expected to provide NIR rest-frame observations of SNe Ia and extend the NIR Hubble diagram from redshifts $\sim$0.3 in $J$ and $\sim$0.1 in $H$ provided by the $F160W$ HST filter, to $\sim$0.7 in $J$ and $\sim$0.3 in $H$ (e.g. see Figure 2 in \citealt{2022ApJ...933..172J}). Roman, together with RAISIN and the 24 very nearby SNe Ia from the {\it Supernovae in the Infrared avec Hubble} (SIRAH; \citealt{2019hst..prop15889J}) HST programme, will provide a full space-based NIR SN Ia Hubble-Lema\^itre diagram. NIR observations of SNe Ia have already been used to measure $H_0$, the most prominent efforts being those of \cite{2018ApJ...869...56B} using all CSP observations and the SuperNovae in object-oriented Python (SNooPy) template fitting, and \citet[][hereafter D18]{2018AA...609A..72D} by directly measuring SN Ia $J$-band peak magnitudes from a literature sample.

In this work, we obtain an updated measurement of $H_0$ with SNe Ia in the NIR. Building from D18, we present the following improvements: 
(i) we collected all available NIR light-curves of SNe Ia to date with data during the rise phase that allows the measurement of their peak $J$ and $H$-band magnitudes, including updated photometry of the CSP from their 3rd data release \citep{2017AJ....154..211K};
(ii) all photometry is put in the CSP photometric system by applying S-corrections;
(iii) the number of SNe Ia in galaxies with Cepheid-based distances is more than doubled from 9 to 19, thanks to the recently increased sample from SHOES;
(iv) the number of SNe Ia in the Hubble flow was also increased from 27 to 52(40) in $J$($H$); and
(v) the analysis is extended to the $H$ band.
Although the distances in the second rung of our distance ladder are based on SHOES distances, with this independent analysis we can test whether SNe Ia in the optical introduce a bias in the $H_0$ measurement, due to systematic uncertainties introduced in their standardization.


\section{Data sample}

To put constraints on the current value of the Hubble expansion using SNe Ia in the NIR, we need (i) a sample of nearby SN Ia observed in the NIR hosted by galaxies whose distance has been independently measured using other techniques; and (ii) a sample of SNe Ia with NIR observations located further in the Hubble flow ($z>0.01$). The peak absolute magnitude of SNe Ia in those nearby galaxies (hereafter {\it calibrators}) can be then determined simply by measuring their peak apparent brightness, and in turn this reference-calibrated magnitude used in the Hubble-flow SNe~Ia to determine distances to their hosts. 

For both distance regimes our criteria to select SNe Ia is the same: SNe~Ia light-curves must be sparsely sampled and have at least a NIR pre-maximum photometric point to allow for a reliable measurement of the peak magnitude. D18 represents our reference work where 36 SNe~Ia were selected based on their high-quality $J$-band photometry. Nine of those SNe Ia exploded in galaxies whose distances were determined independently by the SHOES project \citep{2016ApJ...826...56R}, while 27 were in the Hubble flow. 

In this work, for the calibrator sample we make use of the recently updated sample from SHOES (R22), which has been extended from 19 to 42 SNe~Ia in 38 nearby galaxies with distances measured using Cepheids. We have performed a thorough search of NIR $J$ and $H$-band photometry of all SNe~Ia in the latest SHOES sample of galaxies and found that up to 19 SNe Ia (including those 9 in D18) have NIR light-curves of sufficient quality to be included in this analysis. Three SNe~Ia (2013dy, 2012ht, 2012fr) were initially included in \cite{2016ApJ...826...56R} but their photometry has been published more recently, and other 7 SNe~Ia were in galaxies whose Cepheid distance was presented for the first time in R22. In addition, we have performed a thorough search in the literature for $J$ and $H$-band NIR photometry of SN~Ia in the Hubble flow ($z>0.01$), and found 57 candidate objects.

All 19 SNe~Ia in the calibrator sample have $J$ band light-curves that allows for the peak-brightness determination, however only 16 have light-curves with enough quality in the $H$ band. Similarly, for the Hubble-flow sample, while 55 SNe~Ia have good $J$-band light-curves, 13 have an $H$-band light-curve that does not permit the determination of the peak-brightness. Our initial sample is therefore 19/16 SNe~Ia with $J$/$H$-band light-curves in calibrator galaxies, and 55/44 SNe~Ia in the Hubble flow. The list of SNe~Ia in our calibrator sample (galaxies with Cepheid distances) are listed in Table \ref{tab:calibrators}, while our Hubble-flow SNe Ia sample is presented in Table \ref{tab:hubbleflow}. Photometry for all objects have been obtained from references listed in the last column of these tables.


\section{Methods}

Our approach is based on the assumption that SNe Ia are good natural standard candles in the NIR, so then their peak magnitudes derived directly from the observations are enough to estimate cosmological distances.

\subsection{Distances and redshifts}

Distances $\mu_{\rm Ceph}$ and uncertainties $\sigma_{\rm Ceph}$ to galaxies in the calibrator sample are taken from R22 (listed in Table \ref{tab:calibrators}). Heliocentric redshifts ($z_{\rm helio}$) and their uncertainties for galaxies in the Hubble-flow sample are obtained from the SN host galaxy catalogue provided in \cite{2022PASA...39...46C}, which are usually consistent within $|\Delta z|<0.0005$ (corresponding to 150~km~s$^{-1}$) with the redshifts reported in their reference sources and in the NASA/IPAC Extragalactic Database (NED\footnote{\url{http://ned.ipac.caltech.edu/}}) with a few exceptions: SN~2006kf, SN~2007ba, and SN~2008hs. In addition, five SNe had different redshifts in NED and in the reference sources and \cite{2022PASA...39...46C} differs in one or the other: while for SN~2005M, \cite{2022PASA...39...46C} redshift is similar to that in NED, for SN~2010ai, SN~2008bf, iPTF13asv, and iPTF13azs it is more similar to the reference source. In particular, SN~2008bf has an ambiguous host being in the middle of two nearby galaxies, NGC~4061 and NGC~4065. In our baseline analysis we choose to use redshifts from \cite{2022PASA...39...46C}. These redshifts are then converted to the 3K CMB reference frame ($z_{\rm CMB}$) and corrected for peculiar velocities ($z_{\rm corr}$) induced by visible structures as described in their work. We study the effect of the selection of heliocentric redshifts, CMB frame correction, and peculiar velocity correction in section \ref{sec:vars}.

\subsection{S-corrected photometry}\label{sec:scor}

Published $J$ and $H$-band photometry for both the calibrator and the Hubble-flow sample is S-corrected to the same photometric system using internal routines in SNooPy (v.2.5.3; \citealt{2015ascl.soft05023B}). Given that most of the objects were published by the CSP we use their photometric system as our reference system. More details about the nature of S-corrections can be found in Appendix A of \cite{2007AA...469..645S}. In our Appendix \ref{app:scor}, we show most of the filter transmission used to obtain SN Ia light-curves used in this work, and the magnitude of the S-corrections for tha calibrator sample. S-corrected light-curves to the CSP $J$ and $H$ filters are publicly available at \href{https://github.com/hostflows/H0nirR22}{https://github.com/hostflows/H0nirR22}. 

We note that for SN~2013dy, before S-correction, we had to first convert SN~2013dy published photometry in \cite{2015MNRAS.452.4307P} from the AB to the Vega system following \cite{2007ASPC..364..227M} (their Table 4), 
\begin{eqnarray}
m_J^{\rm Vega}&=&m_J^{\rm AB}+0.894\\
m_H^{\rm Vega}&=&m_H^{\rm AB}+1.368.
\end{eqnarray}

\begin{figure*}[!t]
    \includegraphics[width=\columnwidth]{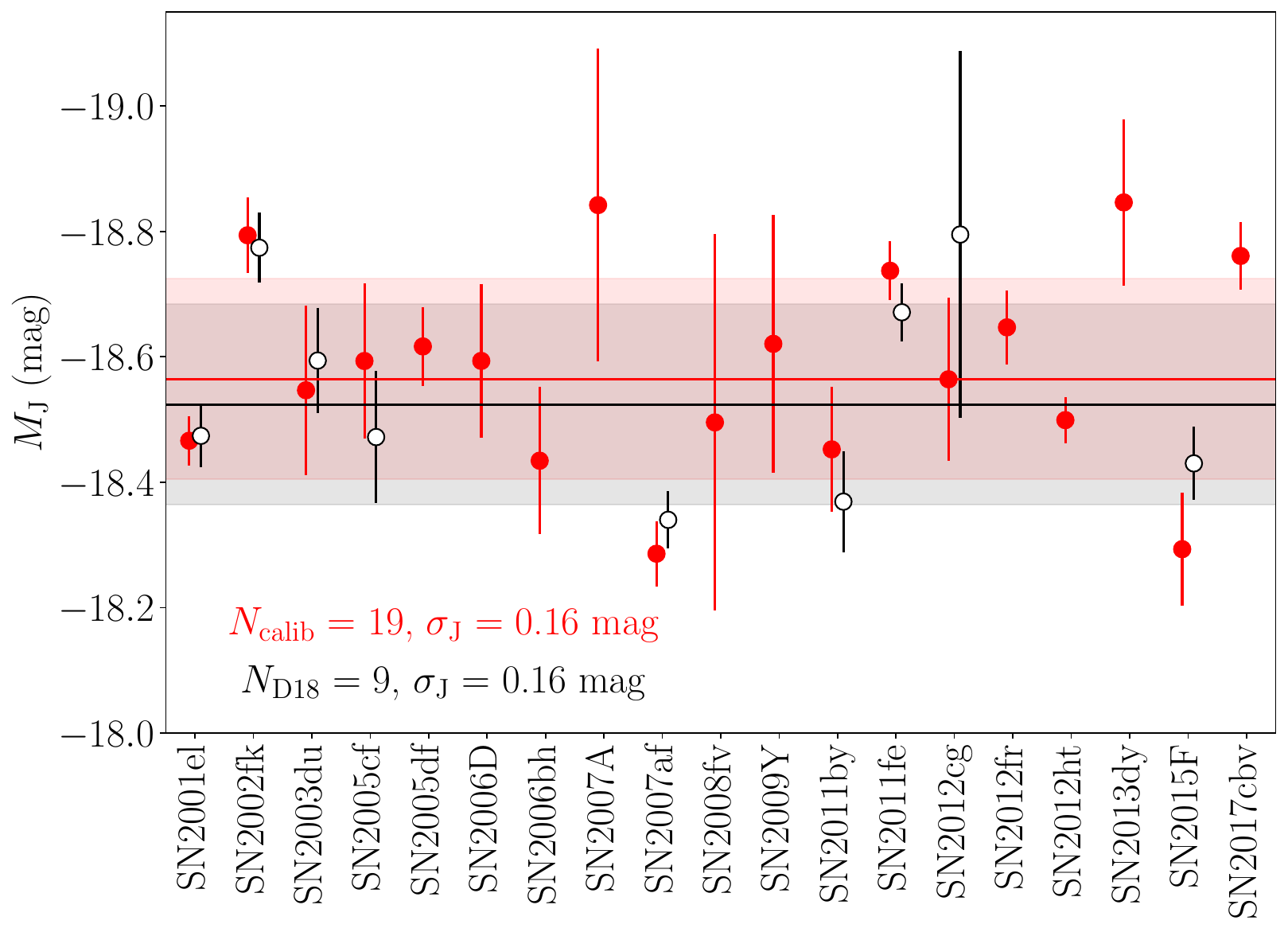}
    \includegraphics[width=\columnwidth]{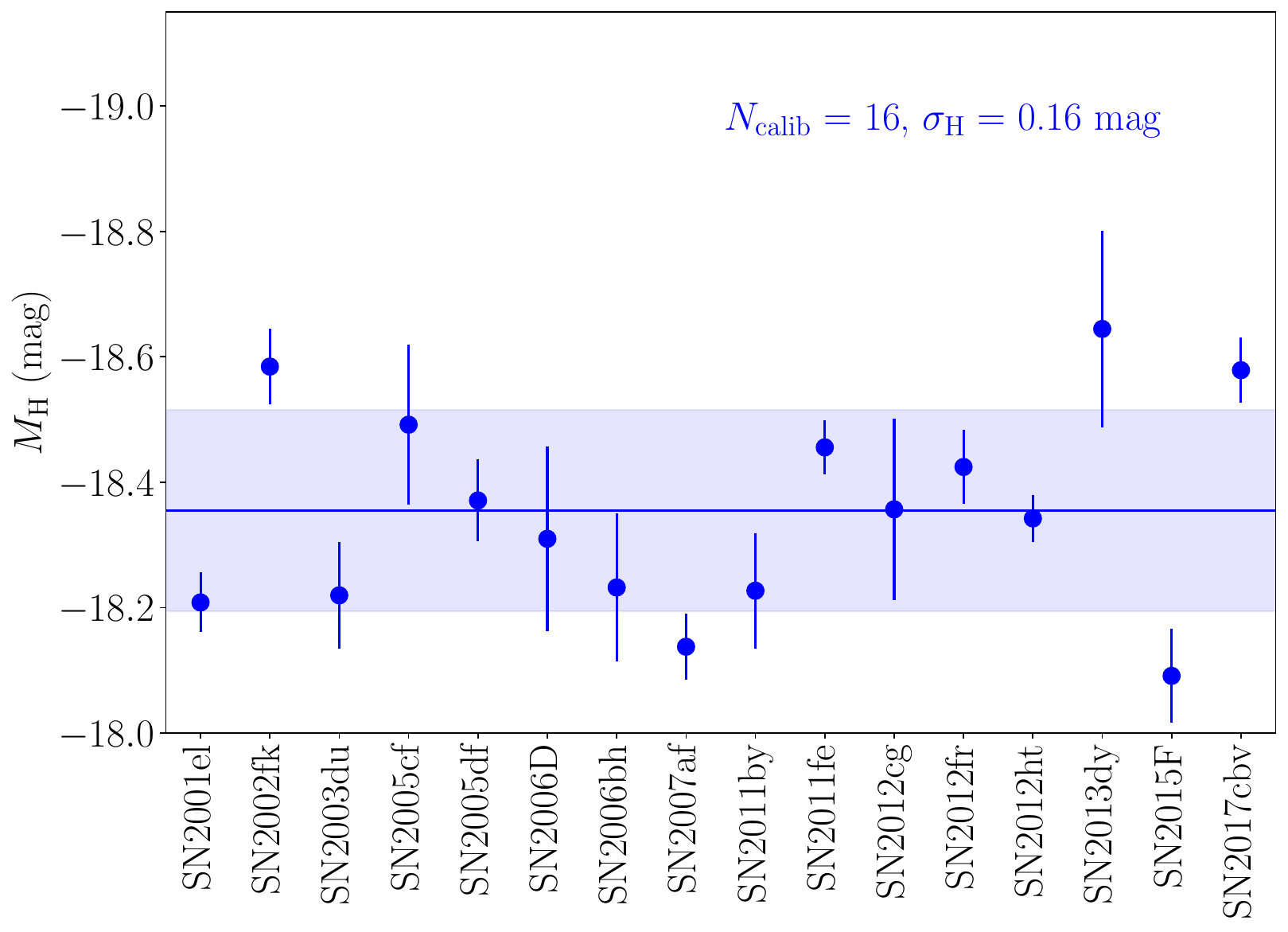}
    \caption{$J$ (left) and $H$ (right) absolute magnitudes of the SNe Ia in the calibrator sample. Horizontal lines represent the weighted average and the strip the standard deviation around that value. In the left panel, we included the 9 SN~Ia included in D18 for reference. Uncertainties correspond to those described in section \ref{sec:absmag} and do not include the $\sigma_{int}$ term added in quadrature.}
    \label{fig:calibrators}
\end{figure*}

\subsection{Multiband SNooPy fits}

We fit UV+optical+NIR light-curves of both the calibrator and the Hubble-flow samples with SNooPy, using the {\it EBV\_model2} and {\it max\_model} models with $\Delta m_{15}$ as a light-curve width parameter. Prior to the template fitting and within the SNooPY framework, the photometric points are corrected for Milky Way extinction using the dust maps from \cite{2011ApJ...737..103S}, and then K-corrected using the \cite{2007ApJ...663.1187H} template. The K-correction involves first color-correcting the Spectral Energy Distribution (SED) by multiplying the original template with a smooth function, which ensures that the observed colors match synthetic colors derived from the corrected SED. Regarding K-corrections, since the redshift range of our sample is quite narrow ($z<0.04$, except 5 objects up to $z\sim$0.08), they are in general small. At the median redshift of our sample $z=0.023$ the K-correction is 0.058 mag in $J$ and 0.030 mag in $K$, and up to 0.156 mag in $J$ and 0.125 mag in $H$ for the SN at the highest redshift ($z=0.08$). 

When using the {\it EBV\_model2}, the fitter provides an estimate of the time of maximum in the $B$-band $T_{\rm max,B}$, the light-curve width parameter $\Delta m_{15}$ in the $B$ band, and the color excess at peak $E(B-V)$. Moreover, we also obtain the $J$ and $H$-band peak magnitude given by the template. However, some of these parameters suffer from covariances among bands that are intrinsic to the model (see more details in e.g. \citealt{2020ApJ...901..143U}). For this reason, in this work we rather use the results from the more versatile {\it max\_model}, which fits each band independently and it is more convenient for the purpose of this work. We get the time of maximum $T_{\rm max,B}$ and light-curve width parameter $\Delta m_{15}$ in the $B$ band, and the $J$ and $H$-band time of maximum $T_{\rm max,X}^{\rm T}$ and peak magnitude $X_{max}^{\rm T}$. All these parameters are listed in Appendix \ref{app:snoopy}.

\subsection{$J$ and $H$ peak magnitudes}

Besides the NIR peak magnitudes from template fitting, following D18, we estimate $J$ and $H$ peak magnitudes through simple interpolation of their light-curves. In this way, we can independently get these values without relying on a particular light-curve template, which include corrections for SN light-curve shape and color.

We interpolate $J$ and $H$ light curves individually by using SNooPy internal routines that rely on either Gaussian Processes (GP) using the {\tt scikit-learn} package \citep{scikit-learn} with a constant plus a Mat\'ern kernel or spline fits using {\sc FITPACK} \citep{1993csfw.book.....D}\footnote{In this work we use version 2.5.3 of SNooPy, which is written in python 3. Previous versions based on python 2 used {\sc pymc} for GP interpolation, but it was replaced in newer versions of SNooPy since it was not compatible with python 3.}. For the GP interpolation we set the time-scale over which the function varies at 10 days, the amplitude of typical function variations as the standard deviation of the photometric points in magnitudes in each light-curve, and a smoothness of $\nu=3.5$. For some objects\footnote{32(56)\% in J(H) for the calibration sample, and 16(30)\% in J(H) for the Hubble flow sample.} the GP is less reliable than a simple spline interpolation, and in our baseline calculation of $H_0$ we choose the fit that provides the lower $\chi^2$ between data and fit in the time range ($-$10,+20) days. Our best light-curve fits are shown in Appendix \ref{app:lcfits} for both the calibrator and the Hubble-flow samples, where GP fits are shown in solid lines and splines in dashed lines.

We obtain the peak magnitude in the $X$ band $m_X$ and its uncertainty $\sigma_{m_X}$ from the interpolated light-curves, which are then corrected for Milky Way reddening using maps of \cite{2011ApJ...737..103S} and a \cite{1999PASP..111...63F} extinction law with $R_V=3.1$, equivalent to $R_J=0.86$ and $R_H=0.53$, and K-corrected using the SN Ia spectral energy distribution models from \cite{2007ApJ...663.1187H}, following the same procedure as in the SNooPy fits. The corrected peak magnitudes, the extinction and K-corrections terms are presented in Table \ref{tab:calibrators} for the calibrator sample and in  Table \ref{tab:hubbleflow} for the Hubble-flow sample.

\subsection{Absolute magnitudes} \label{sec:absmag}

To obtain the absolute magnitudes of SNe~Ia in the calibrator sample we subtract R22 Cepheid distance moduli of their host galaxies from the apparent peak magnitude, 
\begin{equation}
M_X=m_X-\mu_{\rm Ceph},
\end{equation}
and add their uncertainties in quadrature,
\begin{equation}
\sigma_{M_X}^2 = {\sigma_{m_X}^2 + \sigma_{\rm Ceph}^2}.
\end{equation}
The final absolute magnitudes are included in Table \ref{tab:calibrators}.

For the Hubble-flow SNe, we subtracted the distance modulus $\mu(z)$ using a flat $\Lambda$CDM cosmology with $\Omega_\Lambda=0.7$ (equivalent to a deceleration parameter $q_0=\Omega_m/2 - \Omega_\Lambda = -0.55$ and a jerk or prior deceleration $j_0=1.0$; \citealt{2004CQGra..21.2603V}) and $H_0=70$ km s$^{-1}$ Mpc$^{-1}$ from the apparent peak magnitude. For the uncertainty we add the peak magnitude error in quadrature with the redshift ($z_{corr}$) and peculiar velocity uncertainties converted to magnitudes as, 
\begin{equation}
\sigma^2_{M_X} = \sigma^2_{m_{X}} +  \left(\frac{5}{\ln 10}\frac{\sigma_z}{z}\right)^2 + \left(\frac{5}{\ln 10}\frac{v_{pec}}{cz}\right)^2,
\end{equation}
where we adopted a $v_{pec}$ = 250 km s$^{-1}$ \citep{2022MNRAS.514.4620D}. 

\begin{figure*}
    \includegraphics[width=\columnwidth]{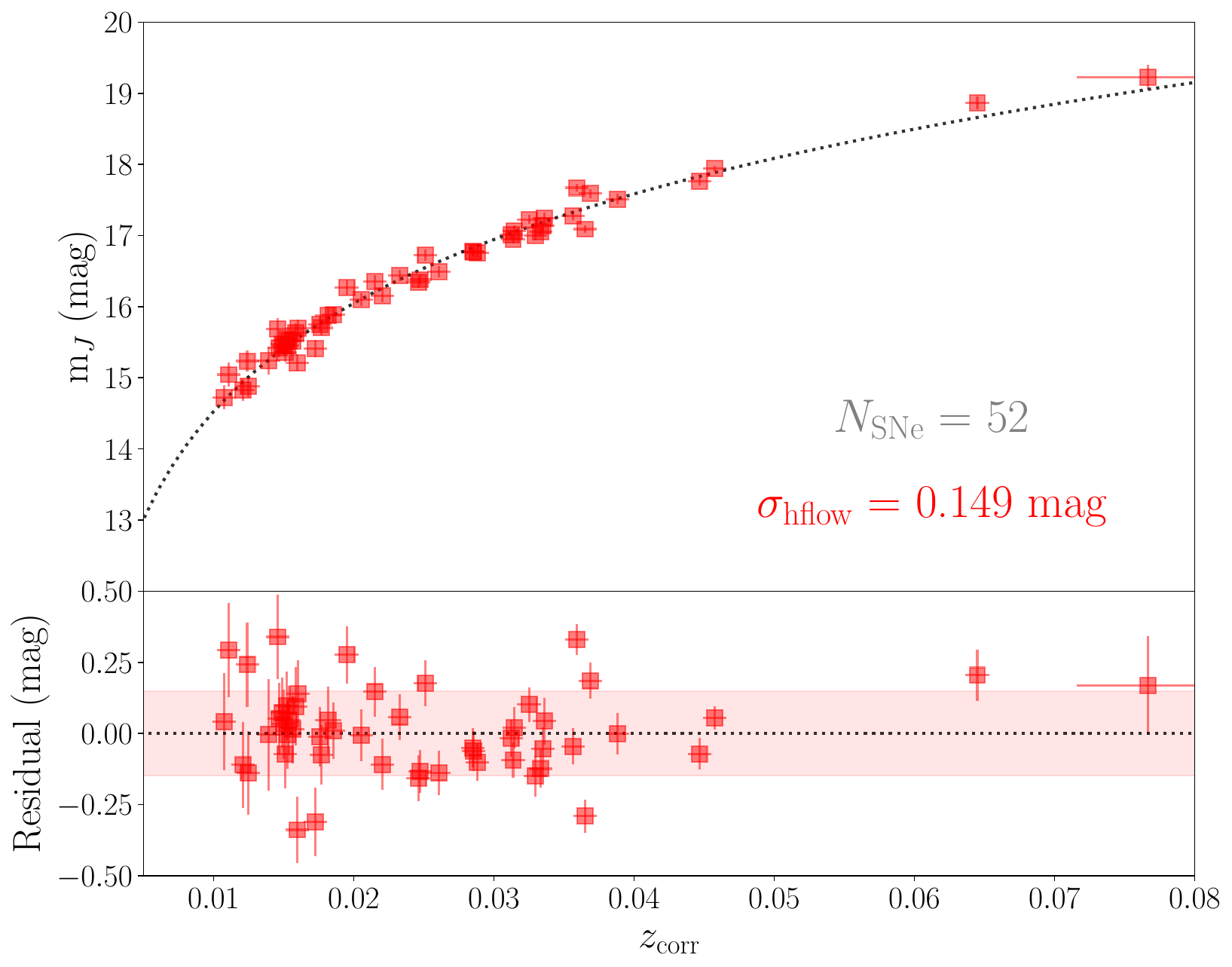} 
    \includegraphics[width=\columnwidth]{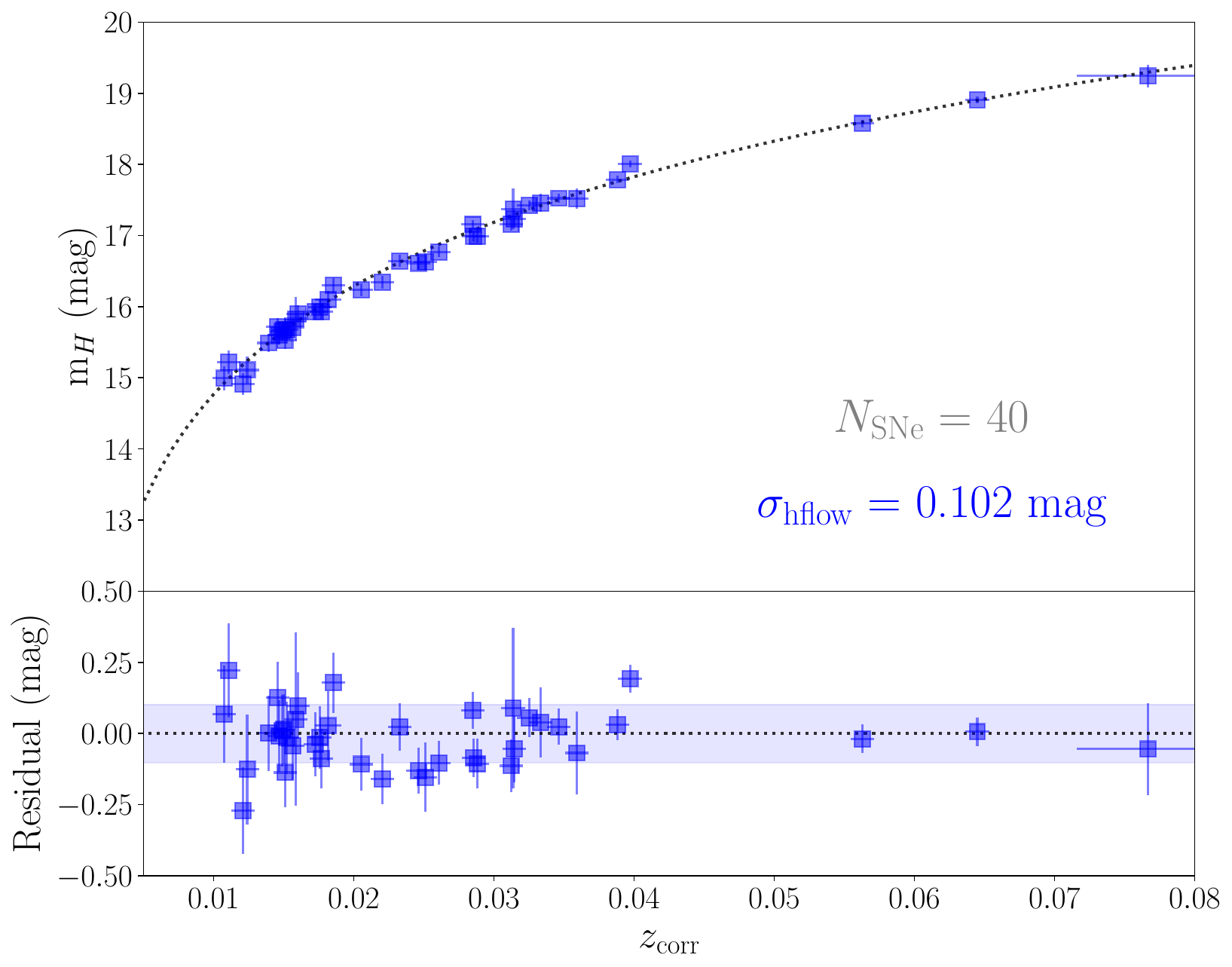}
    \caption{$J$ (left) and $H$ (right) Hubble-Lema\^itre diagram (top panels) and residuals (bottom panels) of our SN~Ia in the Hubble-flow sample.  Uncertainties correspond to those described in section \ref{sec:absmag} and do not include the $\sigma_{int}$ term added in quadrature.}
    \label{fig:hubblediag}
\end{figure*}

\subsection{$H_0$ determination}

To provide an estimate of $H_0$ with SN~Ia in the NIR we need to combine our calibrator sample, which will constrain the absolute magnitude $M_X$, and the Hubble-flow sample, which will determine the zero-point intercept $a_X$ of the NIR SN~Ia magnitude–redshift relation.

We follow here a similar procedure as in D18, \cite{2020MNRAS.496.3402D}, and more recently \cite{2022MNRAS.514.4620D}. Combining the expression of the distance modulus,
\begin{equation}
\mu = m - M = 5 \log_{10}(d_L) + 25,
\end{equation}
and the kinematic expression of the luminosity distance as defined by \cite{2007ApJ...659...98R},
\begin{equation}
d_L \approx \frac{c z}{H_0} \left( 1 + \frac{(1 - q_0) z}{2} + \frac{(1 - q_0 - 3 q_0^2 + j_0) z^2}{6} \right) = \frac{c z}{H_0} K(z),
\end{equation}
we end up with the simple equation,
\begin{equation}
\log_{10} H_0 = \frac{M_X + 5 a_X + 25}{5},
\end{equation}
where $M_X$ is constrained by the calibrator sample, and $a_X$ is the intercept of the distance–redshift relation given for an arbitrary expansion history and for $z > 0$ \cite{2022ApJ...934L...7R}, and it is determined from the Hubble-flow sample by
\begin{equation}
a_X = \log_{10} cz + \log_{10} K(z) - 0.2 m_X.
\end{equation}
To find $H_0$ and $M_X$ we fit a joint Bayesian model to the combined dataset using the Markov chain Monte Carlo (MCMC) sampler of the posterior probability function {\tt emcee} \citep{2013PASP..125..306F} with 200 walkers and 2000 steps each, burning the first 1000 steps per each walker, so with a total of 200000 samples. In addition, we account for an unmodeled intrinsic NIR SN~Ia scatter $\sigma_{int}$, as a nuisance parameter, which is added in quadrature to the calibrator and Hubble-flow peak magnitude uncertainty ($\sigma_{ {\rm Cal},i}^2 = \sigma_{M_{X,i}}^2 + \sigma_{int}^2$; $\sigma_{ {\rm HF},i}^2 = \sigma_{M_{X,i}}^2 + \sigma_{int}^2$), and that we interpret as SN-to-SN variation in the peak luminosity to be constrained by the data and marginalized over. The likelihood we optimize is,
\begin{eqnarray}\label{eq:likelihood}
\nonumber \log \mathcal{L} = -\frac{1}{2} \sum_i \left( \log(2 \pi \sigma_{{\rm Cal},i}^2) + \frac{(M_{{\rm Cal},i} - M_X)^2}{\sigma_{{\rm Cal},i}^2} \right) \\
 -\frac{1}{2} \sum_i \left( \log(2 \pi \sigma_{{\rm HF},i}^2) + \frac{(M_{{\rm HF},i} - M_X + 5  \log(H_0/70))^2}{\sigma_{{\rm HF},i}^2} \right), 
\end{eqnarray}
where the calibrator term penalizes depending on how far the calibrators are to the mean absolute magnitude, and the Hubble-flow term penalizes depending on how close the Hubble-flow objects are to the mean absolute magnitude for the input $H_0$. We use as initial guesses for the walkers a $H_0=70$ km s$^{-1}$ Mpc$^{-1}$, a $M_X$ equal to the average calibrator absolute peak magnitude in each band, and a  $\sigma_{int}=\sqrt{{\rm stddev}(M_X)^2 - {\rm avg}({\sigma_{M_X}})^2}$, and allow them to vary in a scale of 10 km s$^{-1}$ Mpc$^{-1}$, 1 mag, and 0.1 mag, respectively. We also use a single scale-free prior of $\log(p(\sigma_{int})) = -\log \sigma_{int}$ with the conditions $H_0 > 0$ and $\sigma_{int}>0$.


\section{Results}\label{sec:cal}

\subsection{Properties of the calibrator and Hubble-flow samples}\label{sec:prop}

The absolute magnitudes of 19 SNe Ia in the calibrator sample in the $J$ band and 16 in the $H$ band are presented in Figure \ref{fig:calibrators}. The average $J$- and $H$-band calibrator absolute magnitudes are, $\langle M_{J} \rangle=(-18.565\pm0.025)$ and $\langle M_{H} \rangle=(-18.355\pm0.023)$ mag. In the $J$-band the average absolute magnitude is only slightly brighter (by $\sim$0.04 mag) than that presented in D18. These calibrator absolute magnitudes show a dispersion of $\sigma_{calib}$ = 0.16 mag in both bands, which is comparable to the typical scatter found in the optical after  light-curve shape and color corrections. The dispersion is larger than what can be accounted for by the formal uncertainties $\sigma_{M_X}$, with the reduced $\chi^2>5$ in both bands, confirming that an additional intrinsic scatter, $\sigma_{int}$, will be needed in our analysis to account for SN-to-SN luminosity variations. We will discuss below that, once included, the reduced $\chi^2$ are reduced to around unity. 

Figure \ref{fig:hubblediag} presents the Hubble-Lema\^itre diagrams and residuals of our Hubble-flow sample. We used here the redshift corrected for peculiar velocities in the X-axes, and the apparent peak magnitude of the SN in each band in the Y-axes. Hubble residuals are calculated against a flat $\Lambda$CDM cosmology with $H_0=70$ km s$^{-1}$ Mpc$^{-1}$. Five SNe, SN~2008hs, SN~2010ai, PTF10tce (only in $J$), iPTF13asv (only in $H$), and iPTF14bdn (only in $H$) have been removed applying a Chauvenet criterion, which for the sample size is usually around 2.6$\sigma$, leaving the sample with 52 SNe in $J$ and 40 SNe in $H$ bands. The standard deviation of the residuals is $\sigma_{\rm HF}$ = 0.149 and 0.102 mag in the $J$ and $H$ bands, respectively. This dispersion is similar to that found with optical corrected magnitudes by SHOES ($\sigma=0.135$ mag; R22), and comparable to previous works using NIR template fitting ($\sigma=0.116$ mag in $J$ and $\sigma=0.088$ mag in $H$; \citealt{2012MNRAS.425.1007B}), and interpolation ($\sigma=0.106$ mag; D18). 

\begin{figure}
    \includegraphics[width=\columnwidth]{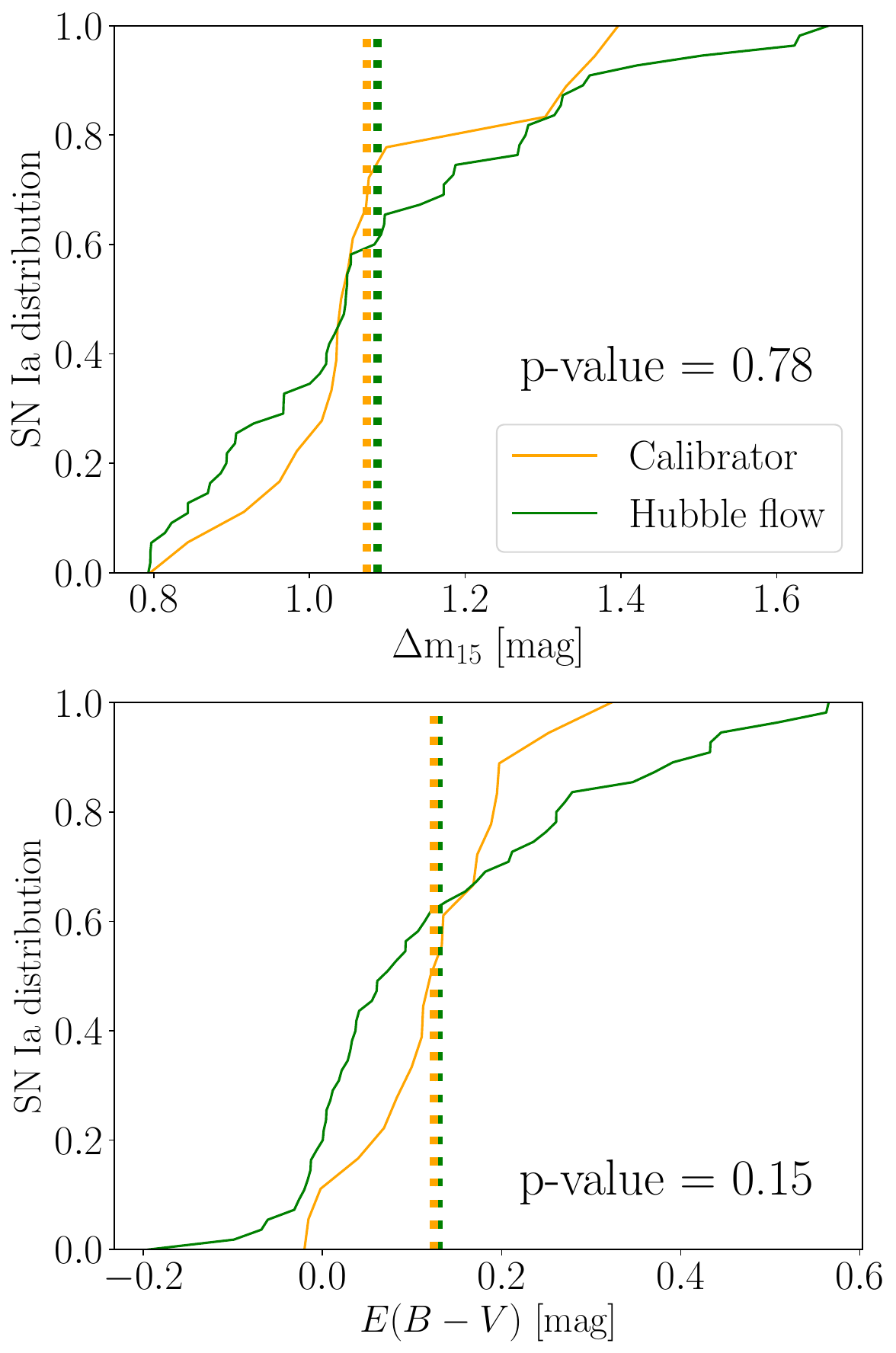} 
    \caption{Distributions of $\Delta m_{15}$ light-curve width parameter (top) and color excess at peak $E(B-V)$ (bottom) of the calibrator (in orange) and Hubble-flow (in green) SN~Ia samples obtained from SNooPY fitting UV, optical and NIR light-curves simultaneously with the $EBV\_model2$. Vertical dashed lines represent the average value of the distributions.
    The p-value of the two-sample Kolmogorov-Smirnov (K-S) test is included in each panel.}
    \label{fig:comp}
\end{figure}

Figure \ref{fig:comp} shows the cumulative distributions of $\Delta m_{15}$ and $E(B-V)$ obtained from SNooPy fits for our calibrator and Hubble-flow samples. The shapes of their distributions and their average values are quite similar at around $\Delta m_{15}$=1.1 mag and $E(B-V)$=0.12 mag. Both samples include objects with $\Delta m_{15}$ larger than 1.2 mag, but none reach the value typically associated to subluminous objects ($>$1.8 mag). Conversely, the color excess distribution of the calibrator sample appears to be slightly narrower than the Hubble-flow objects,  with a larger median value of 0.13 mag compared to 0.07 mag for the Hubble flow sample. We performed a two-sample Kolmogorov-Smirnov (K-S) test using the {\sc scipy} library \citep{2020SciPy-NMeth}, and we obtained a p-value of 0.78 for $\Delta m_{15}$ and 0.15 for $E(B-V)$, respectively, indicating that both distributions are consistent with being drawn from the same sample population. A similar Figure with the cumulative distributions of the color-stretch $s_{BV}$ parameter is included in Appendix \ref{app:sbv}.

\begin{figure*}
    \includegraphics[width=\columnwidth]{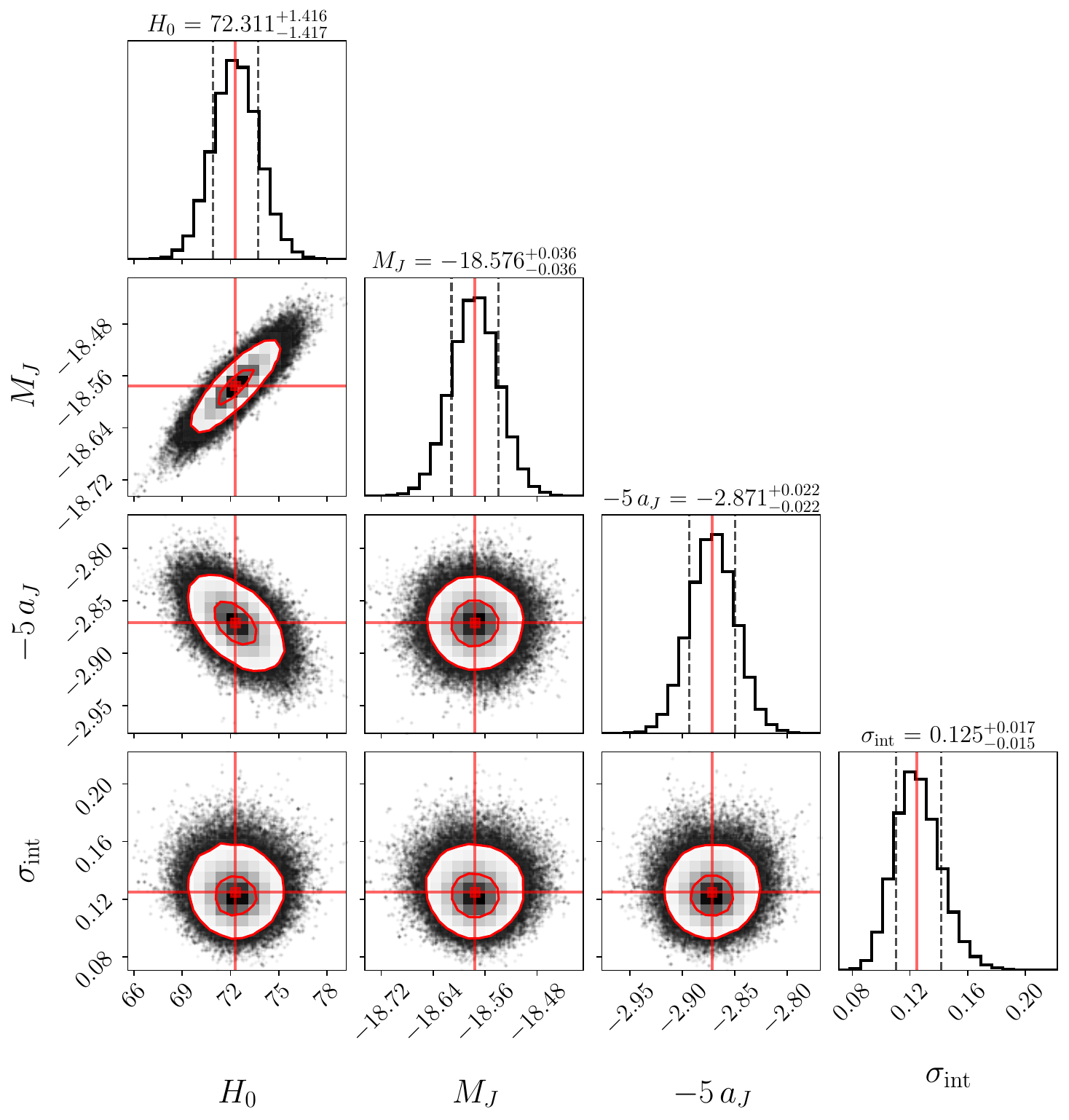} 
    \includegraphics[width=\columnwidth]{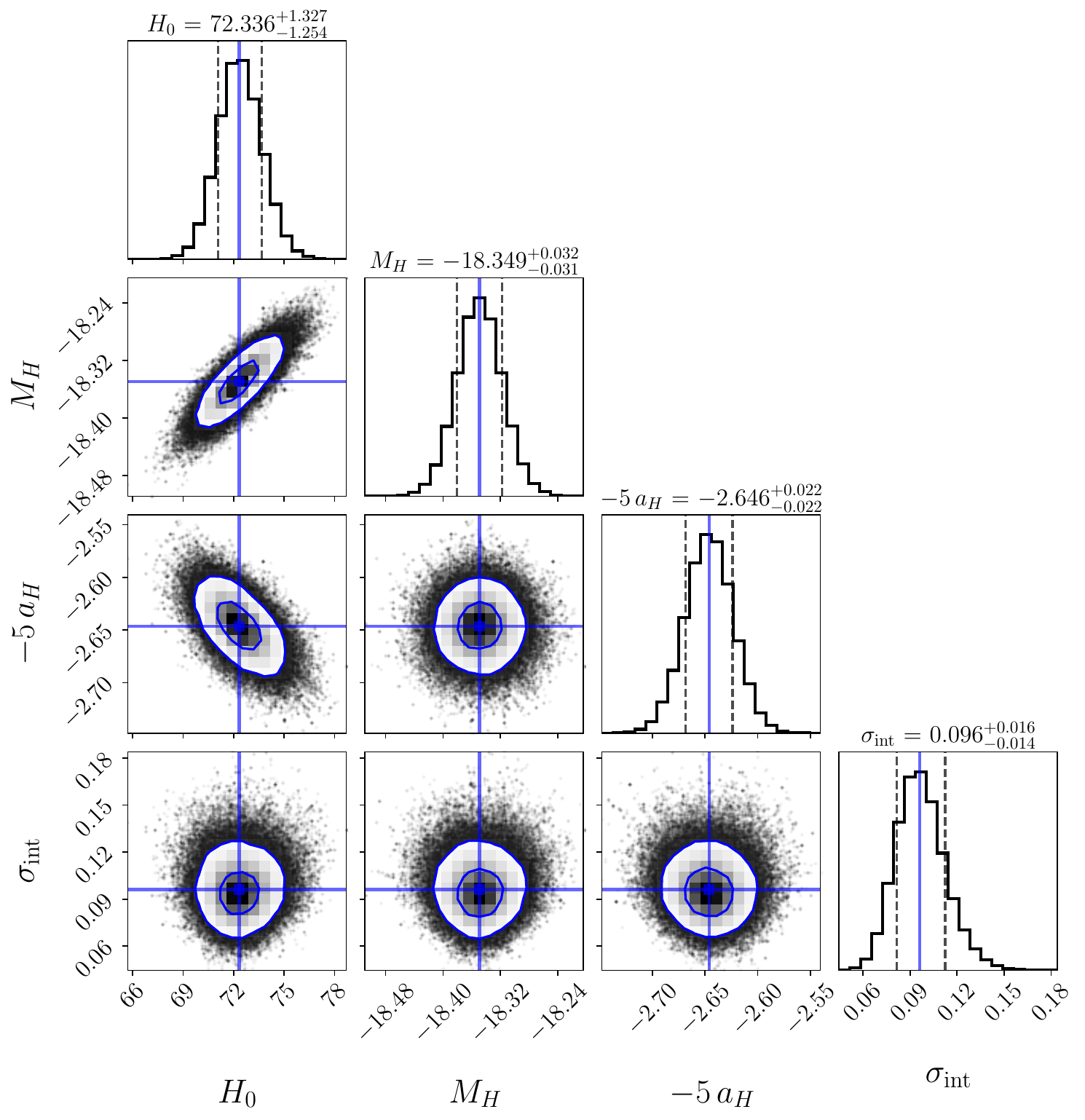}
    \caption{Corner plot with the results of the MCMC posteriors of our baseline analysis. Results for the $J$ band on the left, and for the $H$ band on the right. The red/blue contours on the scatter plots correspond to 1$\sigma$ and 2$\sigma$ of the 2D distributions, and the vertical and horizontal lines the medians of the posteriors.}
    \label{fig:h0}
\end{figure*}

\subsection{$H_0$ measurement}

Our baseline determination of $H_0$ includes 19/16 SNe Ia from the calibrator sample and 52/40 SNe Ia in the Hubble-flow sample for bands $J$/$H$, respectively. We use R22 Cepheid distances for the calibrator sample and peculiar-velocity corrected redshifts for the Hubble-flow sample. The results from 2$\times$10$^5$ posterior samples of the MCMC are shown in Figure \ref{fig:h0} and summarized in Table \ref{tab:h0fits}.

Our baseline result for $H_0$ is 72.31$\pm$1.42 km s$^{-1}$ Mpc$^{-1}$ in the $J$ band and 72.34$^{+1.25}_{-1.33}$ km s$^{-1}$ Mpc$^{-1}$ in the $H$ band, where the errors represent the 16th and 84th percentile range that includes 68\% of the posterior samples and only include statistical uncertainties. This measurement of $H_0$ has a $\sim$1.9\% precision in both bands, respectively, which is lower than that found by D18 (2.2\% in $J$). Both absolute magnitudes $M_X$, $-18.576\pm0.036$ in $J$ and $-18.349\pm0.032$ in $H$, are similar by 0.01 mag to those found just averaging out the calibrator sample magnitudes. The resulting values for the intercept $a_i$, $-2.871\pm0.022$ mag in $J$ and $-2.646\pm0.022$ mag in $H$, contribute less than 2\% to the statistical $H_0$ uncertainty, while the absolute magnitudes contribute around 2.5\%. The additional nuisance parameter introduced in the model to account for remaining scatter $\sigma_{int}$ is found to be of around 0.125 and 0.096 mag in $J$ and $H$, respectively. The presence of this intrinsic scatter increases the uncertainty in the peak absolute magnitude compared to the weighted mean calculated in Section \ref{sec:cal}, from less than 0.01 to about 0.03 mag in both bands. As noted by D18, since the same $\sigma_{int}$ is included in quadrature in the uncertainties of both the calibrator and Hubble-flow samples, $M_X$ and $a_X$ appear to be not correlated because they are constrained separately by each subsample.

\subsection{Distance ladder}

\begin{figure*}
    \includegraphics[width=\columnwidth]{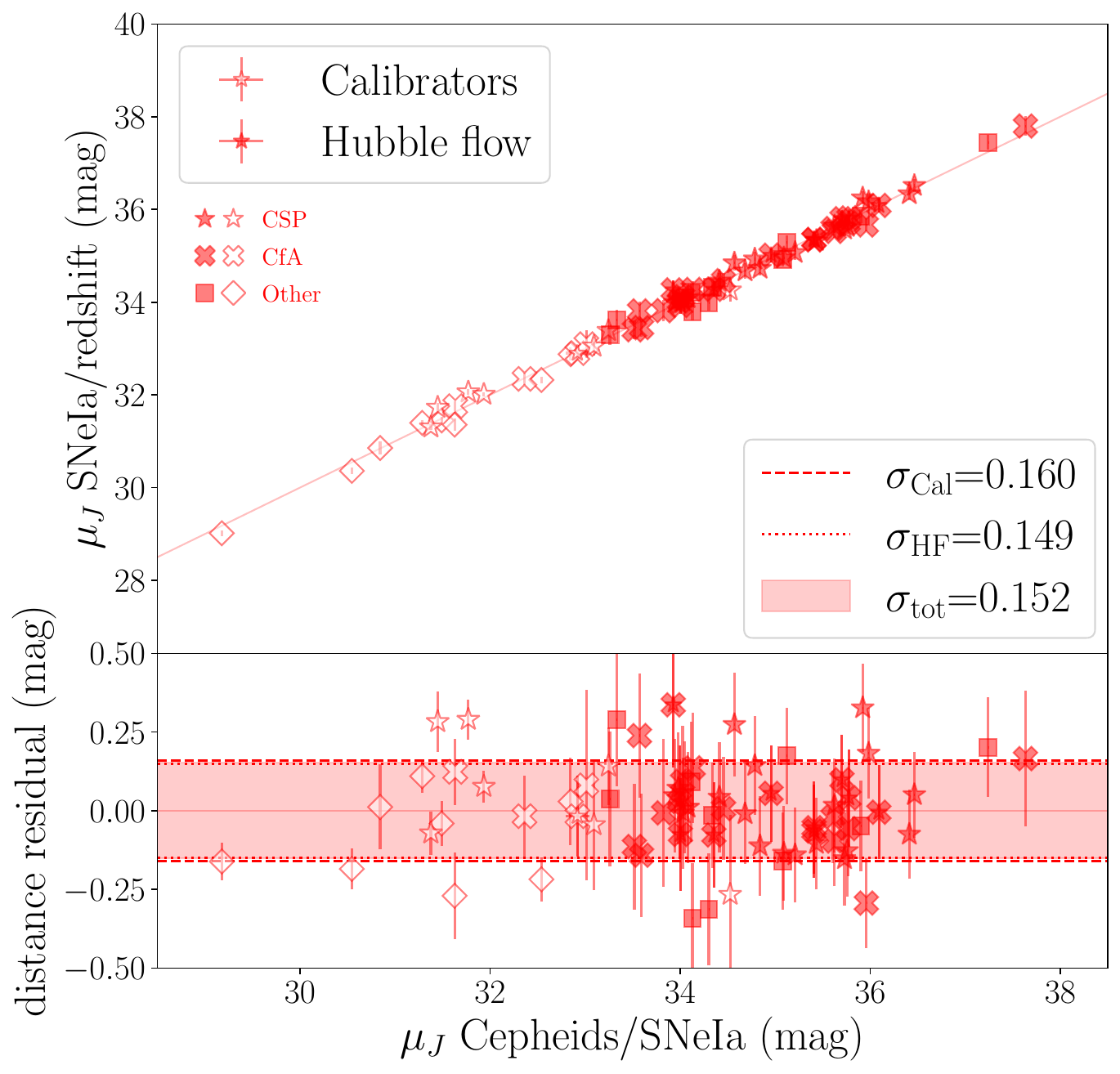} 
    \includegraphics[width=\columnwidth]{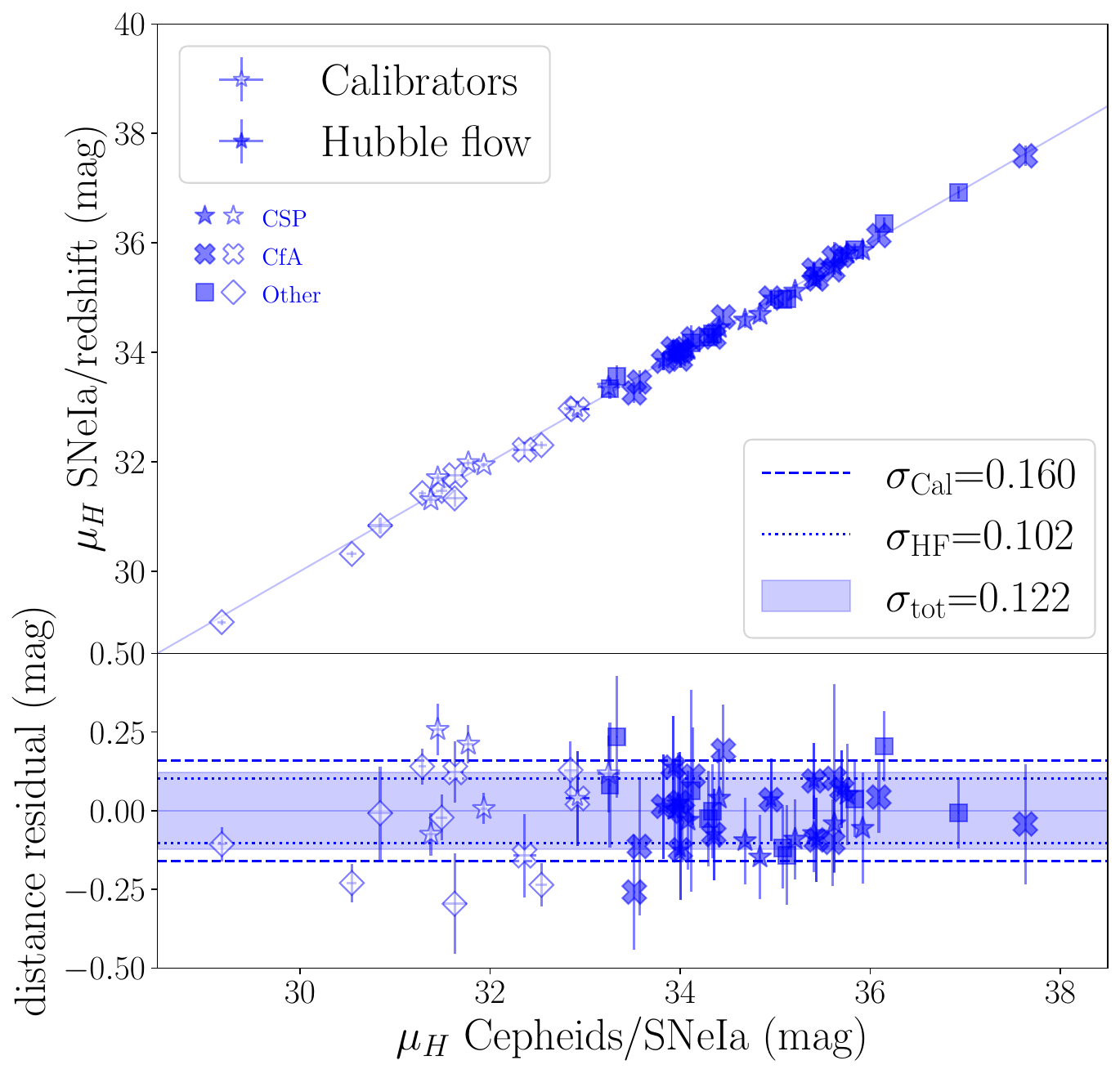}
    \caption{Second and third rungs of the distance ladder for $J$ (left) and $H$ (right) bands. Empty symbols represent SNe Ia in the calibrator sample, corresponding to the second rung where absolute magnitudes were calibrated from Cepheid distances, so X-axis $mu$ is the Cepheid-based distance from SHOES and the Y-axis $mu$ is the SNIa-based distance. Filled symbols correspond to the Hubble-flow sample ($z>0.01$ in our baseline analysis) in the third rung of the distance ladder, where X-axis $mu$ is the SNIa-based distance from SHOES and the Y-axis $mu$ is from the redshift. Symbols are different for thoae SNe Ia from CSP (stars), CfA (crosses), and other SNe Ia from other sources (squares/diamonds).}
    \label{fig:DL}
\end{figure*}

Using the values found for $H_0$ and $M_X$, and once $\sigma_{int}$ is included in quadrature to the absolute magnitudes in both calibrator and Hubble-flow samples, we construct the second and third rungs of the distance ladder of SN~Ia in the NIR in Figure \ref{fig:DL}. We measured the distance modulus for both calibrators and Hubble-flow SNe Ia by subtracting the average $M_X$ found with the MCMC procedure from our measured apparent peak magnitudes (in the Y-axis), against an independent measure of the distance (in the X-axis). For calibrators, we used the R22 Cepheids distances, and for objects in the Hubble flow we used the predicted distance modulus by a flat $\Lambda$CDM cosmology with $\Omega_\Lambda=0.7$ and our baseline $H_0$. The resulting scatter in the full distance ladder is 0.152 mag in the $J$-band and 0.122 mag in the $H$-band.

Once $\sigma_{int}$ is included in the uncertainty budget we have a $\chi^2$=19.7 for 18 degrees of freedom (dof) in $J$ and 26.1 for 15 dof in $H$ for the calibrators, and a $\chi^2$=44.2 for 51 dof in $J$ and 19.1 for 39 dof in $H$ for the Hubble-flow sample, both leading to reduced $\chi^2$ around the unity (0.50 to 1.70). Although the reduced $\chi^2$ for the calibrator sample may seem noisier than the Hubble flow, we caution the reader that any conclusions about the scatter and $\chi^2$ is very sensitive to the exclusion of the outliers based on the Chauvenet criterion, given that our sample size is still small. Once the outliers are included, we get $\chi^2$/dof of 1.2 and 0.9 for calibrator and Hubble flow samples. One of our variations of the analysis presented in section \ref{sec:vars} includes those outliers removed in the baseline analysis.

\section{Discussion}\label{sec:disc}

The main differences between this work and D18 are: 
(i) the third and last data release of the CSP photometry from \cite{2017AJ....154..211K} is used here, while D18 used the at that time available second data release from \cite{2011AJ....142..156S}; 
(ii) we applied S-corrections to put all compiled photometry in the same photometric system, which we chose to be the CSP system;
(iii) NIR luminosity of SNe Ia is calibrated with the NIR Cepheid distances from R22, while D18 used NIR Cepheid distances from \cite{2016ApJ...826...56R}. R22 not only increased the number of distances available but updated those from \cite{2016ApJ...826...56R}; 
(iv) while the method used to determine $J$ and $H$ peak magnitudes is similar between the two works, the actual code for interpolation within the SNooPY framework varied from {\tt pymc} included in the previous version to {\tt scipy} in the current version; and 
(v) here we also extend the analysis to the $H$-band, providing an independent measurement of $H_0$.

\subsection{Analysis variations} \label{sec:vars}
We investigate possible sources of systematic uncertainty in our measurement by performing some variations in the analysis, applying different cuts on the calibrator and Hubble-flow samples, and studying their effects in the determination of $H_0$. All results from these variations are summarized in Table \ref{tab:h0fits}. 

\subsubsection{Peculiar velocities}
The first test is regarding the assumed peculiar velocities uncertainty $v_{pec}$ added in quadrature to the magnitudes of the Hubble-flow sample. Instead of the assumed value of 250 km~s$^{-1}$, we tried to increase the value to 350 km~s$^{-1}$, take a lower value of 150 km~s$^{-1}$ as assumed in other works, and also tried removing this term from the analysis. The net effect of reducing the significance of this term is reducing the value of $H_0$ by 0.15 km~s$^{-1}$~Mpc$^{-1}$ at most, which represent a 0.2\% shift with respect to the fiducial value, and trespassing this uncertainty to the nuisance $\sigma_{int}$ parameter. All other parameters remain mostly unaltered.

\subsubsection{Redshifts}
Next, the analysis is repeated but this time changing the redshift of SNe~Ia in the Hubble-flow sample. We first used all $z_{cmb}$ redshifts instead of those corrected for peculiar velocities $z_{corr}$provided by \cite{2022PASA...39...46C}. Secondly, we repeat the analysis using the host galaxy redshifts reported in NED, converted to the CMB reference frame and with and without corrections for peculiar velocities using the model of \citet{2015MNRAS.450..317C}. Finally, we repeat the process this time starting from the redshifts reported in the reference papers instead, converted to the CMB frame and with and without \citet{2015MNRAS.450..317C} peculiar velocity corrections. The most significant change of all these variations is a reduction in $H_0$ of down to a $\sim$1.6\% in the $H$ band when using $z_{cmb}$ with no peculiar velocity corrections. Since this variation does not affect the calibration sample, we see similar changes of the $a_X$ value, and an increase in $\sigma_{HF}$ and the resulting $\sigma_{int}$. Our result is in agreement with \cite{2022ApJ...938..112P} who found peculiar velocity corrections do not affect significantly the measurement of $H_0$.

\subsubsection{Extinction}
The next test consists in removing all objects with $E(B-V)>0.3$ as measured from SNooPY using the $EBV\_model2$. SHOES sample was selected to have no high-extinction, so this cut results in only one object removed from the calibrator sample (SN~2001el) and eight in the Hubble-flow sample. The effect in $H_0$ is a change of $\sim$0.6\% in opposite directions for each filter. Notably, since the remaining sample is less affected by reddening, $\sigma_{\rm HF}$ is reduced from 0.124 to 0.109 mag in $J$ and unchanged in $H$ (0.096 to 0.097 mag), confirming that $H$ band is less affected by extinction effects.

\subsubsection{Hubble flow cut}
We also tested to remove all objects in the Hubble-flow sample with redshifts $z>0.023$ as in R22. Alternatively we also removed the only object at $z>0.05$ (PTF10ufj) to see their role in the determination of the parameters. These two cuts resulted in a reduction of the Hubble-flow sample size from 52 to 22 and 20, respectively, in the $J$ band, and from 40 to 18 and 15 in the $H$ band. The resulting $H_0$ was not significantly affected in any case for the $H$ band, and increases up to 1.61\% for the $J$ band, corresponding to 01.2 km~s$^{-1}$~Mpc$^{-1}$, becoming even more consistent to the SHOES value.

\subsubsection{Chauvenet criterion}
The next test consists on not applying in any case the Chauvenet criterion and use all SNe~Ia available including clear outliers. Adding those $>$3$\sigma$ outliers produces an increase of $\sigma_{\rm HF}$ from 0.149 to 0.185 mag in $J$ and from 0.102 to 170 in $H$, the largest of all our variations. Also, it produces a 0.9 and a 0.1 km~s$^{-1}$~Mpc$^{-1}$ reduction in $H_0$ in the $J$ and $H$ bands, respectively.

\begin{table*}[t]
\centering
\caption{Results of MCMC posteriors for our baseline analysis and the 21 variations.}
\label{tab:h0fits}
\begin{tabular}{lccccccccc}
\hline
Variation      &$N_{\rm cal}$&$\sigma_{\rm cal}$&$N_{\rm HF}$&$\sigma_{\rm HF}$&$H_0$& $M_X$                     & $-5 a_X$                     & $\sigma_{\rm int}$        & $\Delta H_0$\\
               &               & (mag) &    & (mag) & (km~s$^{-1}$~Mpc$^{-1}$)   & (mag)                         & (mag)                        & (mag)                     & (\%)   \\
\hline
 \multicolumn{10}{c}{$J$-band}\\
\hline
{\bf Baseline}            & 19 & 0.160 & 52 & 0.149 & 72.31$_{-1.42}^{+1.42}$ & $-$18.58$_{-0.04}^{+0.04}$ & $-$2.87$_{-0.02}^{+0.02}$ & 0.125$_{-0.015}^{+0.017}$ &   $...$  \\
$v_{pec}=350$ km s$^{-1}$ & 19 & 0.160 & 52 & 0.149 & 72.36$_{-1.40}^{+1.41}$ & $-$18.57$_{-0.03}^{+0.04}$ & $-$2.87$_{-0.02}^{+0.02}$ & 0.117$_{-0.015}^{+0.018}$ &    0.07  \\
$v_{pec}=150$ km s$^{-1}$ & 19 & 0.160 & 52 & 0.149 & 72.26$_{-1.45}^{+1.51}$ & $-$18.57$_{-0.04}^{+0.04}$ & $-$2.87$_{-0.02}^{+0.02}$ & 0.135$_{-0.014}^{+0.015}$ & $-$0.07  \\
$v_{pec}=0$ km s$^{-1}$   & 19 & 0.160 & 52 & 0.149 & 72.16$_{-1.50}^{+1.55}$ & $-$18.58$_{-0.04}^{+0.04}$ & $-$2.87$_{-0.02}^{+0.02}$ & 0.145$_{-0.013}^{+0.015}$ & $-$0.21  \\
$z_{cmb}$                 & 19 & 0.160 & 52 & 0.152 & 71.53$_{-1.39}^{+1.45}$ & $-$18.57$_{-0.04}^{+0.04}$ & $-$2.85$_{-0.02}^{+0.02}$ & 0.124$_{-0.015}^{+0.017}$ & $-$1.08  \\
$z_{corr}$ from NED+C15   & 19 & 0.160 & 52 & 0.161 & 71.96$_{-1.41}^{+1.42}$ & $-$18.57$_{-0.04}^{+0.04}$ & $-$2.86$_{-0.02}^{+0.02}$ & 0.125$_{-0.016}^{+0.018}$ & $-$0.48  \\
$z_{cmb}$ from NED        & 19 & 0.160 & 52 & 0.145 & 71.62$_{-1.37}^{+1.36}$ & $-$18.57$_{-0.04}^{+0.03}$ & $-$2.85$_{-0.02}^{+0.02}$ & 0.116$_{-0.015}^{+0.016}$ & $-$0.95  \\
$z_{corr}$ from REF+C15   & 19 & 0.160 & 52 & 0.165 & 72.05$_{-1.47}^{+1.49}$ & $-$18.57$_{-0.04}^{+0.04}$ & $-$2.86$_{-0.02}^{+0.02}$ & 0.132$_{-0.016}^{+0.018}$ & $-$0.36  \\
$z_{cmb}$ from REF        & 19 & 0.160 & 52 & 0.148 & 71.60$_{-1.36}^{+1.39}$ & $-$18.57$_{-0.04}^{+0.04}$ & $-$2.85$_{-0.02}^{+0.02}$ & 0.121$_{-0.015}^{+0.017}$ & $-$0.99  \\
$E(B-V)<0.3$              & 18 & 0.162 & 44 & 0.141 & 72.72$_{-1.42}^{+1.44}$ & $-$18.58$_{-0.04}^{+0.04}$ & $-$2.89$_{-0.02}^{+0.02}$ & 0.117$_{-0.015}^{+0.017}$ & $ $0.56  \\
HF $z>0.023$              & 19 & 0.160 & 24 & 0.136 & 72.99$_{-1.63}^{+1.64}$ & $-$18.58$_{-0.04}^{+0.04}$ & $-$2.89$_{-0.03}^{+0.03}$ & 0.132$_{-0.018}^{+0.021}$ &    0.94  \\
HF $0.023<z<0.05$         & 19 & 0.160 & 22 & 0.128 & 73.47$_{-1.64}^{+1.65}$ & $-$18.58$_{-0.04}^{+0.04}$ & $-$2.91$_{-0.03}^{+0.03}$ & 0.131$_{-0.018}^{+0.021}$ &    1.61  \\
no Chauvenet              & 19 & 0.160 & 55 & 0.185 & 71.37$_{-1.65}^{+1.64}$ & $-$18.57$_{-0.04}^{+0.04}$ & $-$2.84$_{-0.03}^{+0.02}$ & 0.157$_{-0.017}^{+0.019}$ & $-$1.31  \\
only best fits            & 17 & 0.167 & 42 & 0.151 & 71.82$_{-1.53}^{+1.57}$ & $-$18.58$_{-0.04}^{+0.04}$ & $-$2.86$_{-0.03}^{+0.03}$ & 0.133$_{-0.017}^{+0.019}$ & $-$0.68  \\
all gp                    & 19 & 0.170 & 50 & 0.152 & 72.90$_{-1.45}^{+1.44}$ & $-$18.57$_{-0.04}^{+0.04}$ & $-$2.88$_{-0.02}^{+0.02}$ & 0.126$_{-0.015}^{+0.017}$ &    0.82  \\
$m_X$ from template       & 19 & 0.170 & 49 & 0.144 & 73.94$_{-1.38}^{+1.38}$ & $-$18.55$_{-0.03}^{+0.03}$ & $-$2.90$_{-0.02}^{+0.02}$ & 0.120$_{-0.015}^{+0.017}$ &    2.25  \\
removing subluminous      & 19 & 0.160 & 49 & 0.144 & 73.20$_{-1.38}^{+1.37}$ & $-$18.57$_{-0.03}^{+0.03}$ & $-$2.90$_{-0.02}^{+0.02}$ & 0.116$_{-0.015}^{+0.017}$ &    1.22  \\
only CSP                  &  8 & 0.176 & 29 & 0.130 & 75.09$_{-2.07}^{+2.07}$ & $-$18.48$_{-0.05}^{+0.05}$ & $-$2.86$_{-0.03}^{+0.03}$ & 0.113$_{-0.019}^{+0.022}$ &    3.85  \\
only CfA                  &  4 & 0.062 & 22 & 0.145 & 73.53$_{-2.35}^{+2.41}$ & $-$18.53$_{-0.07}^{+0.07}$ & $-$2.86$_{-0.02}^{+0.02}$ & 0.060$_{-0.032}^{+0.030}$ &    1.69  \\
only CSP+CfA              & 11 & 0.154 & 40 & 0.127 & 75.07$_{-1.72}^{+1.72}$ & $-$18.48$_{-0.04}^{+0.04}$ & $-$2.86$_{-0.02}^{+0.02}$ & 0.100$_{-0.016}^{+0.018}$ &    3.82  \\
only spirals in HF        & 19 & 0.160 & 39 & 0.135 & 73.43$_{-1.40}^{+1.44}$ & $-$18.57$_{-0.03}^{+0.03}$ & $-$2.90$_{-0.02}^{+0.02}$ & 0.115$_{-0.016}^{+0.018}$ &    1.54  \\
SHOES selection           & 19 & 0.160 & 18 & 0.117 & 74.02$_{-1.67}^{+1.70}$ & $-$18.57$_{-0.04}^{+0.04}$ & $-$2.92$_{-0.03}^{+0.03}$ & 0.123$_{-0.018}^{+0.022}$ &    2.36  \\
\hline
 \multicolumn{10}{c}{$H$-band}\\
\hline
{\bf Baseline}            & 16 & 0.160 & 40 & 0.102 & 72.34$_{-1.25}^{+1.33}$ & $-$18.35$_{-0.03}^{+0.03}$ & $-$2.65$_{-0.02}^{+0.02}$ & 0.096$_{-0.014}^{+0.016}$ &    $...$ \\
$v_{pec}=350$ km s$^{-1}$ & 16 & 0.160 & 40 & 0.102 & 72.35$_{-1.36}^{+1.36}$ & $-$18.35$_{-0.03}^{+0.03}$ & $-$2.65$_{-0.03}^{+0.03}$ & 0.097$_{-0.016}^{+0.018}$ &    0.03  \\
$v_{pec}=150$ km s$^{-1}$ & 16 & 0.160 & 40 & 0.102 & 72.35$_{-1.23}^{+1.26}$ & $-$18.35$_{-0.03}^{+0.03}$ & $-$2.65$_{-0.02}^{+0.02}$ & 0.101$_{-0.013}^{+0.015}$ &    0.01  \\
$v_{pec}=0$ km s$^{-1}$   & 16 & 0.160 & 40 & 0.102 & 72.34$_{-1.33}^{+1.36}$ & $-$18.35$_{-0.04}^{+0.03}$ & $-$2.65$_{-0.02}^{+0.02}$ & 0.113$_{-0.012}^{+0.014}$ &    0.01  \\
$z_{cmb}$                 & 16 & 0.160 & 40 & 0.109 & 71.42$_{-1.30}^{+1.32}$ & $-$18.35$_{-0.03}^{+0.03}$ & $-$2.62$_{-0.02}^{+0.02}$ & 0.100$_{-0.015}^{+0.017}$ & $-$1.27  \\
$z_{corr}$ from NED+C15   & 16 & 0.160 & 40 & 0.128 & 71.98$_{-1.30}^{+1.32}$ & $-$18.35$_{-0.03}^{+0.03}$ & $-$2.63$_{-0.02}^{+0.02}$ & 0.102$_{-0.016}^{+0.018}$ & $-$0.50  \\
$z_{cmb}$ from NED        & 16 & 0.160 & 40 & 0.111 & 71.44$_{-1.29}^{+1.30}$ & $-$18.35$_{-0.03}^{+0.03}$ & $-$2.62$_{-0.02}^{+0.02}$ & 0.102$_{-0.015}^{+0.017}$ & $-$1.24  \\
$z_{corr}$ from REF+C15   & 16 & 0.160 & 40 & 0.130 & 71.77$_{-1.32}^{+1.36}$ & $-$18.35$_{-0.03}^{+0.03}$ & $-$2.63$_{-0.02}^{+0.02}$ & 0.104$_{-0.016}^{+0.018}$ & $-$0.78  \\
$z_{cmb}$ from REF        & 16 & 0.160 & 40 & 0.112 & 71.21$_{-1.31}^{+1.33}$ & $-$18.35$_{-0.03}^{+0.03}$ & $-$2.61$_{-0.02}^{+0.02}$ & 0.102$_{-0.015}^{+0.017}$ & $-$1.56  \\
$E(B-V)<0.3$              & 15 & 0.161 & 32 & 0.105 & 71.88$_{-1.35}^{+1.46}$ & $-$18.36$_{-0.03}^{+0.03}$ & $-$2.65$_{-0.03}^{+0.03}$ & 0.100$_{-0.016}^{+0.018}$ & $-$0.63  \\
HF $z>0.023$              & 16 & 0.160 & 18 & 0.084 & 72.11$_{-1.56}^{+1.60}$ & $-$18.35$_{-0.03}^{+0.03}$ & $-$2.64$_{-0.03}^{+0.03}$ & 0.110$_{-0.018}^{+0.022}$ & $-$0.31  \\
HF $0.023<z<0.05$         & 16 & 0.160 & 15 & 0.091 & 72.22$_{-1.72}^{+1.77}$ & $-$18.35$_{-0.04}^{+0.04}$ & $-$2.64$_{-0.04}^{+0.04}$ & 0.117$_{-0.020}^{+0.023}$ & $-$0.16  \\
no Chauvenet              & 16 & 0.160 & 44 & 0.170 & 72.26$_{-1.51}^{+1.56}$ & $-$18.35$_{-0.04}^{+0.04}$ & $-$2.64$_{-0.03}^{+0.03}$ & 0.130$_{-0.018}^{+0.020}$ & $-$0.10  \\
only best fits            & 15 & 0.162 & 33 & 0.100 & 72.36$_{-1.33}^{+1.36}$ & $-$18.36$_{-0.03}^{+0.03}$ & $-$2.65$_{-0.02}^{+0.02}$ & 0.095$_{-0.016}^{+0.018}$ &    0.03  \\
all gp                    & 16 & 0.176 & 39 & 0.110 & 72.61$_{-1.37}^{+1.42}$ & $-$18.33$_{-0.03}^{+0.03}$ & $-$2.63$_{-0.02}^{+0.02}$ & 0.109$_{-0.015}^{+0.018}$ &    0.38  \\
$m_X$ from template       & 16 & 0.151 & 41 & 0.144 & 73.29$_{-1.50}^{+1.48}$ & $-$18.39$_{-0.04}^{+0.04}$ & $-$2.72$_{-0.02}^{+0.02}$ & 0.124$_{-0.016}^{+0.018}$ &    1.32  \\
removing subluminous      & 16 & 0.160 & 37 & 0.097 & 72.40$_{-1.27}^{+1.30}$ & $-$18.35$_{-0.03}^{+0.03}$ & $-$2.65$_{-0.02}^{+0.02}$ & 0.097$_{-0.015}^{+0.017}$ &    0.08  \\
only CSP                  &  6 & 0.116 & 20 & 0.075 & 75.73$_{-1.48}^{+1.68}$ & $-$18.27$_{-0.03}^{+0.04}$ & $-$2.66$_{-0.03}^{+0.03}$ & 0.058$_{-0.027}^{+0.026}$ &    4.70  \\
only CfA                  &  3 & 0.111 & 20 & 0.194 & 71.50$_{-3.19}^{+3.19}$ & $-$18.32$_{-0.09}^{+0.09}$ & $-$2.59$_{-0.03}^{+0.03}$ & 0.109$_{-0.058}^{+0.053}$ & $-$1.16  \\
only CSP+CfA              &  8 & 0.128 & 29 & 0.096 & 75.10$_{-1.28}^{+1.41}$ & $-$18.28$_{-0.03}^{+0.03}$ & $-$2.66$_{-0.02}^{+0.02}$ & 0.074$_{-0.032}^{+0.025}$ &   3.83   \\
only spirals in HF        & 16 & 0.160 & 30 & 0.101 & 72.37$_{-1.40}^{+1.41}$ & $-$18.35$_{-0.03}^{+0.03}$ & $-$2.65$_{-0.03}^{+0.03}$ & 0.103$_{-0.016}^{+0.018}$ &   0.04   \\
SHOES selection           & 16 & 0.160 & 14 & 0.091 & 72.32$_{-1.72}^{+1.81}$ & $-$18.35$_{-0.04}^{+0.04}$ & $-$2.64$_{-0.04}^{+0.04}$ & 0.117$_{-0.019}^{+0.024}$ & $-$0.02  \\
\hline
\end{tabular}
\end{table*}

\subsubsection{Best light-curve fits}
In order to construct the purest sample, we tried to exclude the few objects that had the interpolated peak magnitude less well constrained by visual inspection, and that could not have passed more restrictive criteria. These include SN~2008fv and SN~2003du in the calibration sample, and SN~2006hx, SN~2007ai, SN~2007bd, SN~2009bv, SN~2010ag, SN2010ai, PTF10mwb, PTF10tce, iPTF13azs, iPTF13dge, iPTF13duj, and iPTF14atg  in the Hubble-flow sample. With these remaining objects, $H_0$ is shifted down for about 0.7\% in the $J$ band and unchanged in the $H$ band.

\subsubsection{GP interpolation and spline}.  
Another test consisted of using all SN~Ia NIR peak magnitudes as obtained from the GP interpolation instead of the best between GP and spline fits. While this may be a more consistent and  systematic method, we decided to choose the best fit in our baseline analysis based on the reduced $\chi^2$. As expected, this choice affects $\sigma_{\rm cal}$ by increasing the scatter up to 0.02 mag of $J$ and $H$-band calibrators, and in turn increase the value of $H_0$ by 0.4-0.8\% in both bands.

\subsubsection{Light-curve templates}
Going one step further, we repeated the analysis but using this time the peak magnitudes obtained from the SNooPy template fitting using the {\it max\_model}. While in the previous test, only those magnitudes that were obtained by the spline interpolation were modified from the baseline analysis, in this case, all magnitudes are different. We find an increased dispersion in $\sigma_{\rm cal}$ in the $J$ band up to 0.17 mag, but a decrease in $H$ to 0.151 mag. Also, a similar $\sigma_{\rm HF}$ in the $J$-band, but an increase in the $H$ band to 0.144 mag. Regarding $H_0$, the $J$-band value is higher by 2.3\% to 73.9 km~s$^{-1}$~Mpc$^{-1}$, and the $H$-band value increased by 1.3\% to 73.3 km~s$^{-1}$~Mpc$^{-1}$.

\subsubsection{Exclude subluminous SNe Ia}
We also tested applying a cut in $\Delta$m$_{15}$ at 1.6 mag, so excluding those objects that present wider light-curves and at at the fainter end of the luminosity-width relation. This is based on previous works who have shown that, similarly to the behaviour in optical bands, NIR absolute magnitudes of fast-declining SNe Ia diverge considerably from their more normal counterparts \citep{2009AJ....138.1584K,2012PASP..124..114K,2017A&A...602A.118D}. This cut affected three objects, SN~2007ba, SN~2010Y, and iPTF13ebh in the Hubble-flow sample. As expected, the $\sigma_{\rm HF}$ is reduced down to 0.144 in $J$ and to 0.097 in $H$. This pulled up the $H_0$ value by 1.2\% in $J$, and 0.1\% in $H$.

\subsubsection{Survey}
The following three tests have to do with restricting the samples by their original source survey. We focused on the CSP and CfA for three reasons:
(i) they contribute to more than half of the total sample;
(ii) their data is the best and systematically well-calibrated and includes both Hubble flow and calibrator objects (which will cancel calibration errors); and (iii) we expect more data from well-calibrated surveys to come in the near-future.
First, we used only SNe Ia observed by the CSP, so their photometry was not S-corrected because they were already in our reference photometric system. This includes SNe from references 6, 10, 11, and 13 in Tables \ref{tab:calibrators} and \ref{tab:hubbleflow}. Then, we repeated the analysis using only those SNe Ia observed by the CfA SN programme (Reference 4 in Tables \ref{tab:calibrators} and \ref{tab:hubbleflow}). Finally, we combined data from these two surveys, discarding observations collected from other sources that have been reduced differently or less systematically than in those two surveys.

When considering each survey independently sample sizes are reduced, especially for the calibrator sample. There are eight calibrator SNe Ia observed by the CSP\footnote{Here we include SN~2012fr, SN~2012ht, and SN~2015F which were observed during the second stage of the CSP (CSP-II), so reduced following the same procedure as for all CSP-I objects, but published in individual papers by the CSP collaboration.} in the $J$ band and six in the $H$ band. For CfA the corresponding numbers are four and three. For the Hubble-flow sample the reduction is not so significant with 29 for the CSP and 22 for CfA of the 52 SNe in the $J$ band, and with 20 from the 40 SNe in $H$ for both surveys. However, when the surveys are combined the numbers increase to 11 and 40 for the calibrator and Hubble flow samples in the $J$ band, and 8 and 29 for the $H$ band, which confirms most of the sample comes from these two surveys (50 to 80\%).

Interestingly, in most cases the scatter of the samples is lower with respect to baseline, with the exception of $J$-band CSP calibrators and the $H$-band CfA objects in the Hubble-flow, confirming the expectation that the more homogeneous data reduces errors and other historic data is likely to be driving the scatter up.
Particularly noteworthy is the scatter of the calibrator sample in the $H$ band where it is reduced from the baseline 0.160 to 0.128 mag, more in line with the Hubble flow.

$H_0$ values increase in all cases, the highest value being 3.4 km~s$^{-1}$~Mpc$^{-1}$ larger, around a 4.7\% from the baseline value, in the case of using only CSP SNe Ia and the $H$ band. The main reason of the higher value of $H_0$ comes is that the mean absolute magnitudes of these samples are 0.1 fainter ($-$18.48 mag in $J$ and $-$18.27 mag in $H$) compared to baseline. We attribute these differences among subsamples to past inhomogeneities of the NIR systems, filters, zero points, etc. As a way around this problem, we highlight that, in the future, better calibration in the NIR will be important to improve upon these constraints.

\subsubsection{Host galaxy type}
Another test consisted in considering only SNe Ia that occurred in spiral galaxies, by excluding those in E and S0 host galaxies as classified in NED. This was driven by the fact that all calibrator galaxies selected by SHOES were star-forming, in order to be able to measure Cepheids stars. In this way, the resulting SNe Ia in the calibrator and Hubble flow samples were hosted by a similar type of galaxies. For those galaxies with no morphological classification in NED, we searched for host galaxy images in PanStarrs and confirmed they were all blue extended objects with structure, so we classified them all as {\it Spiral} and included them in this test. All morphological classification can be found in Table \ref{tab:snoopyres}.

In the $H$ band we find most of the parameters unchanged, with an increase of less than 0.1\% in $H_0$ and an increase of $\sigma_{int}$ to 0.103 mag. In the $J$ band, $H_0$ is increased by 1.5\% from the baseline to 73.43 km~s$^{-1}$~Mpc$^{-1}$, and the $\sigma_{int}$ is reduced to 0.115 mag.

\subsubsection{SHOES selection}
Finally, we tried to mimic as much as possible the cuts and selection done by SHOES, which consisted in keeping only SNe Ia that occurred in star-forming galaxies and increase the redshift cut of the Hubble-flow sample to $z=0.023$. In this way, the resulting numbers would be the most directly comparable to R22. The $H_0$ values of this variation are 74.02$\pm$1.7 and 72.32$\pm$1.8 km~s$^{-1}$~Mpc$^{-1}$ for $J$ and $H$, respectively. Both are fully consistent to R22 $H_0$ value.

\begin{figure}
    \includegraphics[width=\columnwidth]{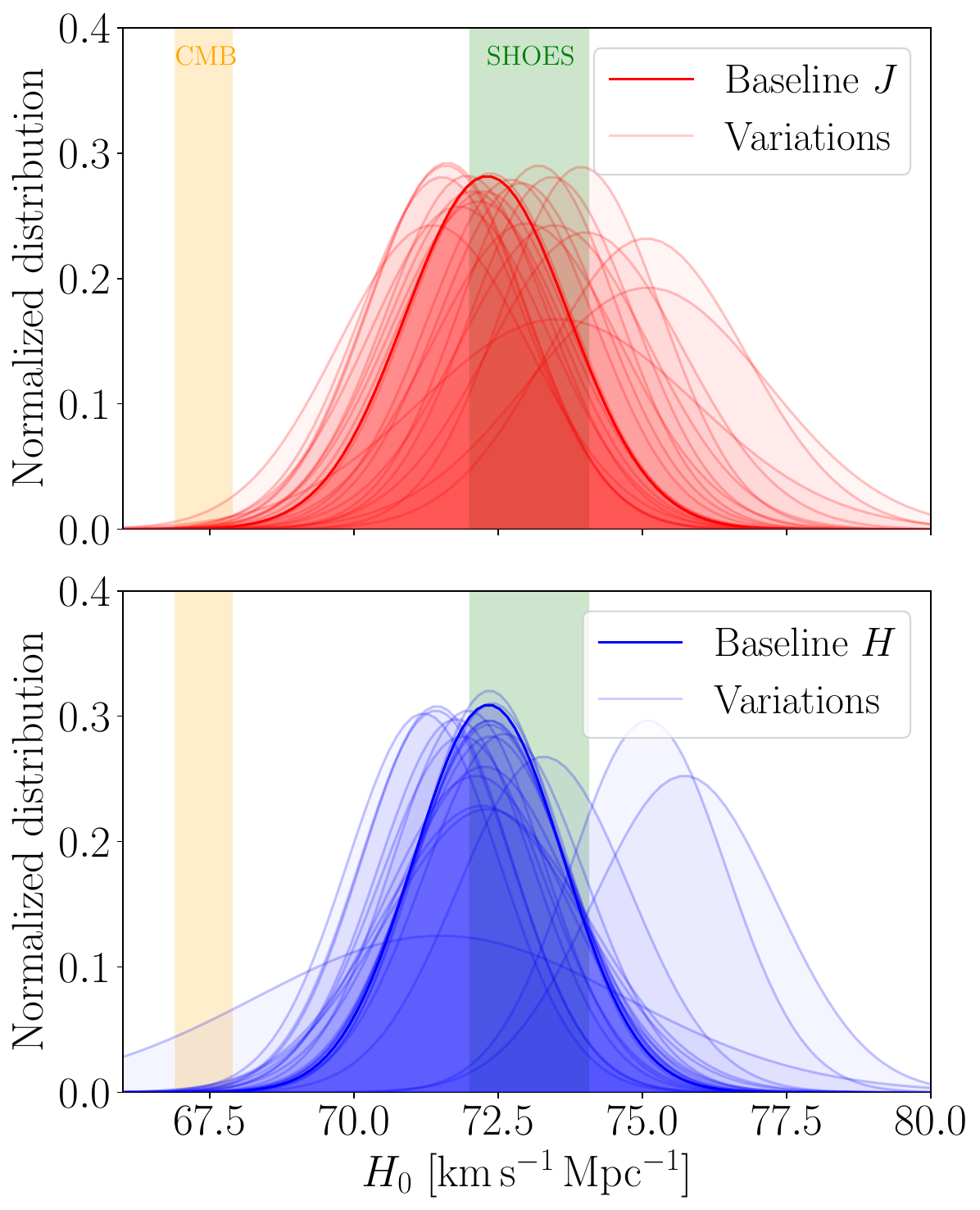} 
    \caption{Probability gaussian densities of our baseline analysis (solid) and all the 21 variations performed in section \ref{sec:vars}. The two vertical strips correspond to the 1$\sigma$ uncertainties around the best $H_0$ value from the \cite{2020A&A...641A...6P} and the SHOES project (R22). Results with the $J$ band in the upper panel and with the $H$ band in the bottom panel.}
    \label{fig:variations}
\end{figure}

\begin{table*}[!ht]
\centering
\caption{Results of testing NIR standardization.}
\label{tab:h0fits2}
\begin{tabular}{lcccccc}
\hline
Variation            & $H_0$                   & $M_X$                      & $-5 a_X$                  & $\sigma_{\rm int}$        & $\alpha$ & $\beta$ \\
                     & (km~s$^{-1}$~Mpc$^{-1}$)& (mag)                      & (mag)                     & (mag)                     &  &  \\
\hline
 \multicolumn{7}{c}{$J$-band}\\
\hline
{\bf Baseline}       & 72.31$_{-1.42}^{+1.42}$ & $-$18.58$_{-0.04}^{+0.04}$ & $-$2.87$_{-0.02}^{+0.02}$ & 0.125$_{-0.015}^{+0.017}$ &  $\cdots$                  & $\cdots$ \\
Stretch corrected    & 72.39$_{-1.27}^{+1.29}$ & $-$18.61$_{-0.03}^{+0.03}$ & $-$2.90$_{-0.02}^{+0.02}$ & 0.105$_{-0.014}^{+0.015}$ &  0.354$_{-0.086}^{+0.086}$ & $\cdots$ \\
Reddening corrected & 72.37$_{-1.31}^{+1.33}$ & $-$18.61$_{-0.04}^{+0.04}$ & $-$2.91$_{-0.02}^{+0.02}$ & 0.112$_{-0.015}^{+0.017}$ &  $\cdots$                  & 0.338$_{-0.115}^{+0.113}$ \\
Both corrections     & 72.36$_{-1.19}^{+1.20}$ & $-$18.64$_{-0.03}^{+0.03}$ & $-$2.94$_{-0.02}^{+0.02}$ & 0.095$_{-0.013}^{+0.015}$ & 0.328$_{-0.081}^{+0.081}$ & 0.297$_{-0.106}^{+0.105}$ \\
\hline
 \multicolumn{7}{c}{$H$-band}\\
\hline
{\bf Baseline}       & 72.34$_{-1.25}^{+1.33}$ & $-$18.35$_{-0.03}^{+0.03}$ & $-$2.65$_{-0.02}^{+0.02}$ & 0.096$_{-0.014}^{+0.016}$ & $\cdots$                  & $\cdots$ \\
Stretch corrected    & 72.22$_{-1.22}^{+1.26}$ & $-$18.36$_{-0.03}^{+0.03}$ & $-$2.66$_{-0.02}^{+0.02}$ & 0.094$_{-0.015}^{+0.017}$ & 0.133$_{-0.097}^{+0.094}$ & $\cdots$ \\
Reddening corrected & 72.44$_{-1.25}^{+1.31}$ & $-$18.36$_{-0.03}^{+0.03}$ & $-$2.66$_{-0.03}^{+0.03}$ & 0.094$_{-0.015}^{+0.017}$ & $\cdots$                  & 0.108$_{-0.121}^{+0.116}$ \\
Both corrections     & 72.32$_{-1.25}^{+1.30}$ & $-$18.37$_{-0.03}^{+0.04}$ & $-$2.67$_{-0.03}^{+0.03}$ & 0.092$_{-0.015}^{+0.017}$ & 0.137$_{-0.092}^{+0.095}$ & 0.115$_{-0.119}^{+0.118}$ \\
\hline
\end{tabular}
\end{table*}

\subsubsection{Summary of all variations}
Summarizing, the largest difference of our 21 variations with respect to baseline analysis is when we used only objects observed by the two main surveys, CSP and CfA, obtaining an $H_0$ of up to 75.7 km~s$^{-1}$~Mpc$^{-1}$, a 4.7\% higher. However, as mentioned above this may be due to the small size of the resulting calibration samples and because those few objects are on average fainter than the full sample. Also, most of the objects followed up by these projects come from targeted searches, which may be biasing SN and host galaxy properties. Future work using SNe Ia from unbiased searches may be able to quantify how important is this bias in this regard. Besides these, the variations that provided the largest difference in $H_0$ were: mimicking the SHOES selection, and using peak magnitudes from template fit. In the $J$ band, the largest change was the SHOES selection increasing the $H_0$ value an 2.4\% up to 74.02 km~s$^{-1}$~Mpc$^{-1}$, which highlights the effect that SNe at redshifts in between the Hubble flow cuts ($0.01<z<0.023$) may be introducing. The largest change in $H$ is when using $z_{cmb}$ redshifts from the reference papers without peculiar velocity corrections, reducing the $H_0$ value in a 1.6\% to 71.2 km~s$^{-1}$~Mpc$^{-1}$. In the $J$ band $H_0$ is reduced a 1\% to 71.6 km~s$^{-1}$~Mpc$^{-1}$. In general, we see that peculiar velocity corrections increase the v$H_0$ value by about 0.5 km~s$^{-1}$~Mpc$^{-1}$ independently of which $z$ we are using. The second largest change in $J$ and $H$ was when using the peak magnitudes from template fitting, increasing the $H_0$ value a 2.3\% up to 73.9 in $J$ and by 1.3\% up to 73.3 km~s$^{-1}$~Mpc$^{-1}$ in the $H$ band. In this case, the difference is highlighting how important are the assumptions of the template fitting when combining optical and NIR data. The larger amount of optical data has more weight in determining light-curve parameters and colors, and may be leaving not enough leverage for the NIR light-curve shapes and peak magnitudes to match well.

All 21 $H_0$ measurements from the aforementioned analysis variants are consistent with our baseline result. They are all plotted in Figure \ref{fig:variations} as individual Gaussian distributions, together with the baseline analysis result and the 1$\sigma$ vertical strips of the Planck and SHOES $H_0$ measurements. The median and standard deviation of all the variants is 72.72$\pm$1.08 km s$^{-1}$ Mpc$^{-1}$ in $J$ and 72.32$\pm$1.08 km s$^{-1}$ Mpc$^{-1}$ in $H$, which corresponds to only 0.41 and 0.02 km s$^{-1}$ Mpc$^{-1}$ different than our fiducial value (29\% in $J$ and 2\% $H$ of the statistical uncertainty). 

\subsection{Systematic uncertainties}

To estimate our systematic uncertainties, we consider four different terms, and add them all in quadrature. First, following the conservative approach of \cite{2019ApJ...876...85R}, our {\it internal} systematic uncertainty is calculated as the standard deviation of our variants. From the 21 variants presented in Table \ref{tab:h0fits}, we obtain a systematic uncertainty of 1.08 km s$^{-1}$ Mpc$^{-1}$ (1.4\%) in both bands. Second, we consider the systematic distance scale error as the mean of the three SH0ES Cepheid anchors in R22 (see their Table 7). This amounts up to 0.7\% of the $H_0$ value, thus 0.51 km s$^{-1}$ Mpc$^{-1}$ also for both bands. Third, we include a photometric zeropoint systematic error between the calibrator and the Hubble flow sample, given the fact that while the Hubble flow sample comes mostly from the CSP and CfA, the calibrator sample is only 11 out of 19 (in $J$) and 8 of 16 ($H$) from these surveys. We consider there can be a 1$\sigma$ ($\sim$0.04 mag) zeropoint difference in the NIR between the well-calibrated systematics of CfA and CSP with respect to literature SNe Ia, but because almost half of the calibrator sample is CfA or CSP (compared to about most of the Hubble Flow) this error reduces to 0.02 mag or 0.7 km s$^{-1}$ Mpc$^{-1}$. Finally, fourth, we budget for additional peculiar velocity uncertainties correlated on larger scales. We calculate the linear power spectrum using CLASS \citep{2011JCAP...07..034B} and cosmological parameters from \cite{2020A&A...641A...6P}. Using the formulae of \cite{2011ApJ...741...67D} we evaluate the correlations of our sample, and find that for our sample geometry and weighting of Hubble flow supernovae, that the additional cosmic variance expected is 1.9\%. Given the results of \cite{2022ApJ...935...83K}, we expect that the use of peculiar velocity corrections based on \cite{2015MNRAS.450..317C} will reduce this systematic by a factor of four, so we consider a 0.5\%, which translates into 0.4 km s$^{-1}$ Mpc$^{-1}$.

Adding these four terms in quadrature we get a systematic error of 1.44 km s$^{-1}$ Mpc$^{-1}$. Including both statistical and systematic uncertainties, our final $H_0$ value is 72.31$\pm$1.42 (stat) $\pm$1.44 (sys) km s$^{-1}$ Mpc$^{-1}$ in $J$, 72.34$_{-1.33}^{+1.25}$ (stat) $\pm$1.44 (sys) km s$^{-1}$ Mpc$^{-1}$ in $H$, or if reported as a single uncertainty, 72.31$\pm$2.02 km s$^{-1}$ Mpc$^{-1}$ on $J$, and 72.34$_{-1.96}^{+1.91}$ km s$^{-1}$ Mpc$^{-1}$ in $H$, representing a 2.8$-$2.7\% uncertainty. Compared to D18, who obtained a statistical uncertainty of 1.6 and a systematic error of 2.7 km s$^{-1}$ Mpc$^{-1}$ in the $J$ band, the measurement found here entails a reduction of 0.2 and 0.7 km s$^{-1}$ Mpc$^{-1}$, respectively, in the $H_0$ uncertainty. This is the most precise $H_0$ value obtained from SNe Ia only with NIR data. Taking into account both sources of uncertainty, our value differs by 2.3-2.4$\sigma$ from the high-redshift result \cite{2020A&A...641A...6P} and by only 0.3$\sigma$ from the local measurement (R22). 

\begin{figure}
    \includegraphics[width=\columnwidth]{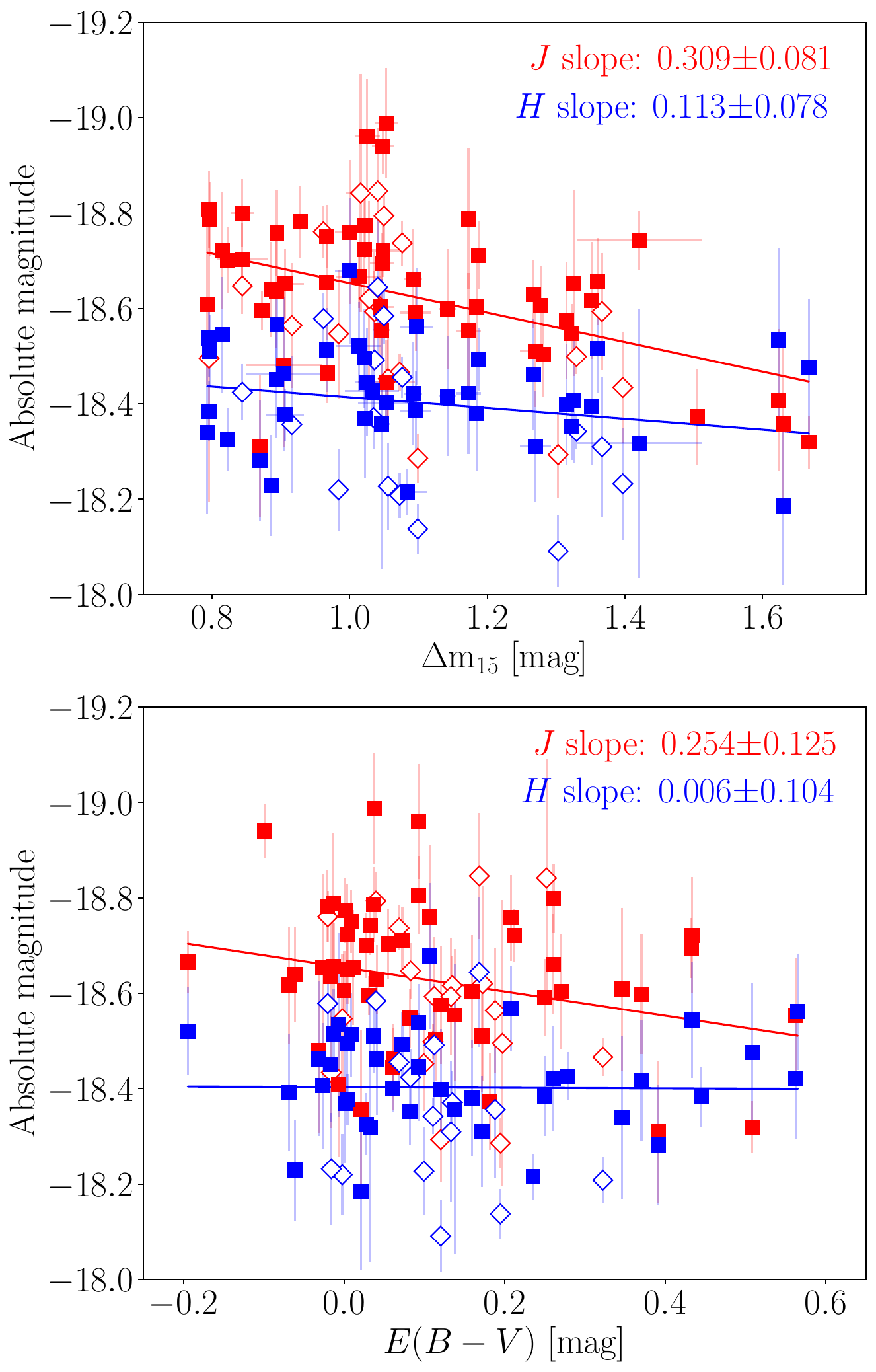} 
    \caption{Dependencies of Hubble residuals on SN Ia light-curve parameters. Open symbols correspond to SNe Ia in the calibrator sample, while closed symbols are for those in the Hubble flow sample.}
    \label{fig:hrlc}
\end{figure}

\subsection{NIR standardization}\label{sec:stand}

Our determination of the NIR ($J$ and $H$) peak magnitude by using GP or spline interpolation, and the subsequent determination of distance, has not been corrected by light-curve and color relations in contrast to what it is common practice when dealing with optical data. To explore whether these corrections may be useful in the NIR to reduce the dispersion, we show in Figure \ref{fig:hrlc} relations between the absolute magnitudes of our sample and light-curve parameters $\Delta$m$_{15}$ and $E(B-V)$, all together for the $J$ and $H$ band. A similar plot with the $s_{BV}$ color-stretch parameter is included in Appendix \ref{app:sbv}.

We note that SN~Ia with higher values of $\Delta$m$_{15}$ tend to be fainter in both bands. A linear regression to all values, calibrators and Hubble-flow SNe Ia, reveals that a 3.8$\sigma$ slope ($0.309\pm0.081$) exists in the absolute magnitude vs. $\Delta$m$_{15}$ relation in the $J$ band, although the significance is reduced to 1.5$\sigma$ (0.113$\pm$0.078) in the $H$ band. For $E(B-V)$ we find a 2$\sigma$ slope (0.254$\pm$0.125) for the $J$ band, and an inexistent relation (0.006$\pm$0.104) for the $H$ band. These two results would be in agreement with these relations being of less importance as redder bands are considered, up to the point where the extinction correction is not needed in the $H$ band.

Applying the stretch correction to the initial absolute magnitude of our calibrators reduces the scatter from 0.160 mag to 0.149 in $J$ and 0.154 mag in $H$. Similarly, applying the reddening  correction, the scatter is mostly unaltered from 0.160 to 0.159 mag in $J$ and 0.160 mag in $H$. Regarding the Hubble flow sample, the stretch correction reduces the scatter from 0.149 to 0.132 mag in $J$, and leaves it at 0.102 mag in $H$. With the reddening correction the reduction of the scatter in $J$ is not so pronounced as with the stretch correction going from 0.149 to 0.142 mag, and it is also unaltered in $H$ at 0.102 mag.

\subsubsection{Applying $\Delta$m$_{15}$ and $E(B-V)$ corrections}
The main assumption in our analysis is that SNe Ia are {\it natural} standard candles in the NIR. However, the relations found in the previous section suggest that SNe Ia  NIR absolute magnitudes can still be corrected using the typical stretch and reddening relations to improve their standardization, especially in the $J$ band. 

First, we explored the inclusion of a stretch correction to the SNe Ia peak magnitudes by adding a term,
\begin{equation}
M_{i}^{corr} = M_{i} - \alpha\times (\Delta m_{15,i} -1),
\end{equation}
in the likelihood presented in equation \ref{eq:likelihood}, where $\alpha$ corresponds to the relation found above between absolute magnitude and stretch, and $\Delta m_{15}$ is found using the $max\_model$ SNooPy fit. We repeated the same analysis with these stretch-corrected magnitudes, using the $\alpha$ values found above as priors in the optimization, and the results are summarized in Table \ref{tab:h0fits2}. The stretch correction is mostly turning the peak magnitudes a bit brighter as it can be seen in the $M_X$ and $-5a_X$ parameters being 0.01-0.03 mag brighter in the two bands. $H_0$ increases by 0.08 km s$^{-1}$ Mpc$^{-1}$ in $J$ and decreases by 0.12 km s$^{-1}$ Mpc$^{-1}$ in $H$, with lower statistical errors in both cases. Interestingly, the intrinsic dispersion is reduced from 0.125 to 0.105 mag in $J$ and from 0.096 to 0.094 mag in $H$. These values are consistent with other previously reported in the literature \citep{2012MNRAS.425.1007B}, and confirms that dispersion is reduced for redder bands. Finally, the best value of the $\alpha$ parameter is 0.354$\pm$0.086 in $J$ and 0.133$^{0.094}_{0.097}$ in $H$, in agreement with the values found in the previous section.

Second, we added instead a reddening correction term to the likelihood,
\begin{equation}
M_{i}^{corr} = M_{i} - \beta\times E(B-V)_i,
\end{equation}
where $\beta$ would correspond to the relation between absolute magnitude and reddening, and $E(B-V)$ is the $EBVhost$ parameter from the $EBV\_model2$ SNooPy fit. The results of this analysis, using as a prior for $\beta$ the value of the relation found above, are also included in Table \ref{tab:h0fits2}. In general, the reddening correction is also turning the peak magnitudes slightly brighter, 0.01-0.03 mag for $M_X$ and $-5a_X$ in both bands, with the exception of just a few objects with negative $EBVhost$ parameter. In turn, the $H_0$ value also increases by 0.06 km s$^{-1}$ Mpc$^{-1}$ with respect to the baseline, and the intrinsic dispersion is reduced from 0.125 to 0.112 mag in $J$. For the $H$ band, the increase in $H_0$ is of 0.10 km s$^{-1}$ Mpc$^{-1}$, and the reduction of the intrinsic dispersion is 0.002 mag, similarly to the stretch correction, to 0.094 mag. In this case the $\beta$ parameter is 0.338$^{+0.113}_{-0.115}$ in $J$ and 0.108$^{+0.116]}_{-0.121}$ in $H$, also consistent with the relation presented in Figure \ref{fig:hrlc}.

Finally, we repeated the analysis adding the two corrections at the same time and minimizing $\alpha$ and $\beta$ simultaneously. Results are also in Table \ref{tab:h0fits2}. In this case, $M_X$ and $-5a_X$ are even brighter with changes up to 0.07 mag in $J$ and 0.02 mag in $H$, while the change is $H_0$ is of only 0.06 and 0.02 km s$^{-1}$ Mpc$^{-1}$ in $J$ and $H$, respectively. The most significant improvement is on the intrinsic scatter, which is reduced to 0.095 mag in $J$ and 0.092 mag in $H$, the lowest of the four analyses. The two nuisance parameters $\alpha$ and $\beta$ take values that are similar to those found when accounted for those corrections separately. 

This test demonstrates that the stretch correction is still needed in the NIR, at least in the $J$ ($\sim 4\sigma$) and $H$ ($\sim 1.5\sigma$) bands, although their importance becomes lower for redder bands. In addition, the reddening correction is still significant in the $J$ band ($\sim 3\sigma$), and starts to be insignificant in the $H$ band ($<1\sigma$), although the intrinsic scatter is still reduced by 0.002 mag when included. It is important to note that, even if these corrections have little effect, they have the virtue of correcting to first order for demographic differences between the calibrator and Hubble flow samples.


\section{Summary and conclusions}

In this work, we present an updated measurement of the Hubble constant $H_0$ using a compilation of published SNe Ia observations in the NIR. All SN in our sample were observed before their maximum, so to estimate their peak magnitude we performed Gaussian process and spline interpolations. Combining SNe Ia in nearby galaxies whose distance has already been determined by the SHOES team using the Cepheid period-luminosity relation, with SNe Ia at further distances, we obtain an $H_0$ of 72.31$\pm$1.42 km s$^{-1}$ Mpc$^{-1}$ in the $J$ band and 72.34$^{+1.33}_{-1.25}$ km s$^{-1}$ Mpc$^{-1}$ in the $H$ band, where all uncertainties are statistical.

\begin{figure}
    \includegraphics[width=\columnwidth]{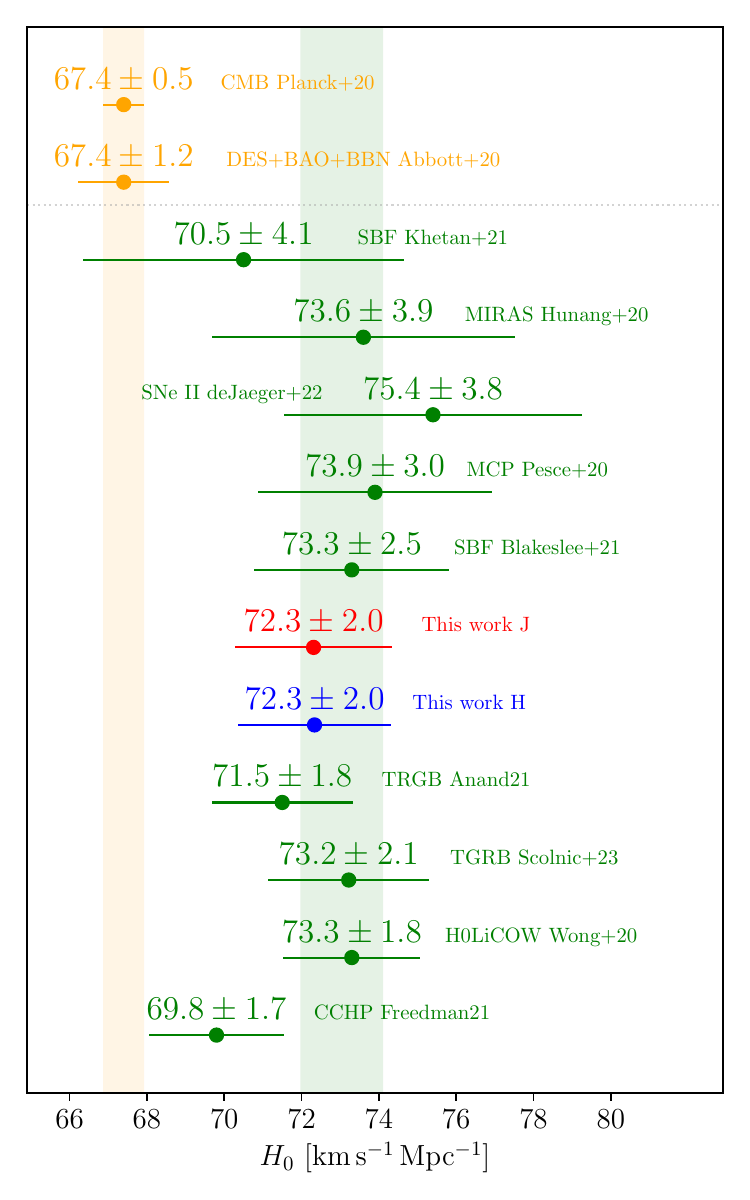} 
    \caption{Summary plot of the latest measurements of $H_0$ using several different techniques from the early (in orange) and late (in green) Universe. Vertical colored strips represent the reference early Universe value by the Planck satellite, and the late Universe value from SHOES. 
    Late Universe measurements are sorted by the size of their uncertainty from top to bottom. Our measurements are included in red (for $J$) and blue (for $H$).}
    \label{fig:hubbletension}
\end{figure}

We have performed up to 21 variations to our baseline analysis to estimate systematic uncertainties, and obtained all consistent with our baseline analysis. The median $H_0$ value and dispersion of the 21 variations is 72.72$\pm$1.08 km s$^{-1}$ Mpc$^{-1}$ in $J$ and 72.32$\pm$1.08 km s$^{-1}$ Mpc$^{-1}$ in $H$, which only differs by less than 0.4 km s$^{-1}$ Mpc$^{-1}$ from the baseline. The largest differences in $H_0$ of these variations with the value from the baseline analysis come from using data from a single survey, varying $z_{cmb}$ values (directly from publications or from large databases), applying peculiar velocity corrections, or when using template fitting instead of direct interpolation, with differences in $H_0$ of up to 4.7\%. 

Taking into account up to four sources of systematic uncertainty added in quadrature, namely the dispersion of the 21 variations, the distance scale error of the three SH0ES Cepheid anchors in R22, a photometric zeropoint error between the calibrator and the Hubble flow sample, and additional peculiar velocity uncertainties correlated on larger scales, our final result of $H_0$ is 72.31$\pm$2.02 km s$^{-1}$ Mpc$^{-1}$ on $J$ (2.8\% uncertainty), and 72.34$_{-1.96}^{+1.91}$ km s$^{-1}$ Mpc$^{-1}$ in $H$ (2.7\% uncertainty), both below the 3\% precision.

Our measurement is in agreement with R22 at 0.3$\sigma$, which used the same Cepheid-based distances but used optical SN Ia data in the third rung of the distance scale, and disagrees with the \cite{2020A&A...641A...6P} value at 2.3-2.4$\sigma$. This independent analysis confirms both that SNe Ia in the optical do not introduce any bias in the $H_0$ measurement due to systematic uncertainties introduced in their standardization, and that SNe Ia in the NIR are a powerful tool for cosmological analysis.

Figure \ref{fig:hubbletension} shows the reference $H_0$ measurements from the \cite{2020A&A...641A...6P} and R22 as vertical strips, together with our results with SNe Ia in the NIR with the $J$ and $H$ bands, and a summary of other recent independent measurements obtained with the Tip of the Red Giant Branch (TRGB; \citealt{2021ApJ...919...16F,2021arXiv210800007A,2023arXiv230406693S}), type II supernovae (SNe II; \citealt{2022MNRAS.514.4620D}), Surface Brightness Fluctuations (SBF; \citealt{2021ApJ...911...65B}), MIRAS \citep{2020ApJ...889....5H}, strong lenses (HOLiCOW;  \citealt{2020MNRAS.498.1420W}), the Megamaser Cosmology Project (MCP;  \citealt{2020ApJ...891L...1P}), and by the Dark Energy Survey combining clustering and weak lensing data with baryon acoustic oscillations and Big Bang nucleosynthesis  \citep{2018MNRAS.480.3879A}. It can clearly be seen how the precision of our measurement is competitive compared to other probes.

We explored the standardization of SN~Ia NIR absolute magnitudes by including a term that accounts for the light-curve width and color excess (obtained from optical+NIR SNooPy fits). All tests point into the direction of reducing the uncertainty in $H_0$, the dispersion of the absolute magnitudes and the intrinsic scatter when performing the $H_0$ minimization. The nuisance parameters $\alpha$ and $\beta$ that describe the relations between absolute magnitudes and light-curve parameters are much smaller but still significant (specially for $J$) compared to optical bands. Even if these corrections have little effect, they have the virtue of correcting to first order for demographic differences between the calibrator and Hubble flow samples.

Based on our results, in order to improve the precision in $H_0$ we will need to: 
(i) increase the number of calibrators, which translates into obtaining high quality NIR data of SNe Ia occurring in very nearby galaxies for which any of these independent techniques can be used to determine its distance. In particular, the James Webb Space Telescope will naturally take over from the effort made by the HST in obtaining NIR Cepheid imaging of more in number and also further galaxies to increase independent Cepheid-based distances of SN Ia hosts; 
(ii) increase the number of well-observed Hubble-flow SNe Ia. In this regard, the future Nancy Roman Space Telescope, with its wide field of view and the F213 filter, will provide a large number of NIR light-curves of SNe Ia at higher redshift, allowing a full $JH$ rest-frame NIR Hubble-Lema\^itre diagram of SN Ia up to redshifts of 0.7; and 
(iii) further study the NIR standardization of SNe Ia light-curves in order to reduce the scatter of their absolute peak magnitudes and therefore their distance estimation. As suggested by our test in section \ref{sec:stand}, some improvement is possible by taking into account light-curve parameters, but definitely more work is needed to determine this more reliably;
and (iv) other improvements include, to name a few, better NIR spectral templates to obtain more precise K-corrections \citep{2019PASP..131a4002H,2020hst..prop16234J}, more standard filter transmissions in the NIR to improve the accuracy of S-corrections to data obtained from different instruments, and consider variable or individual reddening law affecting each SN individually \citep{2021MNRAS.508.4656G}. 

\begin{acknowledgements}
We acknowledge the referee for useful comments that have improved the first version of this manuscript.
L.G. acknowledges financial support from the Spanish Ministerio de Ciencia e Innovaci\'on (MCIN), the Agencia Estatal de Investigaci\'on (AEI) 10.13039/501100011033, and the European Social Fund (ESF) "Investing in your future" under the 2019 Ram\'on y Cajal program RYC2019-027683-I and the PID2020-115253GA-I00 HOSTFLOWS project, from Centro Superior de Investigaciones Cient\'ificas (CSIC) under the PIE project 20215AT016, and the program Unidad de Excelencia Mar\'ia de Maeztu CEX2020-001058-M.
KM is funded by the EU H2020 ERC grant no. 758638. M.D.S. acknowledges funding from the Independent Research Fund Denmark  (IRFD, grant number 10.46540/2032-00022B ).
Ara i aqu\'i \'es que veig, d'entre una grisor espectral, una vaga lluïssor que s'enc\'en amb suavitat (adapted from PR).
\end{acknowledgements}
\bibliographystyle{aa}
\bibliography{h0nir} 

\begin{appendix}

\begin{table*}[h]\footnotesize
\centering
\caption{Summary of telescopes and instruments used to obtain NIR imaging of SNe Ia used in this work.}
\label{tab:infra}
\begin{tabular}{clllcccc} 
\hline
Reference & Observatory & Telescope & Instrument & \multicolumn{2}{c}{$\lambda_{\rm eff}$} & \multicolumn{2}{c}{Effective Width [\AA]} \\ 
  &   &   &   & $J$ & $H$ & $J$ & $H$ \\ 
\hline
 $\left[ 1 \right]$ & Cerro Tololo & 1m YALO & ANDICAM & 12399.85 & 16152.77 & 1517.49 & 2871.15 \\ 
 $\left[ 1,15,16 \right]$   & Las Campanas & 1m Swope & Rockwell & $-$ & $-$ & $-$ & $-$ \\ 
 $\left[ 2,6 \right]$  & Las Campanas & 2.5m du Pont & WIRC & 12368.16 &  16158.61 & 1781.80  & 2398.73 \\
 $\left[ 2 \right]$ & Las Campanas & 6.5m Magellan-Baade & Classic-Cam & $-$ & $-$ & $-$ & $-$ \\ 
 $\left[ 3 \right]$ & Roque Muchachos & 3.5m TNG & Nics & 12564.50 &  16128.27 & 2704.28 &  2852.22 \\
 $\left[ 3 \right]$ & Roque Muchachos & 2.5m NOT & NOTCam & 12390.26  & 16130.87 & 1532.41 &  2847.57 \\
 $\left[ 4,9 \right]$ & Fred L. Whipple & 1.3m PAIRITEL & 2MASS & 12350.00  & 16620.00 & 1624.32  & 2509.40\\ 
 $\left[ 5,14 \right]$ & Cerro Tololo & 1.3m CTIO & ANDICAM & 12399.85 &  16152.77 & 1517.49  & 2871.15\\
 $\left[ 6 \right]$    & Las Campanas & 1m Swope     & RetroCam  (up to 2009) & 12443.03 &  16172.77 & 1676.83  & 2448.11 \\ 
                       &              &              & RetroCam  (since 2009) & 12372.44 &  16172.77 & 1734.26  & 2448.11 \\ 
 $\left[ 7 \right]$ & Campo Imperatore & 1.1m AZT-24 & SWIRCAM  & $-$ & $-$ & $-$ & $-$ \\ 
 $\left[ 8 \right]$ & Kitt Peak & 3.5m WIYN & WHIRC & 12439.66 &  16498.66 & 1597.91 &  2820.70 \\ 
 $\left[ 10 \right]$   & Las Campanas & 2.5m du Pont & RetroCam & 12380.83  & 16136.70 & 1729.40  & 2581.01\\ 
 $\left[ 10,11 \right]$ & Las Campanas & 6.5m Magellan-Baade & FourStar & 12287.26  &  16039.55 & 2214.62  & 2769.45\\
 $\left[ 12,19 \right]$ & San Pedro Martir & 1.5m Johnson & RATIR \\ 
 $\left[ 13 \right]$ & La Silla & 3.5m NTT & SOFI & 12427.53  & 16365.26 & 1521.59  & 2878.91 \\ 
 $\left[ 15,16 \right]$   & Las Campanas & 2.5m du Pont & CIRSI & $-$ & $-$ & $-$ & $-$ \\ 
 $\left[ 16 \right]$ & Fred L. Whipple & 1.2m FLWO Telescope & (not reported) & $-$ & $-$ & $-$ & $-$ \\
 $\left[ 16 \right]$ & Kitt Peak & 2.3m Steward Observatory & HgCdTe NICMOS & $-$ & $-$ & $-$ & $-$ \\ 
 $\left[ 17,18,19 \right]$ & Paranal & 8.1m VLT &  HAWK-I & 12521.99  & 16051.69 & 1524.24  & 2861.22 \\
 $\left[ 17,18 \right]$ & Mauna Kea & 8.2m Gemini North &  NIRI & 12457.75  & 16327.42 & 1576.49  & 2811.51 \\
 $\left[ 19 \right]$ & Paranal & 4.1m VISTA & VIRCAM & 12485.17  & 16354.32 & 1554.12  & 2719.57\\
 $\left[ 19 \right]$ & Palomar & 5.1m Telescope & WIRC & 12403.18  & 16158.90 & 1521.69 &  2884.16 \\
\hline
\end{tabular}
\end{table*}

\section{Near-infrared filters and transmissions} \label{app:scor}

\begin{figure}
\includegraphics[trim=0.cm 0.cm 0.cm 0.cm, clip=True,width=\columnwidth]{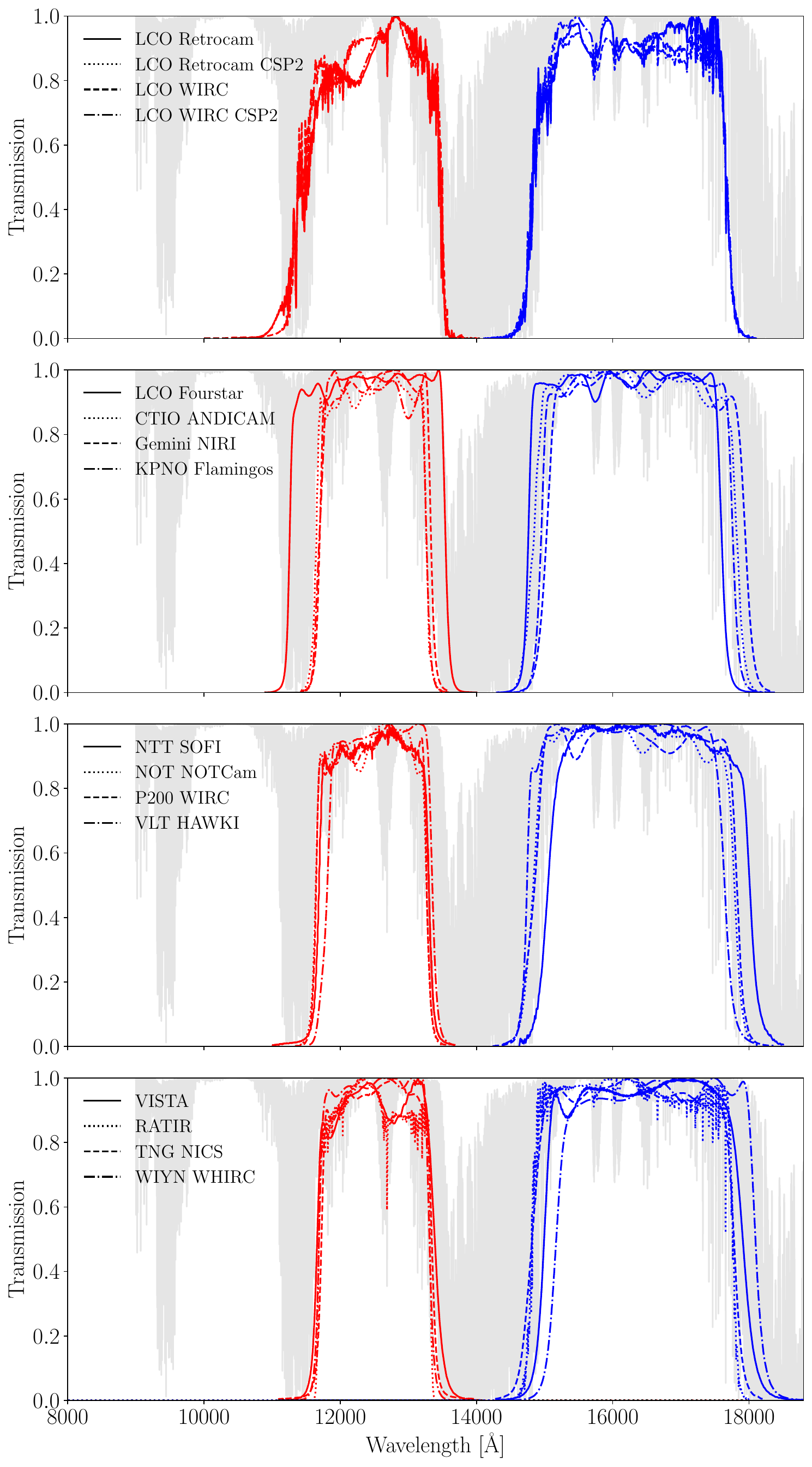}
\caption{Responses of some of the $J$- (red) and $H$-band (blue) filters used in our sample. All transmissions have been obtained from the  Spanish Virtual Observatory filter profile service (SVO; \href{https://svo.cab.inta-csic.es}{https://svo.cab.inta-csic.es}; \protect\cite{2020sea..confE.182R}). }
\label{fig:filters}
\end{figure}

All SNIa observations used in this work have been obtained from the ground at several observatories in the world using different combinations of telescopes, instruments and filters (see Table \ref{tab:infra} for a list), with different characteristics (e.g. weather conditions, sky contamination, detector sensitivity, detector pixel size), and reduced using different methods, software and procedures. Our starting point for the analysis presented in this paper is the SN Ia photometry as it was published in each reference, sometimes in the natural system and other times calibrated to a known sytem (e.g. CSP, 2MASS). As described in section \ref{sec:scor}, we have applied S-corrections in most cases to convert instrumental photometry or photometry in other systems to the CSP photometric system, since that was the system with more data available.

\begin{figure}[t]
\includegraphics[trim=0.cm 0.cm 0.cm 0.cm, clip=True,width=\columnwidth]{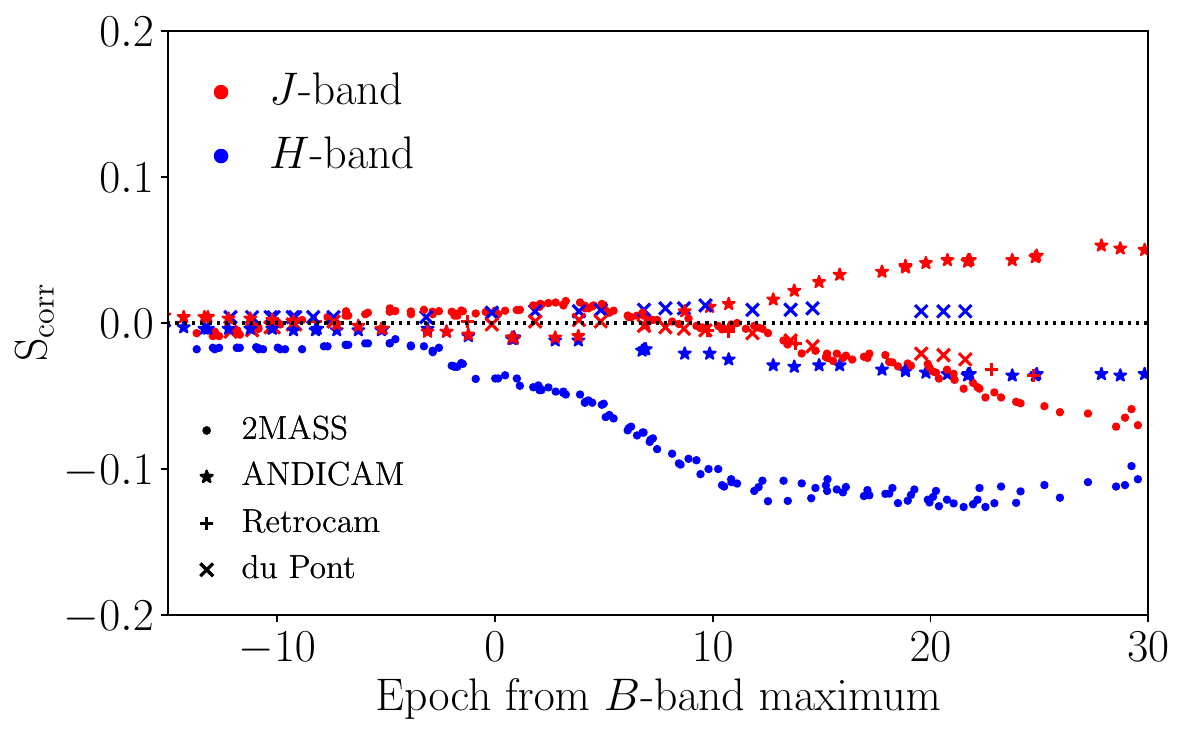}
\caption{S-corrections of all our SN Ia in teh calibrator sample that was not observed in the CSP system. }
\label{fig:scor}
\end{figure}

In Figure \ref{fig:filters} we show 16 of the 24 $J$- and $H$-band transmissions listed in Table \ref{tab:infra}, that were available through the Spanish Virtual Observatory filter profile service (SVO; \citealt{2020sea..confE.182R}. Each panel shows only four trasmissions per band for the sake of clarity. It is evident that transmission functions show quite a variety of shapes. In the top panel the four filters correspond to those used by the CSP collaboration, together with the Fourstar filters in the second panel, for the first (CSP-I) and second (CSP-II) stages of the project. The observed high level of homogeneity within the filters, coupled with the fact that a substantial portion of our SN data comes from this survey, justifies our selection of these bands as the reference in our study. To give an idea of the magnitude of this correction, in Figure \ref{fig:scor} we show the S-corrections applied within SNooPY to the 11 SNe Ia in the calibration sample that were not observed in the CSP system. The largest correction is of about 0.12 mag after 10 days from $B$-band maximum for 2MASS bands, but around $J$ and $H$ maxima the S-correction is as big as 0.01 mag in $J$ and 0.04 mag in $H$, exactly the number reported by \cite{2010AJ....139..519C} to convert between the 2MASS and CSP systems, and consistent with the recalculation done in \cite{2015ApJS..220....9F}.

\section{The $s_{BV}$ parameter} \label{app:sbv}

\begin{figure*}[!th]
\includegraphics[trim=0.cm 0.cm 0.cm 0.cm, clip=True,width=0.44\textwidth]{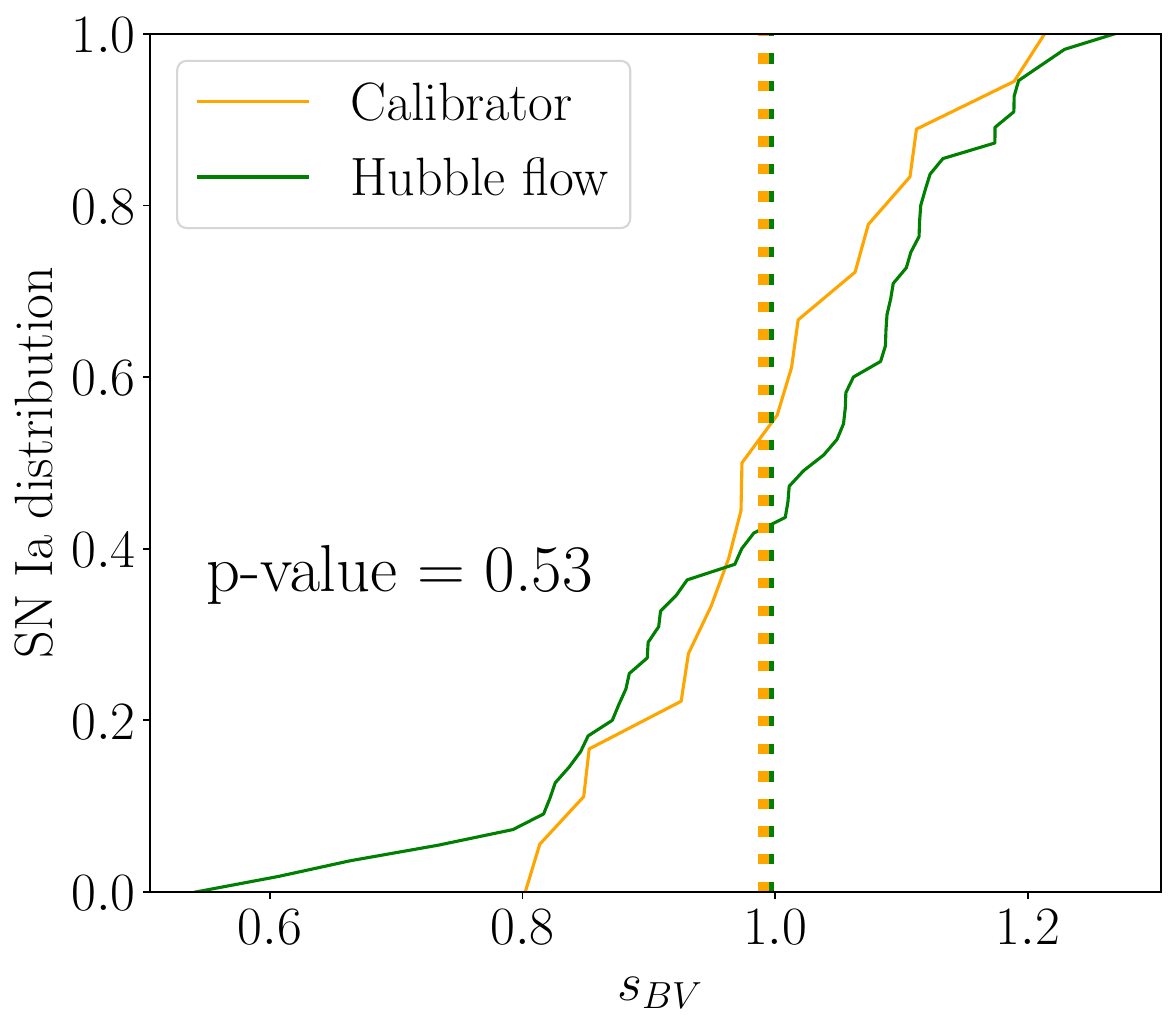}
\includegraphics[trim=0.cm 0.cm 0.cm 0.cm, clip=True,width=0.55\textwidth]{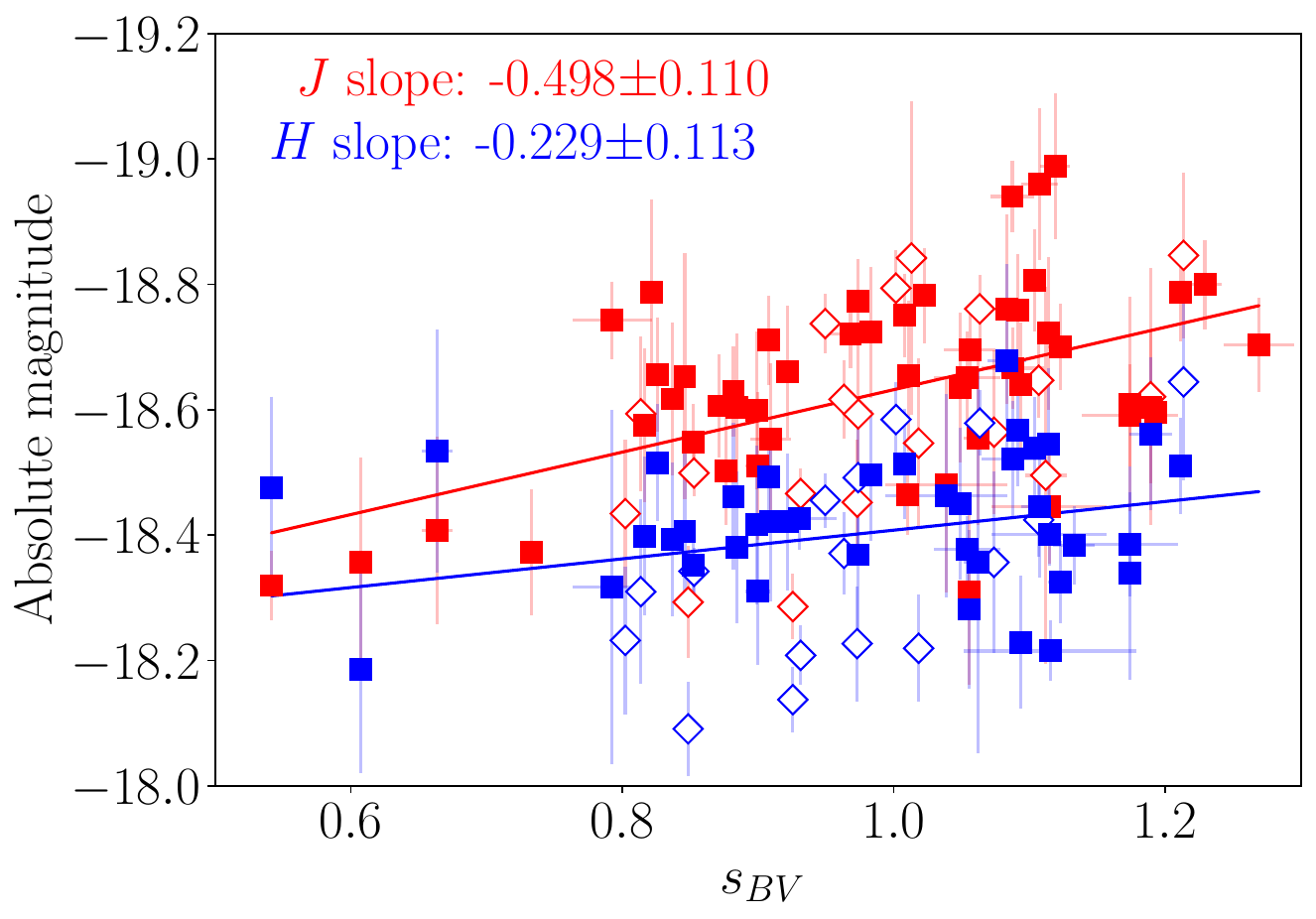}
\caption{Dependencies of Hubble residuals on SN Ia light-curve parameters. Open symbols correspond to SNe Ia in the calibrator sample, while closed symbols are for those in the Hubble flow sample.}
\label{fig:sBV}
\end{figure*}

The $s_{BV}$ parameter combines aspects of the typical stretch and color parameters because it is calculated as the stretch factor of the $B-V$ color-curve relative to the standard behavior of a SN Ia with a $t_{max}^B - t_{max}^{B-V}$ of 30 days. It offers a more meaningful interpretation of SN Ia light-curves as it measures the time at which iron recombines from [Fe\textsuperscript{III}] to [Fe\textsuperscript{II}] \citep{2014ApJ...789...32B}. Importantly, it resolves the degeneracy in the relation between peak brightness and light-curve width for the fainter SNe Ia \citep{2019A&A...630A..76G}. The $s_{BV}$ values for the entire sample used in this paper are provided in Table \ref{tab:snoopyres}.

For completeness, we examined whether the calibrator and Hubble flow SN Ia samples are compatible in terms of this parameter, as we did with $\Delta m_{15}$ and $E(B-V)$ in section \ref{sec:stand}. The right panel of Figure \ref{fig:sBV} shows the cumulative distribution of the color-stretch parameter $s_{BV}$ for the calibrator and Hubble flow samples. Both samples exhibit a similar average $s_{BV}$ around 1.0, and the two-sample K-S test yields a p-value of 0.53, indicating that the distributions are consistent with being drawn from the same population.

Similarly to the test performed in section \ref{sec:prop}, we also examined whether a relation between the NIR peak magnitude and $s_{BV}$ could be used to reduce the dispersion in peak magnitudes and thereby standardize SN Ia in the NIR. The right panel of Figure \ref{fig:sBV} illustrates this relation, showing that brighter SNe Ia tend to have larger $s_{BV}$ values, meaning the peak in the $B-V$ color-curve occurs after 30 days from peak brightness. Performing a linear regression for each band independently results in a slope of -0.498$\pm$0.110 in $J$ and -0.229$\pm$0.113 in $H$. These values correspond to 4.5$\sigma$ and 2.0$\sigma$, respectively. This confirms that in $J$ there is still a significant benefit in standardizing SN Ia, while in $H$, it is less critical.

By linearly applying the color-stretch correction to the absolute magnitudes of our calibrators, we observed a reduction in scatter from 0.160 mag to 0.149 mag in $J$ and from 0.154 mag to 0.146 mag in $H$. This reduction is approximately 0.010 mag better compared to applying the $\Delta m_{15}$ correction. For the Hubble flow sample, the stretch correction resulted in a scatter reduction from 0.149 mag to 0.130 mag in $J$, which is only 0.002 mag less than the $\Delta m_{15}$ correction. In $H$ band, the scatter remained at 0.102 mag after applying the $s_{BV}$ correction. Overall, applying the color-stretch correction led to improvements in scatter reduction for both the calibrators and the Hubble flow sample, particularly in $J$ band, confirming it is a valuable standardization method. 

\onecolumn

\begin{landscape}
\section{SNooPy results}\label{app:snoopy}
\begin{longtable}{lllcccccccc}
\caption{\label{tab:snoopyres} SNooPy $max\_model$ and $EBV\_model2$ (only $E(B-V)$ parameter) template fitting results.}\\
\hline\hline
SN name   & Host galaxy             & Morphology     &T$_{\rm max,B}$&$\Delta m_{15}$&   $E(B-V)$   &  $s_{BV}$   &T$_{\rm max,J}^{\rm T}$& $J_{\rm max}^{\rm T}$&T$_{\rm max,H}^{\rm T}$& $H_{\rm max}^{\rm T}$\\
\hline
\endfirsthead
\caption{continued.}\\
\hline\hline
SN name  & Host galaxy             & Morphology      &T$_{\rm max,B}$&$\Delta m_{15}$&   $E(B-V)$   &  $s_{BV}$   &T$_{\rm max,J}^{\rm T}$& $J_{\rm max}^{\rm T}$&T$_{\rm max,H}^{\rm T}$& $H_{\rm max}^{\rm T}$\\
\hline
\endhead
\hline
\endfoot
\multicolumn{10}{c}{Calibrators}\\
\hline
2001el  & NGC 1448                 & SAcd            & 52182.57 (07) & 1.073 (014) &    0.322 (007) & 0.931 (007) & 52178.67 & 12.761 (027) & 52178.53 & 12.935 (032) \\
2002fk  & NGC 1309                 & SA(s)bc         & 52547.86 (06) & 1.050 (009) &    0.039 (005) & 1.002 (005) & 52543.94 & 13.825 (015) & 52543.81 & 14.047 (020) \\
2003du  & UGC 9391                 & SBdm            & 52766.05 (03) & 0.984 (006) & $-$0.002 (003) & 1.018 (004) & 52762.06 & 14.216 (043) & 52761.93 & 14.514 (037) \\
2005cf  & NGC 5917                 & Sb              & 53534.15 (02) & 1.036 (005) &    0.112 (002) & 0.974 (003) & 53530.22 & 13.773 (020) & 53530.09 & 13.980 (023) \\
2005df  & NGC 1559                 & SB(s)cd         & 53599.06 (07) & 1.035 (010) &    0.134 (011) & 0.963 (007) & 53595.13 & 12.874 (014) & 53595.00 & 13.139 (014) \\
2006D   & Mrk 1337                 & SAB(s)ab pec?   & 53758.00 (03) & 1.366 (004) &    0.133 (003) & 0.814 (003) & 53754.18 & 14.391 (040) & 53753.37 & 14.565 (029) \\
2006bh  & NGC 7329                 & SB(r)b          & 53834.11 (04) & 1.397 (005) & $-$0.015 (006) & 0.802 (003) & 53830.29 & 14.782 (015) & 53829.39 & 14.968 (014) \\
2007A   & NGC 0105                 & Sab             & 54113.62 (08) & 1.016 (021) &    0.252 (008) & 1.013 (011) & 54109.67 & 15.700 (023) & 54109.55 & 16.000 (059) \\
2007af  & NGC 5584                 & SAB(rs)cd       & 54174.96 (03) & 1.099 (007) &    0.194 (002) & 0.926 (003) & 54171.07 & 13.469 (007) & 54170.91 & 13.631 (011) \\
2008fv  & NGC 3147                 & SA(rs)bc        & 54749.67 (13) & 0.796 (012) &    0.197 (012) & 1.112 (015) & 54745.47 & 14.604 (033) & 54745.38 & 14.989 (028) \\
2009Y   & NGC 5728                 & SAB(r)a?        & 54877.08 (05) & 1.028 (013) &    0.172 (006) & 1.189 (008) & 54873.15 & 14.503 (018) & 54873.02 & 14.732 (025) \\
2011by  & NGC 3972                 & SA(s)bc         & 55690.81 (04) & 1.056 (009) &    0.099 (005) & 0.973 (004) & 55686.90 & 13.287 (033) & 55686.77 & 13.504 (027) \\
2011fe  & M101                     & SAB(rs)cd       & 55814.82 (03) & 1.076 (006) &    0.068 (003) & 0.949 (003) & 55810.93 & 10.474 (010) & 55810.79 & 10.745 (010) \\
2012cg  & NGC 4424                 & SB(s)a          & 56082.10 (04) & 0.916 (007) &    0.188 (005) & 1.074 (005) & 56078.02 & 12.346 (020) & 56077.86 & 12.583 (022) \\
2012fr  & NGC 1365                 & SB(s)b          & 56243.33 (04) & 0.844 (005) &    0.082 (005) & 1.107 (005) & 56239.10 & 12.735 (015) & 56238.95 & 12.974 (017) \\
2012ht  & NGC 3447                 & SAB(s)m pec     & 56295.85 (03) & 1.330 (004) &    0.110 (004) & 0.853 (002) & 56292.03 & 13.436 (010) & 56291.33 & 13.627 (012) \\
2013dy  & NGC 7250                 & Sdm?            & 56501.64 (10) & 1.041 (012) &    0.168 (008) & 1.213 (008) & 56497.71 & 13.799 (063) & 56497.59 & 14.409 (062) \\
2015F   & NGC 2442                 & SAB(s)bc        & 57106.86 (03) & 1.303 (006) &    0.120 (004) & 0.848 (004) & 57103.04 & 13.098 (027) & 57102.42 & 13.215 (025) \\
2017cbv & NGC 5643                 & SAB(rs)c        & 57840.51 (03) & 0.962 (006) & $-$0.020 (003) & 1.063 (005) & 57836.50 & 11.917 (015) & 57836.36 & 12.037 (014) \\ 
\hline
\multicolumn{10}{c}{Hubble flow}\\
\hline
1999ee    & IC 5179                  & SA(rs)bc        & 51469.01 (03) & 0.793 (005) &    0.346 (004) & 1.174 (005) & 51464.81 & 14.771 (017) & 51464.71 & 14.958 (015) \\
1999ek    & UGC 03329                & Sbc             & 51482.03 (08) & 1.092 (013) &    0.261 (011) & 0.922 (007) & 51478.14 & 16.104 (013) & 51477.99 & 16.183 (018) \\
2004eo    & NGC 6928                 & SB(s)ab         & 53278.90 (04) & 1.315 (006) &    0.120 (006) & 0.817 (005) & 53275.08 & 15.482 (022) & 53274.42 & 15.667 (032) \\
2004ey    & UGC 11816                & SB(rs)c         & 53304.78 (02) & 0.967 (005) &    0.011 (001) & 1.011 (003) & 53300.77 & 15.508 (007) & 53300.64 & 15.852 (022) \\
2005M     & NGC 2930                 & S?              & 53405.90 (04) & 0.797 (005) &    0.036 (003) & 1.211 (003) & 53401.70 & 16.510 (037) & 53401.60 & 16.659 (032) \\
2005el    & NGC 1819                 & SB0             & 53647.35 (03) & 1.351 (004) & $-$0.068 (005) & 0.837 (003) & 53643.54 & 15.503 (012) & 53642.77 & 15.622 (015) \\
2005eq    & MCG -01-09-06            & SB(rs)cd?       & 53654.89 (06) & 0.823 (006) &    0.028 (003) & 1.123 (009) & 53650.67 & 16.817 (014) & 53650.54 & 17.038 (025) \\
2005hc    & MCG +00-06-003           & Sa              & 53667.70 (07) & 0.872 (009) &    0.030 (005) & 1.193 (006) & 53663.53 & 17.875 (038) & 53663.37 & 17.961 (107) \\
2005iq    & ESO 538-G013             & Sa              & 53688.15 (05) & 1.278 (009) &    0.000 (007) & 0.871 (004) & 53684.33 & 17.236 (030) & 53683.79 & 17.343 (052) \\
2005kc    & NGC 7311                 & Sab             & 53698.31 (04) & 1.142 (013) &    0.370 (007) & 0.899 (006) & 53694.45 & 15.442 (016) & 53694.24 & 15.616 (022) \\
2005ki    & NGC 3332                 & (R)SA0          & 53706.00 (03) & 1.360 (004) & $-$0.012 (004) & 0.826 (003) & 53702.18 & 16.081 (013) & 53701.39 & 16.258 (020) \\
2006ax    & NGC 3663                 & SA(rs)bc        & 53827.74 (03) & 1.022 (007) &    0.003 (002) & 0.983 (003) & 53823.80 & 15.703 (008) & 53823.68 & 15.999 (012) \\
2006et    & NGC 232                  & SB(r)a?         & 53994.39 (05) & 0.894 (008) &    0.207 (004) & 1.092 (008) & 53990.26 & 16.146 (015) & 53990.10 & 16.375 (018) \\
2006hx    & PGC 73820                & S0              & 54022.30 (11) & 1.049 (024) &    0.211 (015) & 0.968 (018) & 54018.38 & 17.799 (071) & 54018.25 & 17.835 (058) \\
2006kf    & UGC 02829                & S0              & 54041.80 (04) & 1.505 (007) &    0.181 (011) & 0.733 (004) & 54037.99 & 16.371 (013) & 54036.87 & 16.484 (032) \\
2006le    & UGC 3218                 & SAb             & 54047.92 (05) & 0.886 (009) & $-$0.061 (008) & 1.094 (008) & 54043.77 & 16.186 (015) & 54043.61 & 16.460 (036) \\
2006lf    & UGC 3108                 & S?              & 54045.05 (09) & 1.172 (008) & $-$0.013 (012) & 0.822 (008) & 54041.20 & 15.630 (031) & 54040.93 & 15.734 (049) \\
2007S     & UGC 5378                 & Sb              & 54144.52 (05) & 0.815 (005) &    0.433 (003) & 1.114 (008) & 54140.30 & 15.346 (010) & 54140.18 & 15.505 (012) \\
2007ai    & ESO 584-G007             & Sc              & 54174.03 (19) & 0.844 (016) &    0.261 (008) & 1.229 (012) & 54169.80 & 17.152 (023) & 54169.65 & 17.242 (039) \\
2007as    & PGC 026840               & SB(rs)c         & 54181.91 (07) & 1.185 (005) &    0.159 (006) & 0.885 (005) & 54178.06 & 15.891 (017) & 54177.78 & 16.120 (032) \\
2007ba    & UGC 09798                & S0/a            & 54197.87 (11) & 1.667 (010) &    0.508 (032) & 0.541 (006) & 54194.20 & 17.527 (057) & 54193.06 & 17.527 (153) \\
2007bc    & UGC 06332                & (R)SBa          & 54200.79 (08) & 1.281 (011) &    0.114 (006) & 0.876 (006) & 54196.96 & 16.242 (025) & 54196.42 & 16.351 (063) \\
2007bd    & UGC 4455                 & SB(r)a          & 54207.40 (06) & 1.267 (011) &    0.040 (008) & 0.882 (005) & 54203.58 & 17.017 (050) & 54203.07 & 17.173 (080) \\
2007ca    & MCG -02-34-61            & Sc              & 54227.80 (53) & 0.870 (009) &    0.391 (006) & 1.056 (007) & 54223.63 & 15.660 (015) & 54223.47 & 15.729 (019) \\
2008bc    & PGC 90108                & S               & 54550.11 (04) & 0.893 (005) & $-$0.016 (004) & 1.049 (004) & 54545.98 & 15.699 (016) & 54545.82 & 15.844 (024) \\
2008bf    & NGC 4061 / NGC 4065      & E: / E          & 54555.19 (04) & 0.928 (006) & $-$0.020 (003) & 1.023 (004) & 54551.13 & 16.402 (012) & 54550.98 & 16.821 (049) \\
2008gl    & UGC 00881                & E               & 54768.82 (05) & 1.323 (006) &    0.081 (005) & 0.852 (004) & 54765.00 & 17.109 (028) & 54764.32 & 17.282 (053) \\
2008gp    & CGCG 390-078             & (R)SAB(r)a pec? & 54779.63 (03) & 1.022 (007) &    0.001 (004) & 0.974 (005) & 54775.68 & 17.091 (020) & 54775.56 & 17.429 (043) \\
2008hj    & MCG -02-01-014           & SB(rs)c?        & 54802.12 (05) & 0.968 (010) &    0.061 (006) & 1.010 (006) & 54798.11 & 17.478 (029) & 54797.97 & 17.613 (072) \\
2008hs    & NGC 0919                 & Sab             & 54813.18 (08) & 1.605 (012) &    0.030 (012) & 0.637 (008) & 54809.38 & 16.315 (041) & 54808.24 & 16.511 (057) \\
2008hv    & NGC 2765                 & S0              & 54817.64 (01) & 1.325 (004) & $-$0.026 (004) & 0.846 (003) & 54813.82 & 15.238 (013) & 54813.14 & 15.461 (017) \\
2009aa    & ESO 570-G020             & Sbc             & 54878.80 (03) & 1.188 (005) &    0.072 (005) & 0.908 (004) & 54874.95 & 16.699 (013) & 54874.66 & 16.891 (019) \\
2009ad    & UGC 3236                 & Sbc             & 54886.89 (04) & 0.966 (009) &    0.008 (006) & 1.008 (006) & 54882.88 & 16.792 (016) & 54882.75 & 17.052 (029) \\
2009al    & NGC 3388                 & S0              & 54897.44 (22) & 1.096 (023) &    0.249 (009) & 1.174 (035) & 54893.55 & 16.336 (019) & 54893.40 & 16.540 (034) \\
2009bv    & MCG +06-29-39            & S               & 54926.93 (17) & 0.906 (027) &    0.004 (014) & 1.054 (025) & 54922.83 & 17.449 (041) & 54922.67 & 17.510 (048) \\
2010Y     & NGC 3392                 & E               & 55247.96 (08) & 1.623 (009) & $-$0.007 (013) & 0.664 (011) & 55244.17 & 15.246 (036) & 55243.08 & 15.386 (136) \\
2010ag    & UGC 10679                & Sb(f)           & 55270.95 (34) & 0.844 (036) &    0.055 (026) & 1.269 (026) & 55266.72 & 17.134 (023) & 55266.57 & 17.619 (068) \\
2010ai    & SDSS J125925.04+275948.2 & E               & 55277.46 (05) & 1.339 (006) &    0.040 (012) & 0.833 (006) & 55273.65 & 16.464 (024) & 55272.92 & 16.466 (042) \\
2010ju    & UGC 03341                & SBab            & 55524.64 (11) & 1.172 (018) &    0.562 (011) & 0.909 (015) & 55520.79 & 15.807 (015) & 55520.52 & 15.673 (024) \\
2010kg    & NGC 1633                 & SAB(s)ab        & 55543.97 (12) & 1.270 (022) &    0.171 (013) & 0.899 (009) & 55540.15 & 15.775 (021) & 55539.64 & 15.856 (031) \\
2011ao    & IC 2973                  & SB(s)d          & 55639.78 (08) & 1.000 (022) &    0.106 (006) & 1.084 (014) & 55635.81 & 14.861 (025) & 55635.68 & 14.970 (019) \\
PTF10bjs  & MCG +09-21-83            & Sb              & 55261.94 (24) & 1.013 (032) & $-$0.194 (039) & 1.088 (023) & 55257.99 & 16.884 (041) & 55257.86 & 17.162 (035) \\
PTF10hmv  & SDSSJ121133.31+471628.6  & Spiral          & 55351.70 (17) & 0.796 (009) &    0.445 (025) & 1.133 (015) & $\cdots$ & $\cdots$     & 55347.40 & 17.426 (021) \\
PTF10mwb  & SDSS J171750.05+405252.5 & S(r)c           & 55391.54 (22) & 1.420 (090) &    0.032 (028) & 0.792 (029) & 55387.73 & 17.097 (037) & 55386.77 & 17.286 (047) \\
PTF10nlg  & 2MASSJ16503452+6016345   & Spiral          & 55392.04 (31) & 1.032 (039) &    0.279 (036) & 0.930 (027) & 55388.10 & 18.581 (072) & 55387.98 & 18.567 (048) \\
PTF10tce  & Anon.                    & Spiral          & 55442.92 (30) & 1.083 (030) &    0.236 (020) & 1.115 (064) & 55439.03 & 17.764 (102) & 55438.89 & 17.947 (097) \\
PTF10ufj  & 2MASX J02253767+2445579  & S0/a            & 55457.34 (33) & 0.905 (055) & $-$0.031 (020) & 1.038 (045) & 55453.23 & 19.195 (056) & 55453.07 & 19.176 (034) \\
PTF10wnm  & Anon.                    & Spiral          & 55476.97 (25) & 1.053 (041) &    0.060 (022) & 1.114 (046) & 55473.05 & 18.590 (084) & 55472.92 & 18.844 (090) \\
iPTF13asv & SDSSJ162243.02+185733.8  & Spiral          & 56429.94 (15) & 1.048 (016) & $-$0.099 (011) & 1.087 (016) & 56426.02 & 17.038 (046) & 56425.9  & 17.234 (086) \\
iPTF13azs & LEDA5068207              & Spiral          & 56436.84 (14) & 1.047 (018) &    0.433 (027) & 1.056 (019) & 56432.91 & 17.211 (044) & 56432.79 & 17.116 (093) \\
iPTF13dge & NGC1762                  & SA(rs)c         & 56556.67 (06) & 1.046 (013) &    0.138 (008) & 1.062 (010) & 56552.7  & 15.602 (063) & 56552.5  & 15.829 (133) \\
iPTF13duj & NGC7042                  & Sb              & 56601.77 (07) & 1.025 (018) &    0.092 (007) & 1.108 (013) & 56597.83 & 15.550 (038) & 56597.70 & 15.852 (043) \\
iPTF13ebh & NGC0890                  & SAB(r)0         & 56623.19 (06) & 1.630 (007) &    0.021 (011) & 0.607 (006) & 56619.4  & 15.037 (040) & 56618.31 & 15.175 (052) \\
iPTF14atg & IC0831                   & E               & 56799.33 (16) & 1.097 (023) &    0.565 (016) & 1.189 (016) & 56795.44 & 16.527 (165) & 56795.29 & 16.788 (273) \\
iPTF14bdn & UGC08503                 & Im              & 56822.28 (08) & 1.053 (017) &    0.038 (010) & 1.119 (011) & 56818.36 & 15.117 (028) & 56818.24 & 15.366 (047) \\
iPTF16abc & NGC 5221                 & Sb?             & 57498.67 (04) & 0.796 (004) &    0.093 (004) & 1.104 (006) & 57494.46 & 17.242 (026) & 57494.37 & 17.852 (033) \\
iPTF16auf & ECO3633                  & Spiral          & 57538.05 (08) & 1.044 (021) &    0.271 (012) & 1.189 (012) & 57534.12 & 15.434 (050) & 57534.0  & 15.692 (075) \\
\end{longtable}

\end{landscape}

\section{LC} \label{app:lcfits}

\begin{figure*}[!ht]
\centering
    \includegraphics[trim=0.4cm 0.2cm 2cm 1cm, clip=True,width=0.32\textwidth]{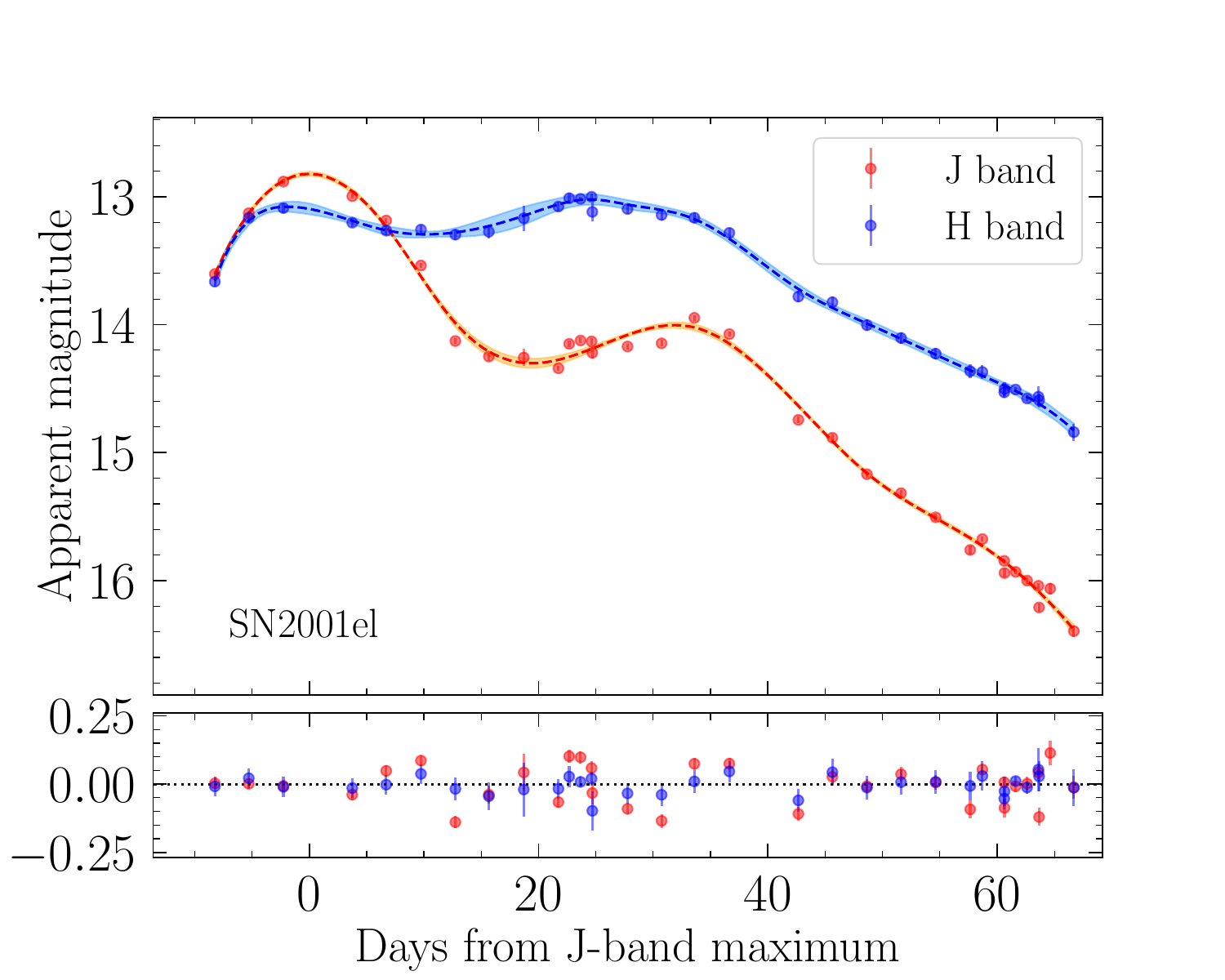}
    \includegraphics[trim=0.4cm 0.2cm 2cm 1cm, clip=True,width=0.32\textwidth]{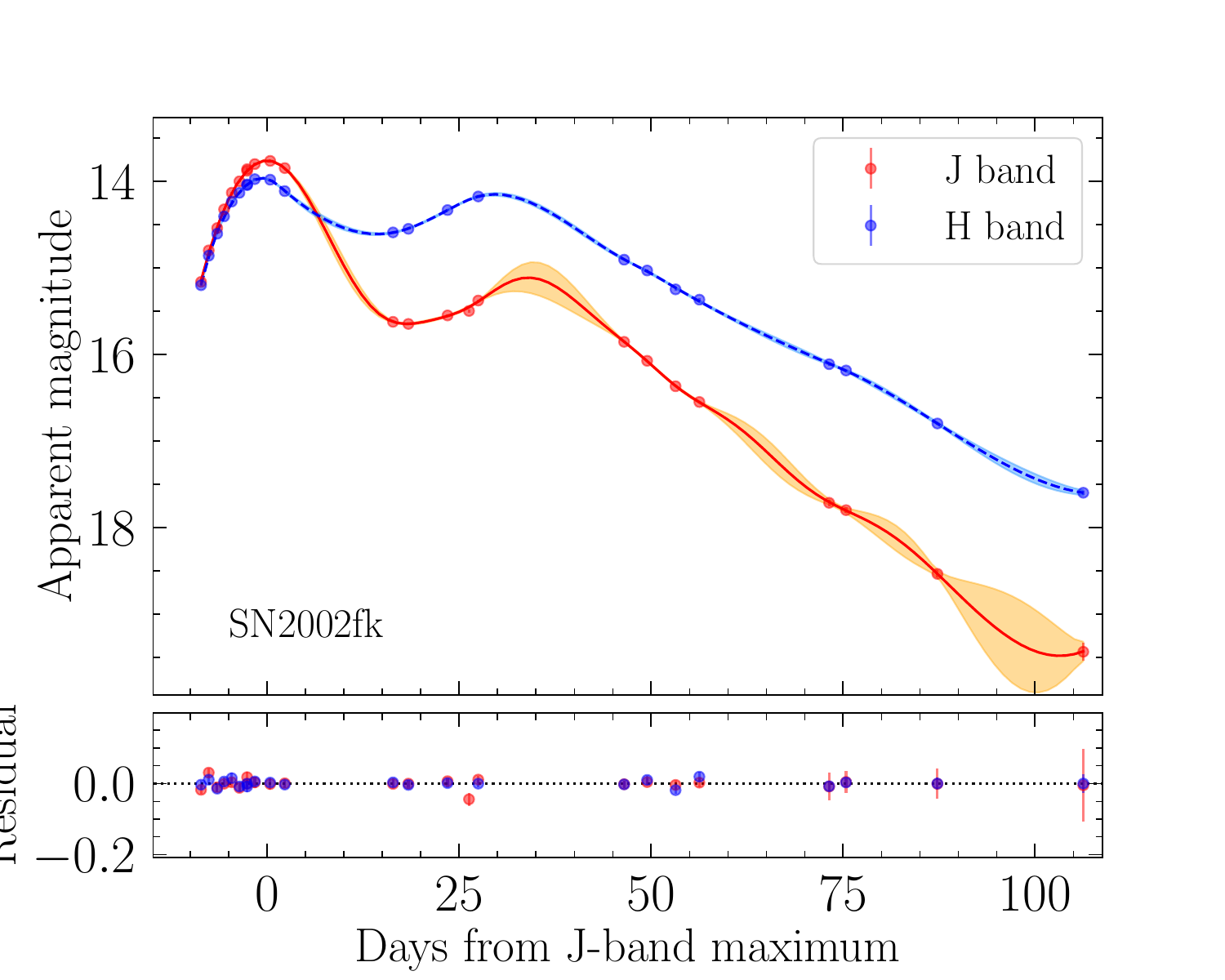}
    \includegraphics[trim=0.4cm 0.2cm 2cm 1cm, clip=True,width=0.32\textwidth]{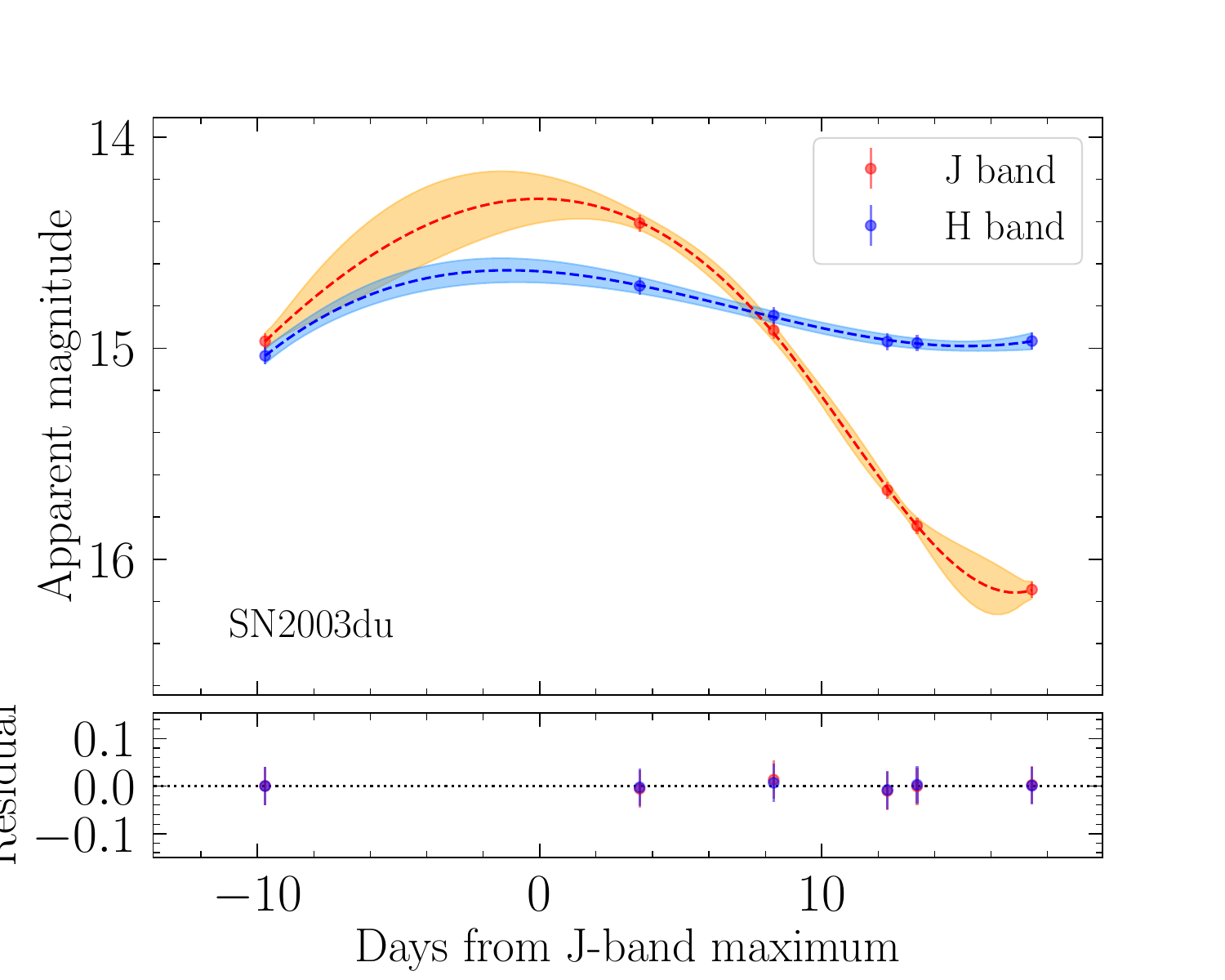}
    \includegraphics[trim=0.4cm 0.2cm 2cm 1cm, clip=True,width=0.32\textwidth]{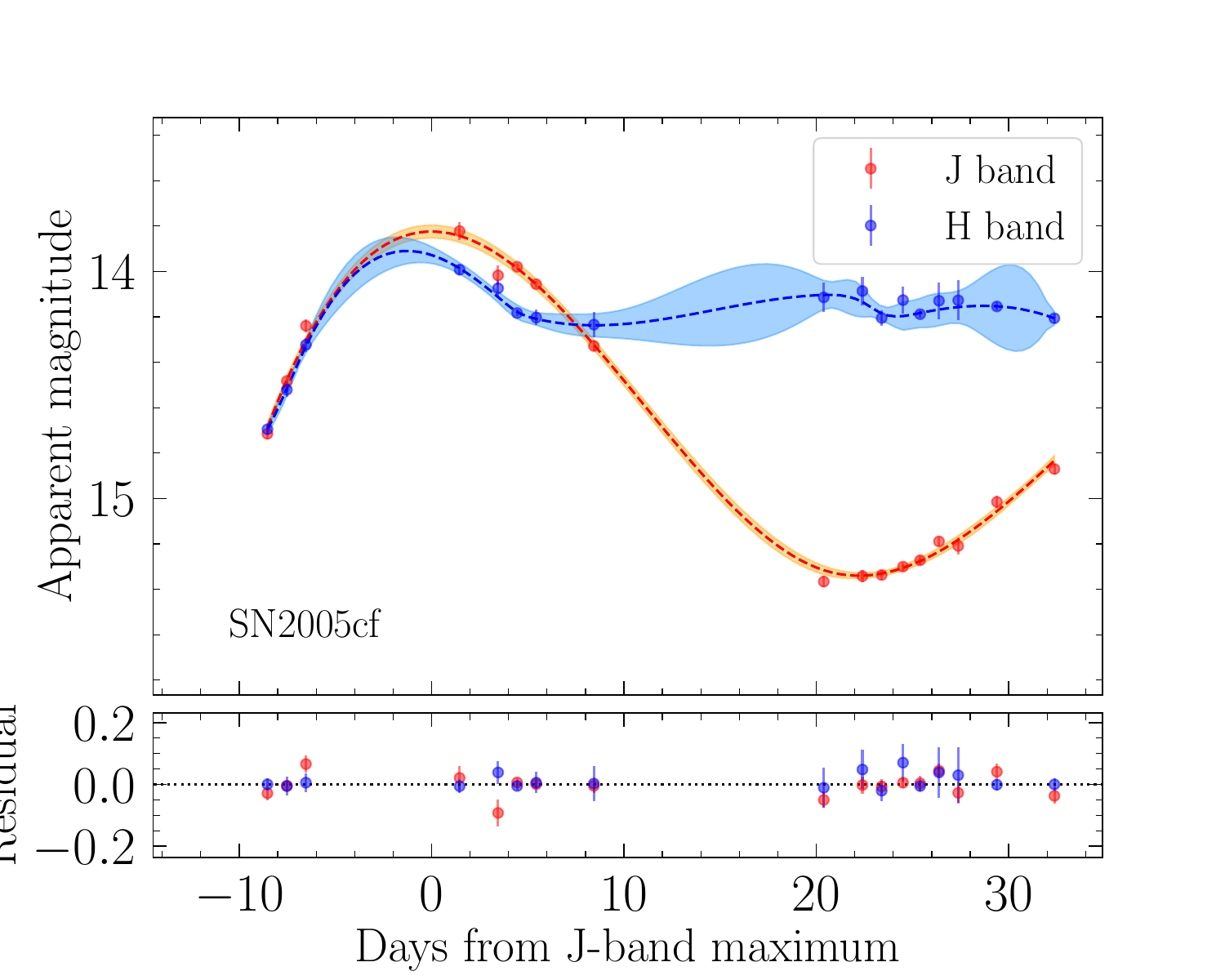}
    \includegraphics[trim=0.4cm 0.2cm 2cm 1cm, clip=True,width=0.32\textwidth]{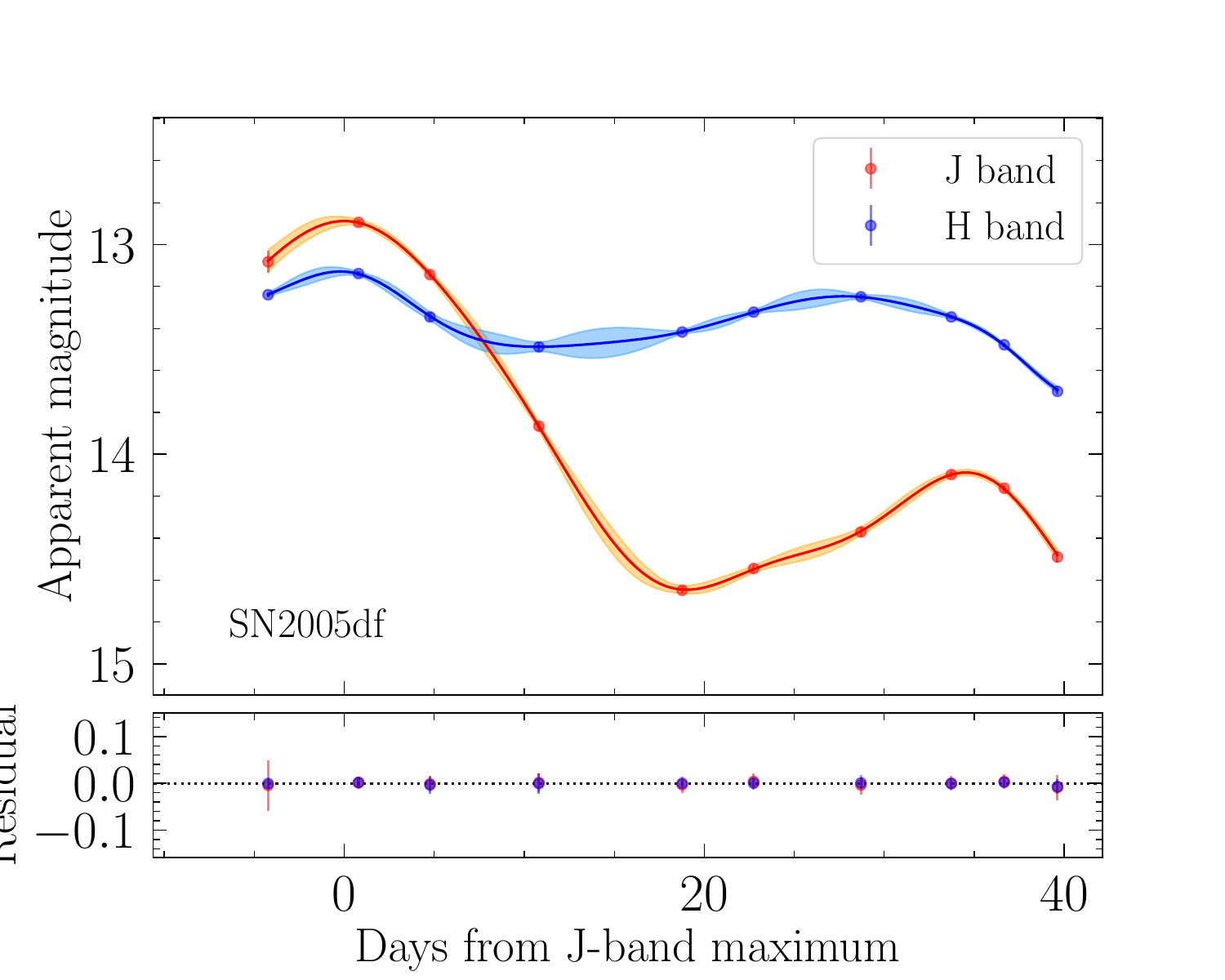}
    \includegraphics[trim=0.4cm 0.2cm 2cm 1cm, clip=True,width=0.32\textwidth]{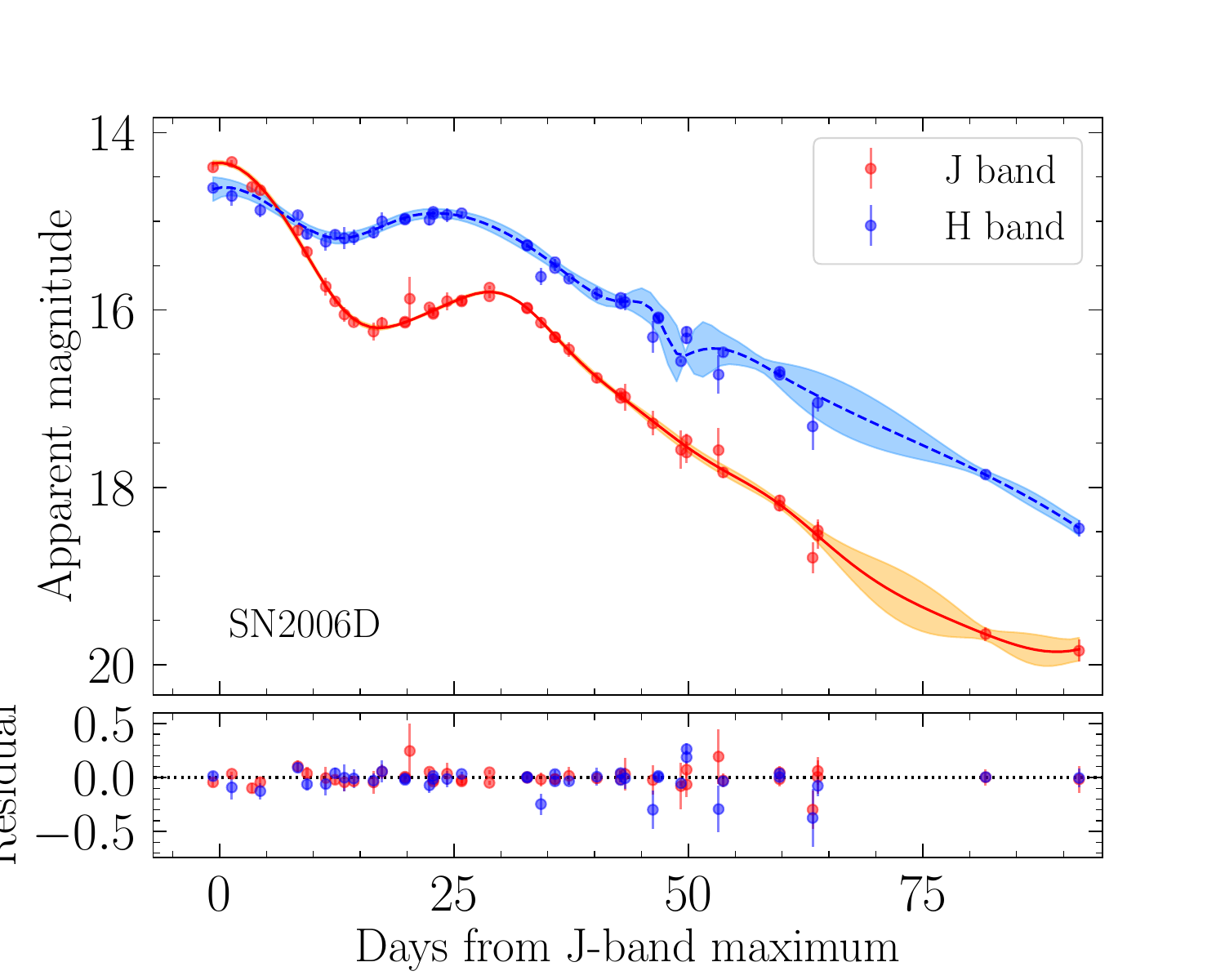}
    \includegraphics[trim=0.4cm 0.2cm 2cm 1cm, clip=True,width=0.32\textwidth]{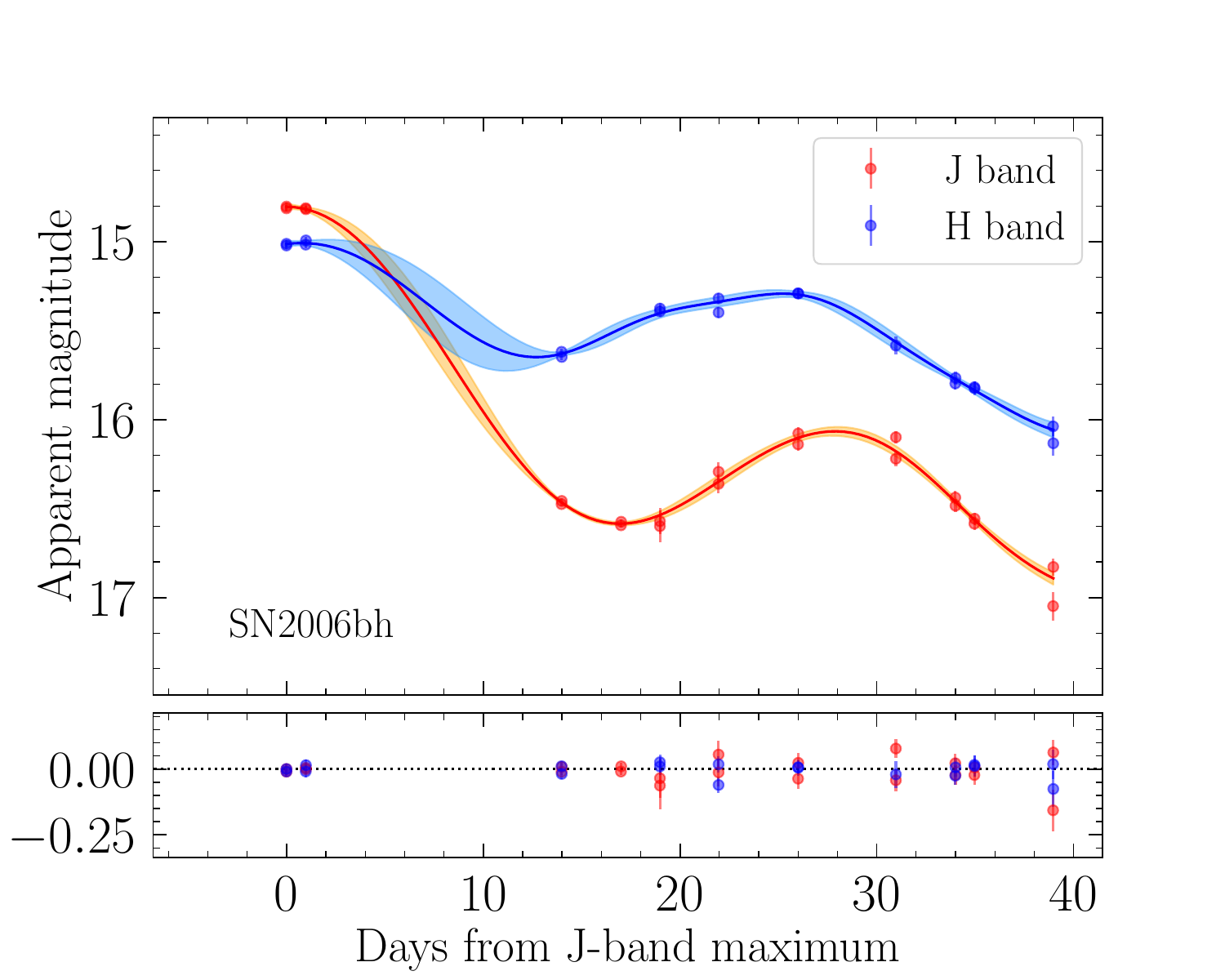}
    \includegraphics[trim=0.4cm 0.2cm 2cm 1cm, clip=True,width=0.32\textwidth]{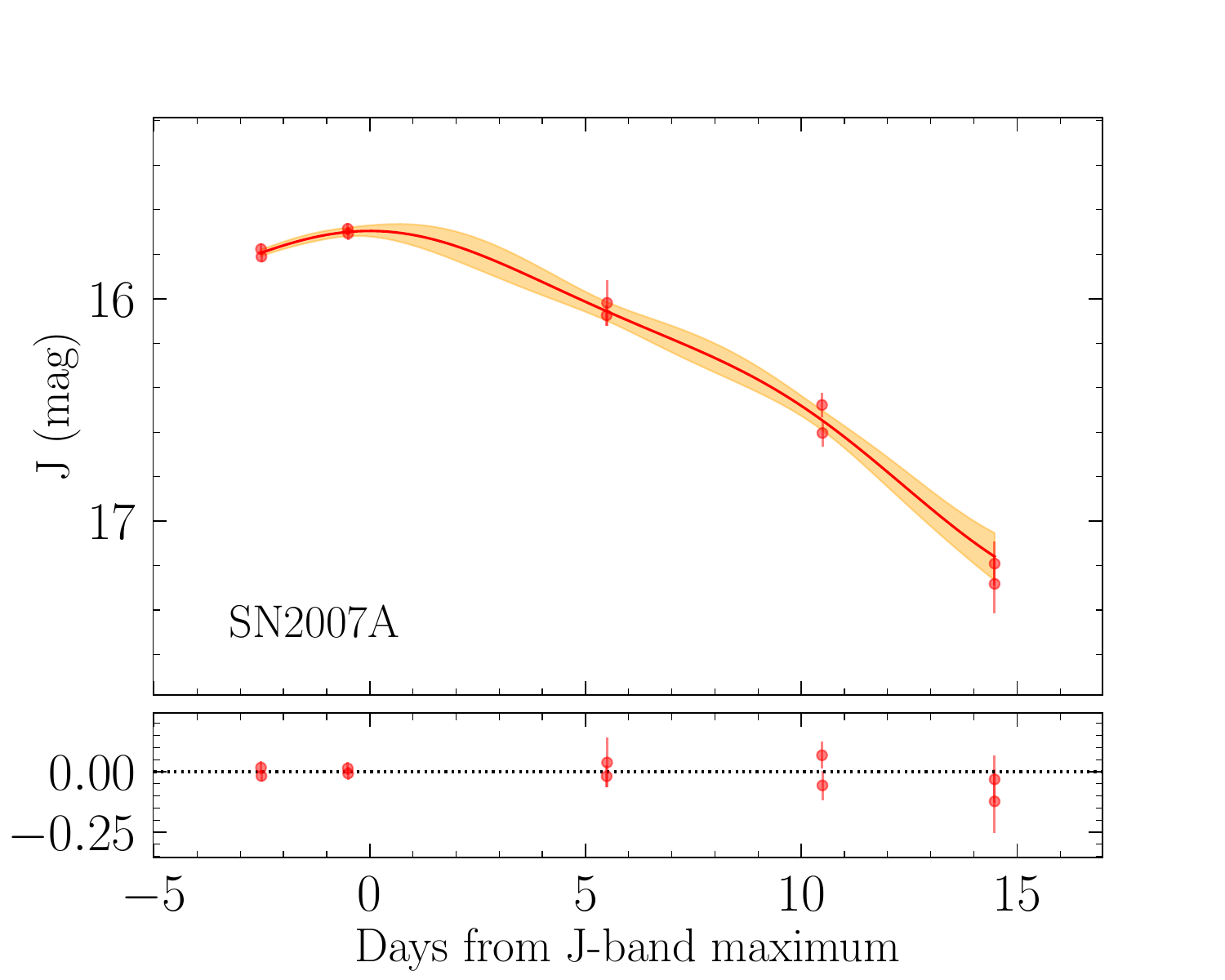}
    \includegraphics[trim=0.4cm 0.2cm 2cm 1cm, clip=True,width=0.32\textwidth]{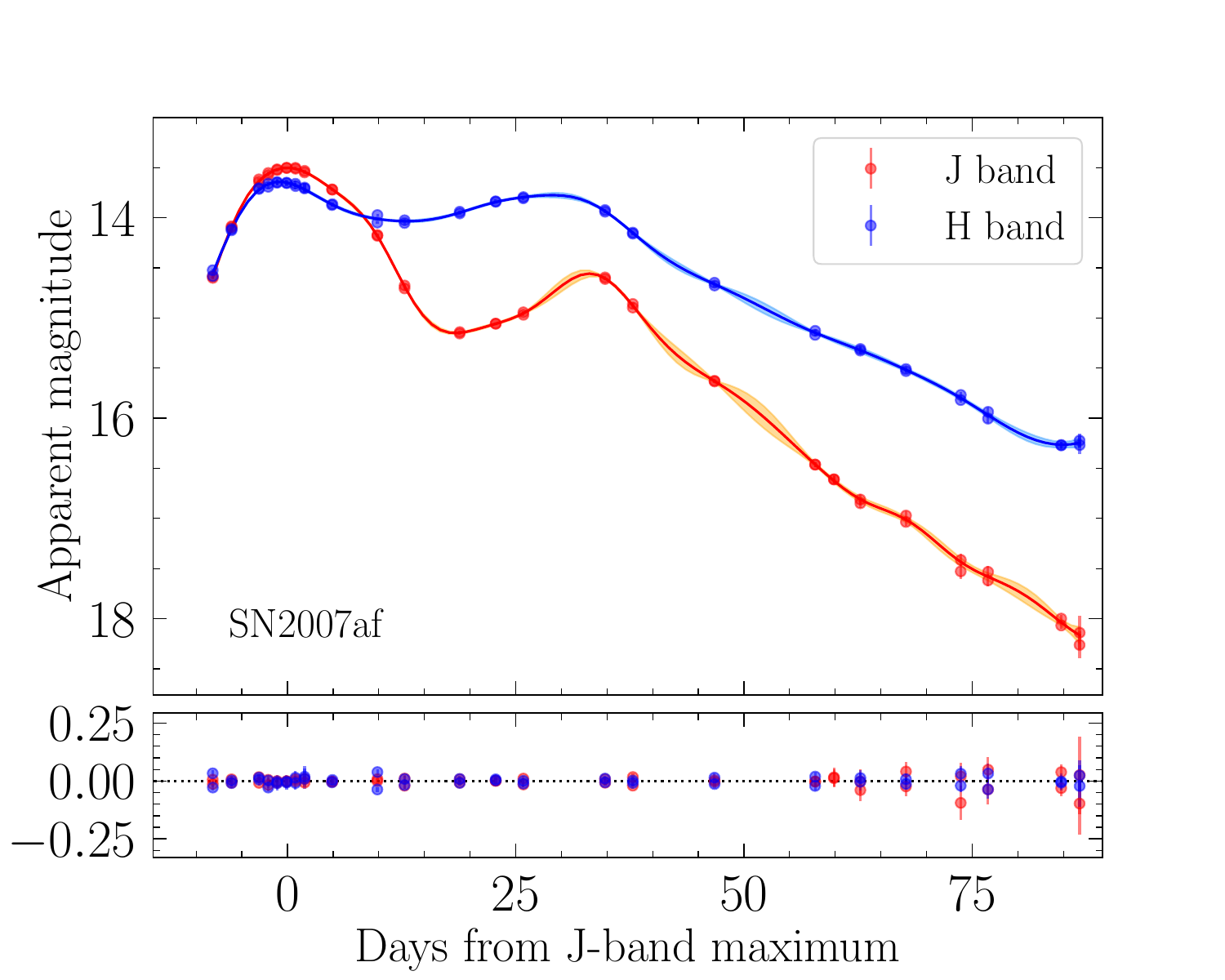}
    \includegraphics[trim=0.4cm 0.2cm 2cm 1cm, clip=True,width=0.32\textwidth]{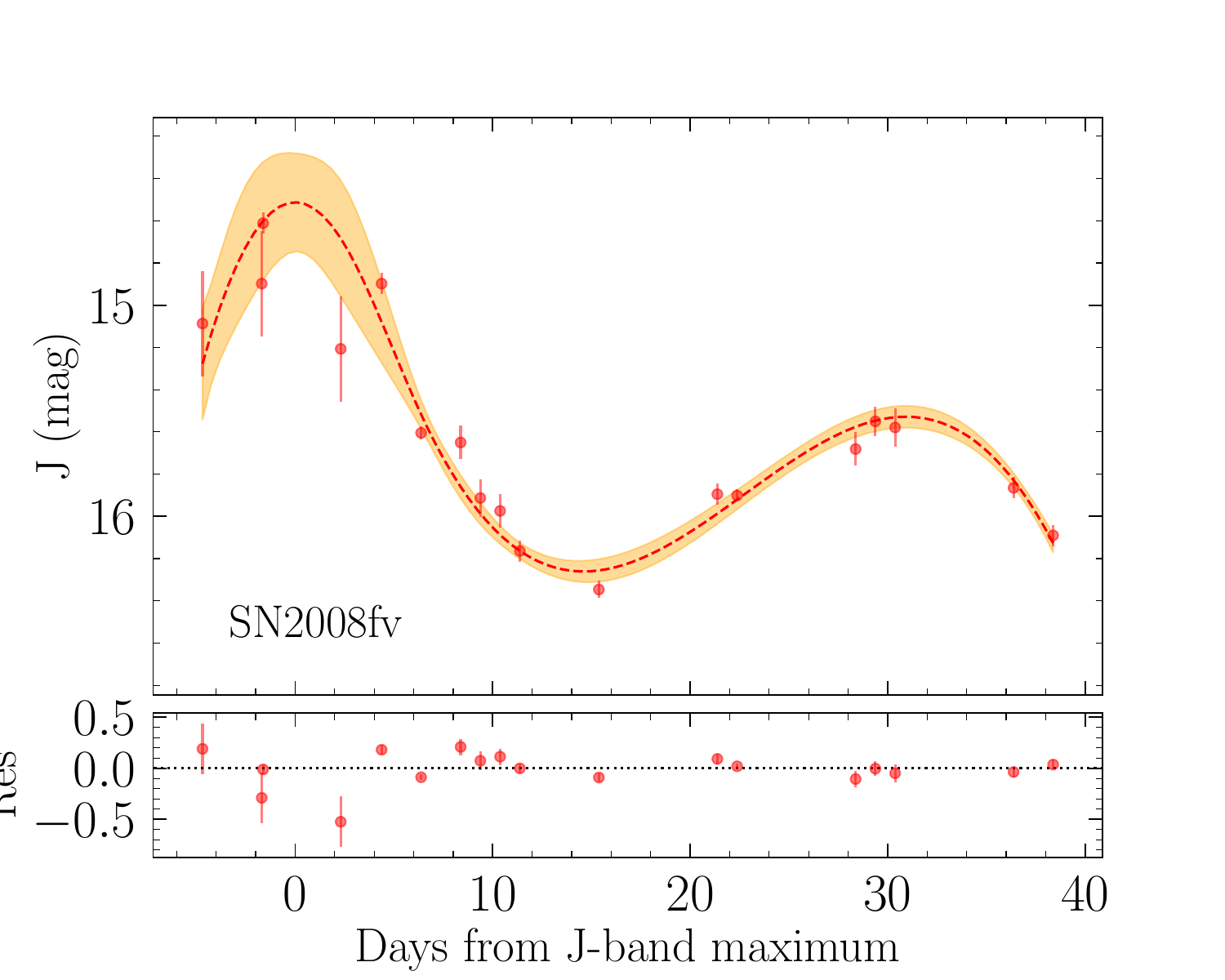}
    \includegraphics[trim=0.4cm 0.2cm 2cm 1cm, clip=True,width=0.32\textwidth]{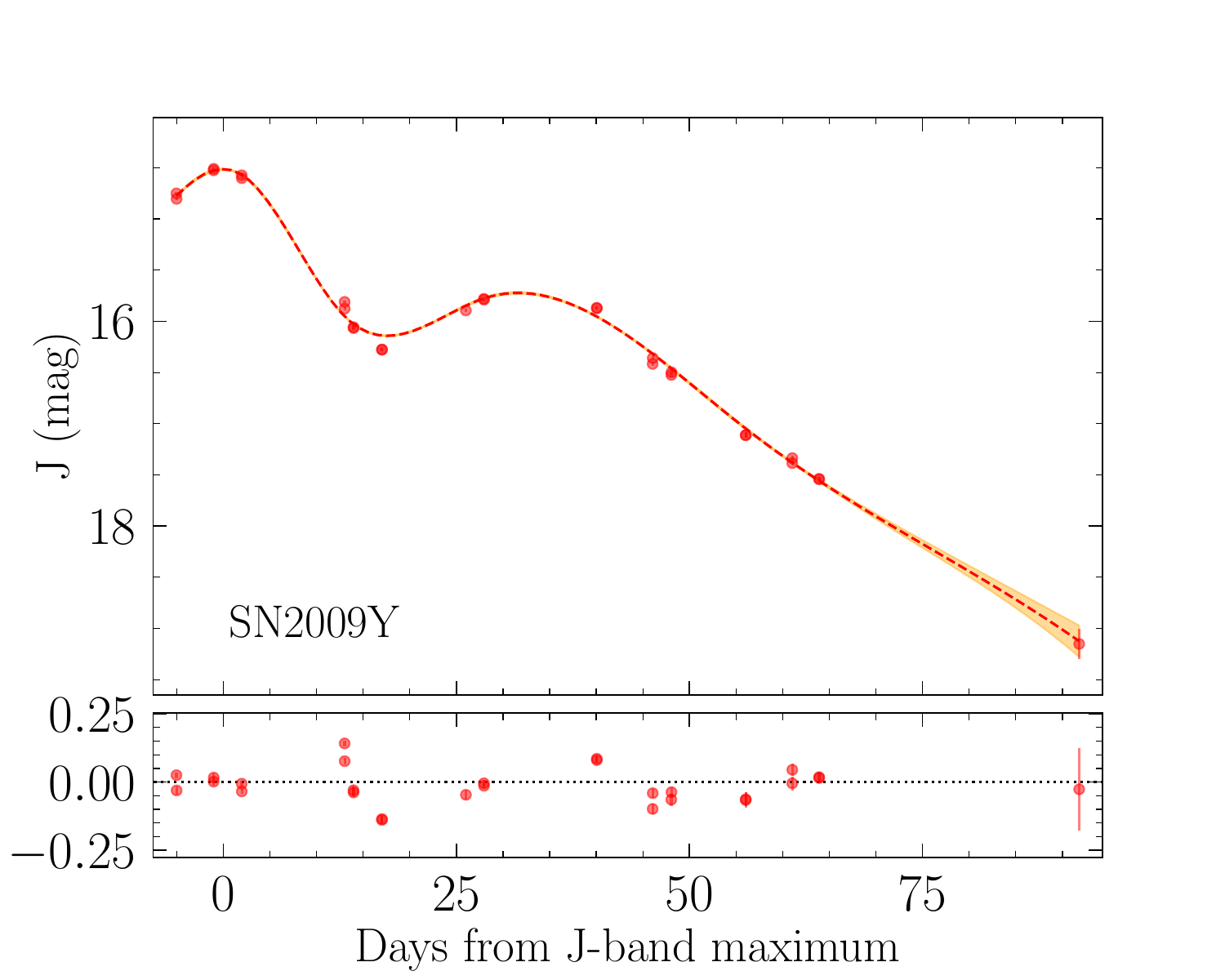}
    \includegraphics[trim=0.4cm 0.2cm 2cm 1cm, clip=True,width=0.32\textwidth]{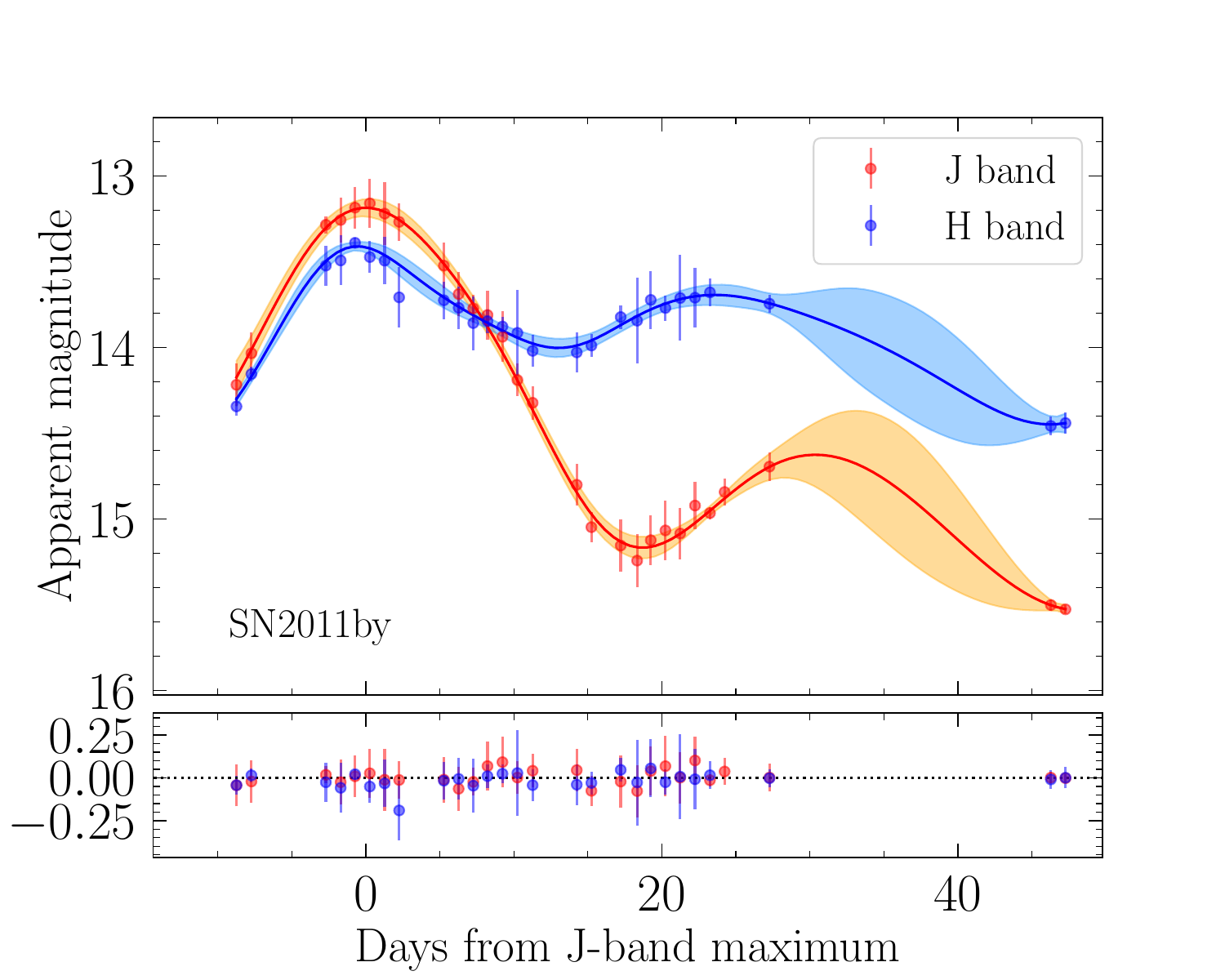}
    \caption{All calibrator gaussian process  (solid lines) or spline fits (dashed lines) in $J$ (red) and $H$ (blue) bands.}
    \label{fig:calibratorslc}
\end{figure*}

Figure \ref{fig:calibratorslc} shows the $J$ and $H$-band light-curves of the 19 SNe in our calibration sample along with the Gaussian process (solid lines) or spline best fits (dashed lines) used to derive peak magnitudes. In Figure \ref{fig:hubbleflowlc} similar plots are presented for the Hubble-flow sample.

\begin{figure*}[!ht]
\centering
    \includegraphics[trim=0.4cm 0.2cm 2cm 1cm, clip=True,width=0.32\textwidth]{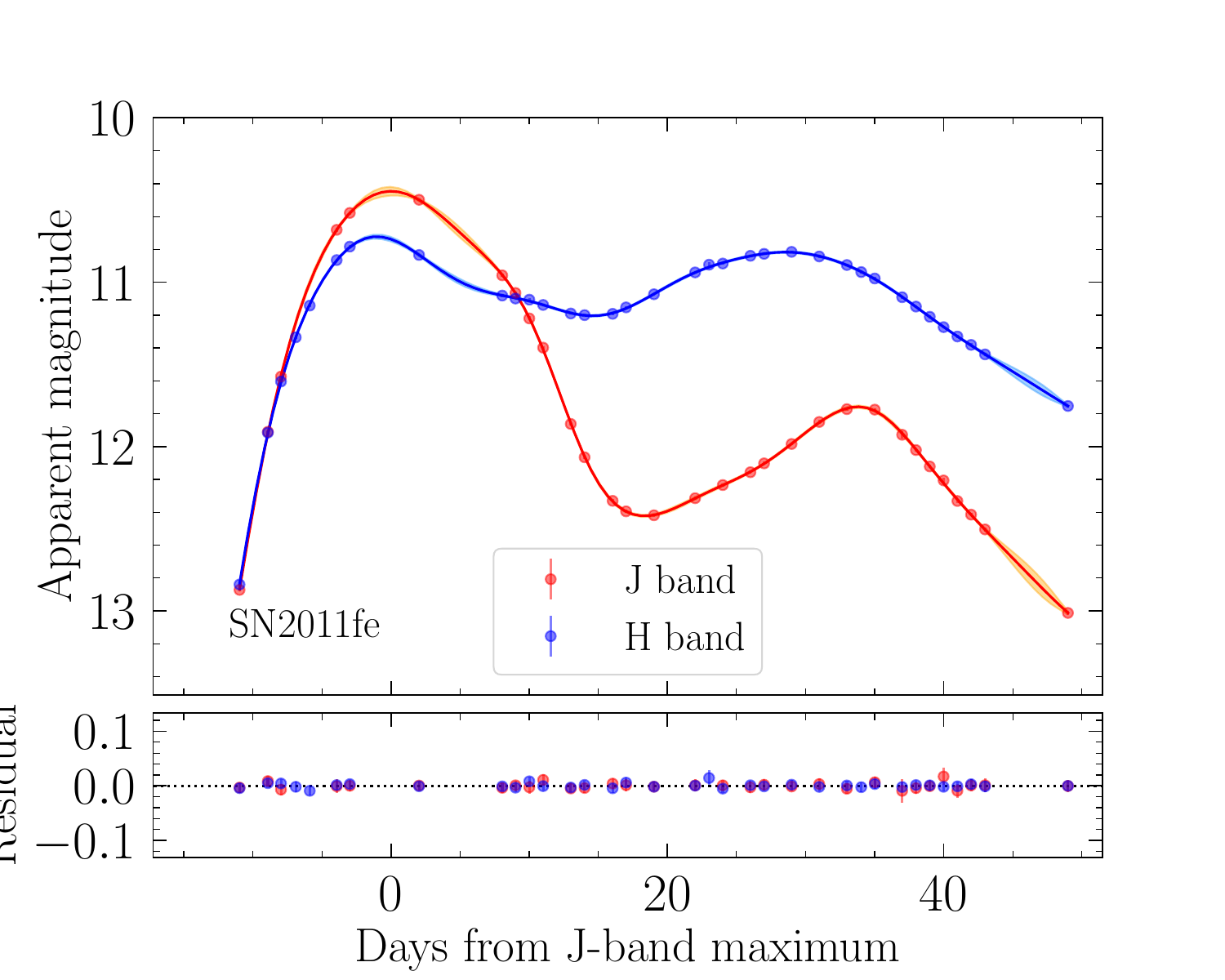}
    \includegraphics[trim=0.4cm 0.2cm 2cm 1cm, clip=True,width=0.32\textwidth]{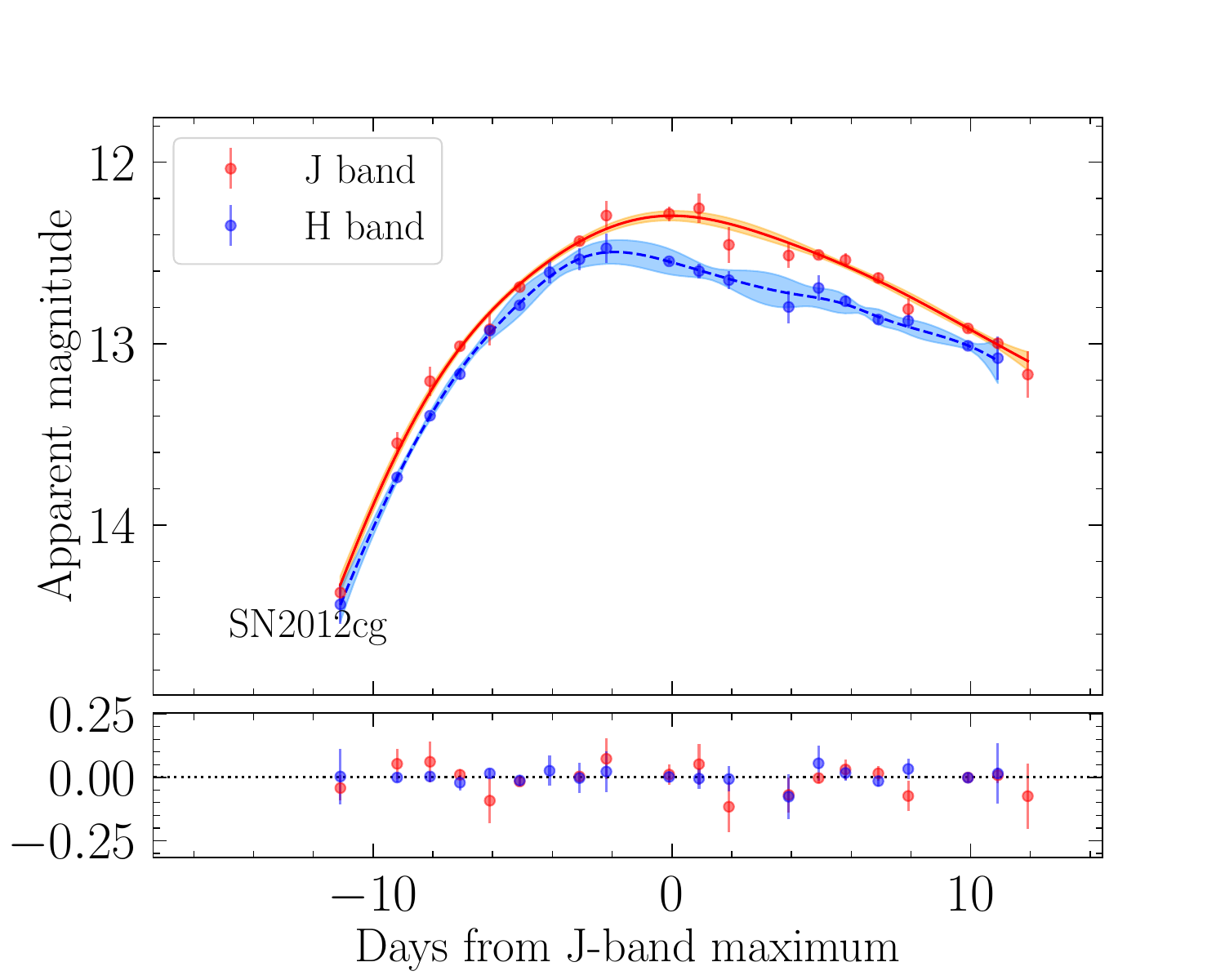}
    \includegraphics[trim=0.4cm 0.2cm 2cm 1cm, clip=True,width=0.32\textwidth]{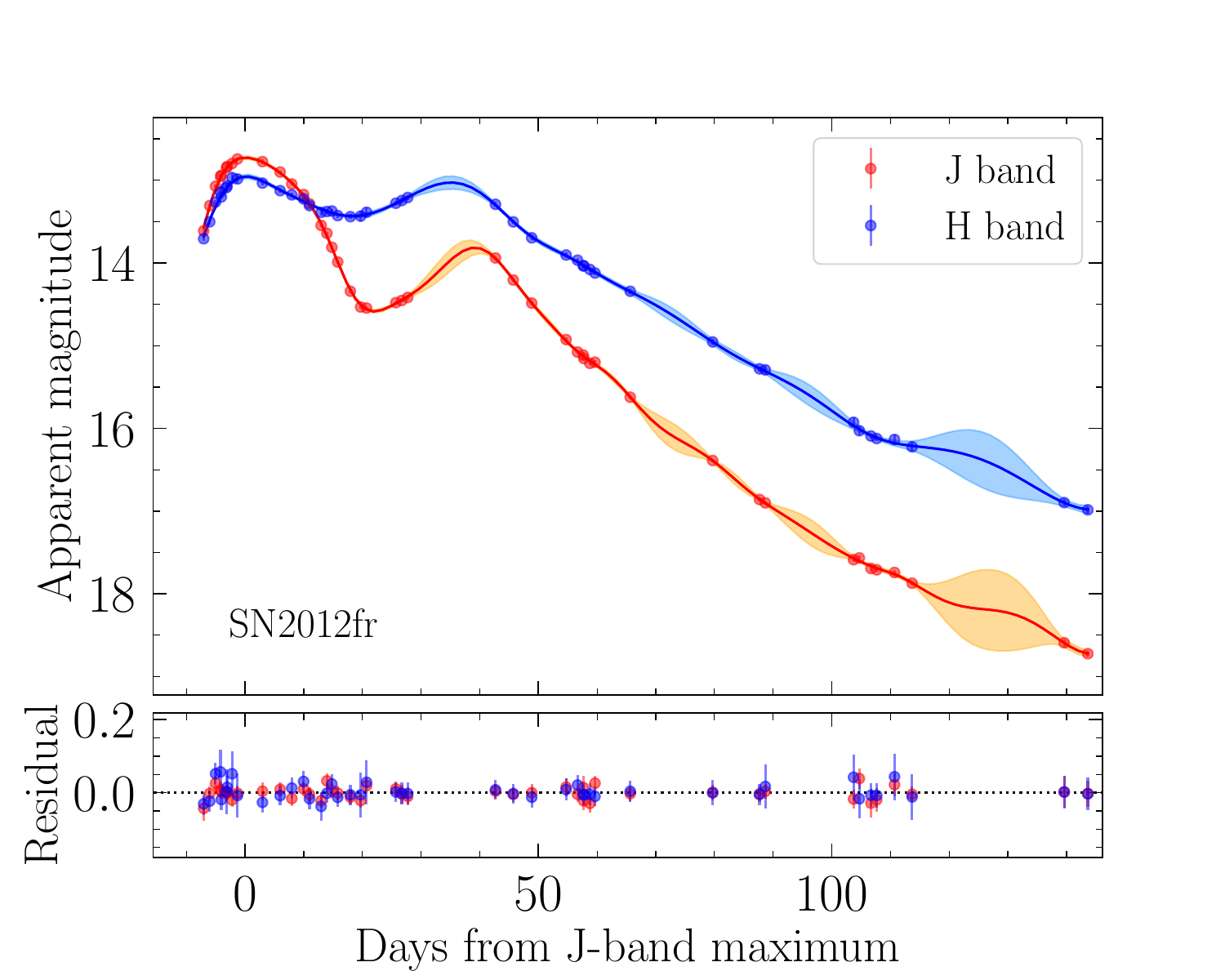}
    \includegraphics[trim=0.4cm 0.2cm 2cm 1cm, clip=True,width=0.32\textwidth]{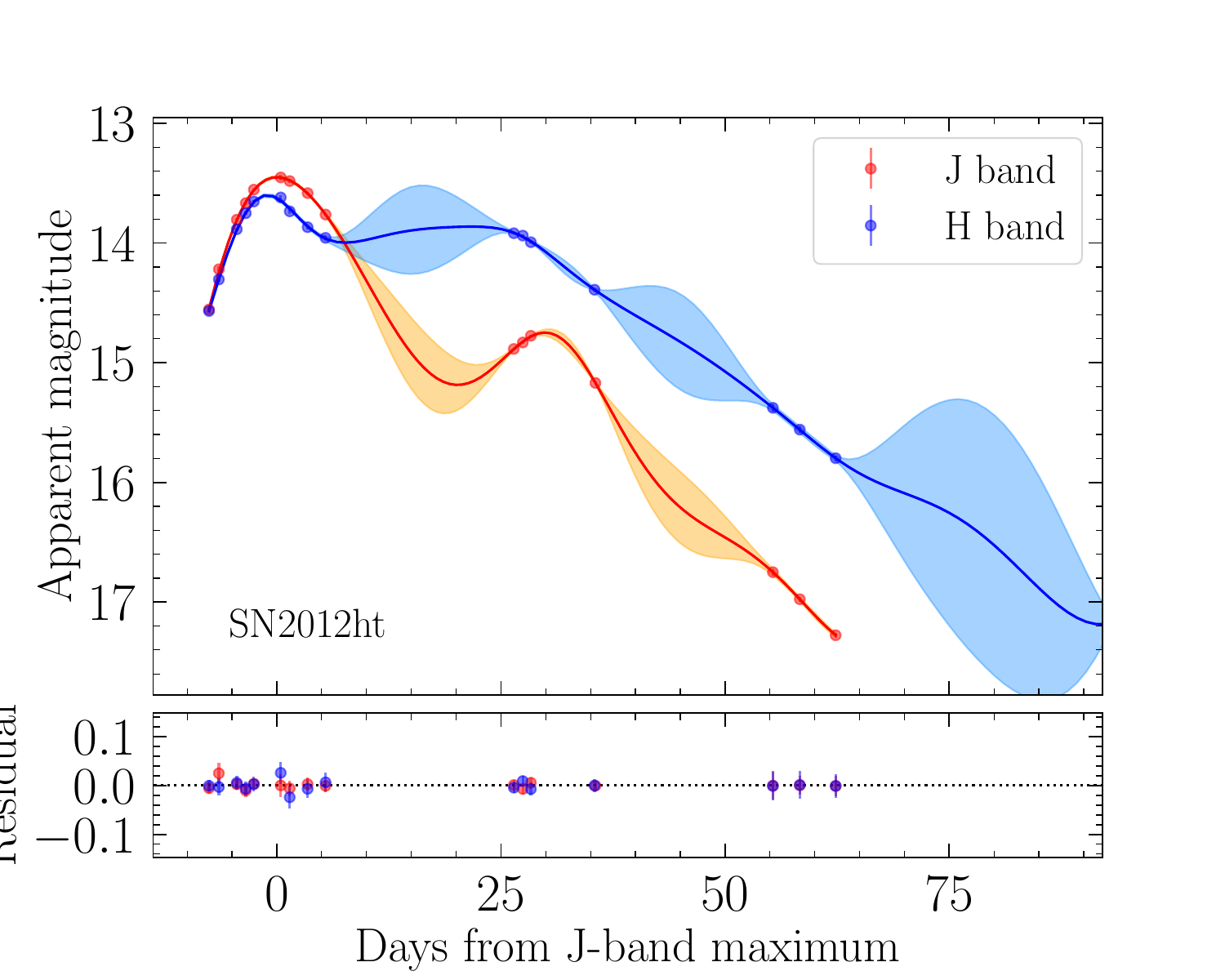}
    \includegraphics[trim=0.4cm 0.2cm 2cm 1cm, clip=True,width=0.32\textwidth]{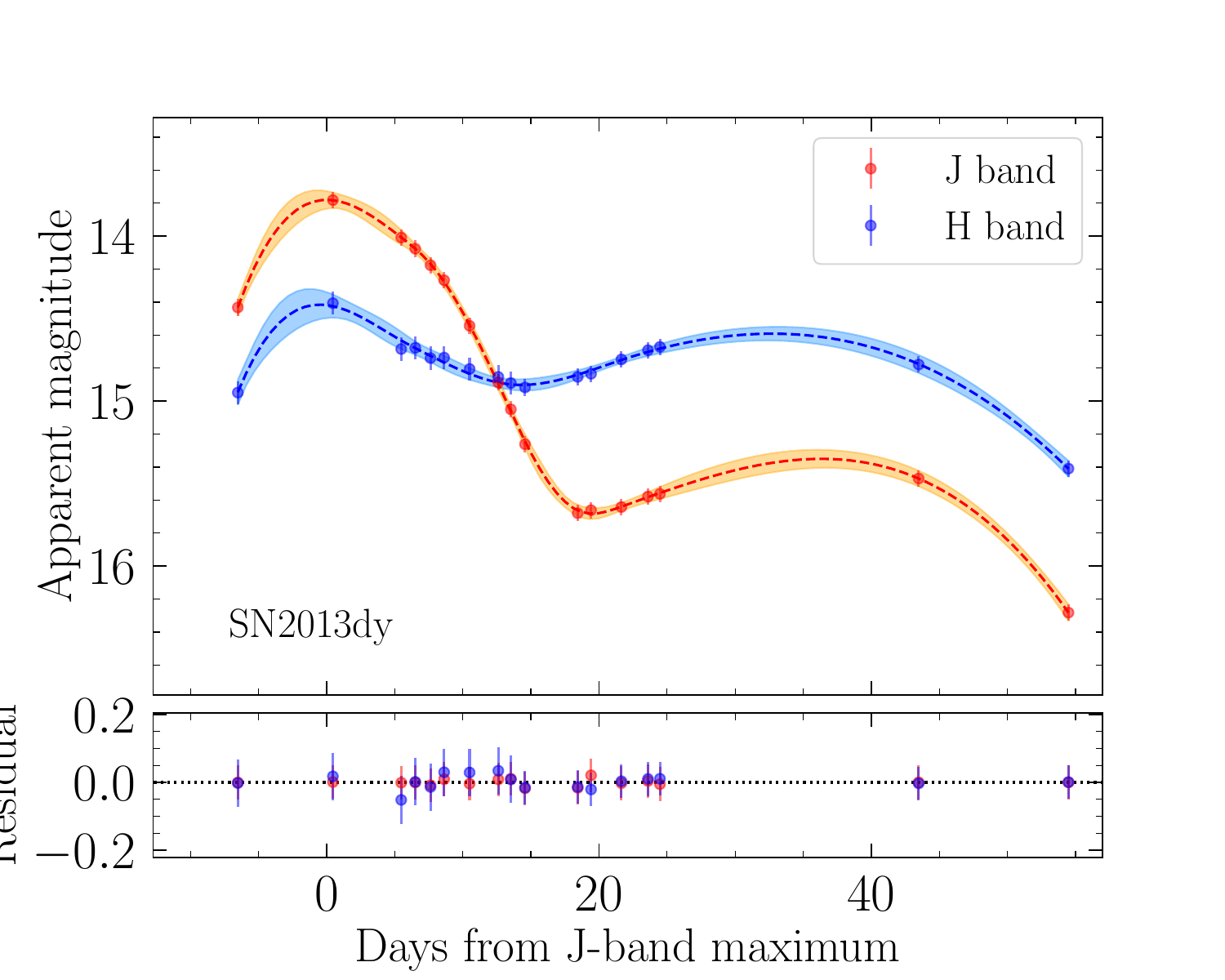}
    \includegraphics[trim=0.4cm 0.2cm 2cm 1cm, clip=True,width=0.32\textwidth]{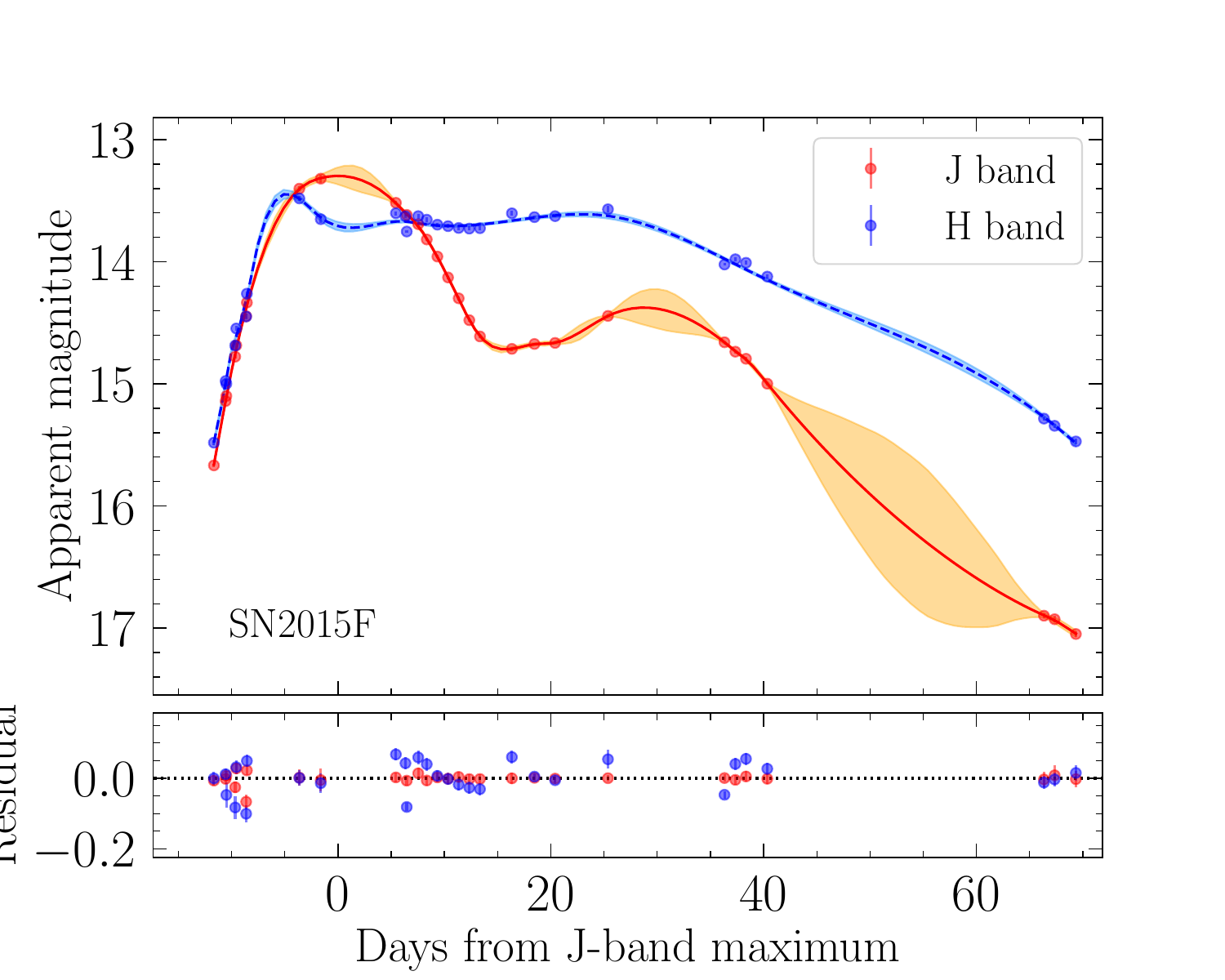}
    \includegraphics[trim=0.4cm 0.2cm 2cm 1cm, clip=True,width=0.32\textwidth]{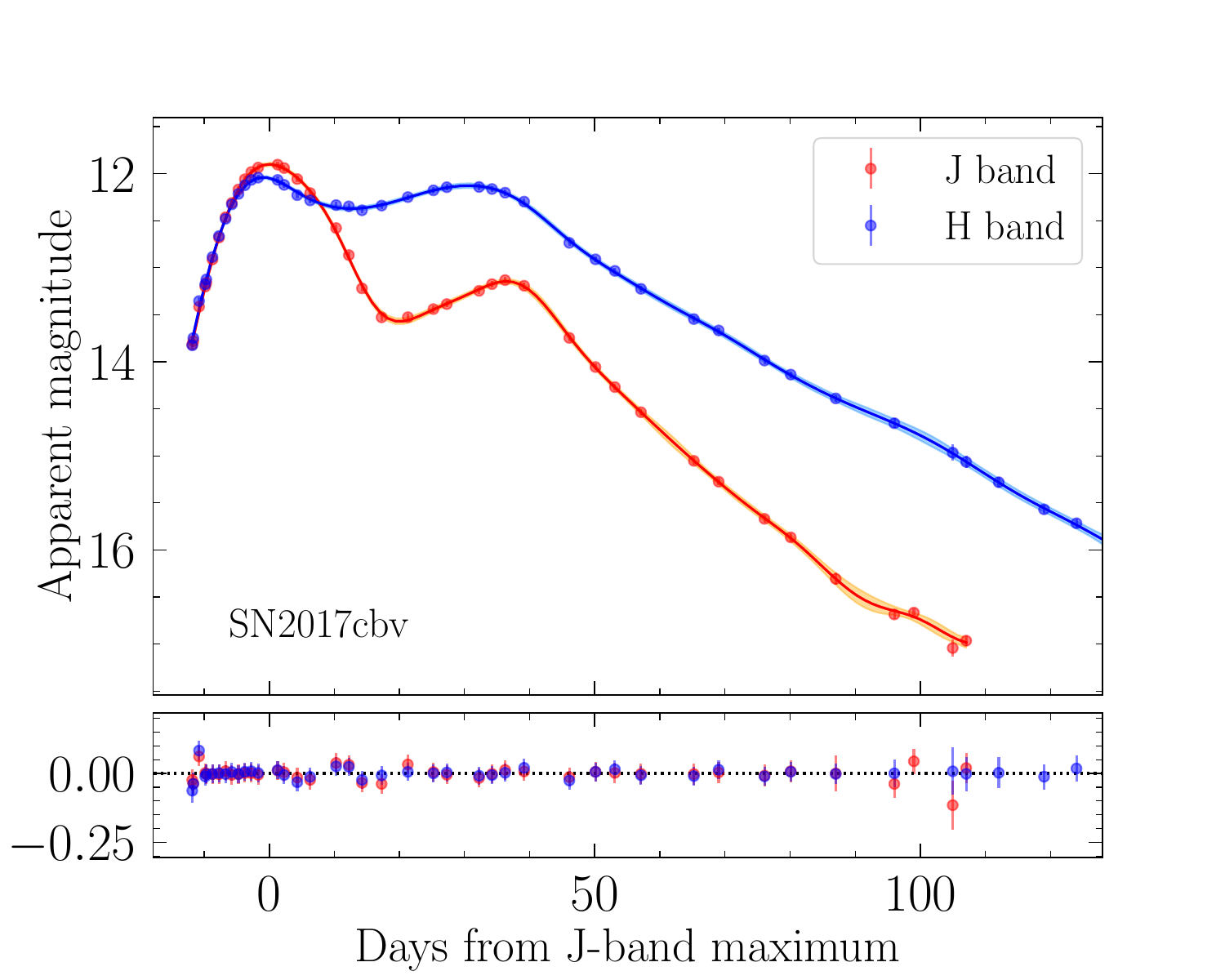}
    \caption{(continuing form Fig D1).}
\end{figure*}

\begin{figure*}
\centering
\includegraphics[trim=0cm 0.2cm 2cm 1cm, clip=True,width=0.32\textwidth]{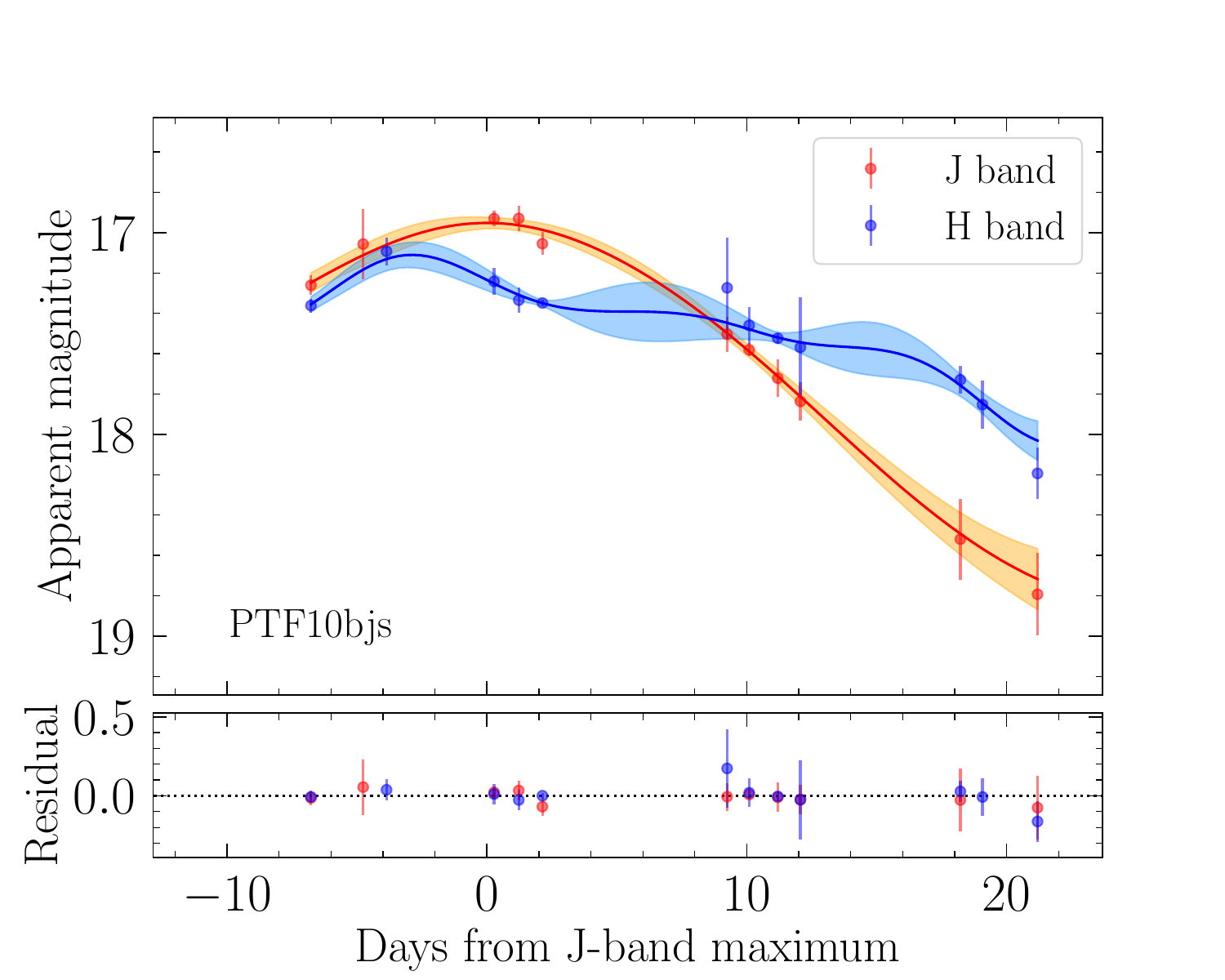}
\includegraphics[trim=0cm 0.2cm 2cm 1cm, clip=True,width=0.32\textwidth]{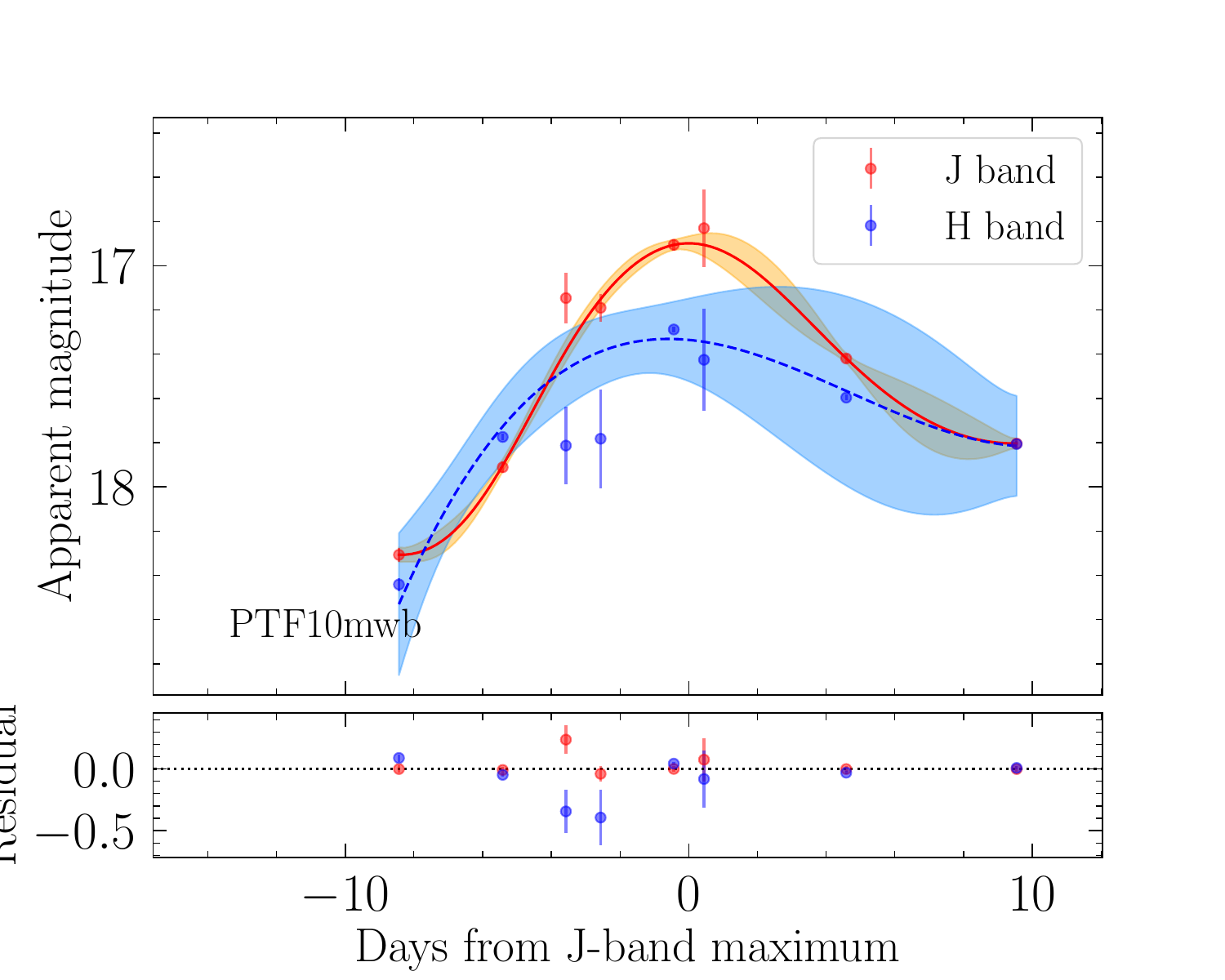}
\includegraphics[trim=0cm 0.2cm 2cm 1cm, clip=True,width=0.32\textwidth]{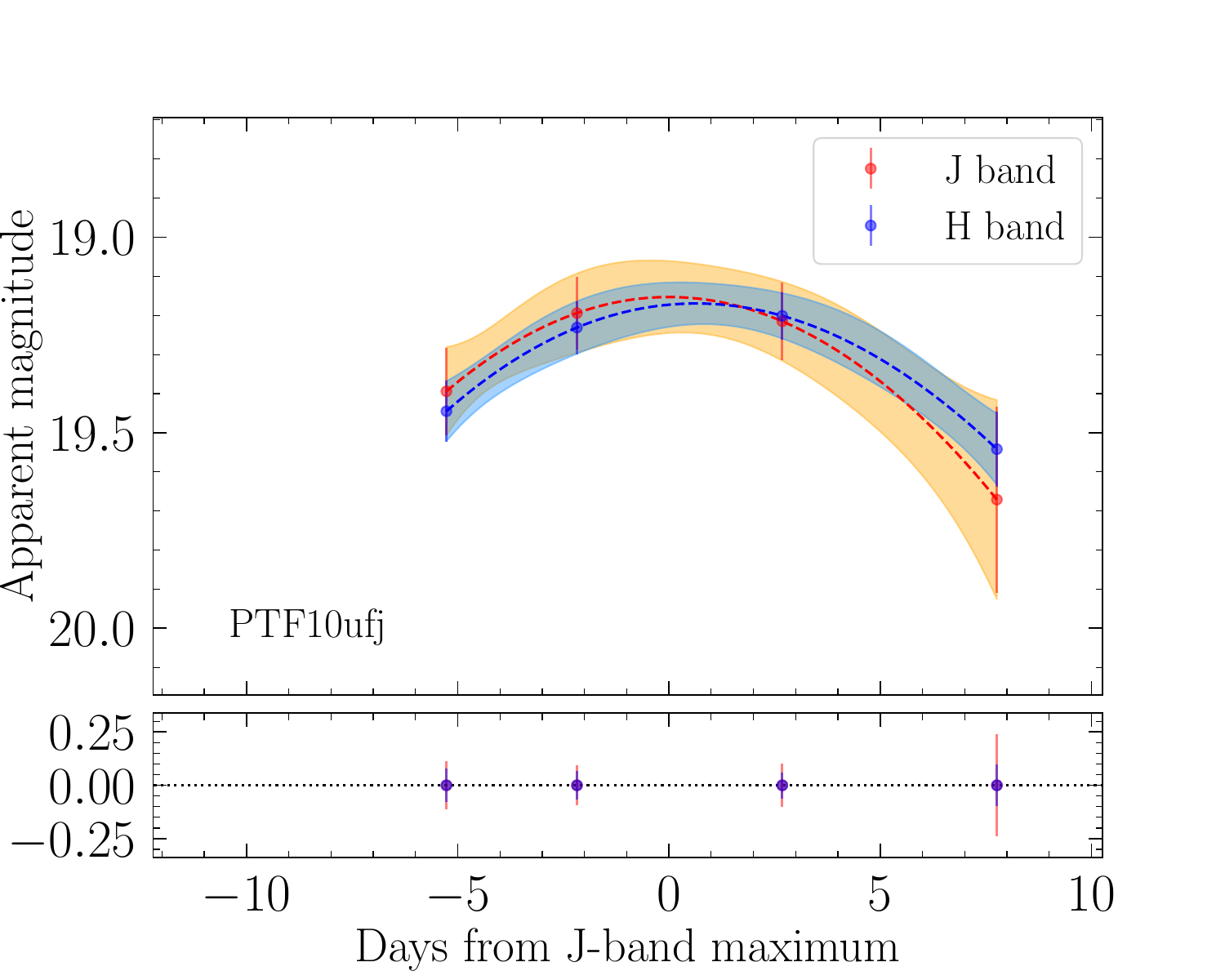}
\includegraphics[trim=0cm 0.2cm 2cm 1cm, clip=True,width=0.32\textwidth]{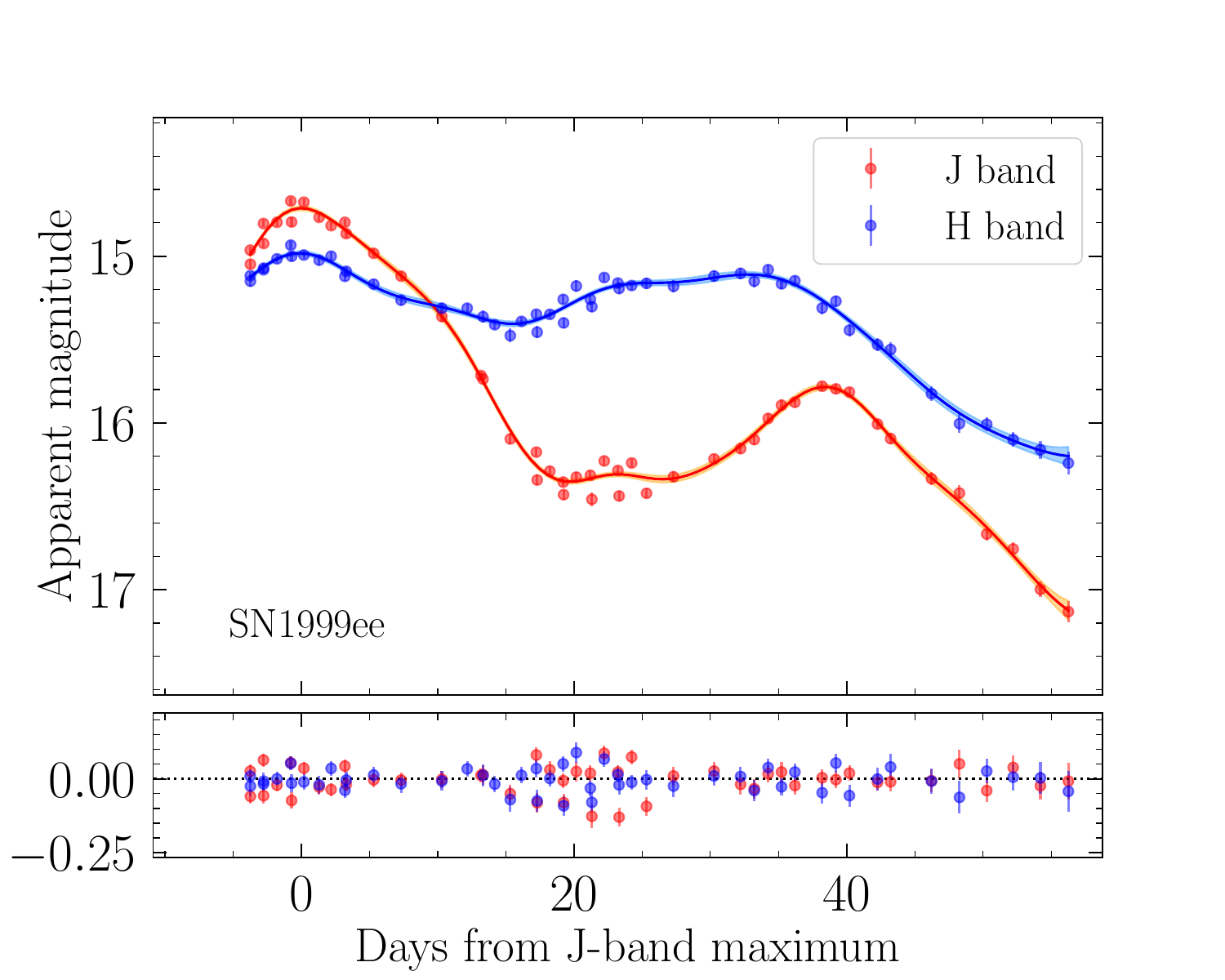}
\includegraphics[trim=0cm 0.2cm 2cm 1cm, clip=True,width=0.32\textwidth]{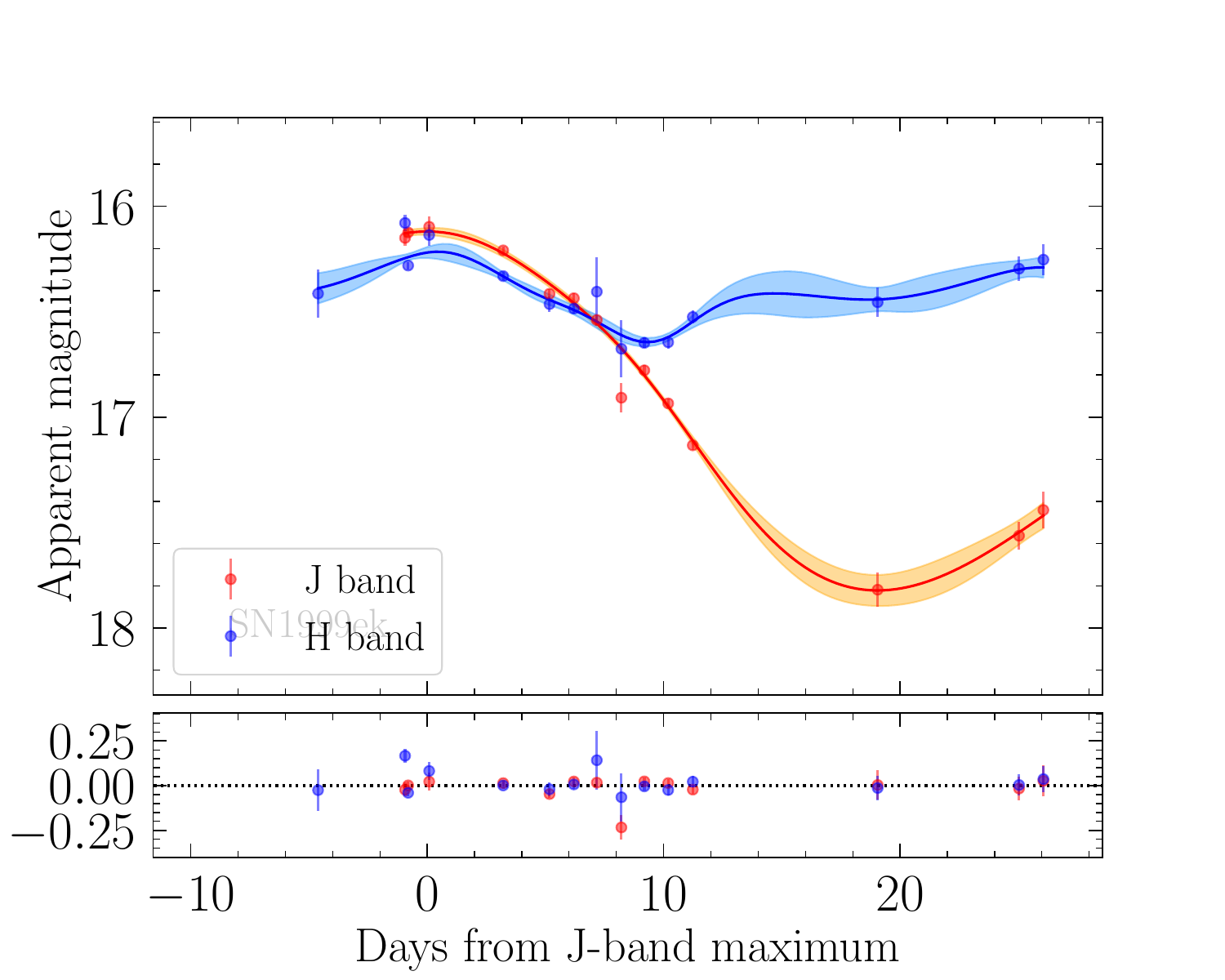}
\includegraphics[trim=0cm 0.2cm 2cm 1cm, clip=True,width=0.32\textwidth]{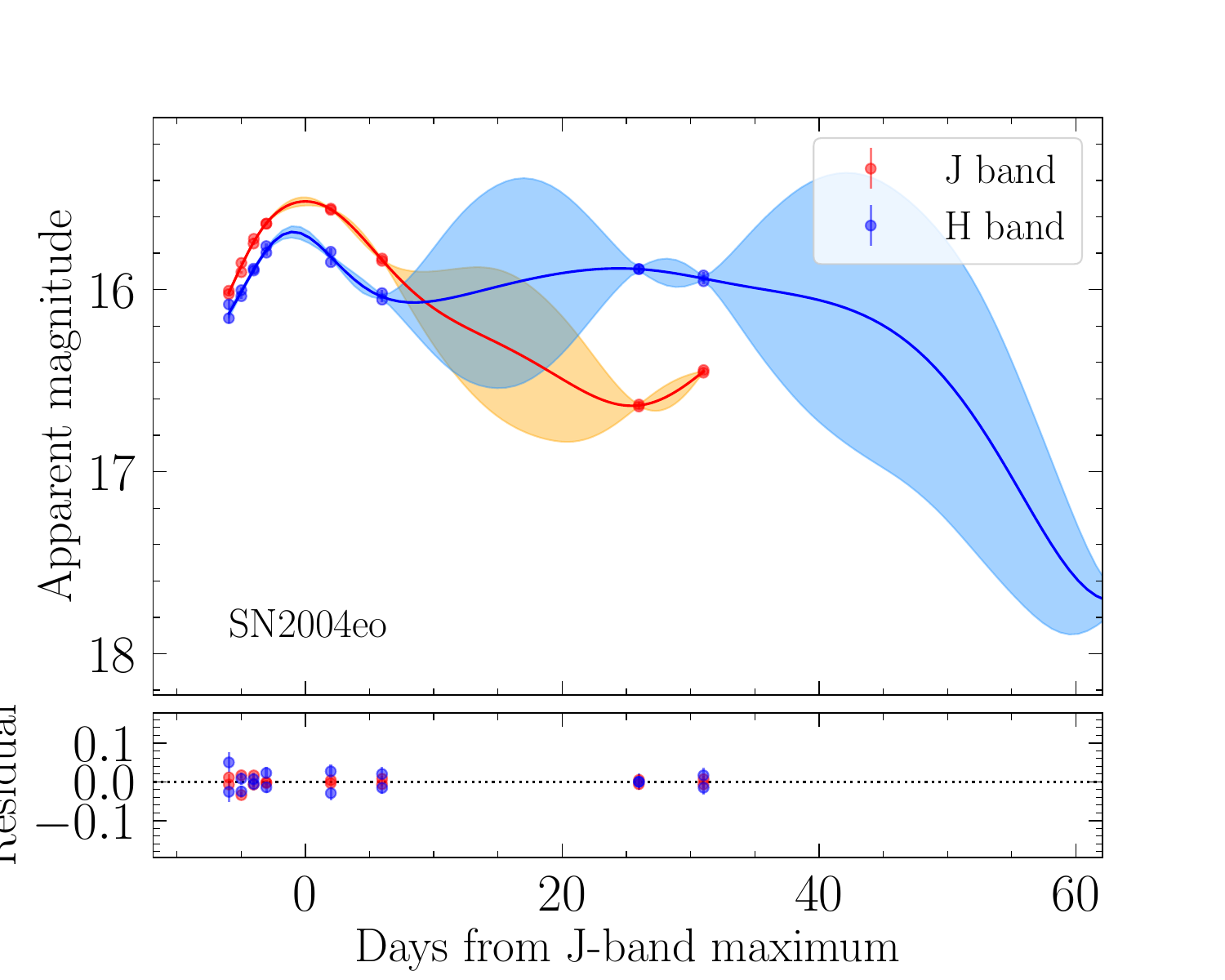}
\includegraphics[trim=0cm 0.2cm 2cm 1cm, clip=True,width=0.32\textwidth]{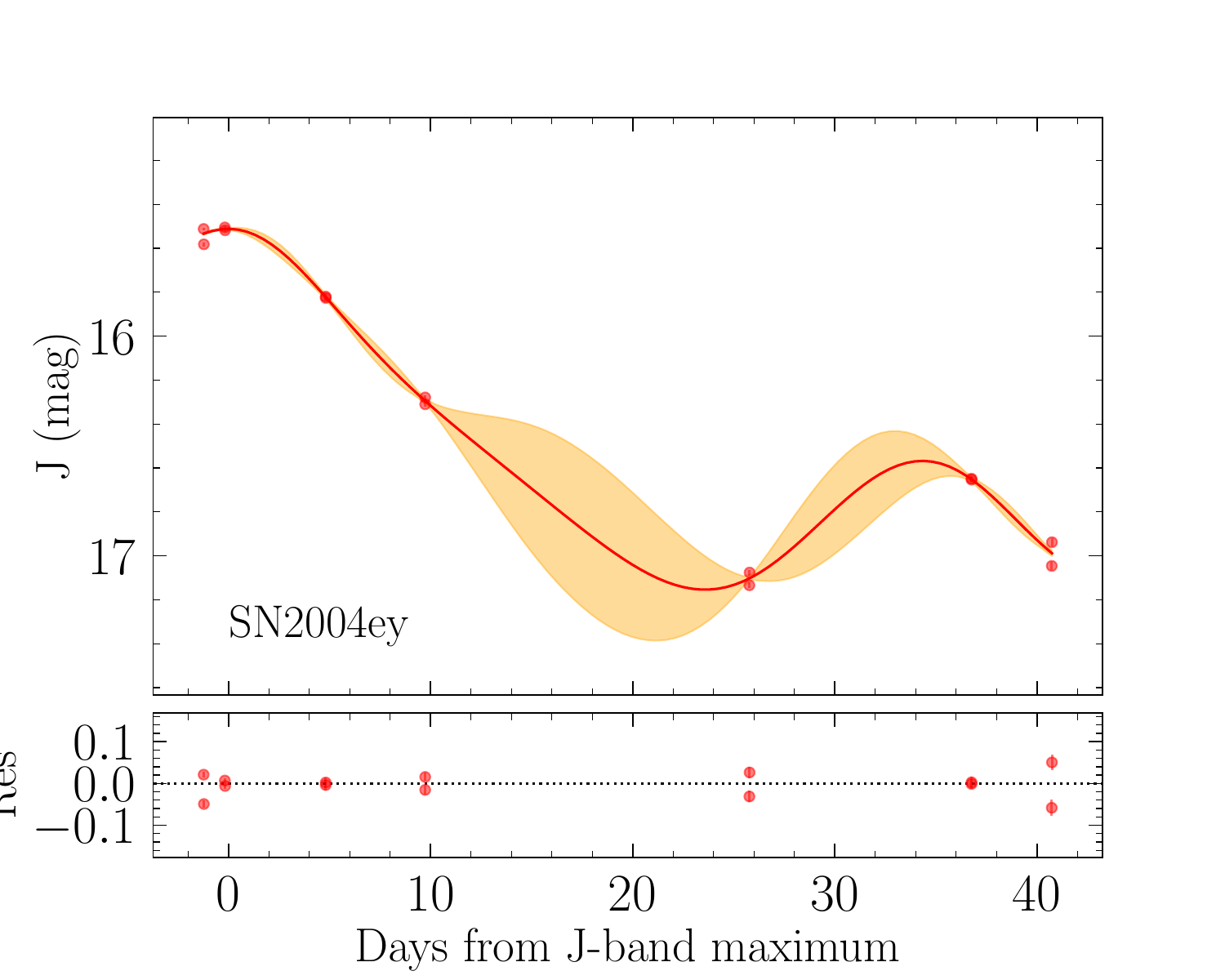}
\includegraphics[trim=0cm 0.2cm 2cm 1cm, clip=True,width=0.32\textwidth]{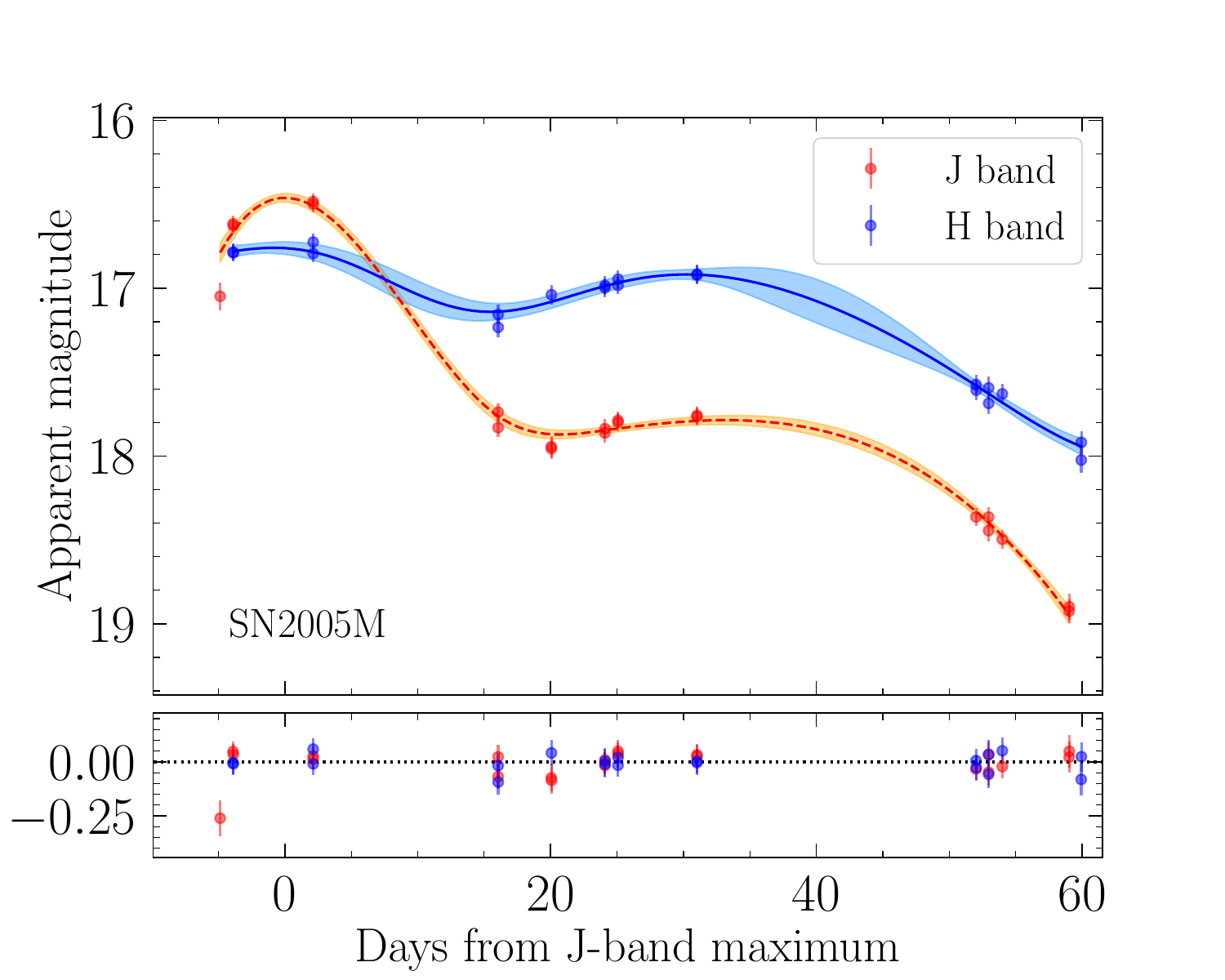}
\includegraphics[trim=0cm 0.2cm 2cm 1cm, clip=True,width=0.32\textwidth]{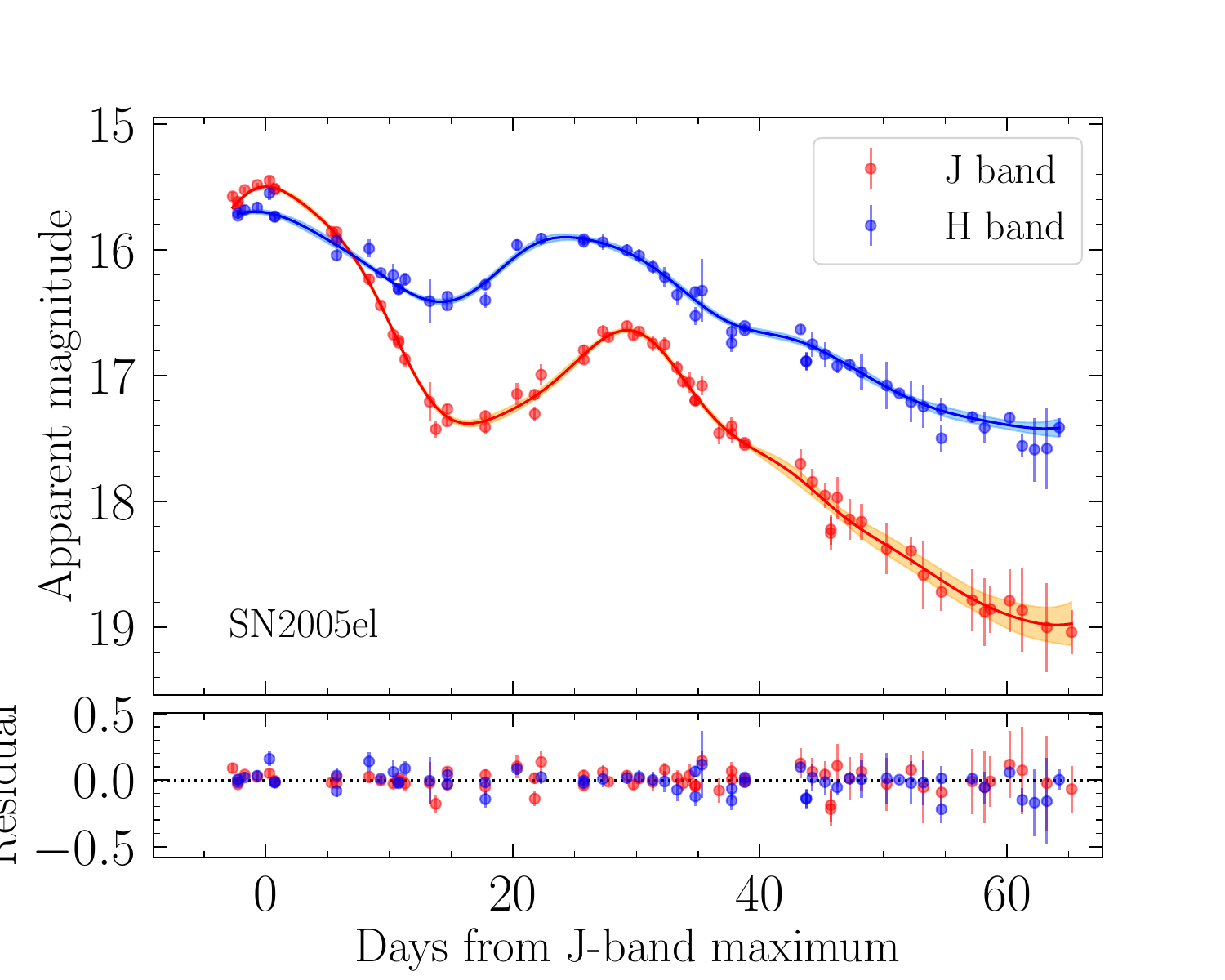}
\includegraphics[trim=0cm 0.2cm 2cm 1cm, clip=True,width=0.32\textwidth]{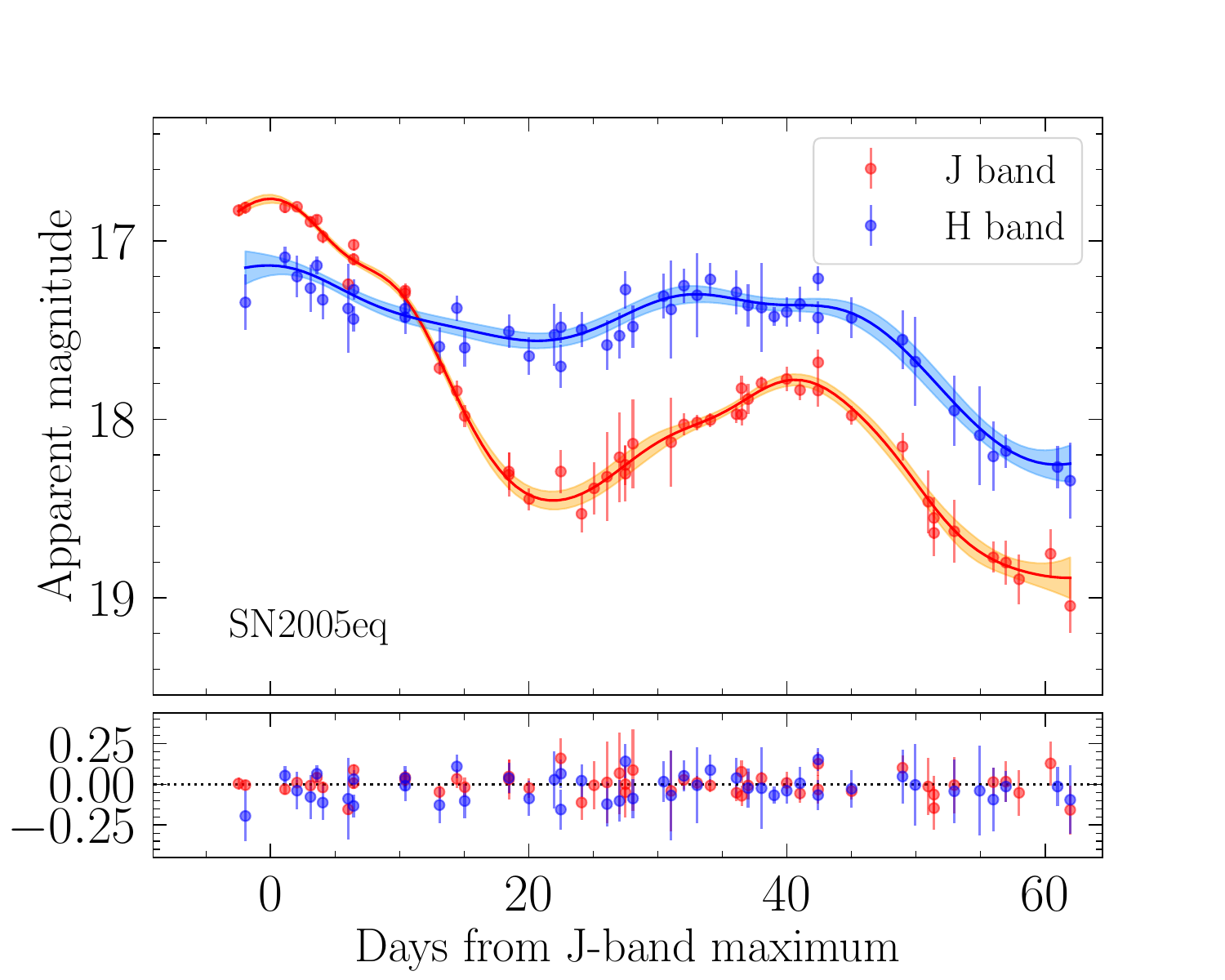}
\includegraphics[trim=0cm 0.2cm 2cm 1cm, clip=True,width=0.32\textwidth]{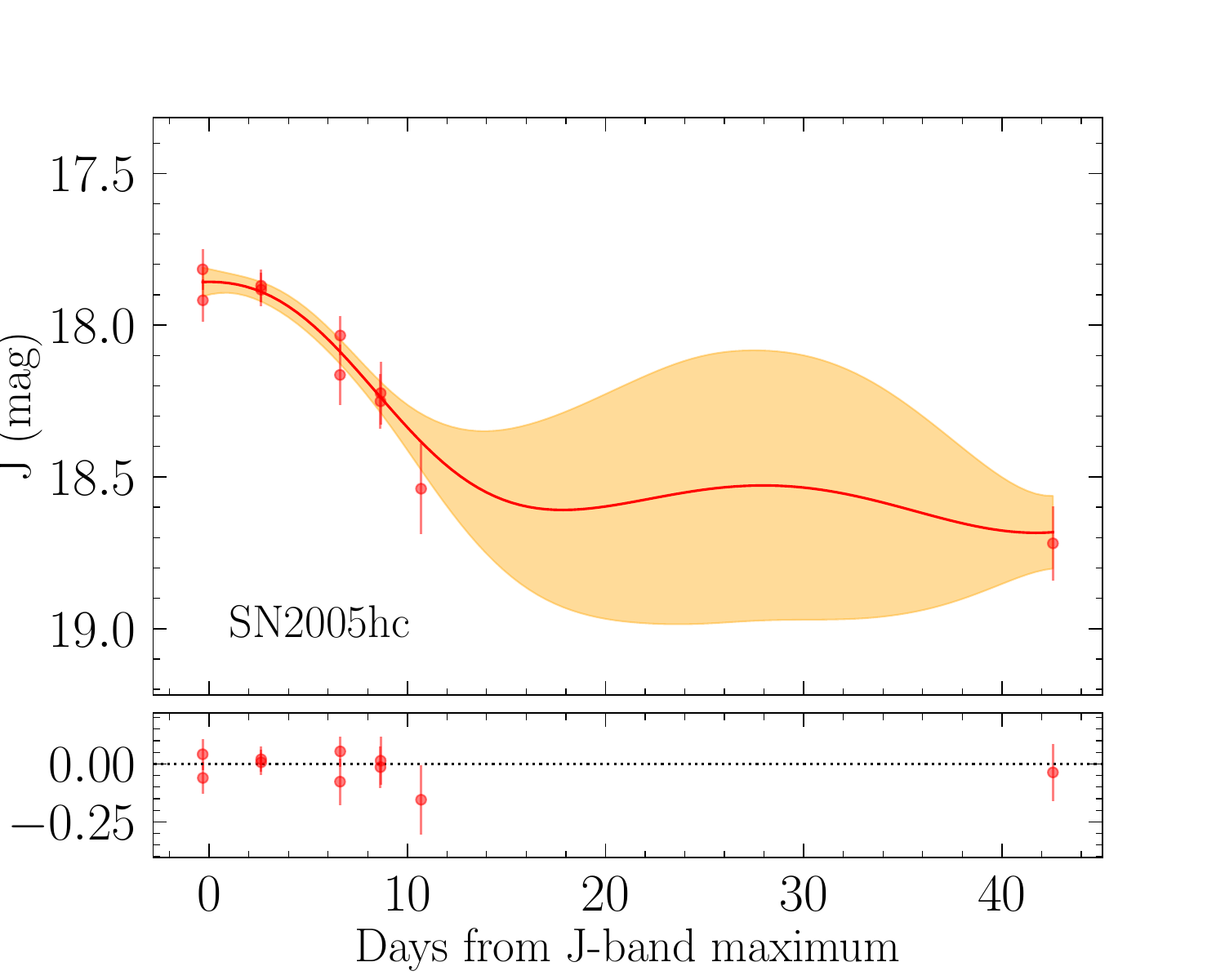}
\includegraphics[trim=0cm 0.2cm 2cm 1cm, clip=True,width=0.32\textwidth]{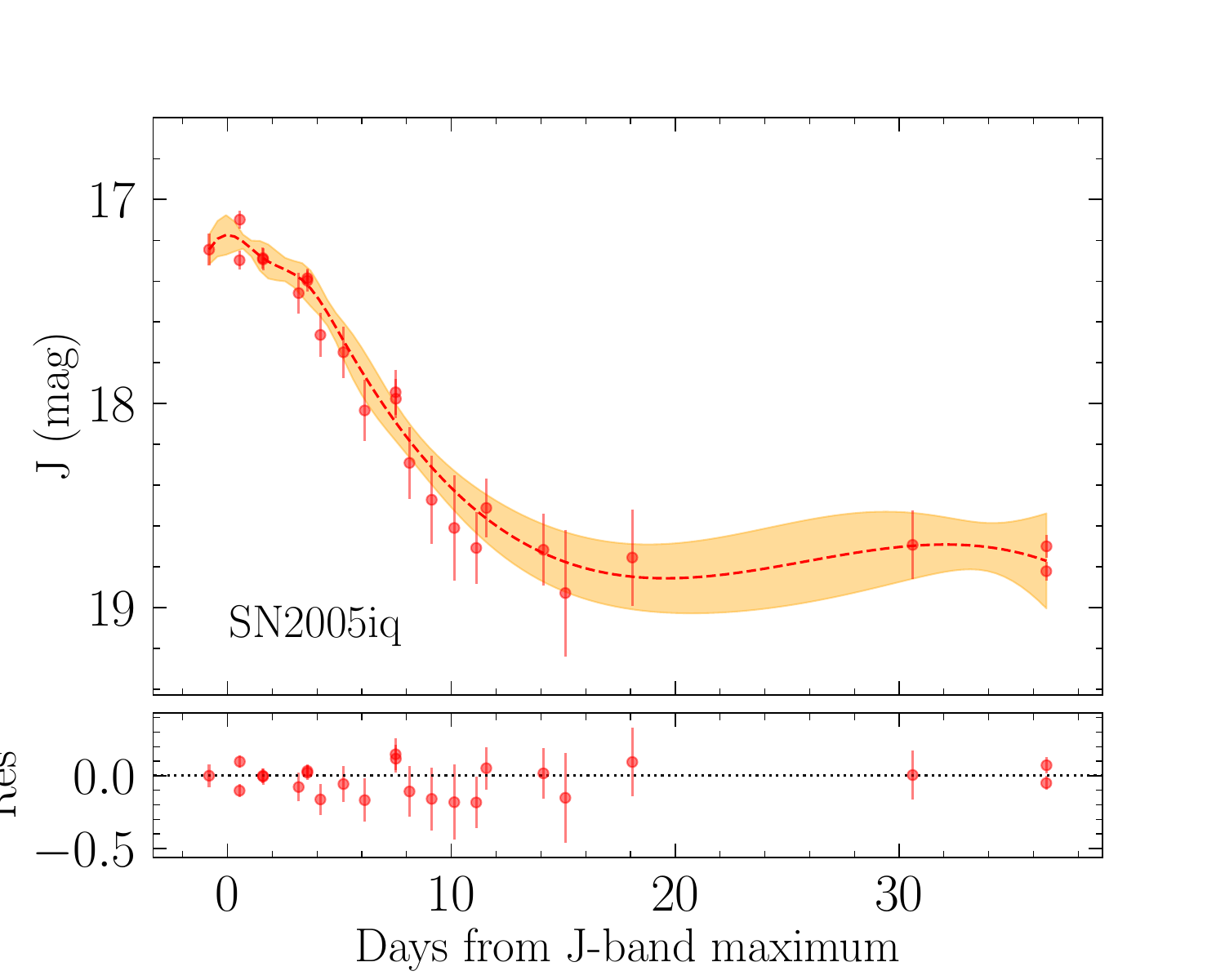}
\includegraphics[trim=0cm 0.2cm 2cm 1cm, clip=True,width=0.32\textwidth]{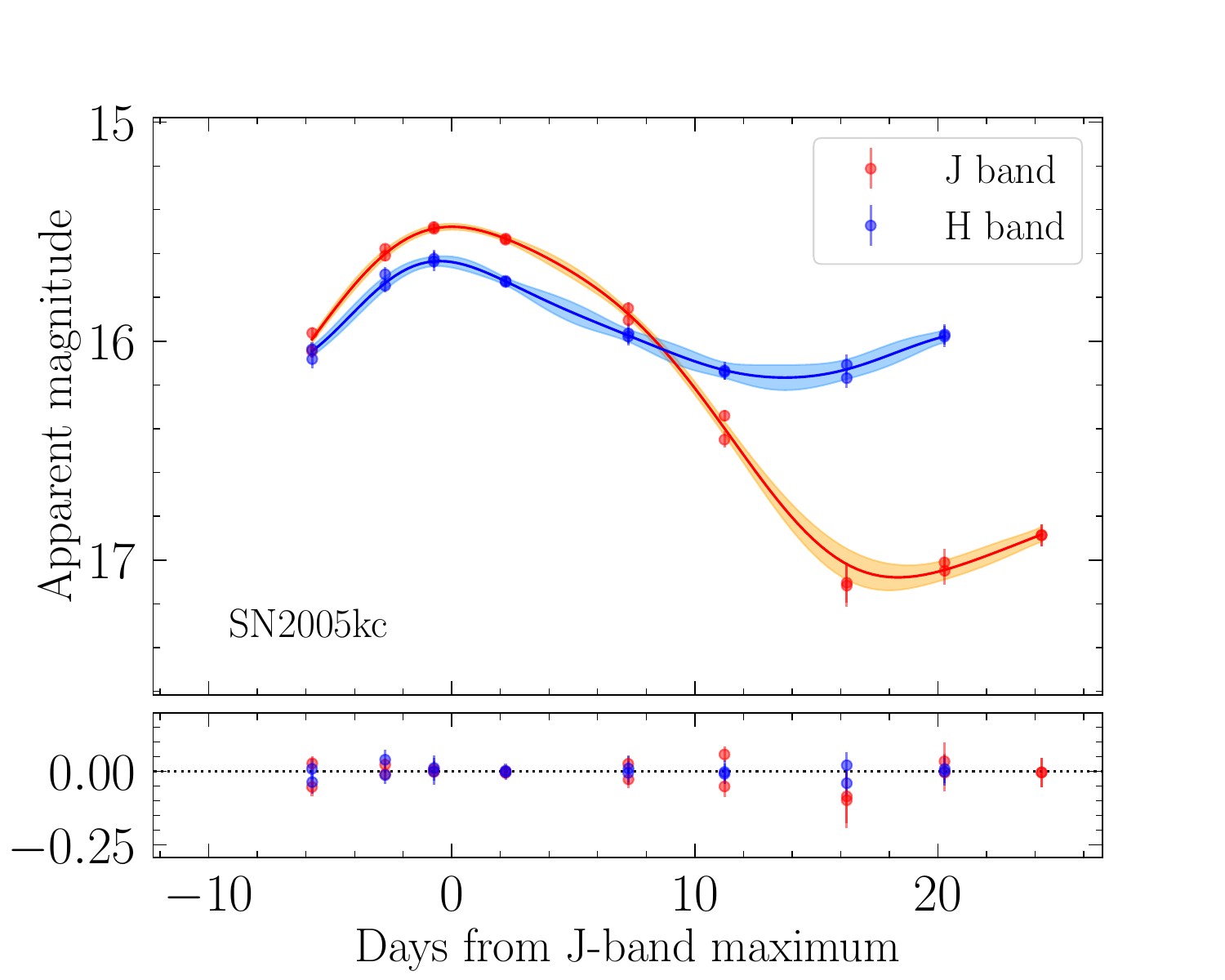}
\includegraphics[trim=0cm 0.2cm 2cm 1cm, clip=True,width=0.32\textwidth]{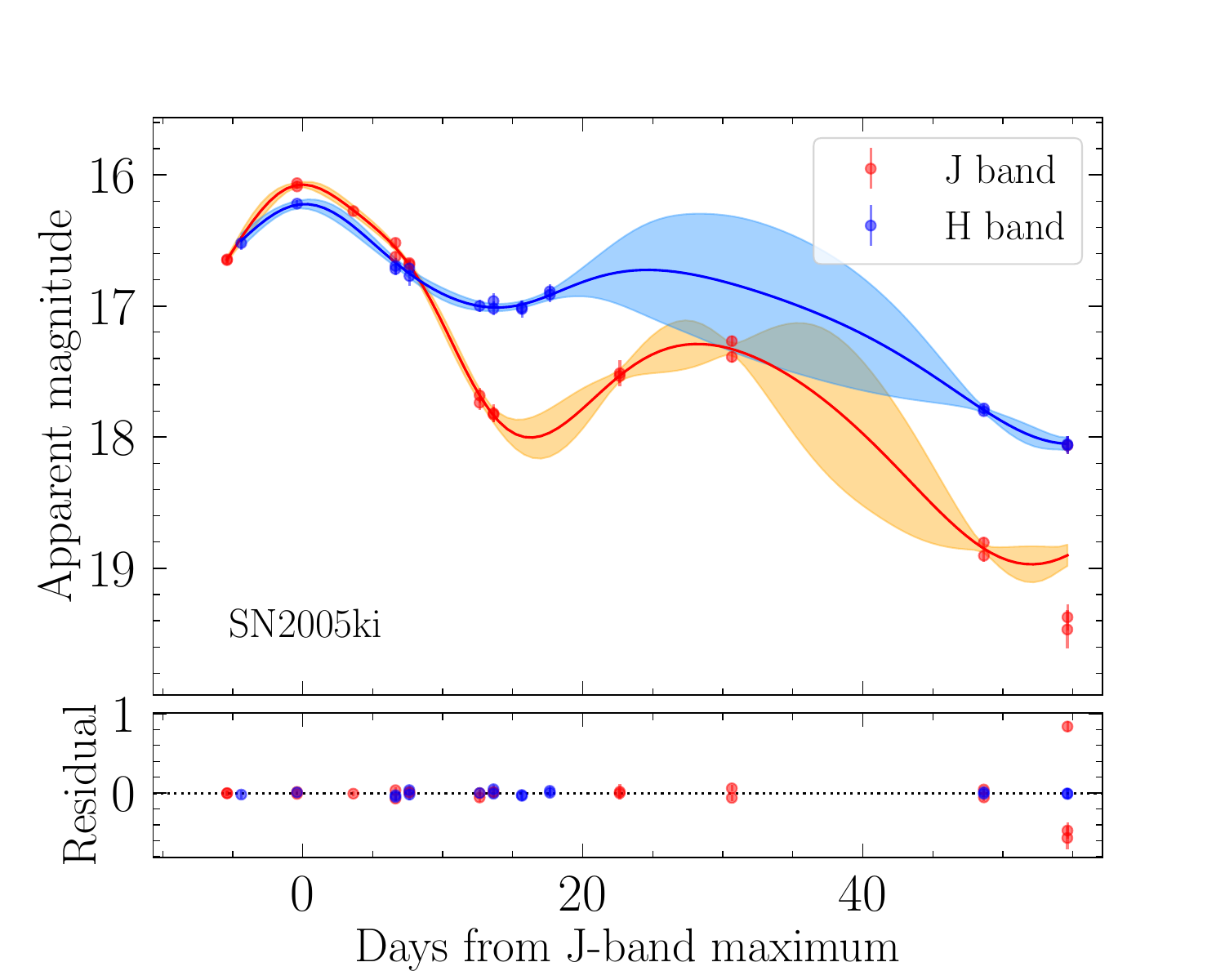}
\includegraphics[trim=0cm 0.2cm 2cm 1cm, clip=True,width=0.32\textwidth]{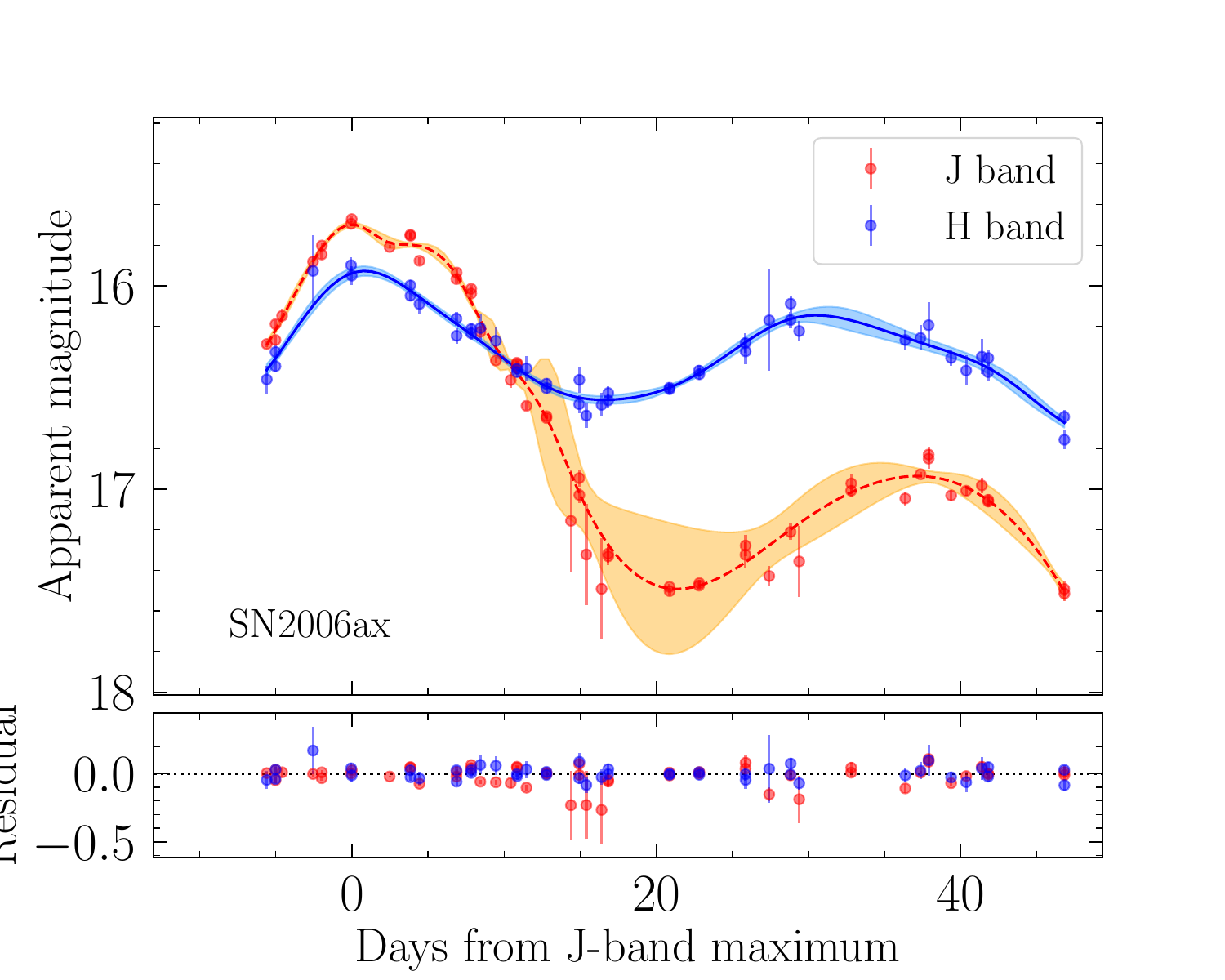}
\caption{All Hubble-flow gaussian process (solid lines) or spline fits (dashed lines) in $J$ (red) and $H$ (blue) bands.}
\label{fig:hubbleflowlc}
\end{figure*}
\begin{figure*}[!ht]
\centering
\includegraphics[trim=0cm 0.2cm 2cm 1cm, clip=True,width=0.32\textwidth]{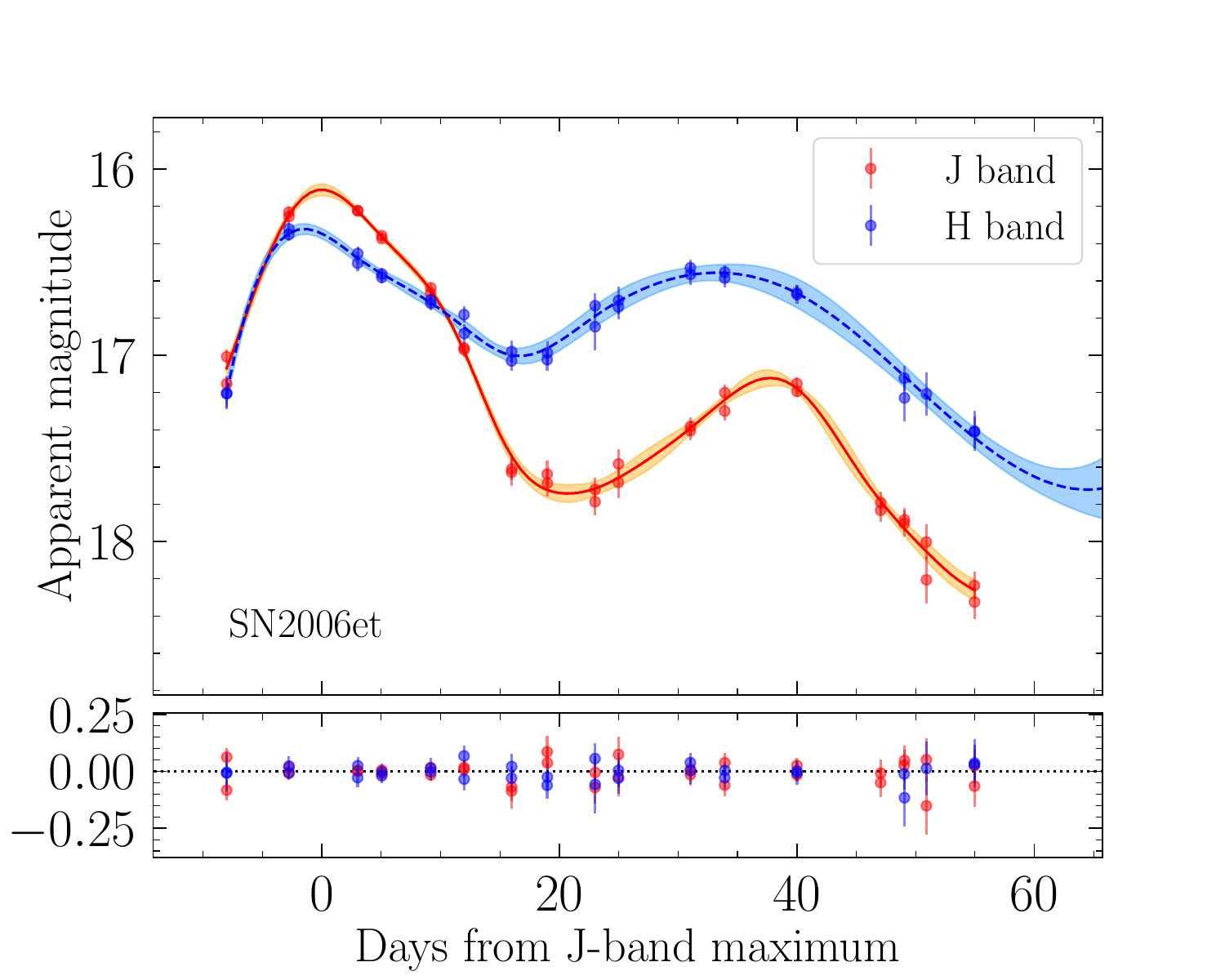}
\includegraphics[trim=0cm 0.2cm 2cm 1cm, clip=True,width=0.32\textwidth]{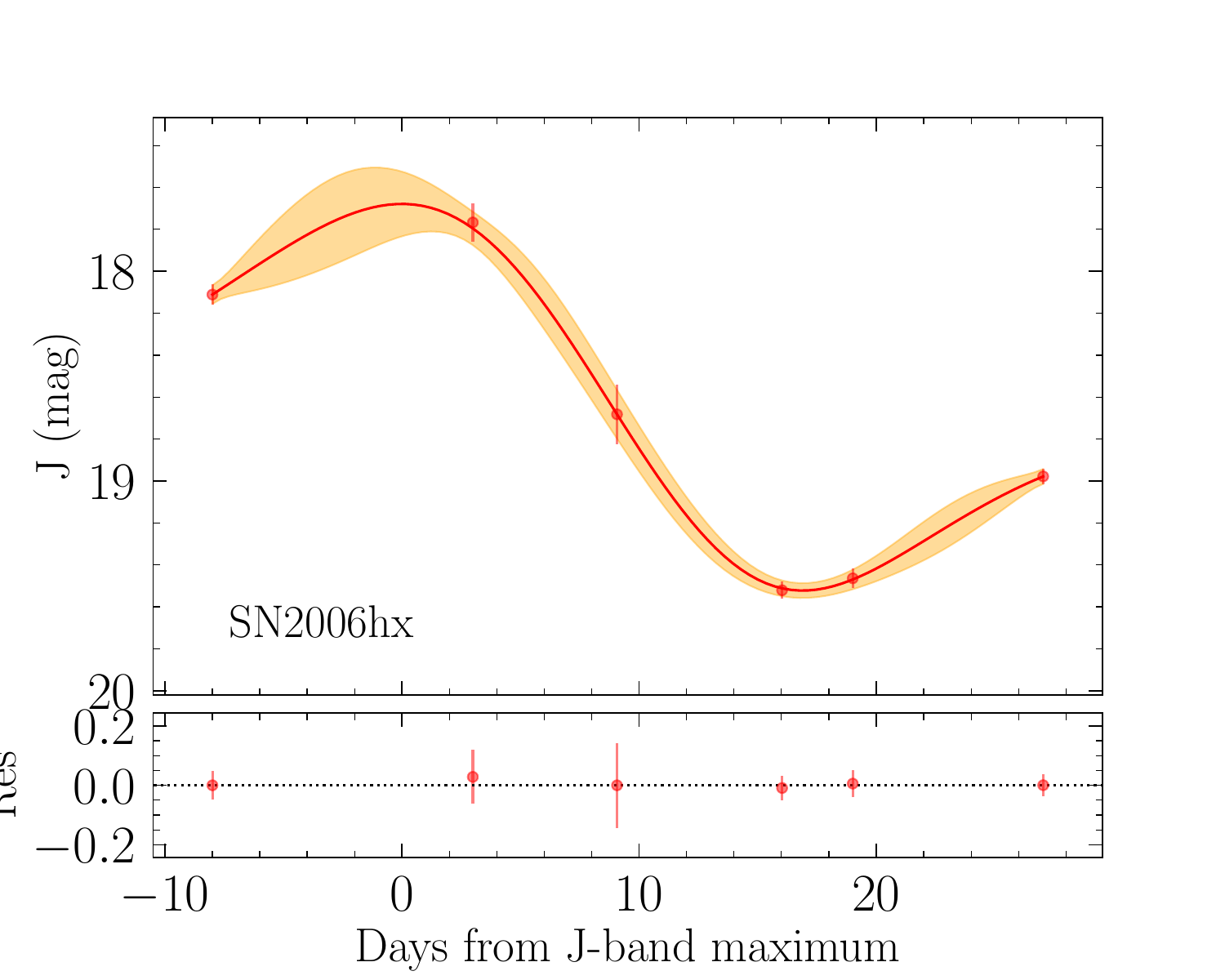}
\includegraphics[trim=0cm 0.2cm 2cm 1cm, clip=True,width=0.32\textwidth]{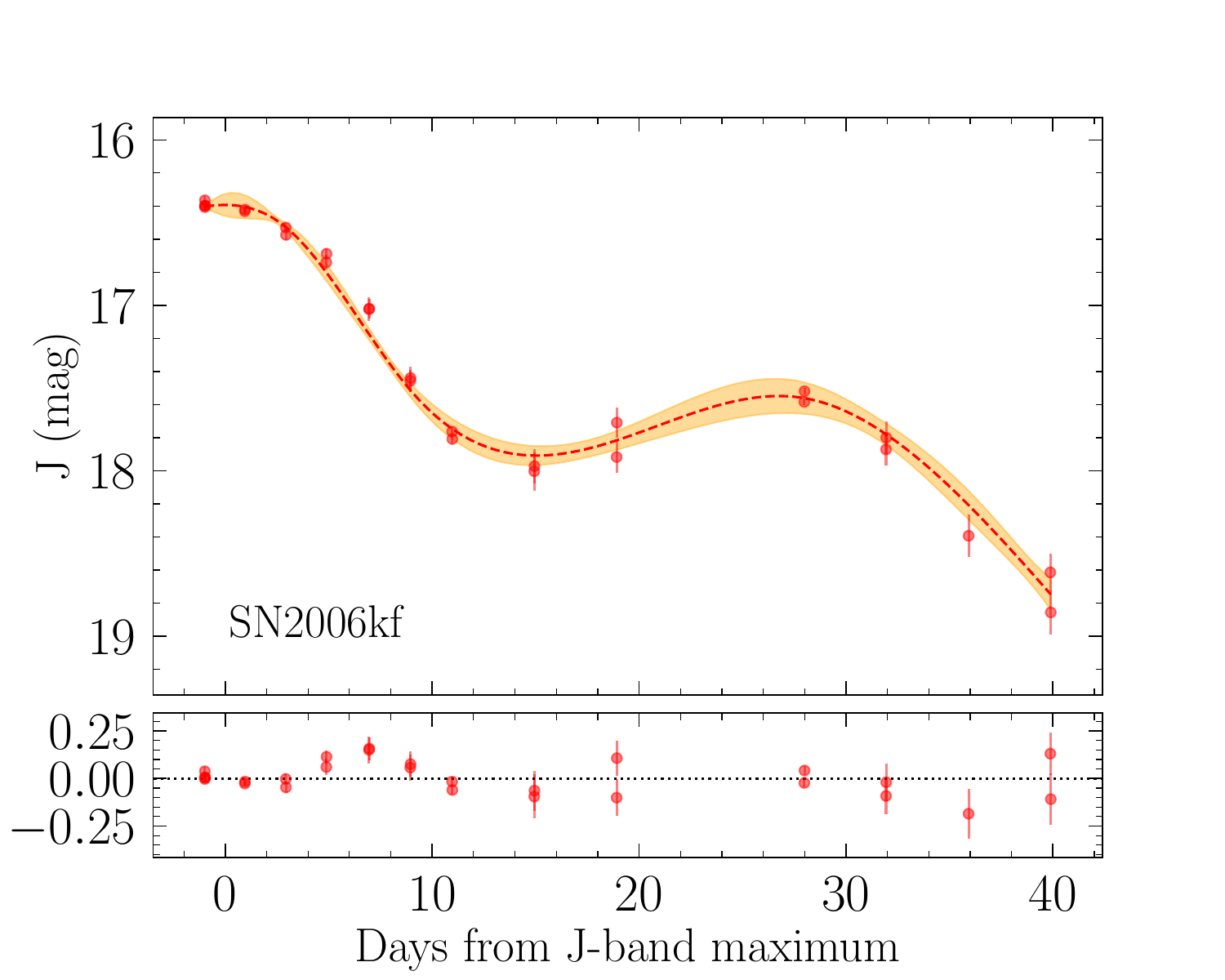}
\includegraphics[trim=0cm 0.2cm 2cm 1cm, clip=True,width=0.32\textwidth]{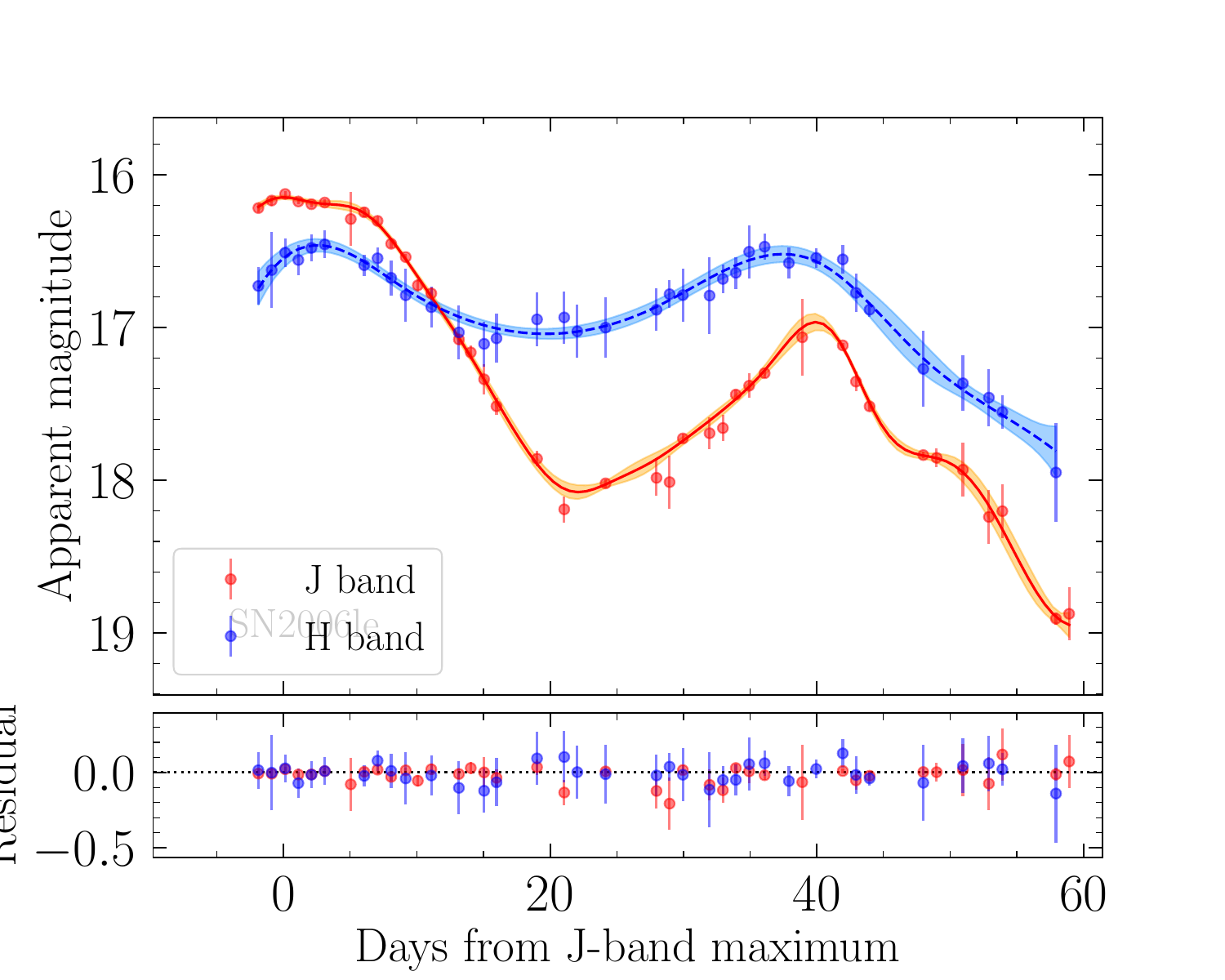}
\includegraphics[trim=0cm 0.2cm 2cm 1cm, clip=True,width=0.32\textwidth]{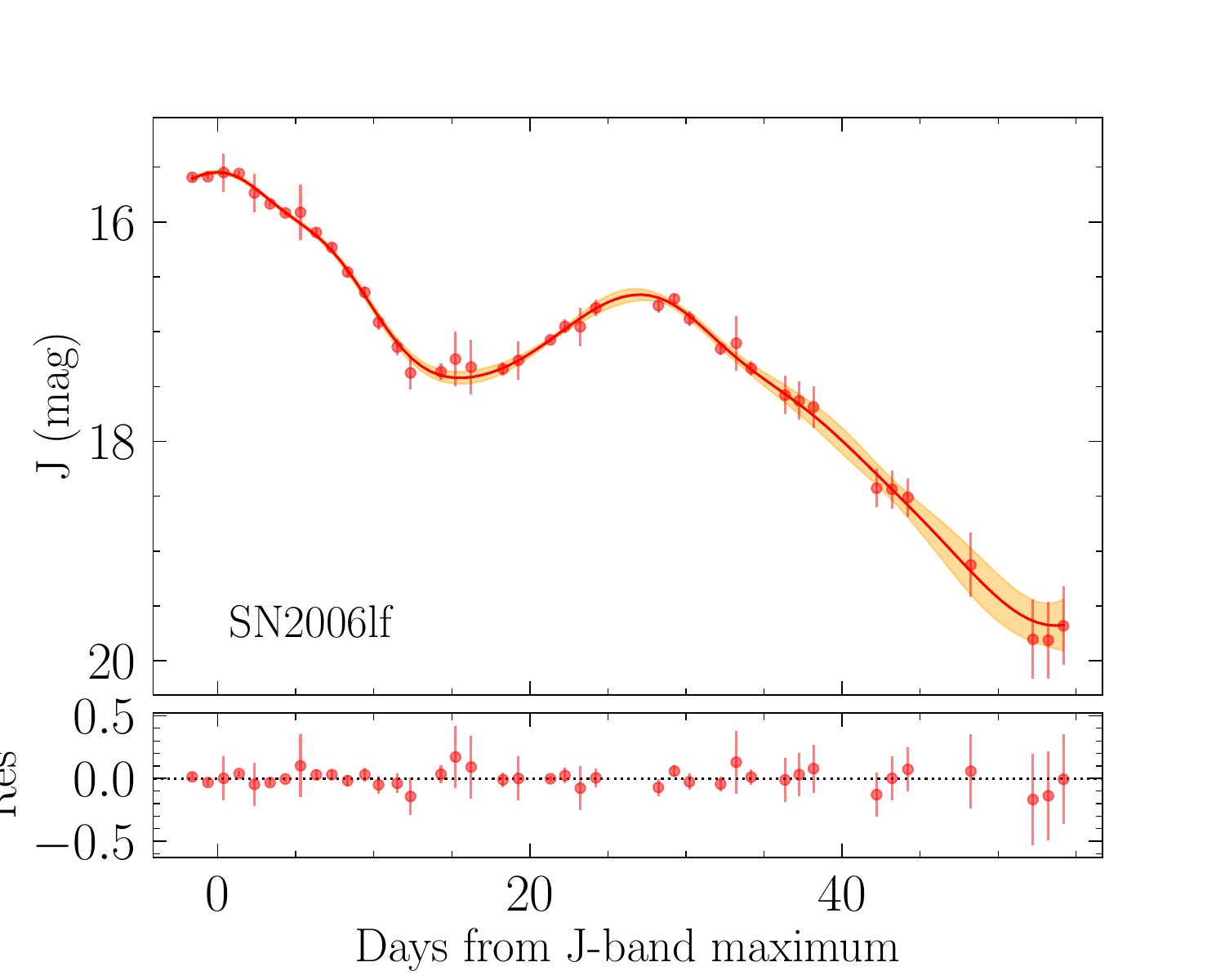}
\includegraphics[trim=0cm 0.2cm 2cm 1cm, clip=True,width=0.32\textwidth]{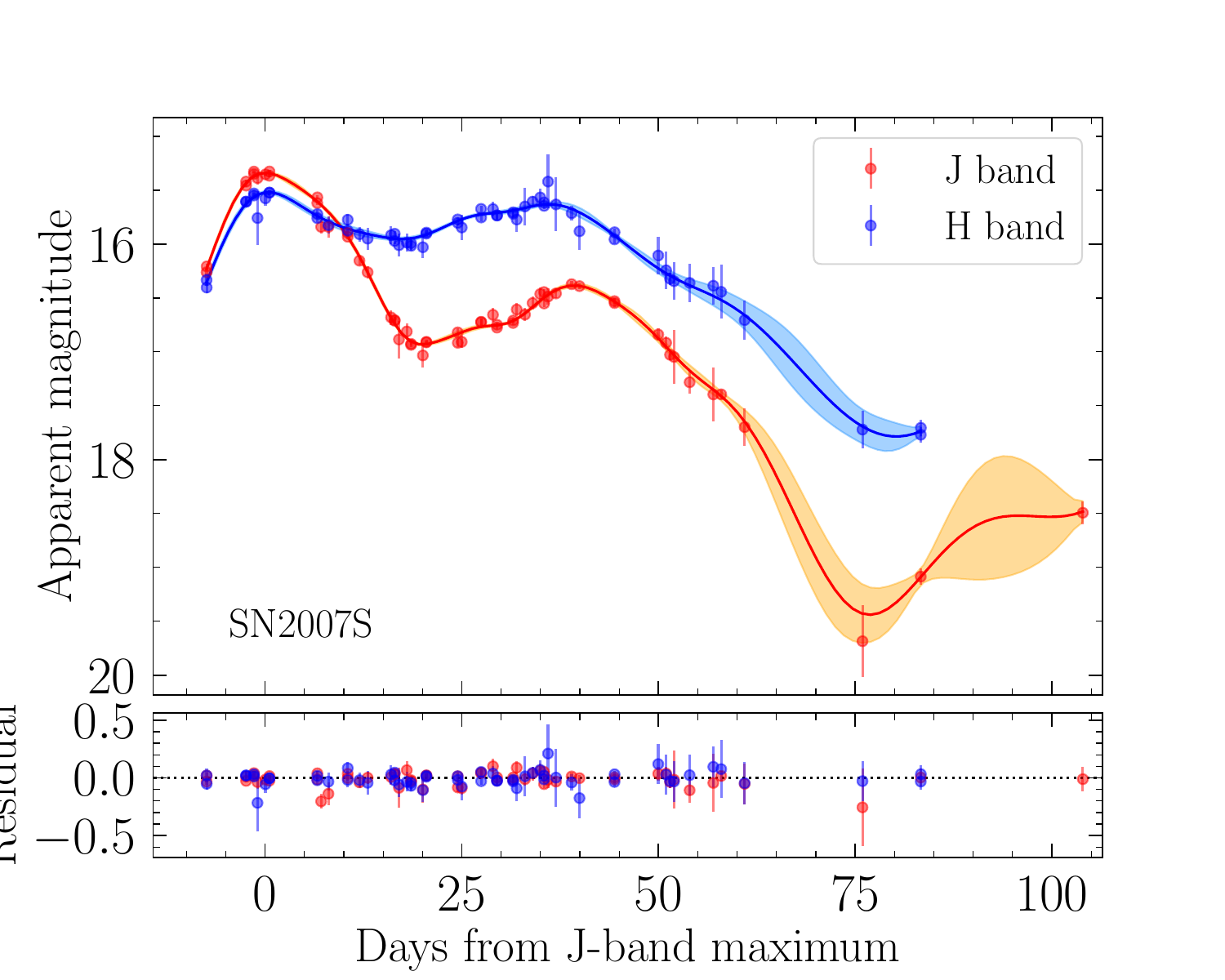}
\includegraphics[trim=0cm 0.2cm 2cm 1cm, clip=True,width=0.32\textwidth]{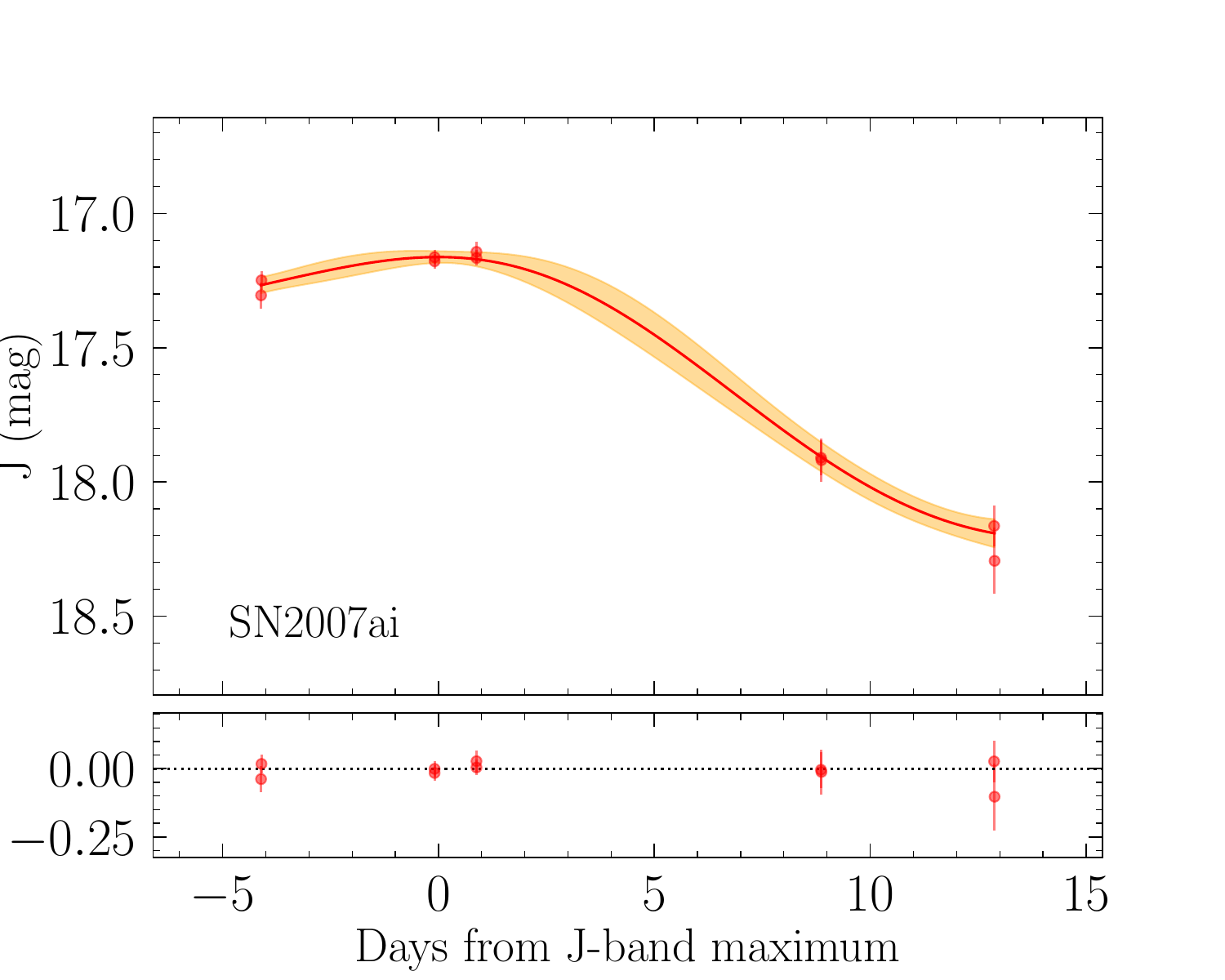}
\includegraphics[trim=0cm 0.2cm 2cm 1cm, clip=True,width=0.32\textwidth]{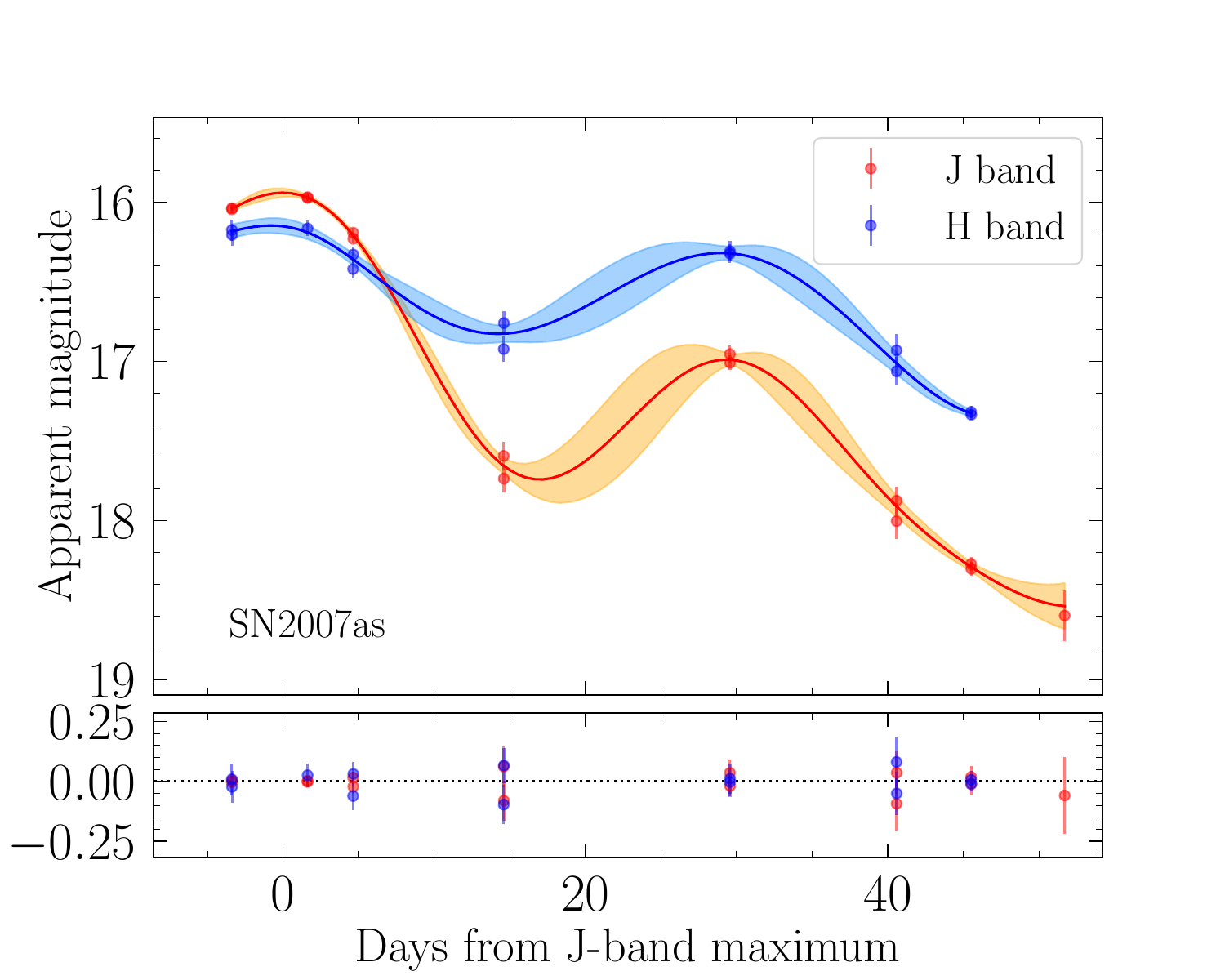}
\includegraphics[trim=0cm 0.2cm 2cm 1cm, clip=True,width=0.32\textwidth]{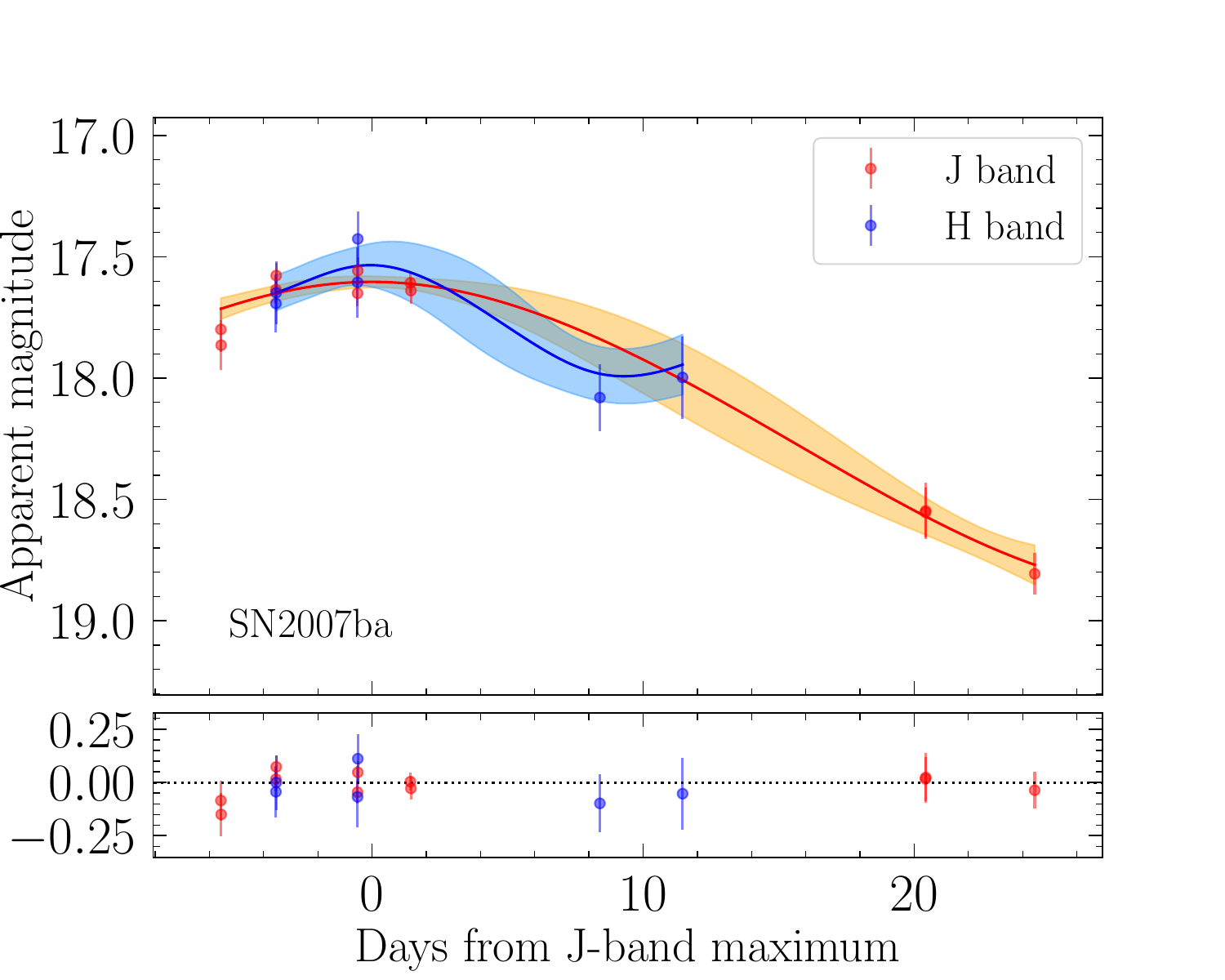}
\includegraphics[trim=0cm 0.2cm 2cm 1cm, clip=True,width=0.32\textwidth]{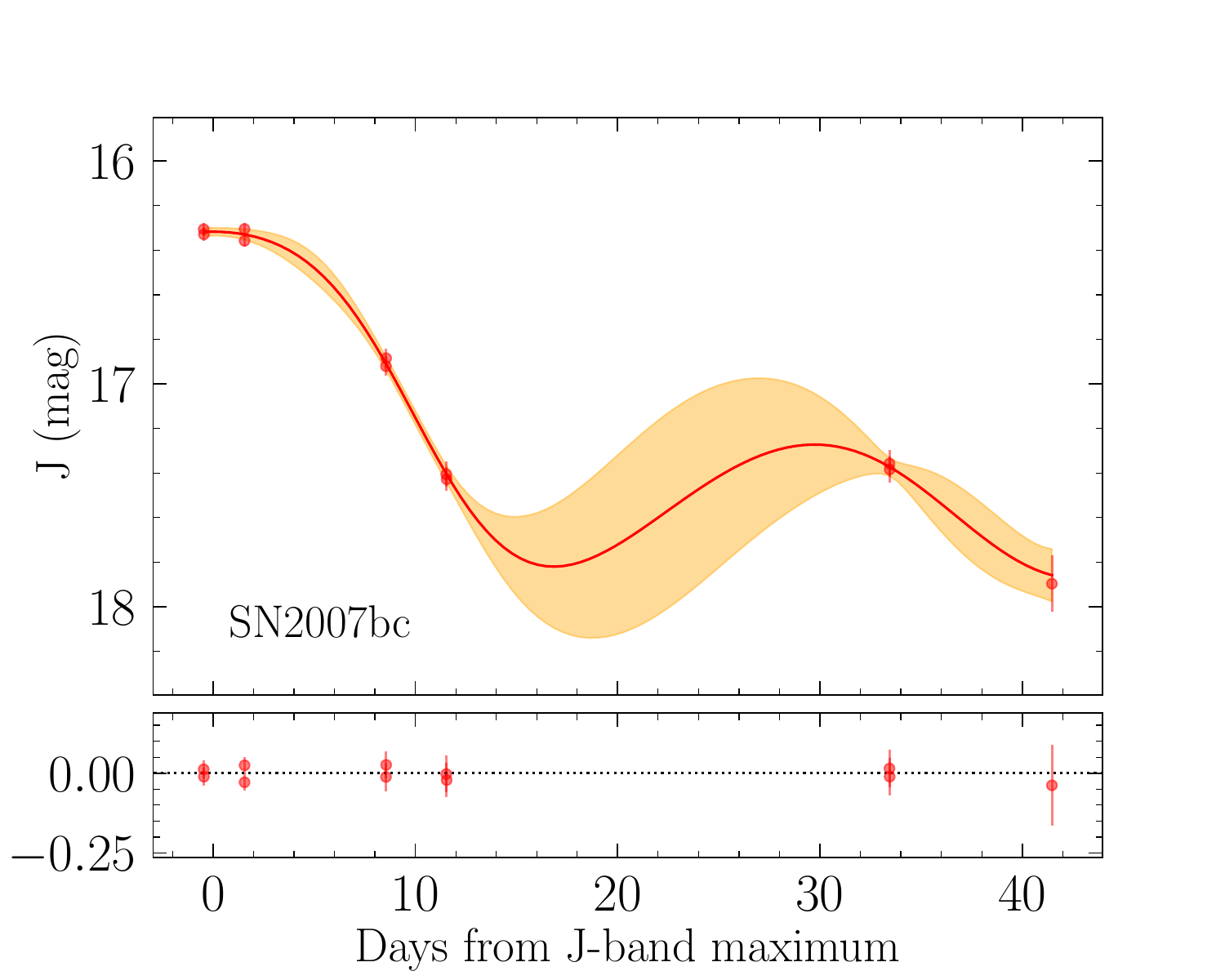}
\includegraphics[trim=0cm 0.2cm 2cm 1cm, clip=True,width=0.32\textwidth]{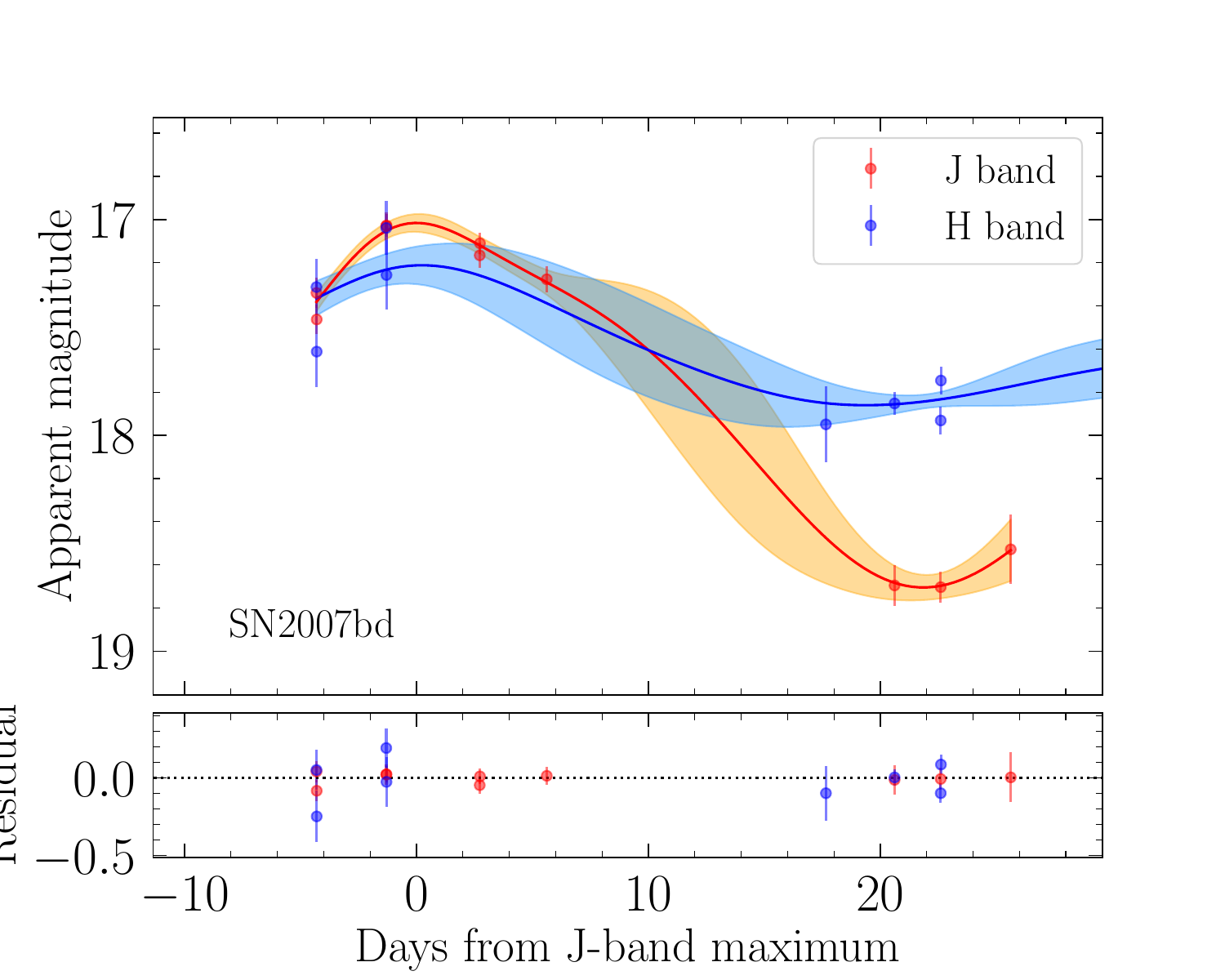}
\includegraphics[trim=0cm 0.2cm 2cm 1cm, clip=True,width=0.32\textwidth]{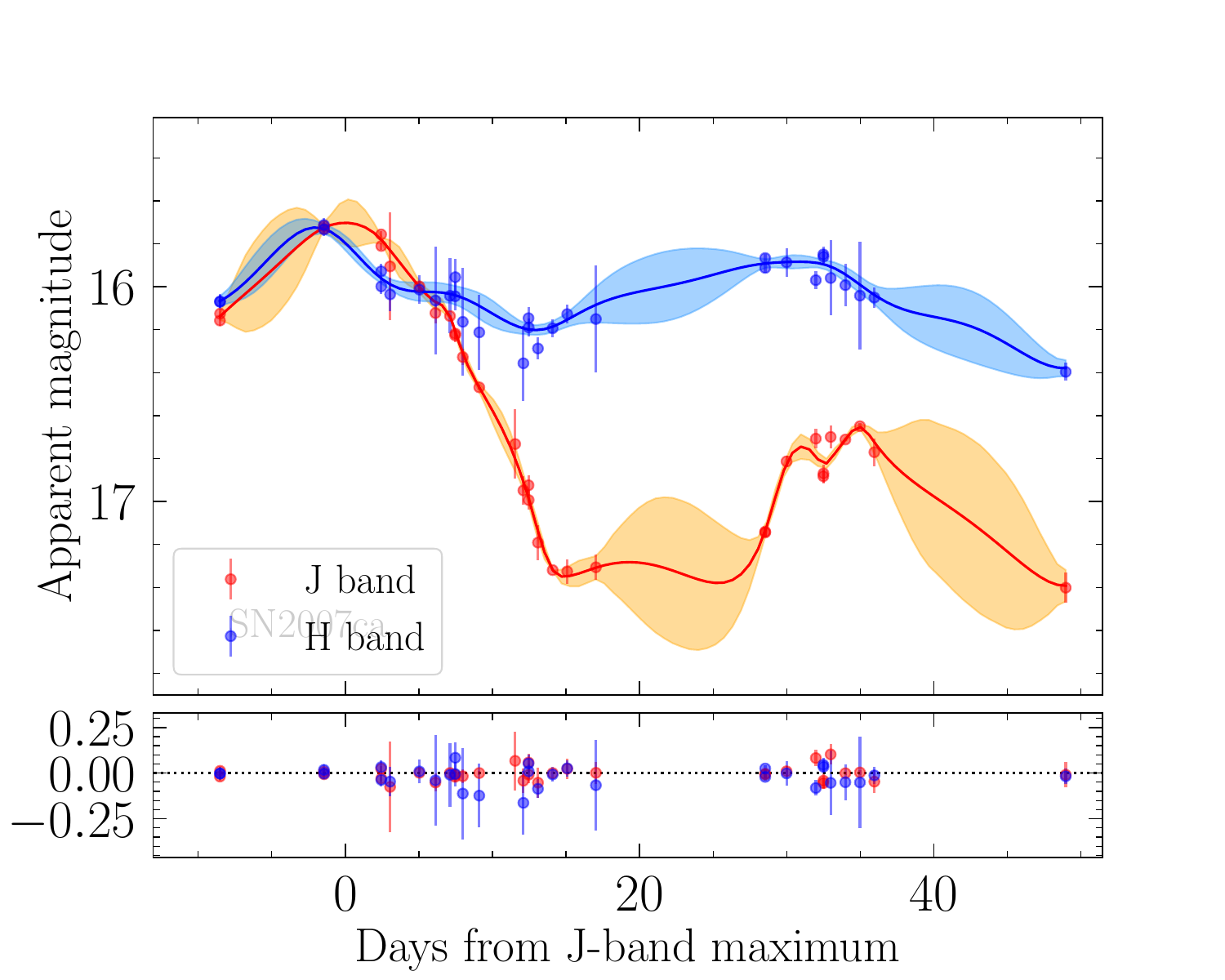}
\includegraphics[trim=0cm 0.2cm 2cm 1cm, clip=True,width=0.32\textwidth]{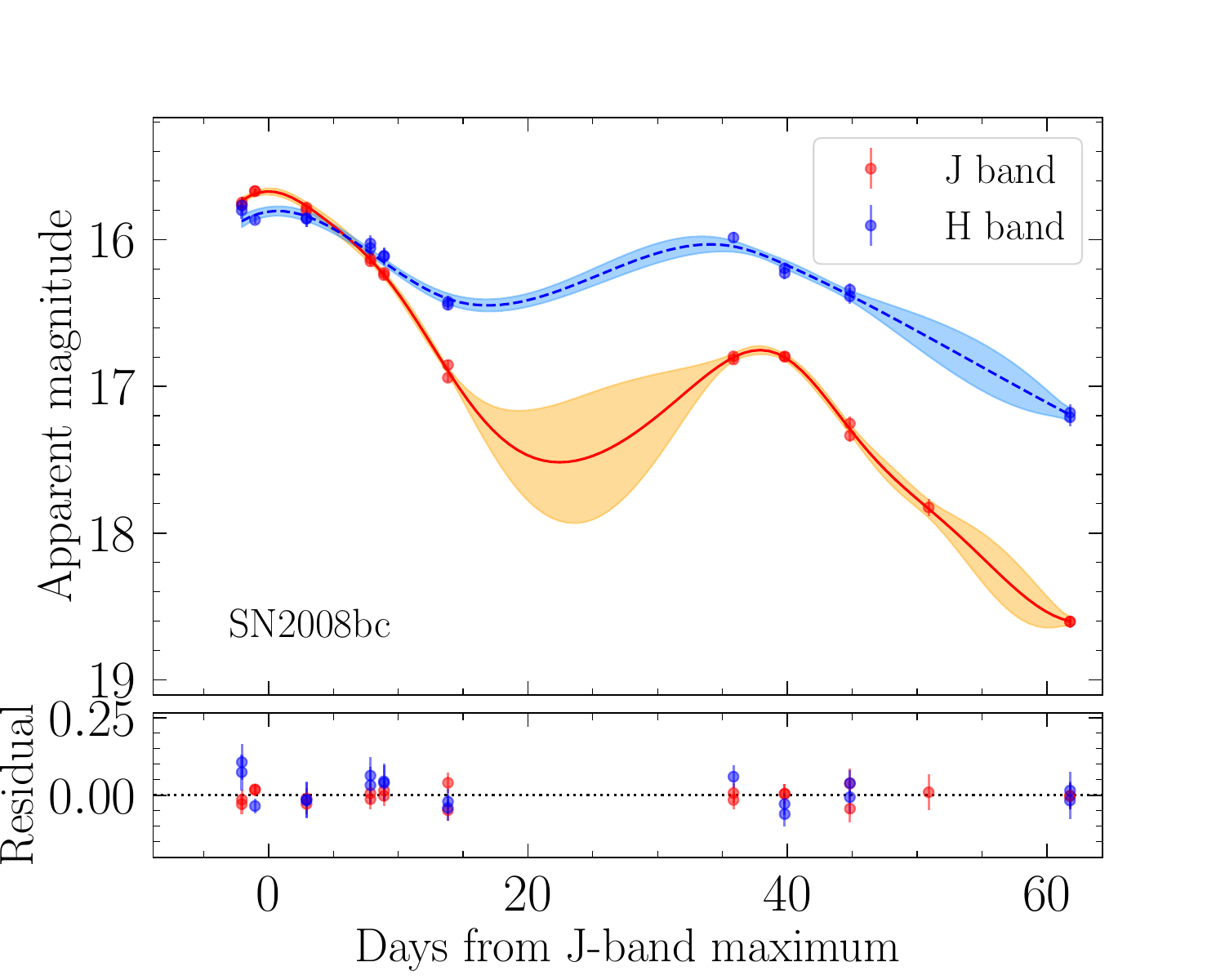}
\includegraphics[trim=0cm 0.2cm 2cm 1cm, clip=True,width=0.32\textwidth]{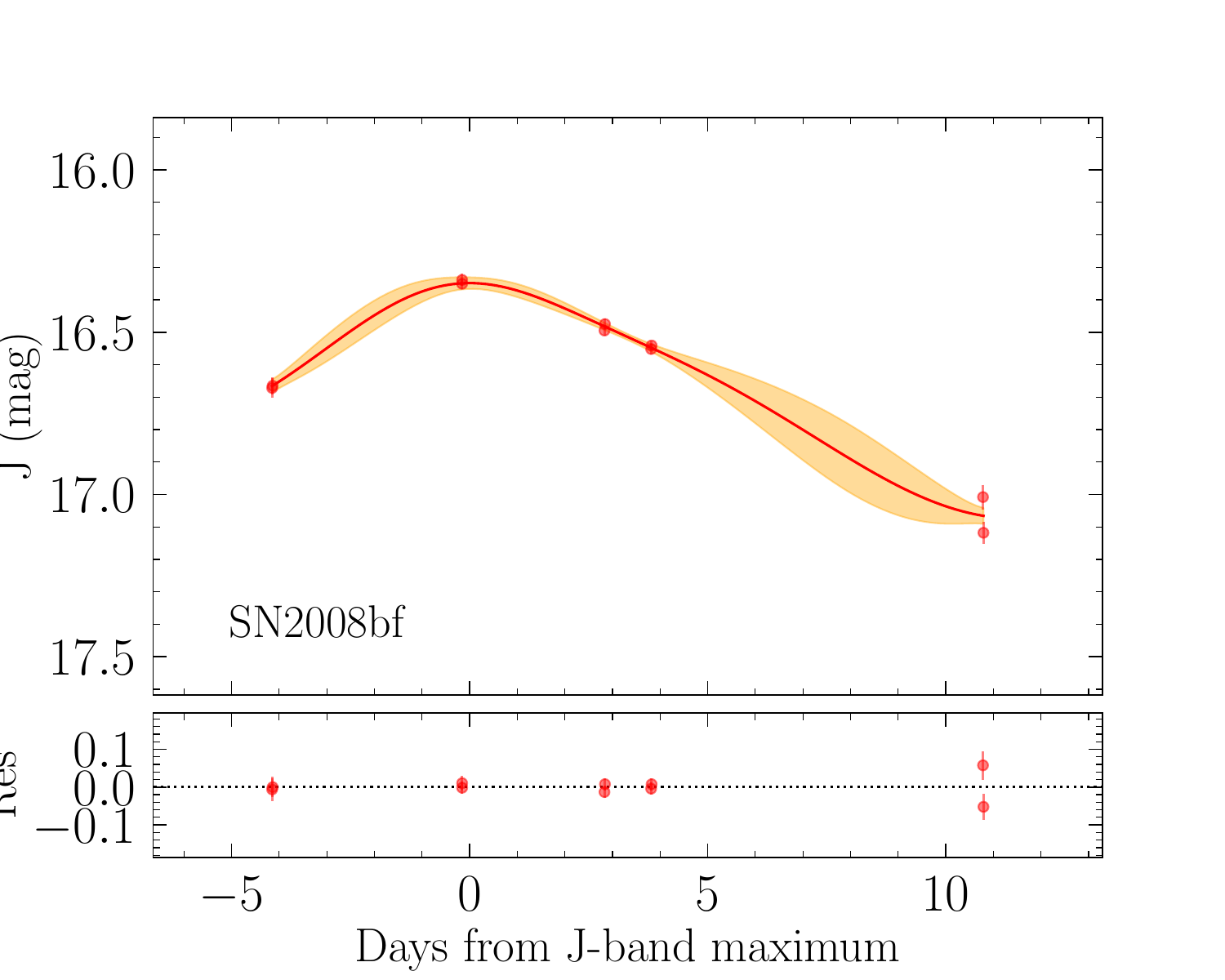}
\includegraphics[trim=0cm 0.2cm 2cm 1cm, clip=True,width=0.32\textwidth]{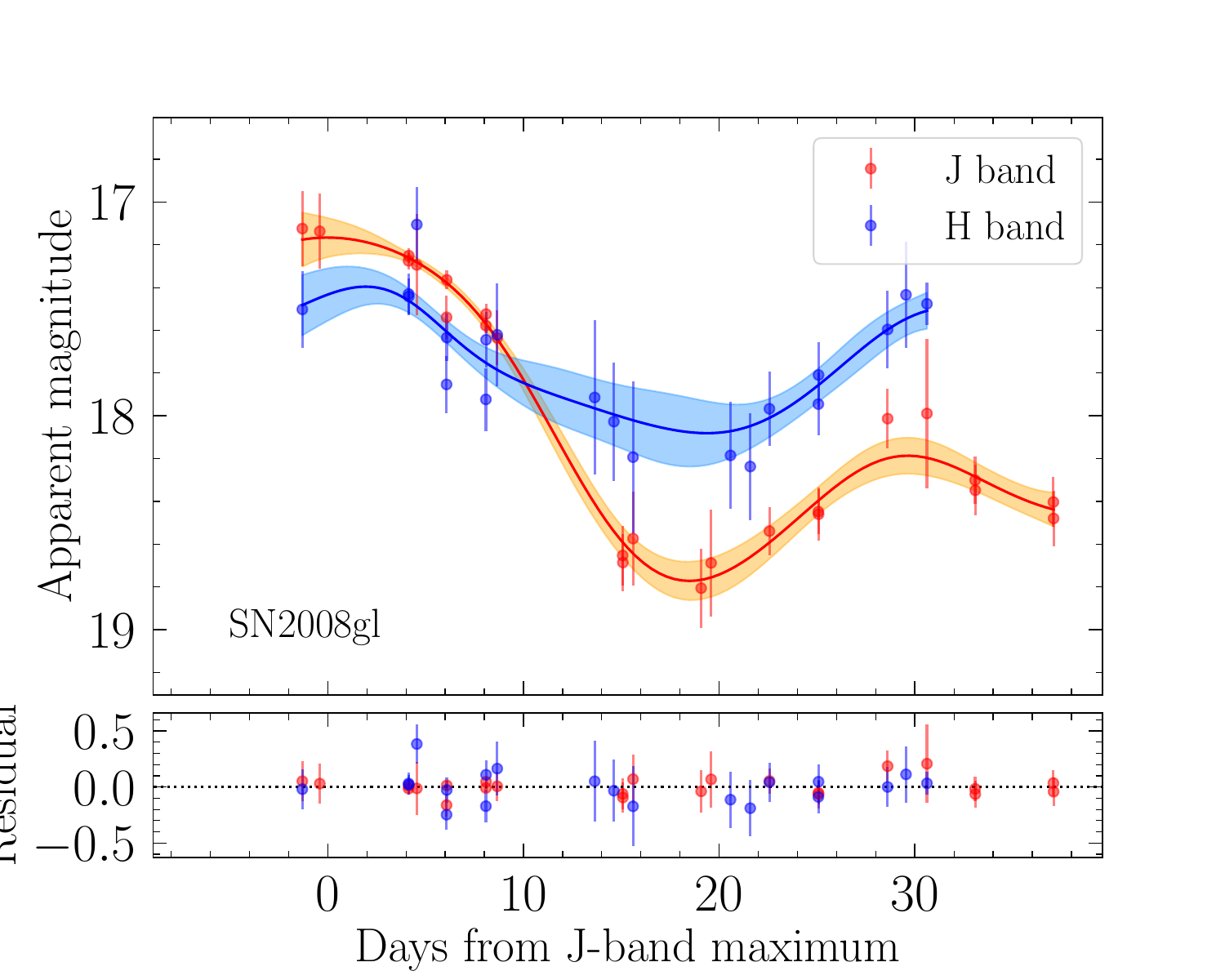}
\caption{(continuing form Fig D3)}
\end{figure*}
\begin{figure*}
\includegraphics[trim=0cm 0.2cm 2cm 1cm, clip=True,width=0.32\textwidth]{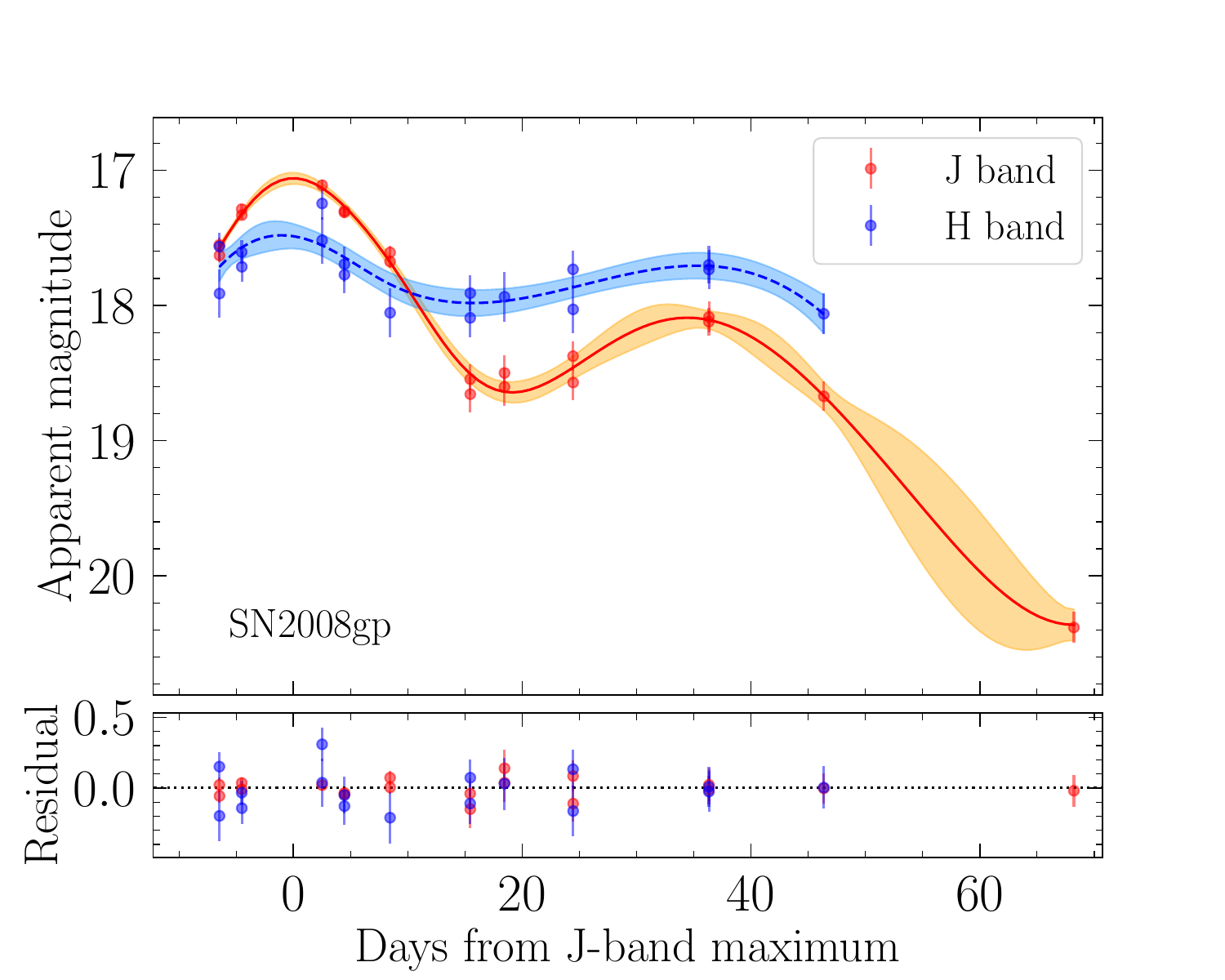}
\includegraphics[trim=0cm 0.2cm 2cm 1cm, clip=True,width=0.32\textwidth]{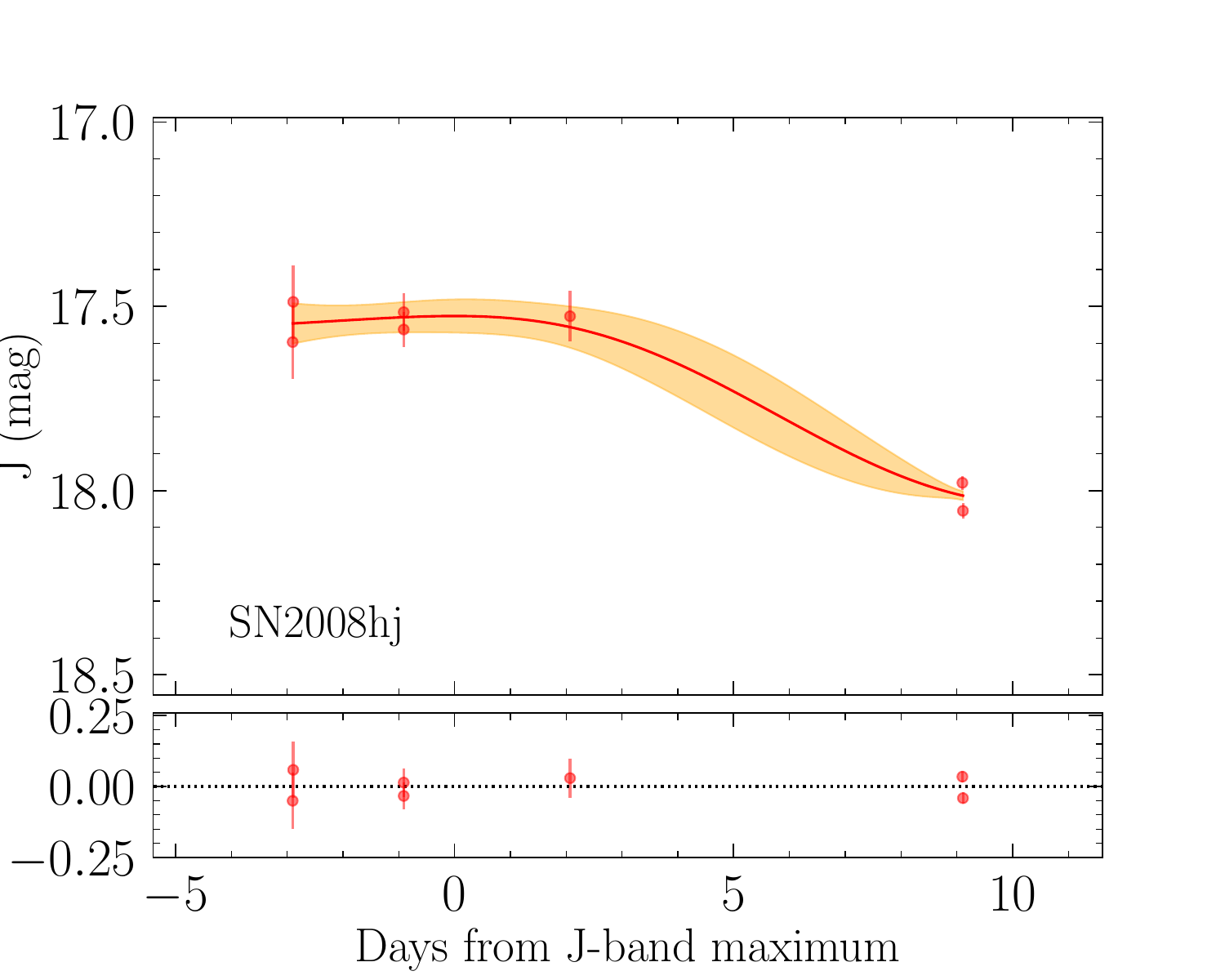}
\includegraphics[trim=0cm 0.2cm 2cm 1cm, clip=True,width=0.32\textwidth]{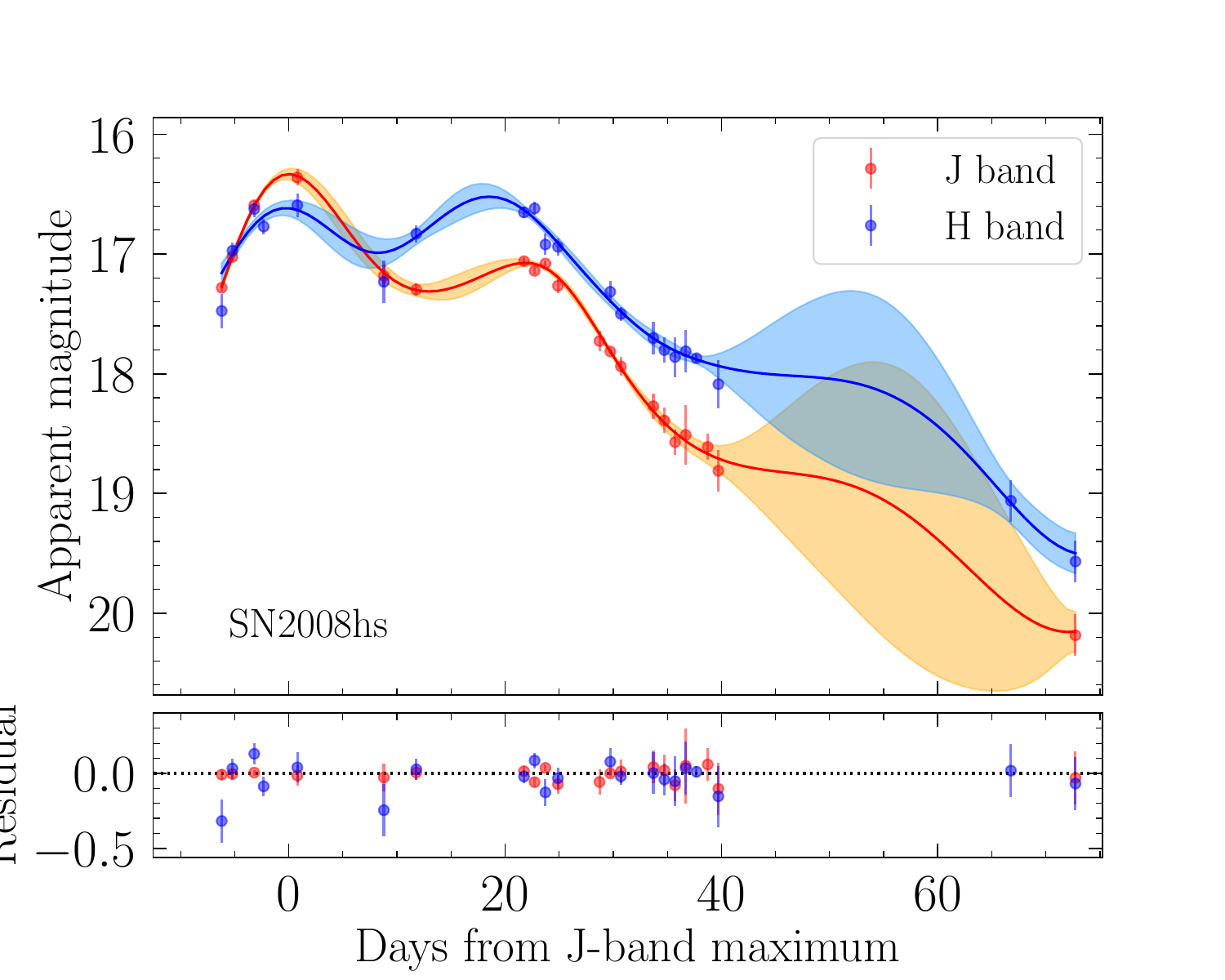}
\includegraphics[trim=0cm 0.2cm 2cm 1cm, clip=True,width=0.32\textwidth]{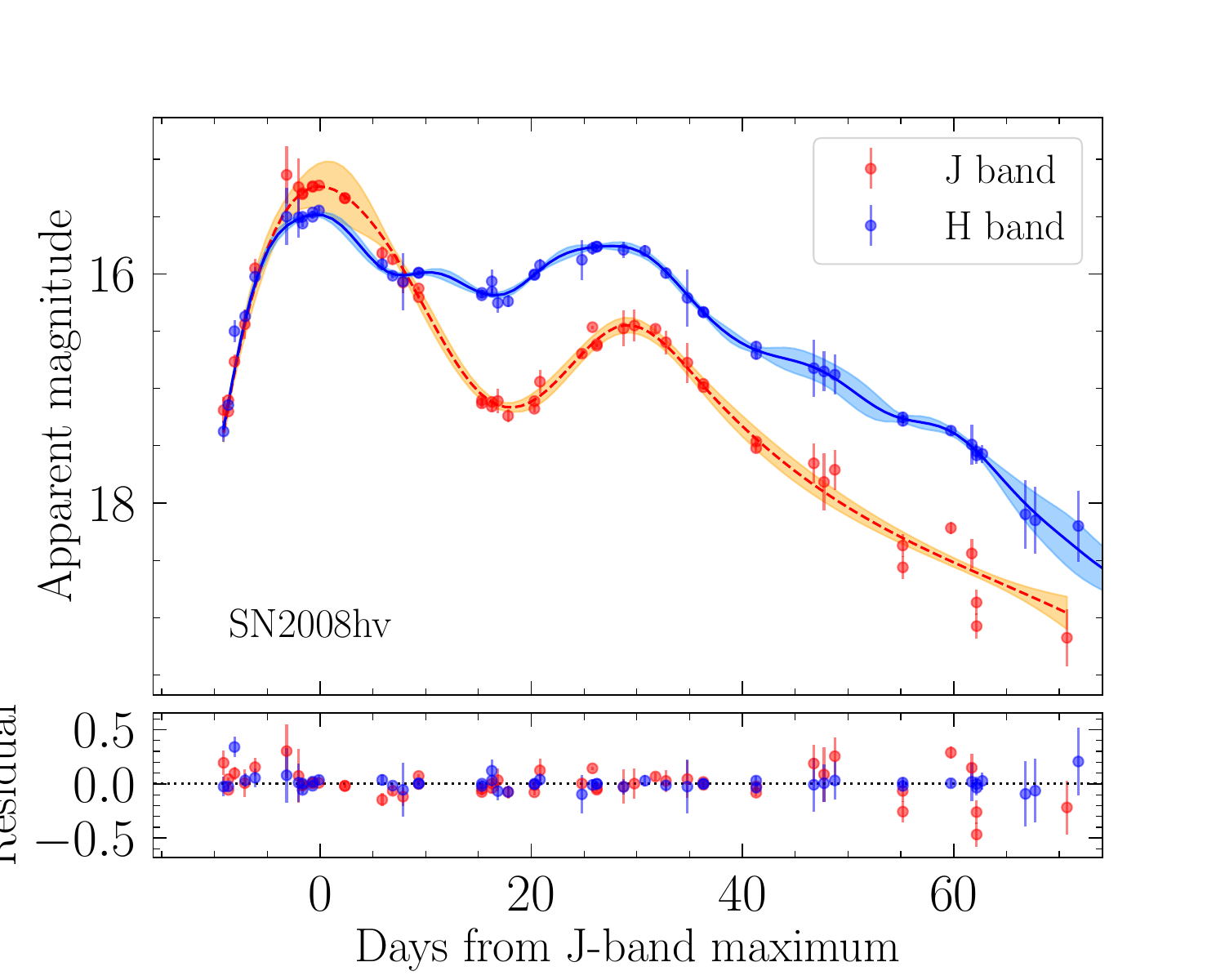}
\includegraphics[trim=0cm 0.2cm 2cm 1cm, clip=True,width=0.32\textwidth]{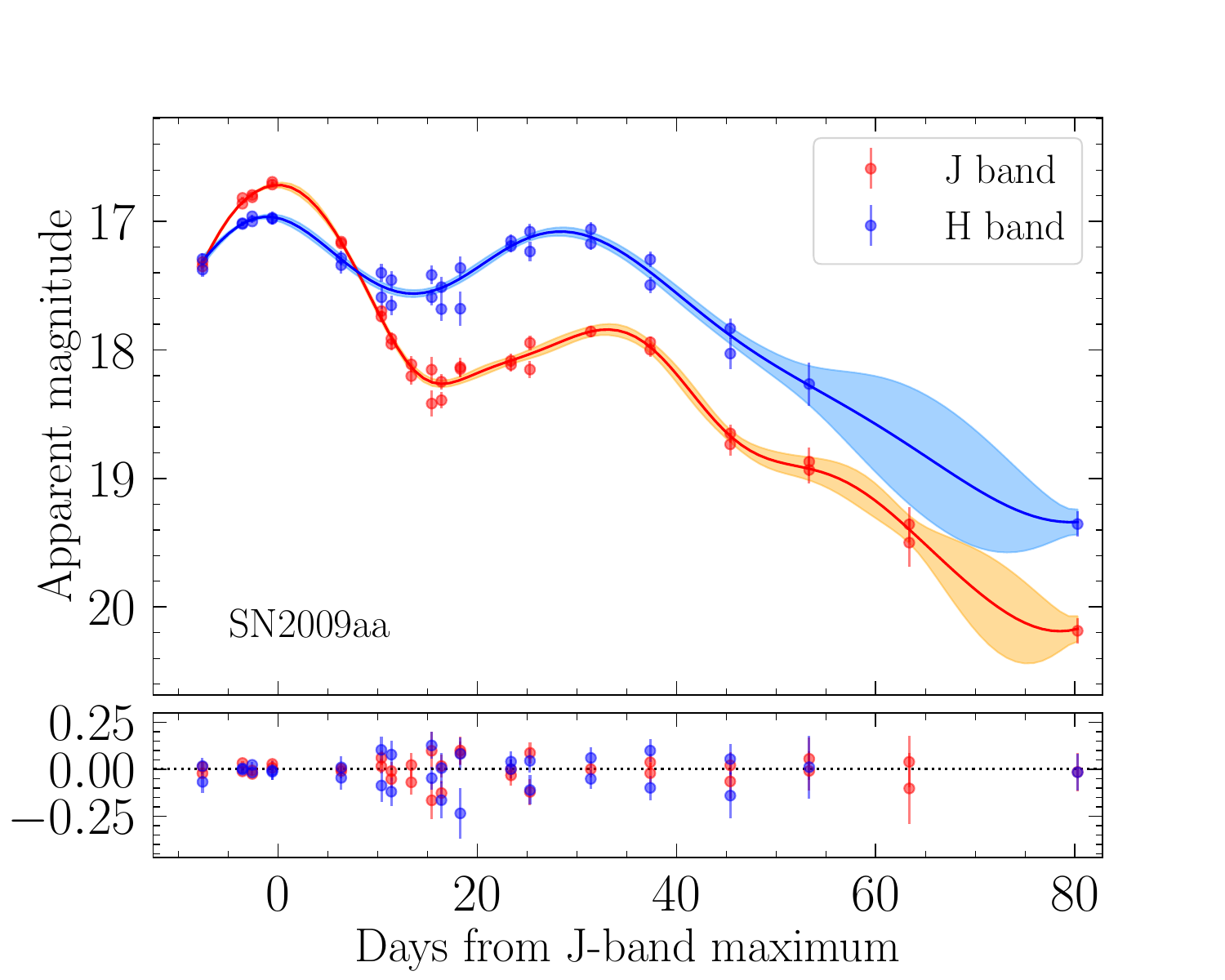}
\includegraphics[trim=0cm 0.2cm 2cm 1cm, clip=True,width=0.32\textwidth]{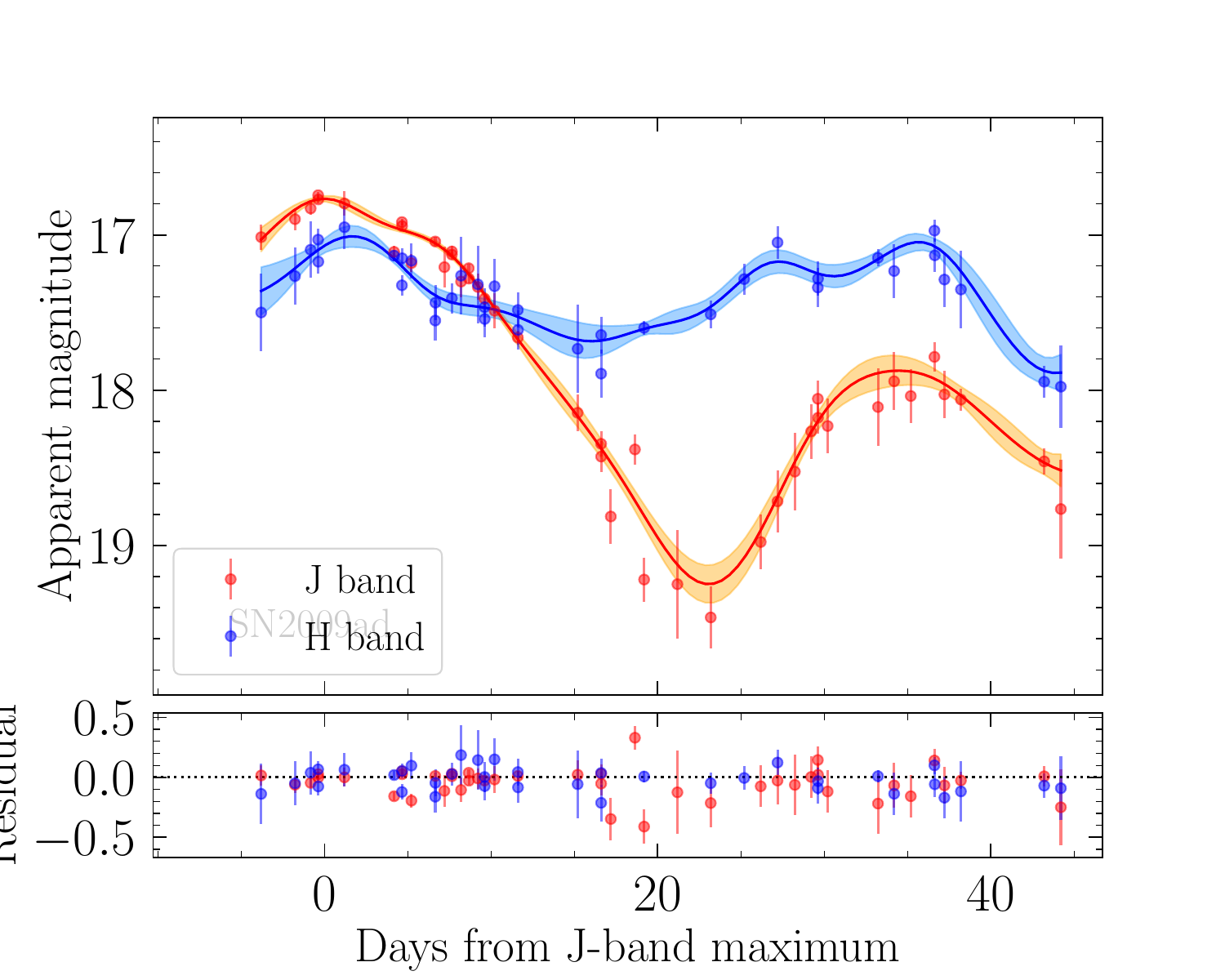}
\includegraphics[trim=0cm 0.2cm 2cm 1cm, clip=True,width=0.32\textwidth]{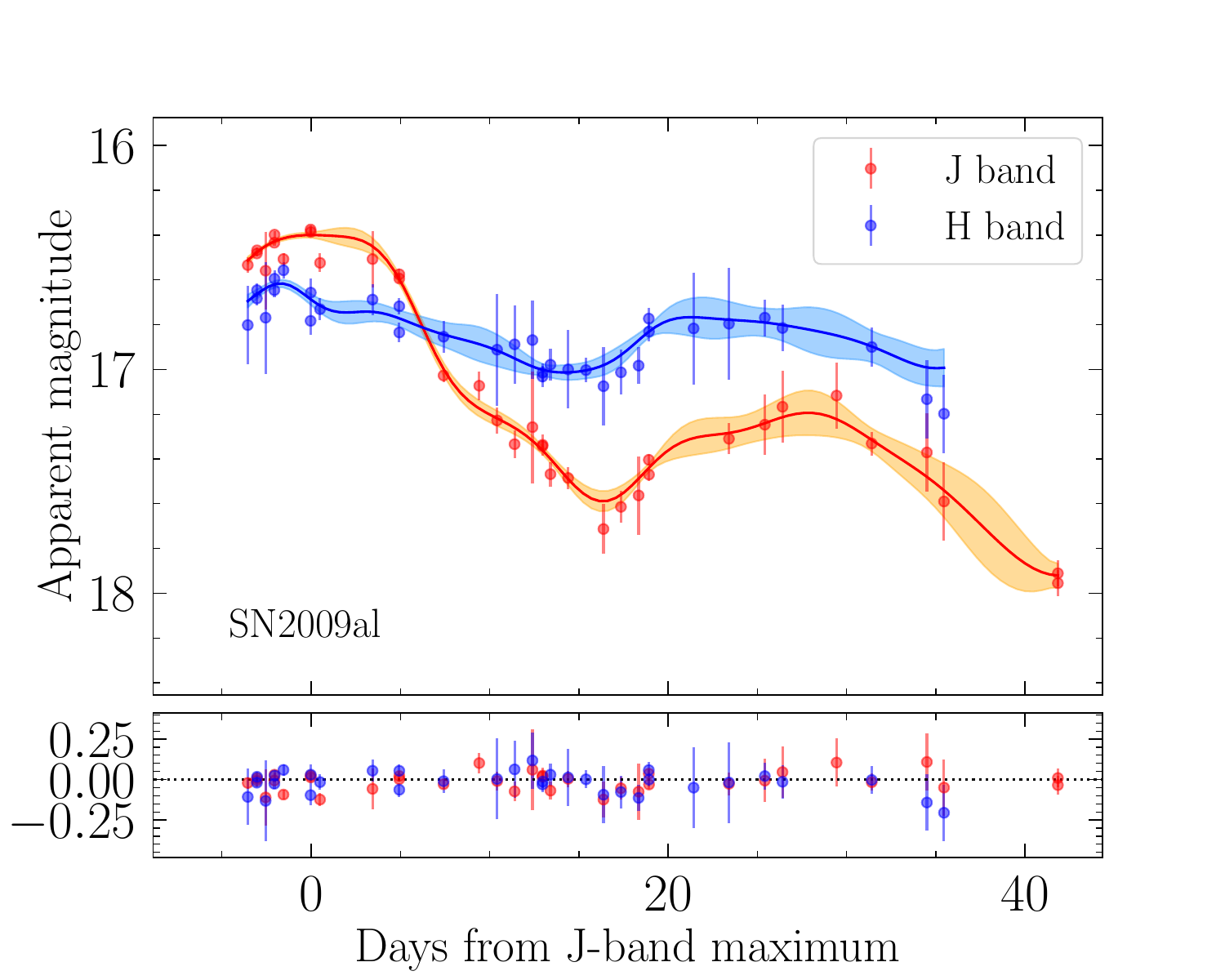}
\includegraphics[trim=0cm 0.2cm 2cm 1cm, clip=True,width=0.32\textwidth]{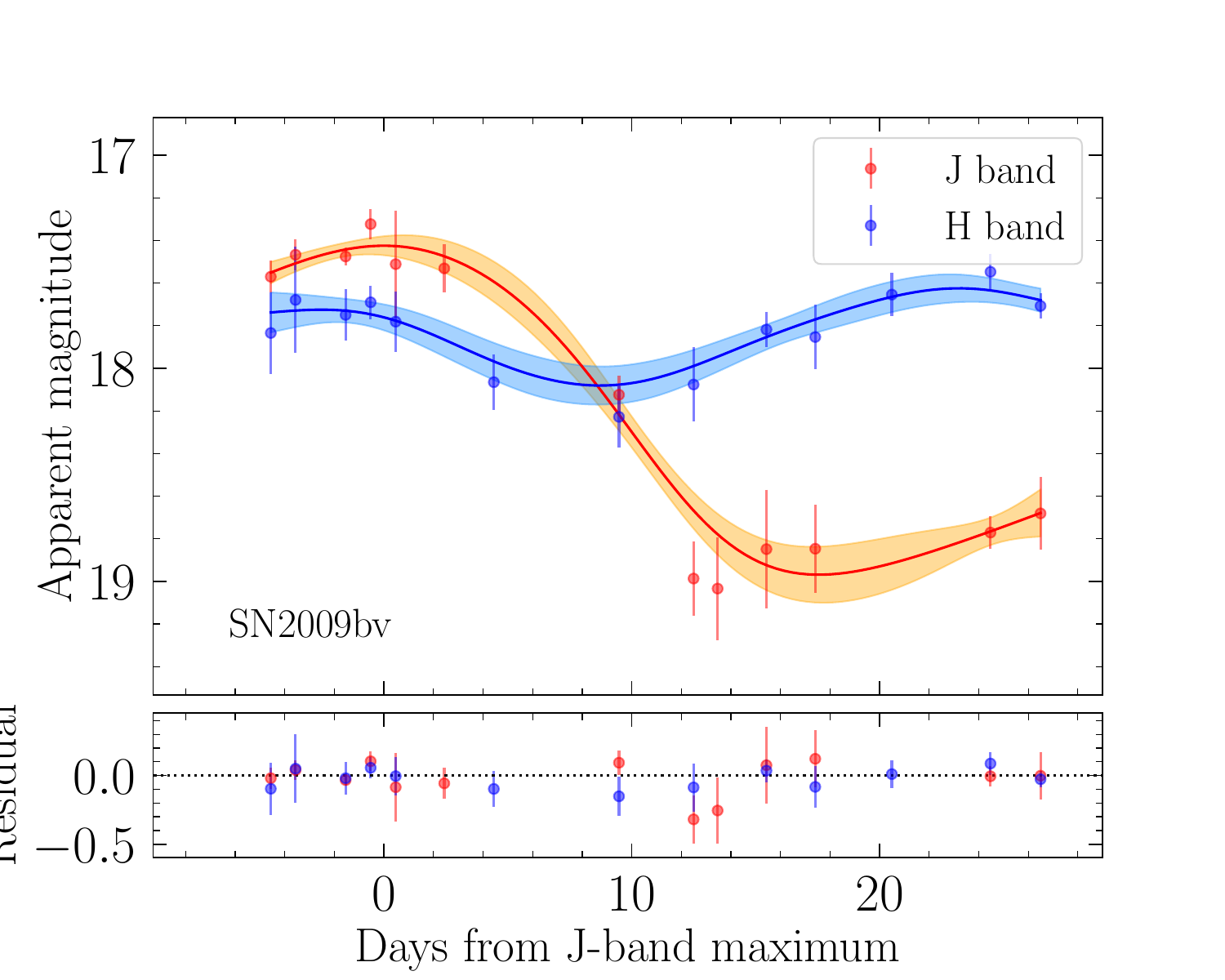}
\includegraphics[trim=0cm 0.2cm 2cm 1cm, clip=True,width=0.32\textwidth]{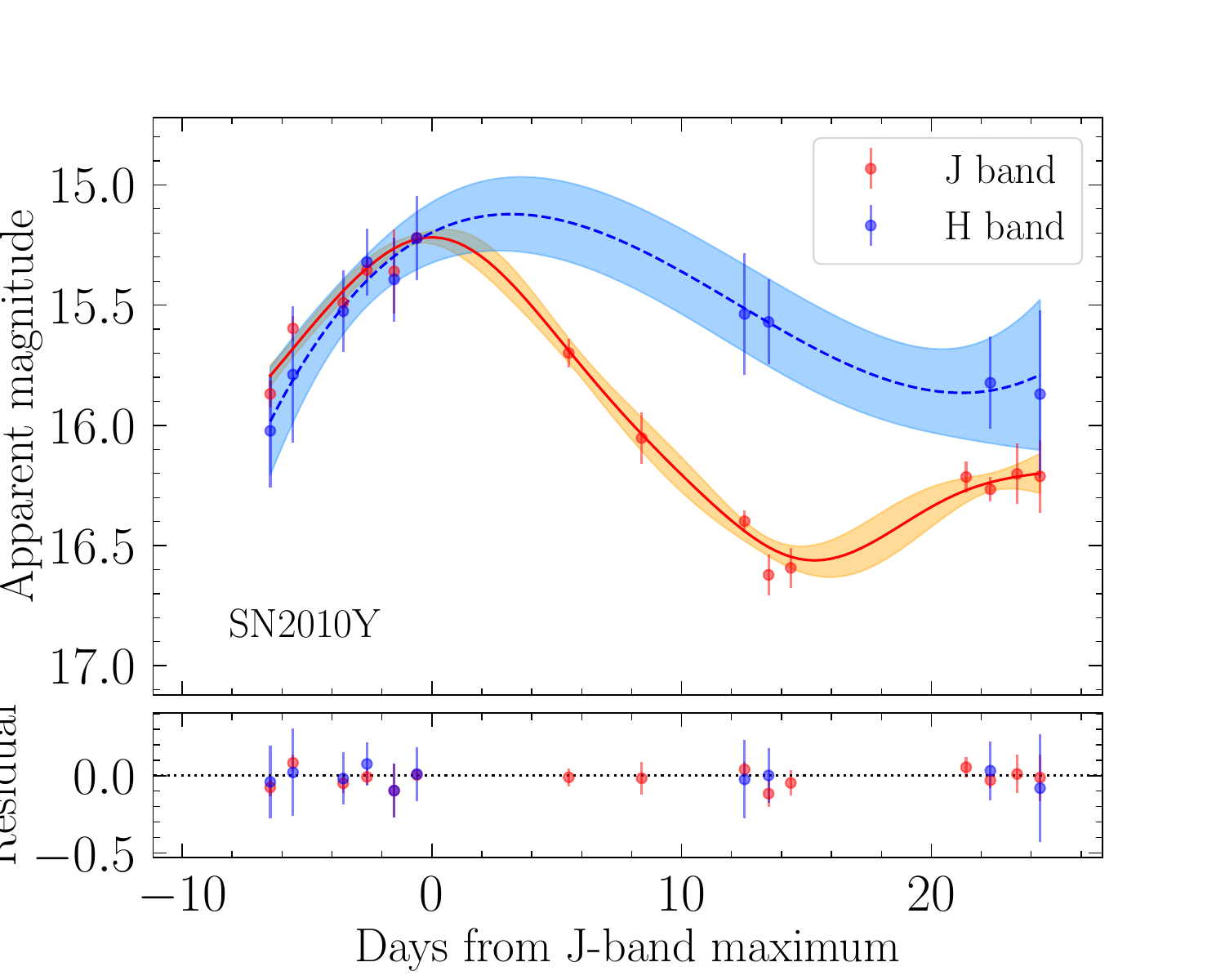}
\includegraphics[trim=0cm 0.2cm 2cm 1cm, clip=True,width=0.32\textwidth]{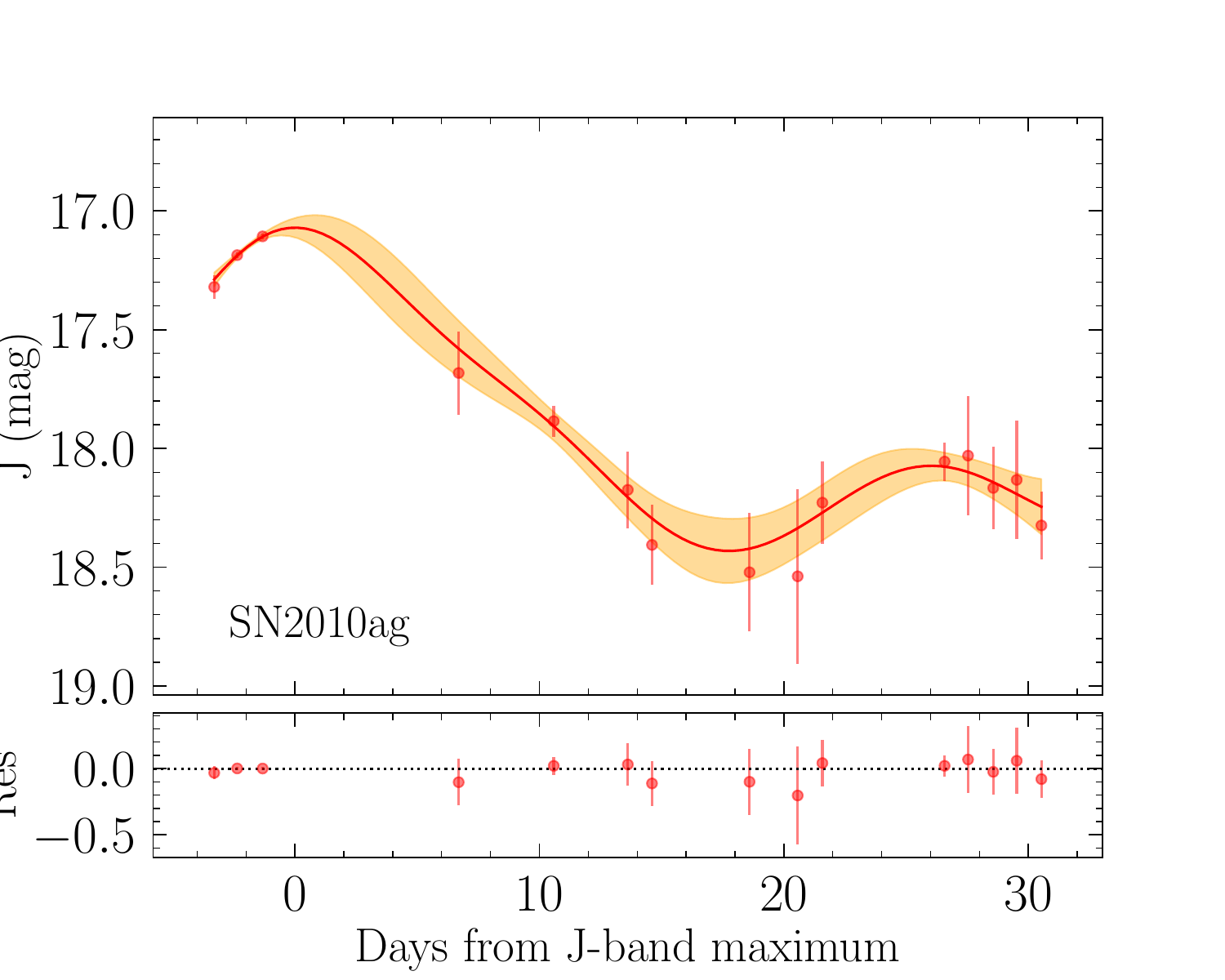}
\includegraphics[trim=0cm 0.2cm 2cm 1cm, clip=True,width=0.32\textwidth]{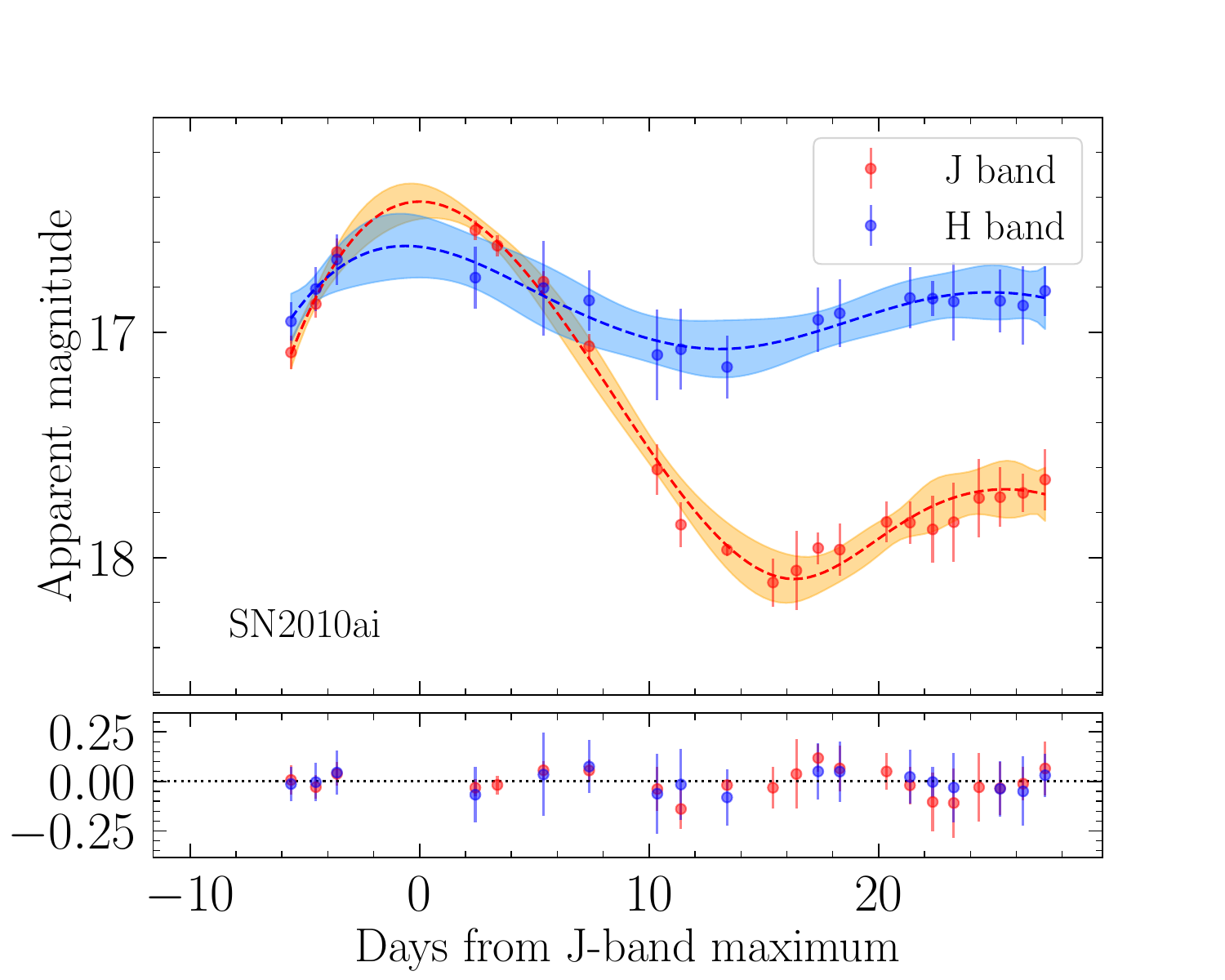}
\includegraphics[trim=0cm 0.2cm 2cm 1cm, clip=True,width=0.32\textwidth]{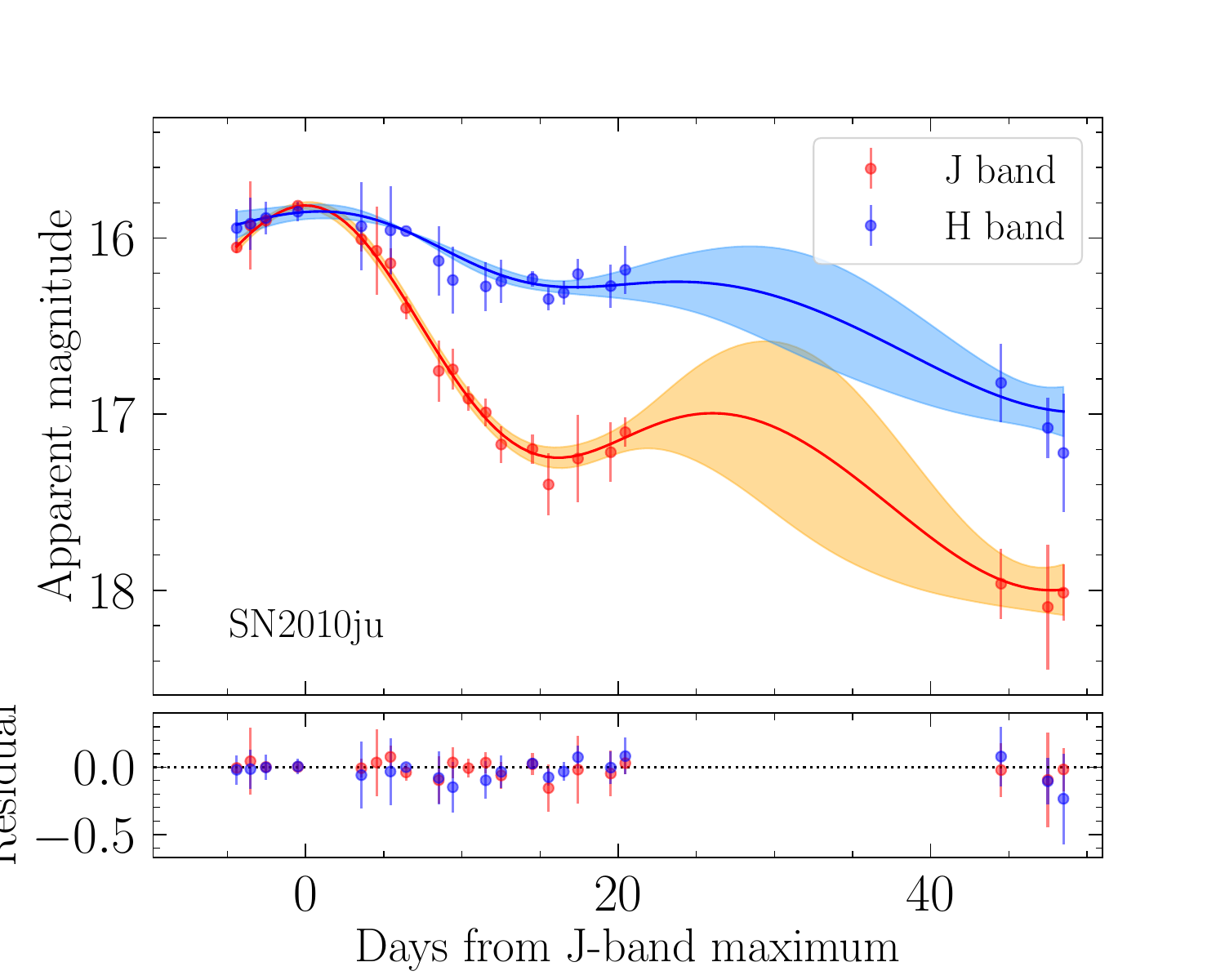}
\includegraphics[trim=0cm 0.2cm 2cm 1cm, clip=True,width=0.32\textwidth]{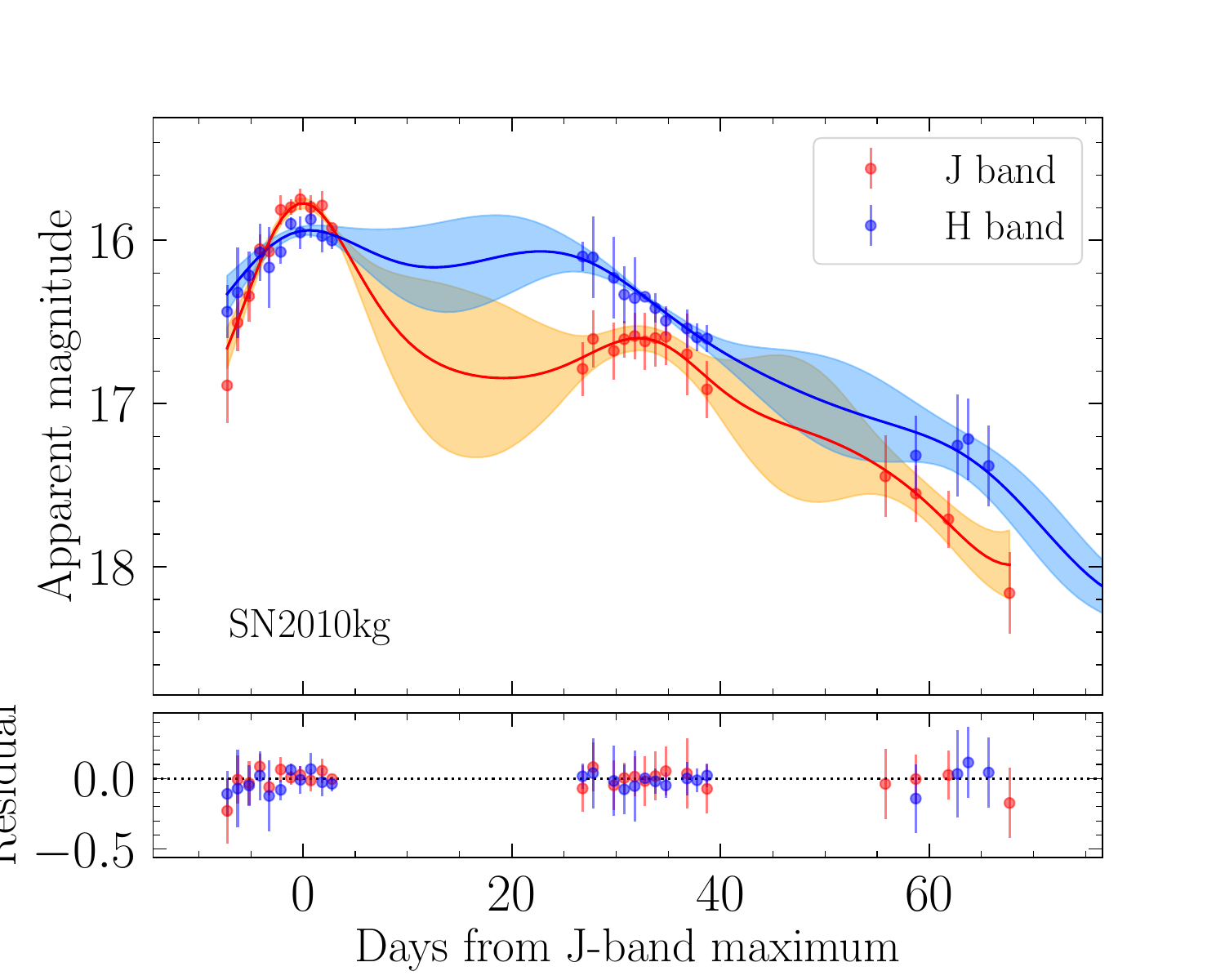}
\includegraphics[trim=0cm 0.2cm 2cm 1cm, clip=True,width=0.32\textwidth]{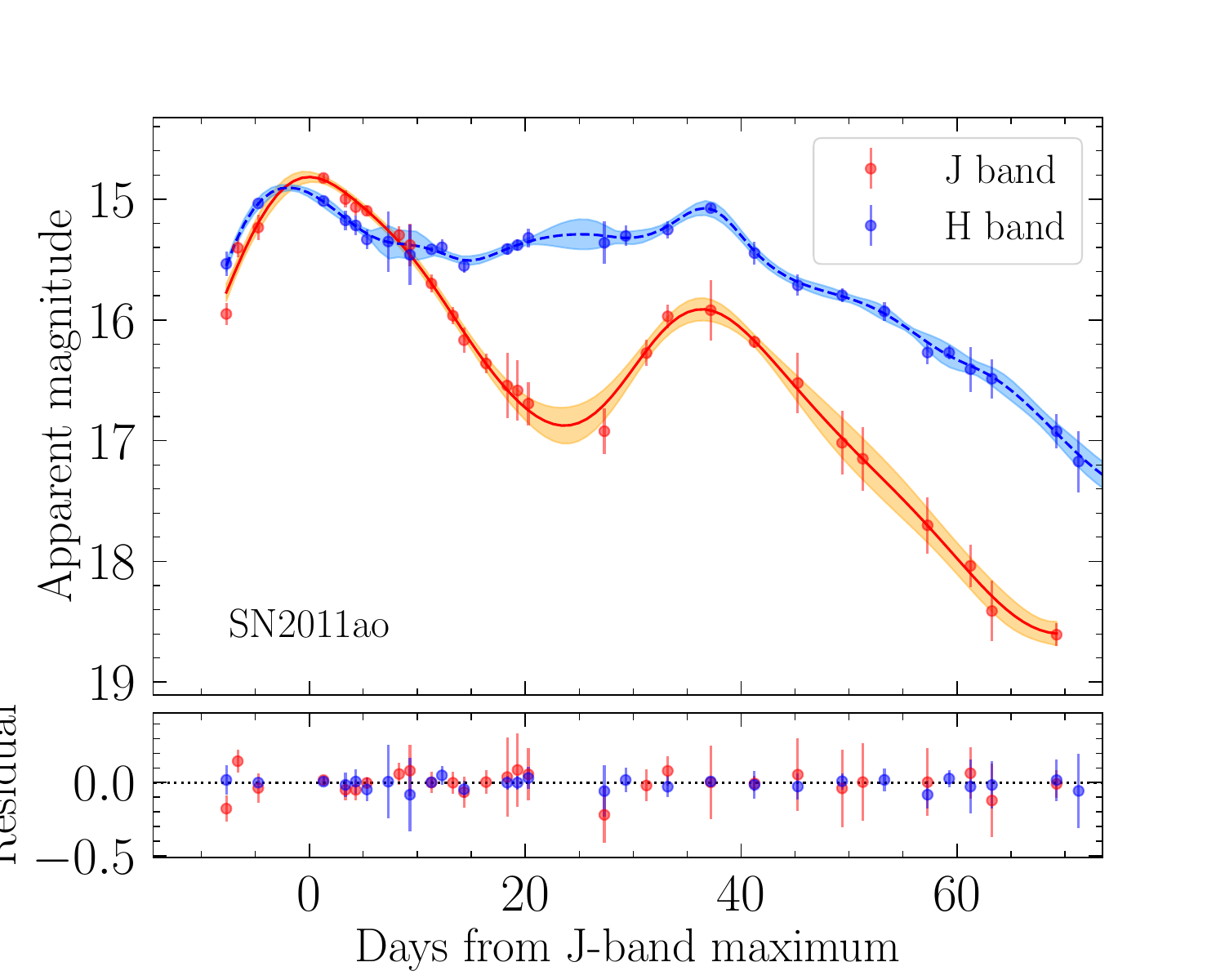}
\includegraphics[trim=0cm 0.2cm 2cm 1cm, clip=True,width=0.32\textwidth]{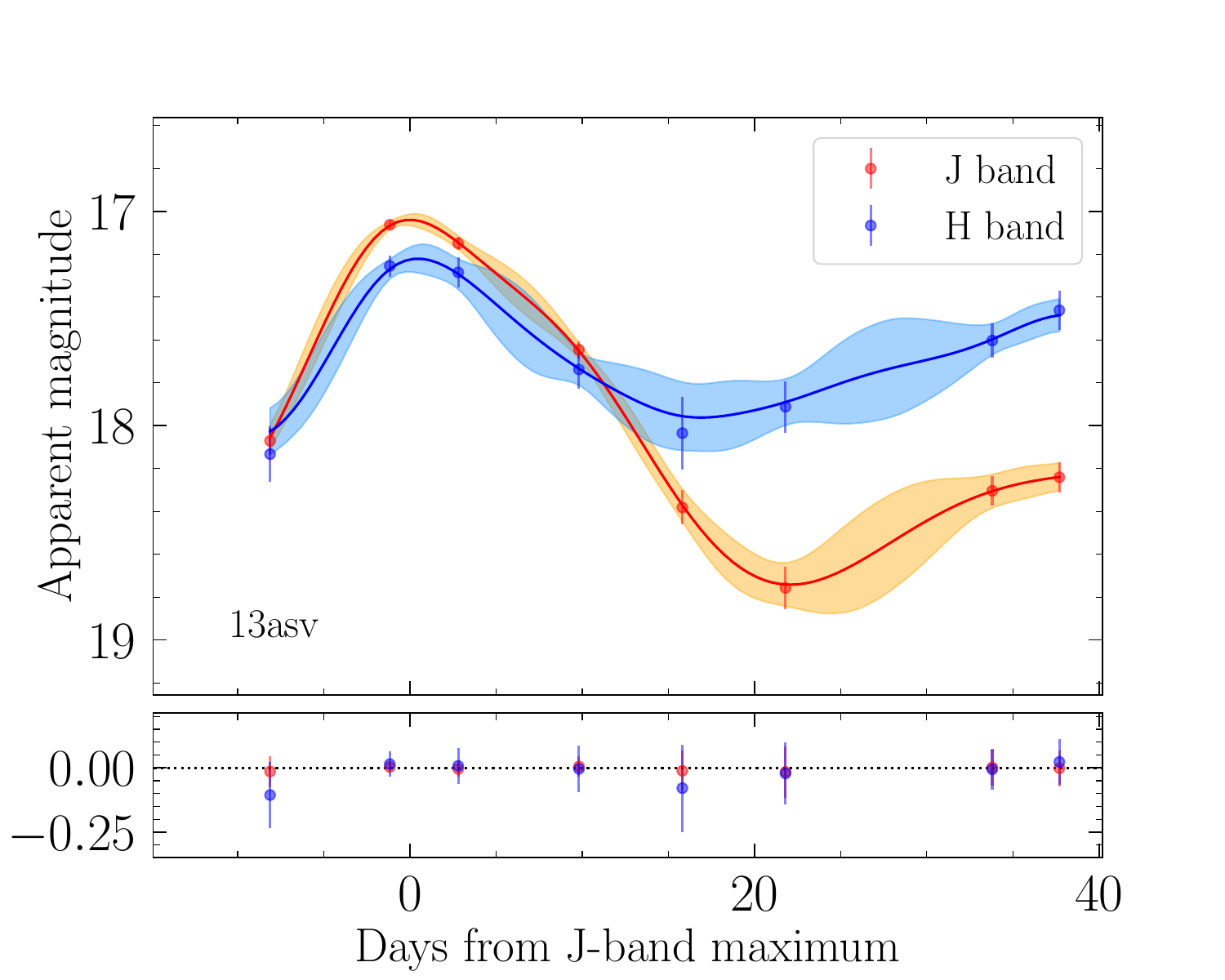}
\caption{(continuing form Fig D3)}
\end{figure*}
\begin{figure*}
\includegraphics[trim=0cm 0.2cm 2cm 1cm, clip=True,width=0.32\textwidth]{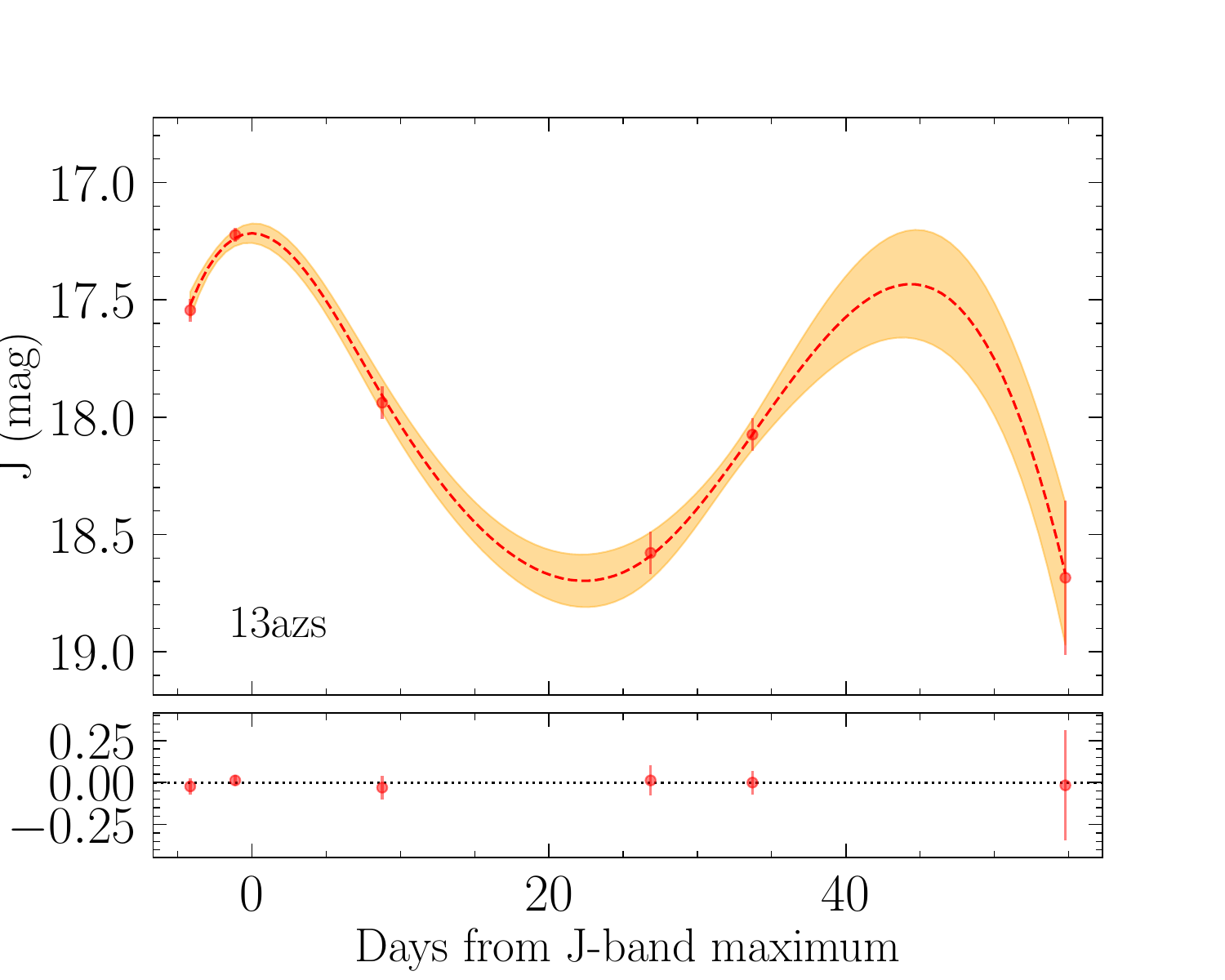}
\includegraphics[trim=0cm 0.2cm 2cm 1cm, clip=True,width=0.32\textwidth]{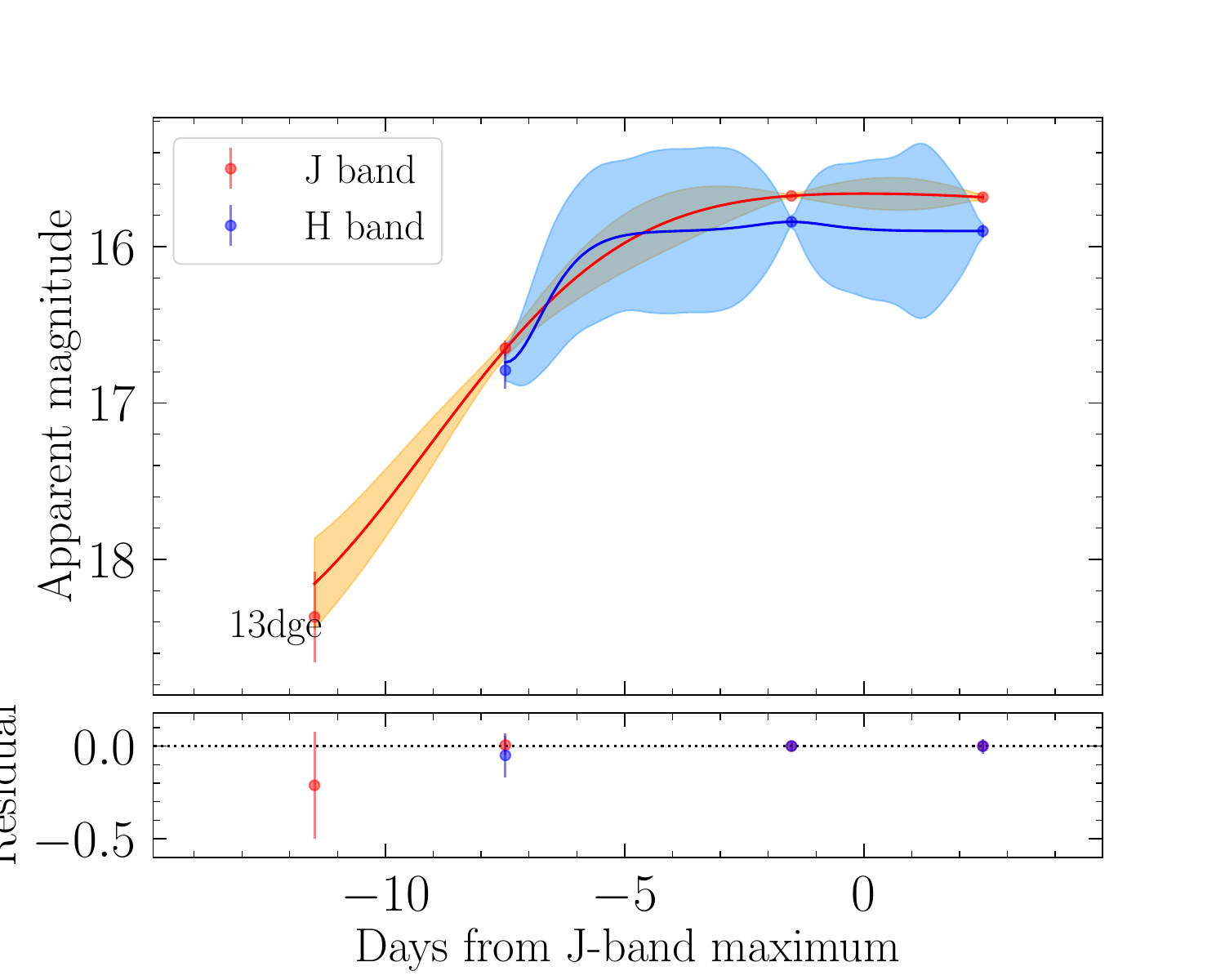}
\includegraphics[trim=0cm 0.2cm 2cm 1cm, clip=True,width=0.32\textwidth]{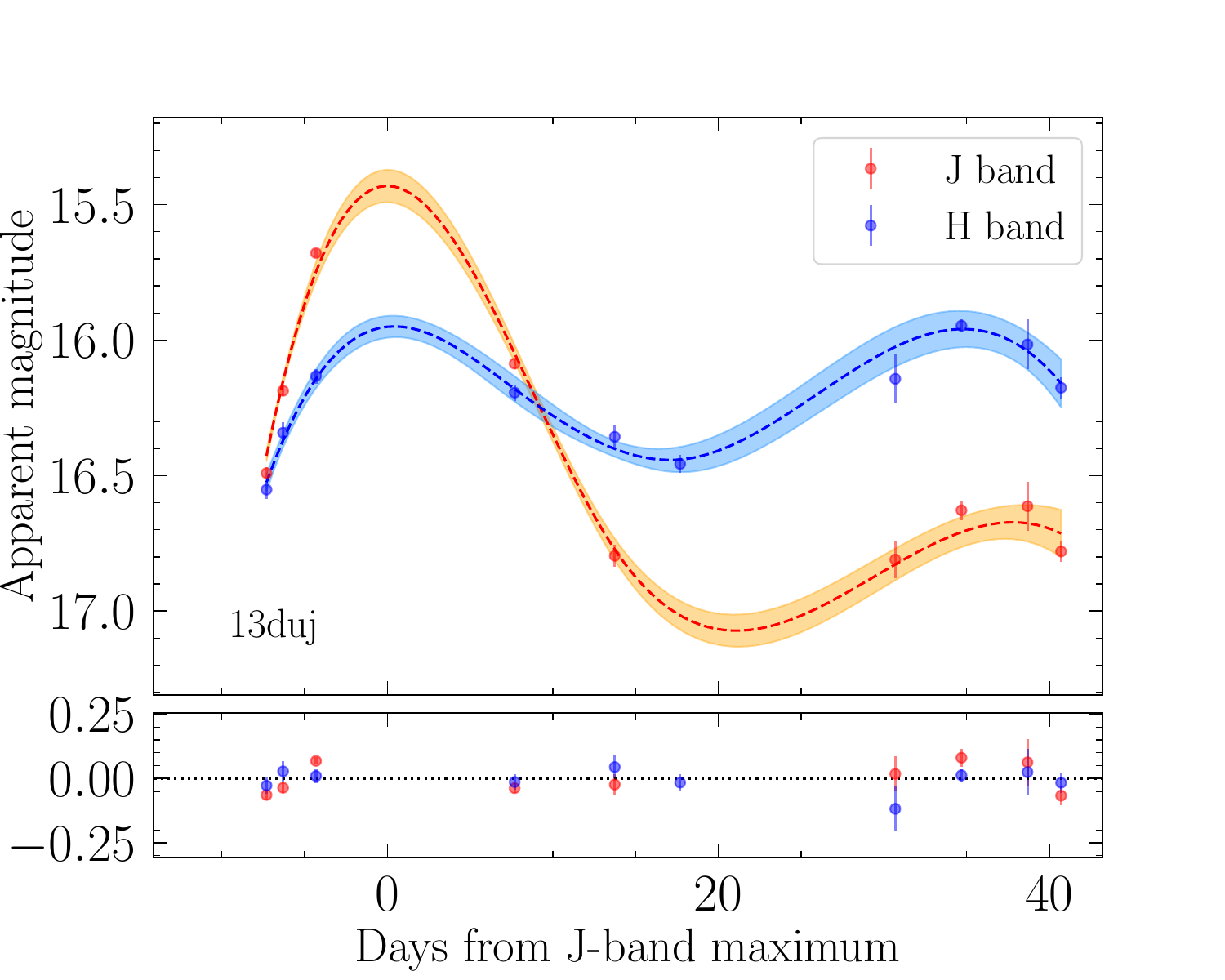}
\includegraphics[trim=0cm 0.2cm 2cm 1cm, clip=True,width=0.32\textwidth]{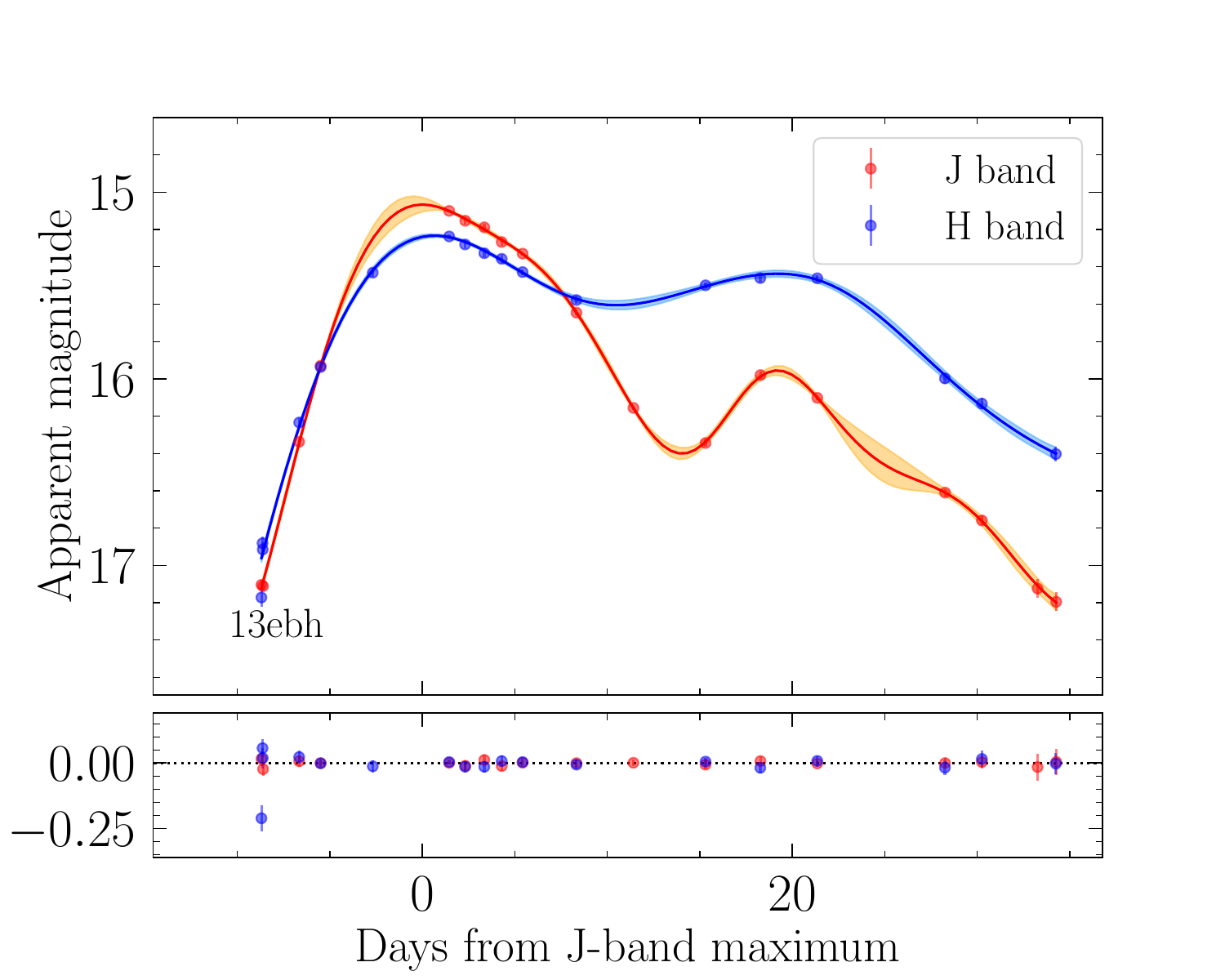}
\includegraphics[trim=0cm 0.2cm 2cm 1cm, clip=True,width=0.32\textwidth]{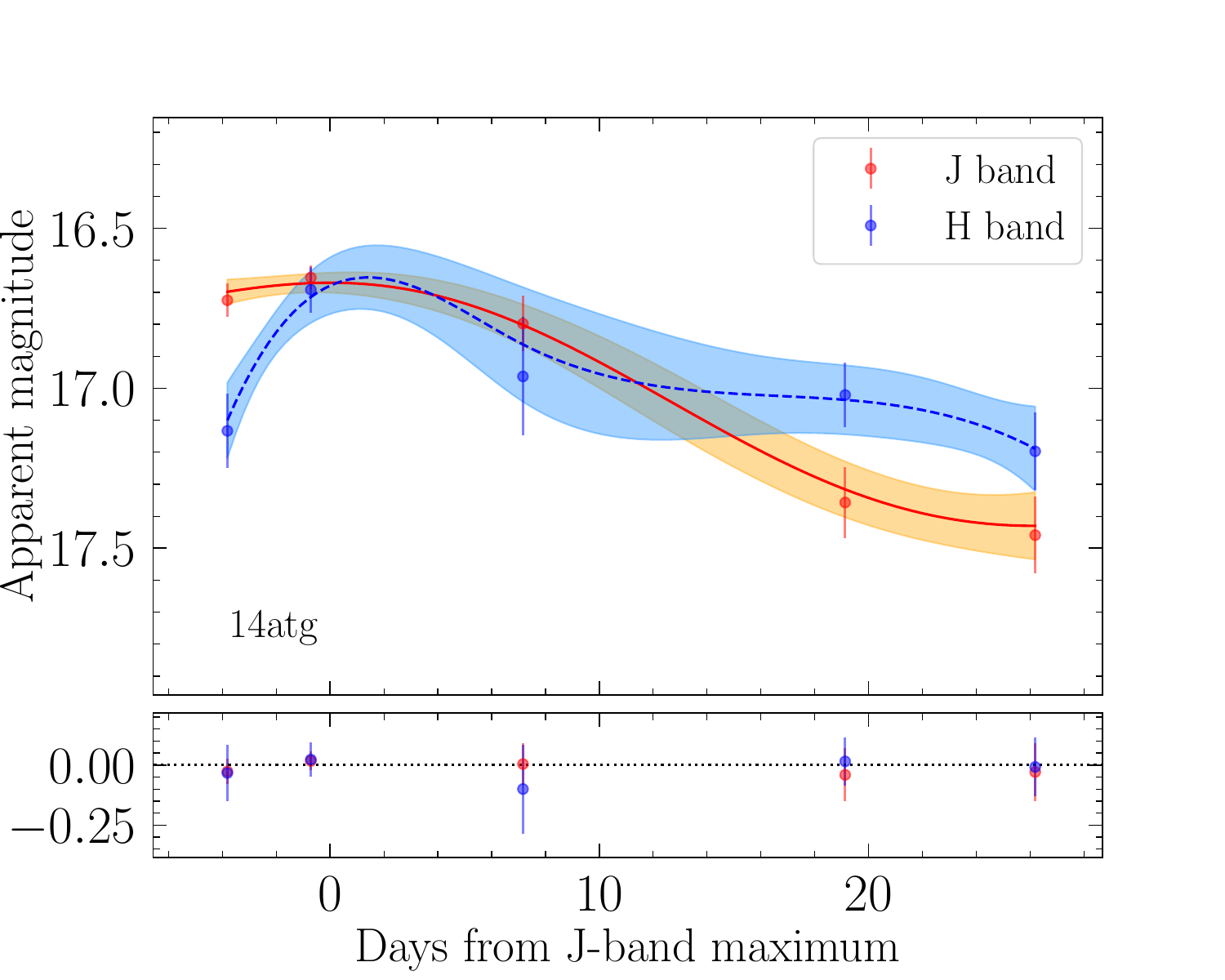}
\includegraphics[trim=0cm 0.2cm 2cm 1cm, clip=True,width=0.32\textwidth]{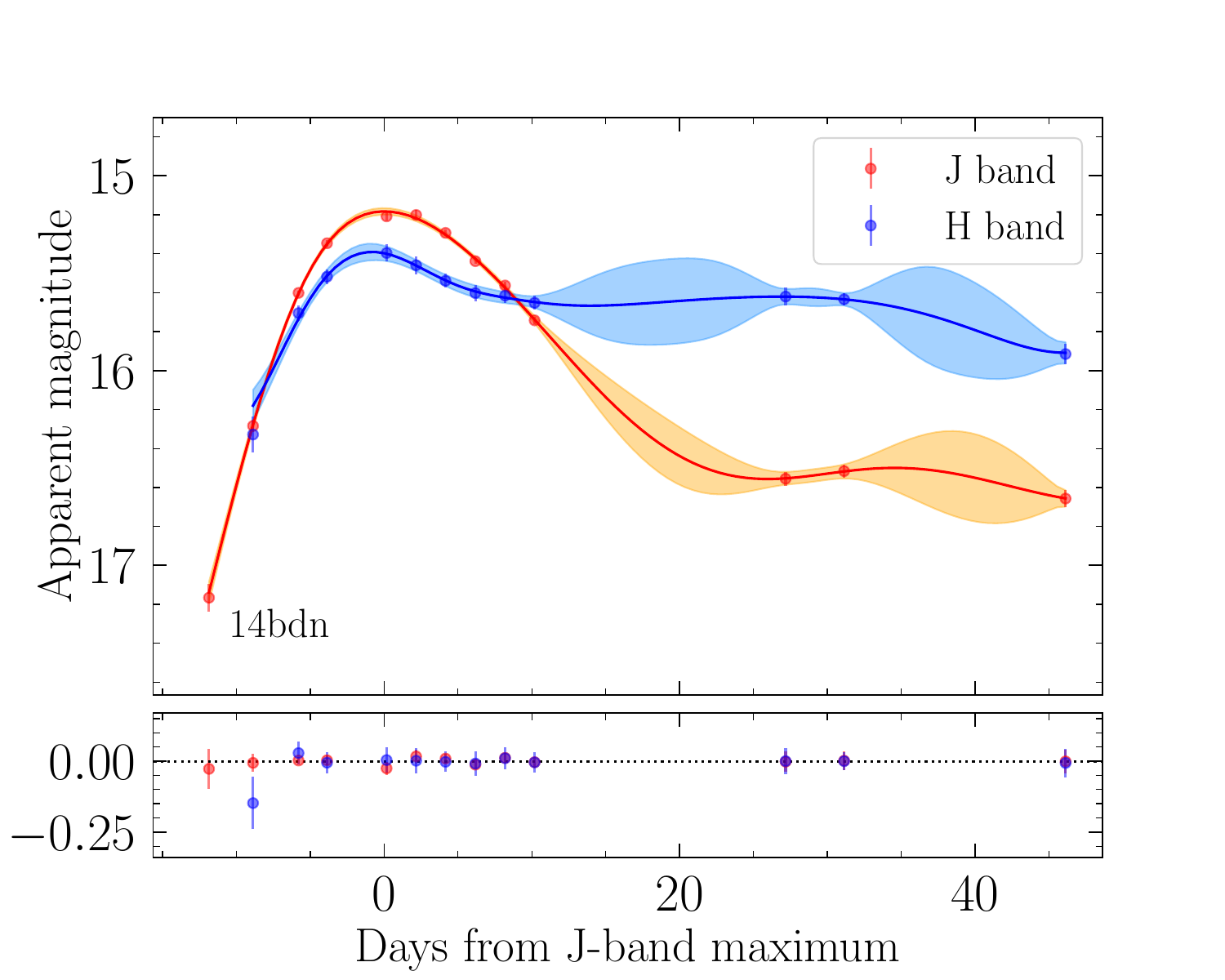}
\includegraphics[trim=0cm 0.2cm 2cm 1cm, clip=True,width=0.32\textwidth]{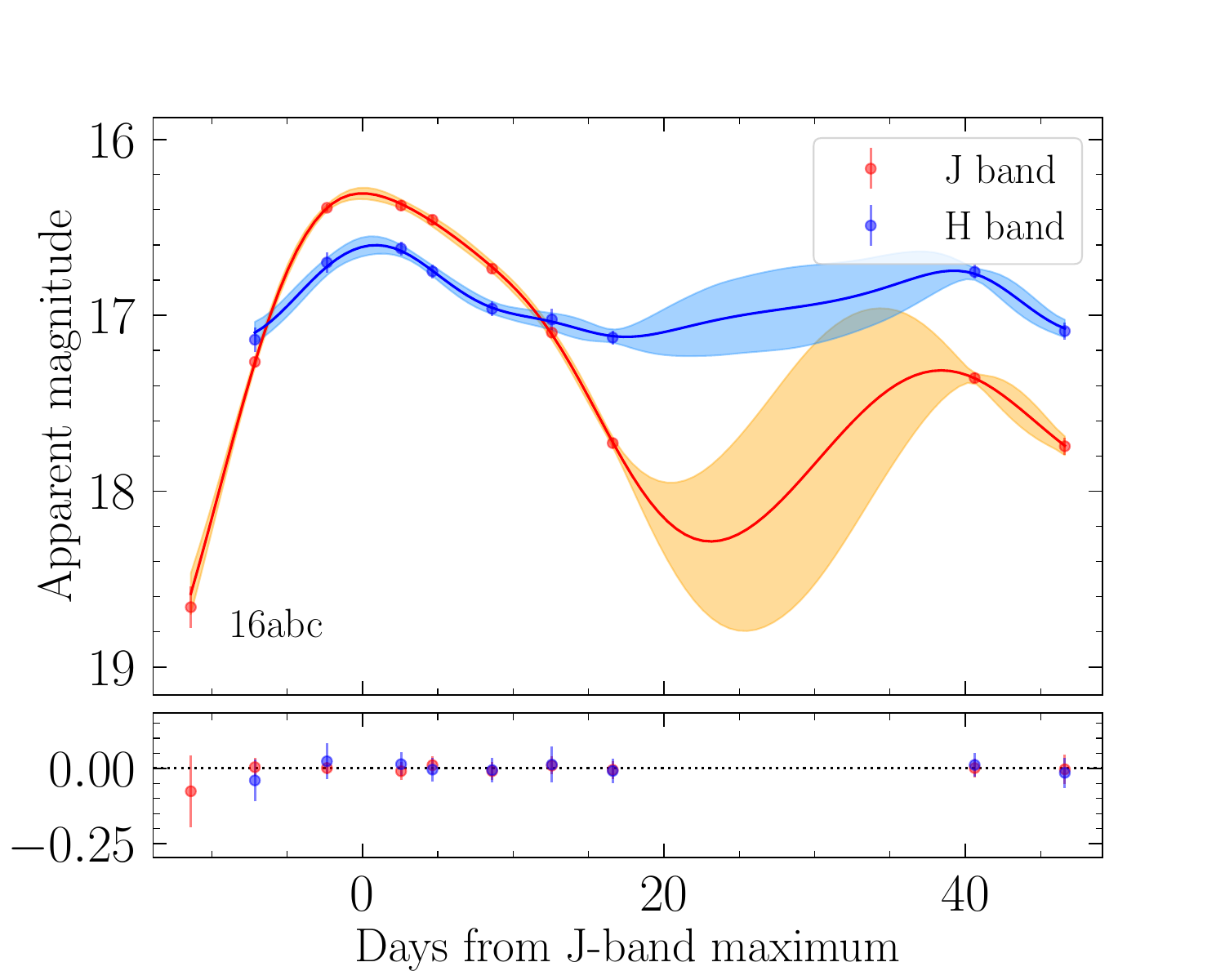}
\includegraphics[trim=0cm 0.2cm 2cm 1cm, clip=True,width=0.32\textwidth]{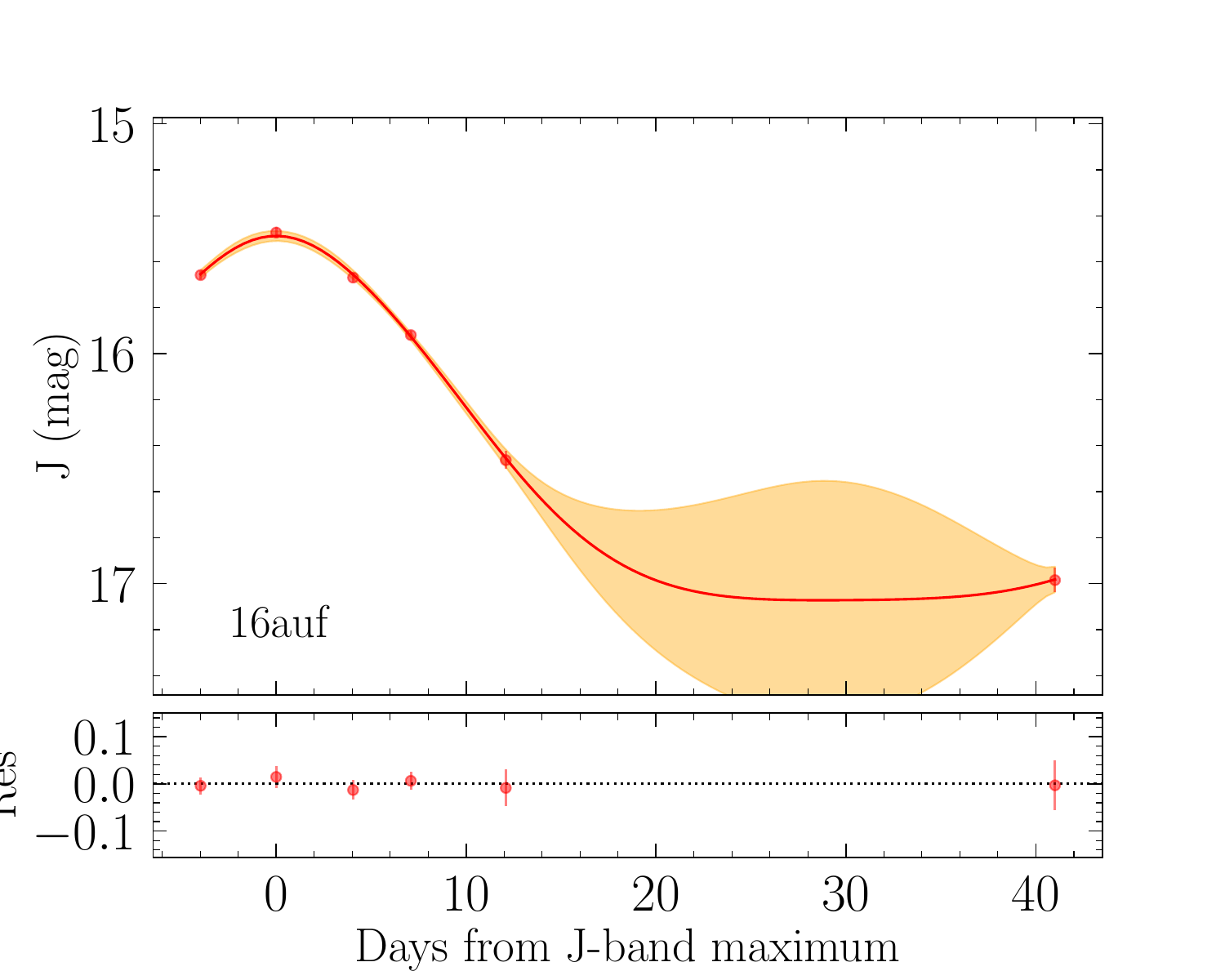}
\includegraphics[trim=0cm 0.2cm 2cm 1cm, clip=True,width=0.32\textwidth]{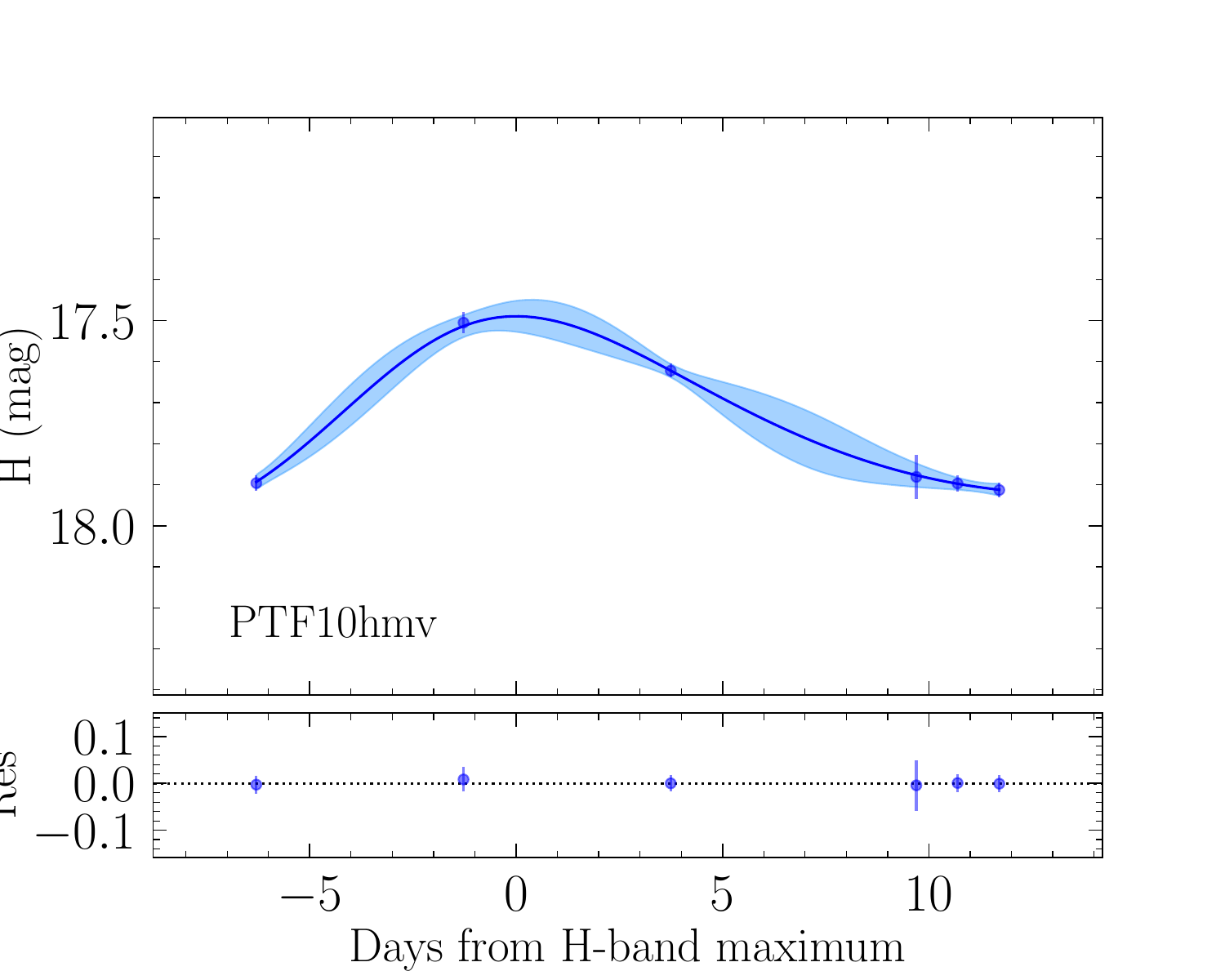}
\includegraphics[trim=0cm 0.2cm 2cm 1cm, clip=True,width=0.32\textwidth]{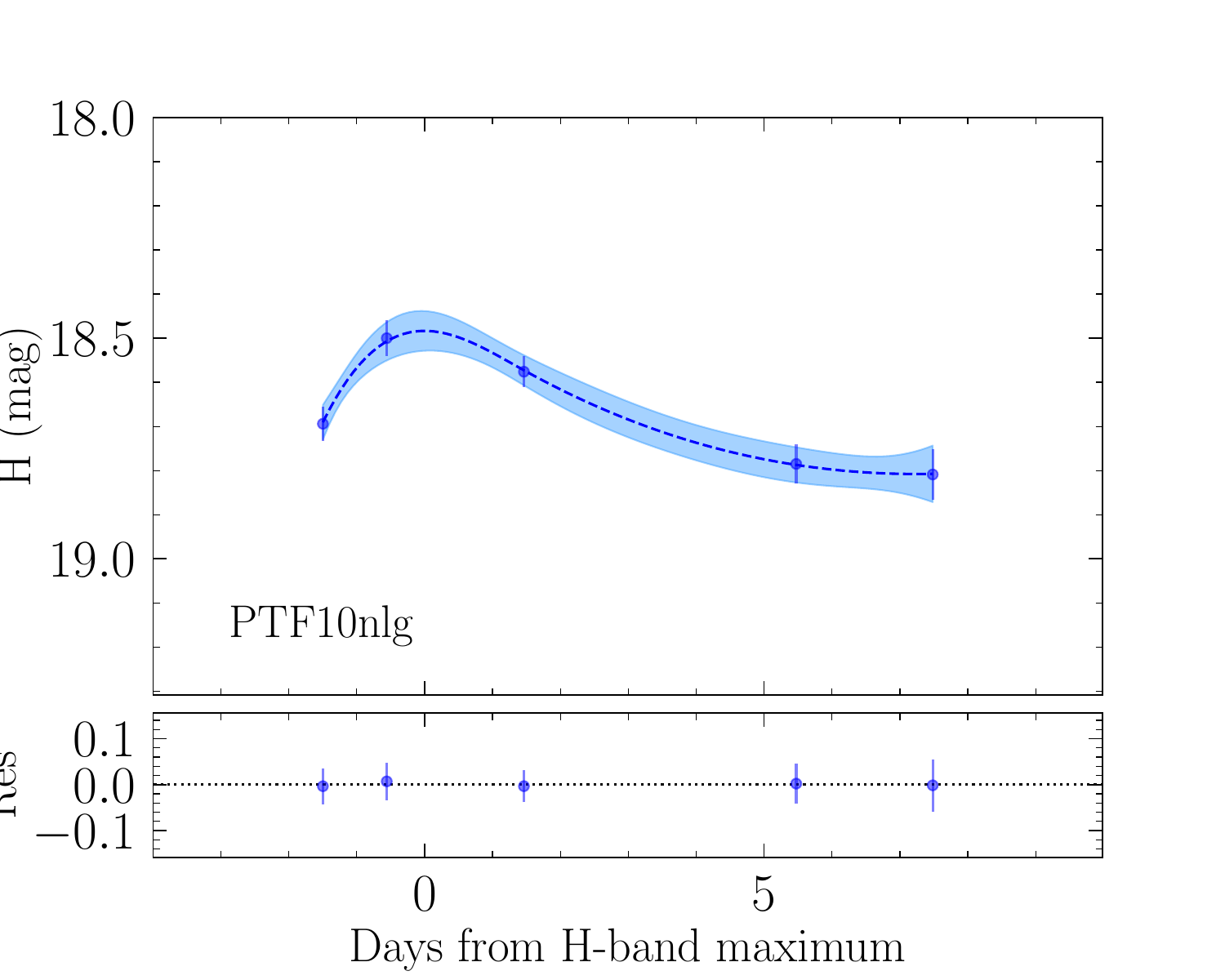}
\includegraphics[trim=0cm 0.2cm 2cm 1cm, clip=True,width=0.32\textwidth]{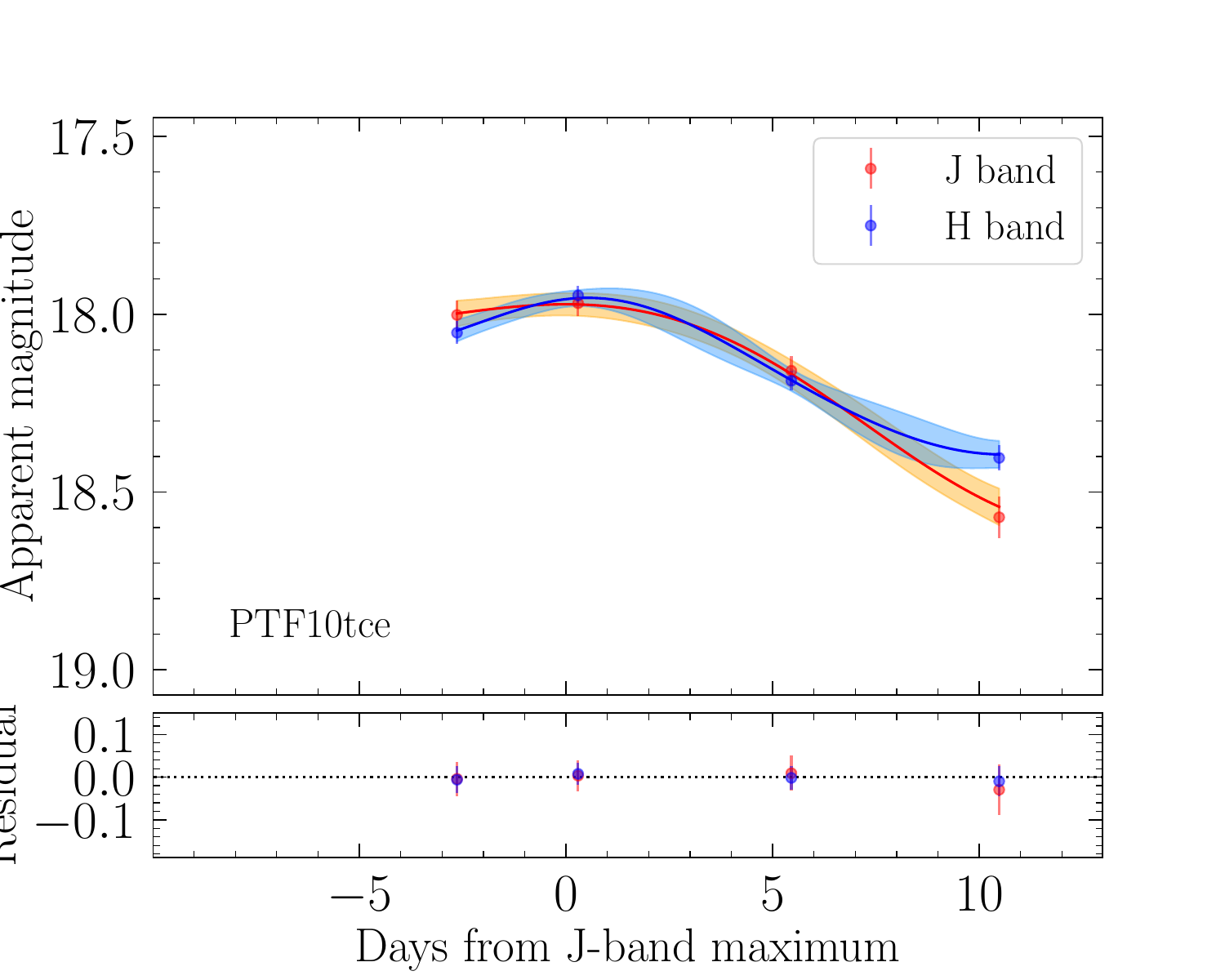}
\includegraphics[trim=0cm 0.2cm 2cm 1cm, clip=True,width=0.32\textwidth]{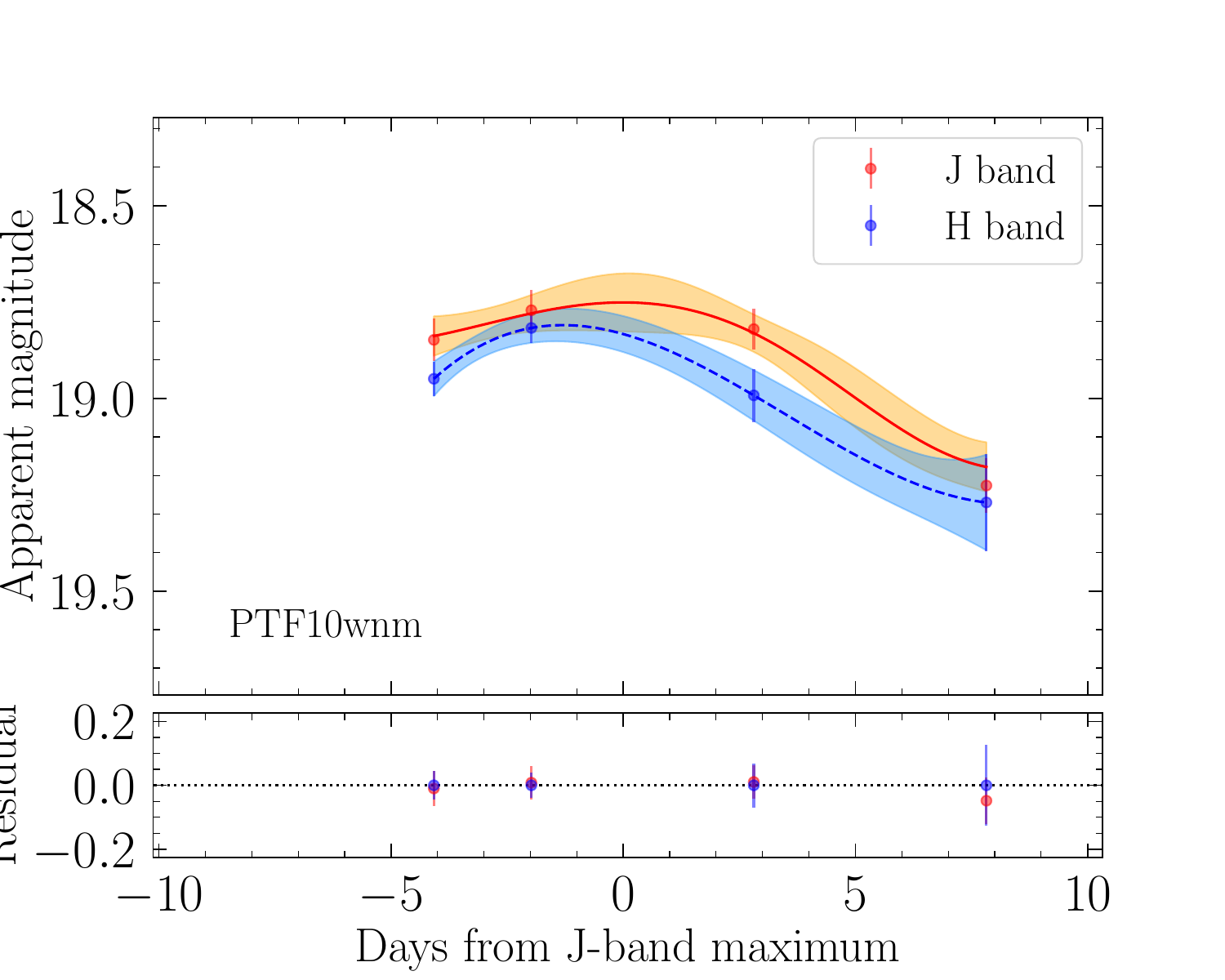}
\caption{(continuing form Fig D3)}
\end{figure*}

\end{appendix}
\end{document}